\documentclass{scrbook} 
\usepackage[english,bibtex]{osm-thesis} 

\newcommand{\hpitype}{Master's Thesis}
\newcommand{\hpiauthor}{Alexander Löser}
\newcommand{\hpititle}{Automatic Clustering in Hyrise}
\newcommand{\hpititleother}{Automatisches Clustering in Hyrise} 
\newcommand{\hpisupervisor}{Dr.\,Michael Perscheid, Jan Koßmann, Martin Boissier}
\newcommand{\hpichair}{Chair for Enterprise Platforms and Integration Concepts}
\newcommand{\hpiexternalsupervisor}{}
\newcommand{\hpiexternal}{}
\SetDate[01/10/2020]
\newcommand{\hpidate}{\today}

\bibliography{thesis}

\begin{document}

	\pagenumbering{alph}
	\ifisbook\begin{titlepage}
	\setlength{\evensidemargin}{0.5\evensidemargin+0.5\oddsidemargin}
	\setlength{\oddsidemargin}{\evensidemargin}

	\centering

	\raisebox{-0.5\height}{\includegraphics[width=5.5cm]{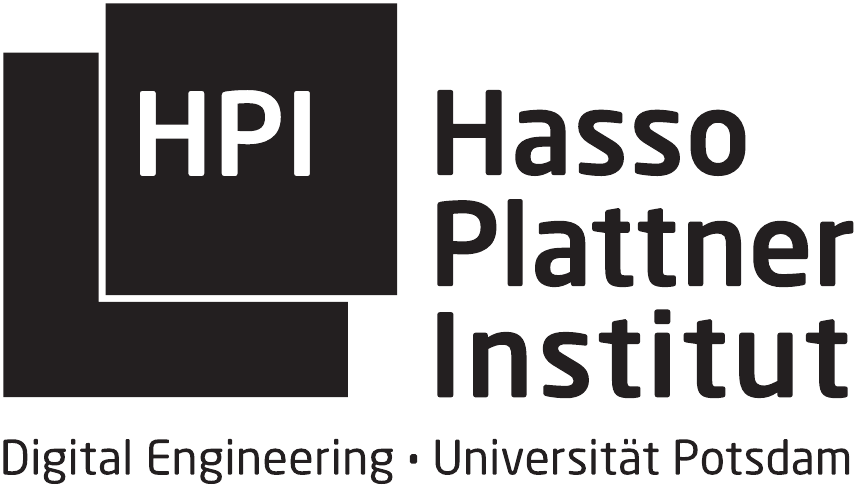}}
	\hspace*{.2\textwidth}
	\raisebox{-0.5\height}{\includegraphics[width=4cm]{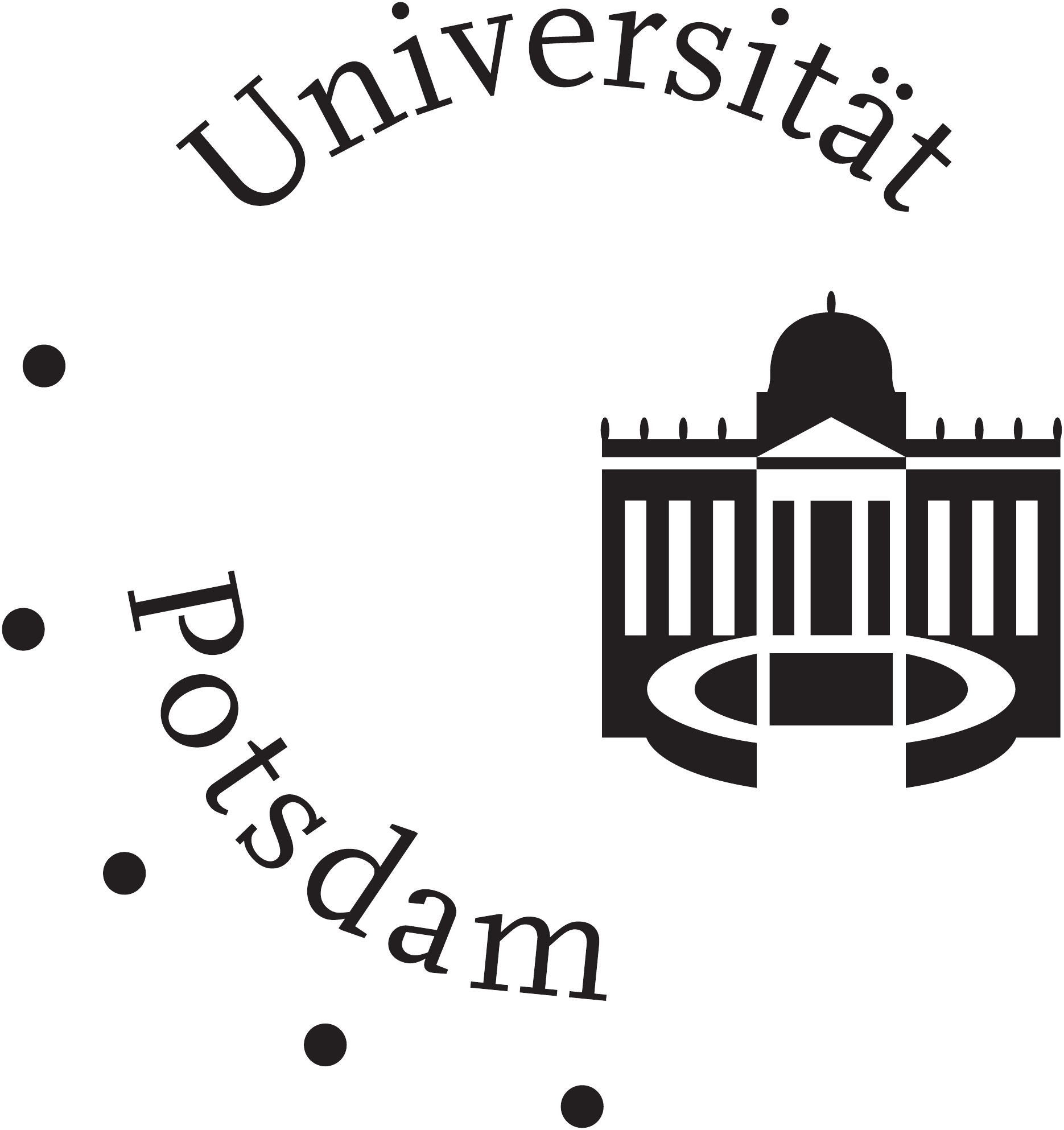}}
	
	\vspace*{4\baselineskip}
	{\usekomafont{subject}\hpitype}\par
	
	\vfill
	{\usekomafont{title}\hpititle\par}
	\vspace*{\baselineskip}
	{\usekomafont{subtitle}\hpititleother}\par
	
	\vfill
	{\textbf{\phantom{\iflanguage{ngerman}{von}{by}}} \\
	 \smallskip\usekomafont{author}\hpiauthor}\par
	
	\vfill
	\phantom{\begin{minipage}{\textwidth}
	{\textbf{\iflanguage{ngerman}{Betreuung}{Supervisors}}\\
	\usekomafont{publishers}\smallskip\hpisupervisor\\ \textit{\hpichair}\\ \smallskip\textbf{\normalfont\hpiexternalsupervisor}\\ \textit{\hpiexternal}}
	\end{minipage}}
	
	\vfill
	{\usekomafont{date}\iflanguage{ngerman}{Hasso-Plattner-Institut an der Universität Potsdam}{Hasso Plattner Institute at the University of Potsdam}}\par
	\vspace*{\baselineskip}
	{\usekomafont{date}\hpidate}\par
	
\end{titlepage}\fi
	\ifisbook\cleardoubleemptypage\fi
	
	\pagenumbering{roman}
	\begin{titlepage}
	\centering

	\raisebox{-0.5\height}{\includegraphics[width=5.5cm]{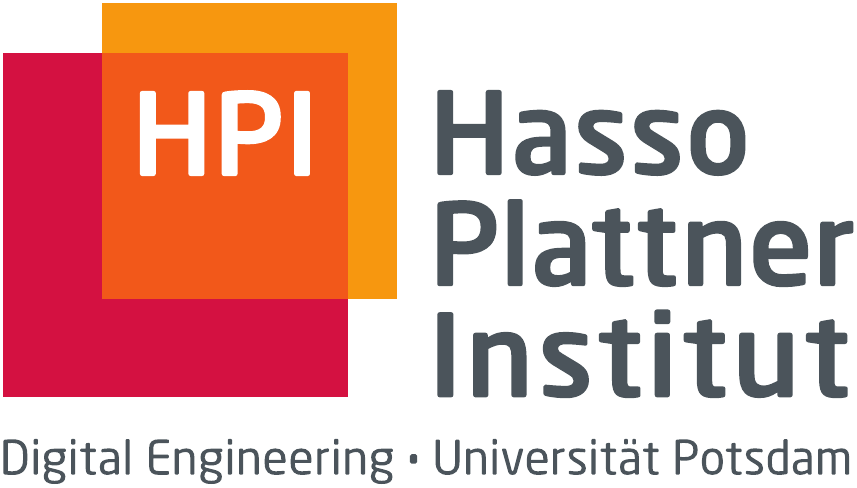}}
	\hspace*{.2\textwidth}
	\raisebox{-0.5\height}{\includegraphics[width=4cm]{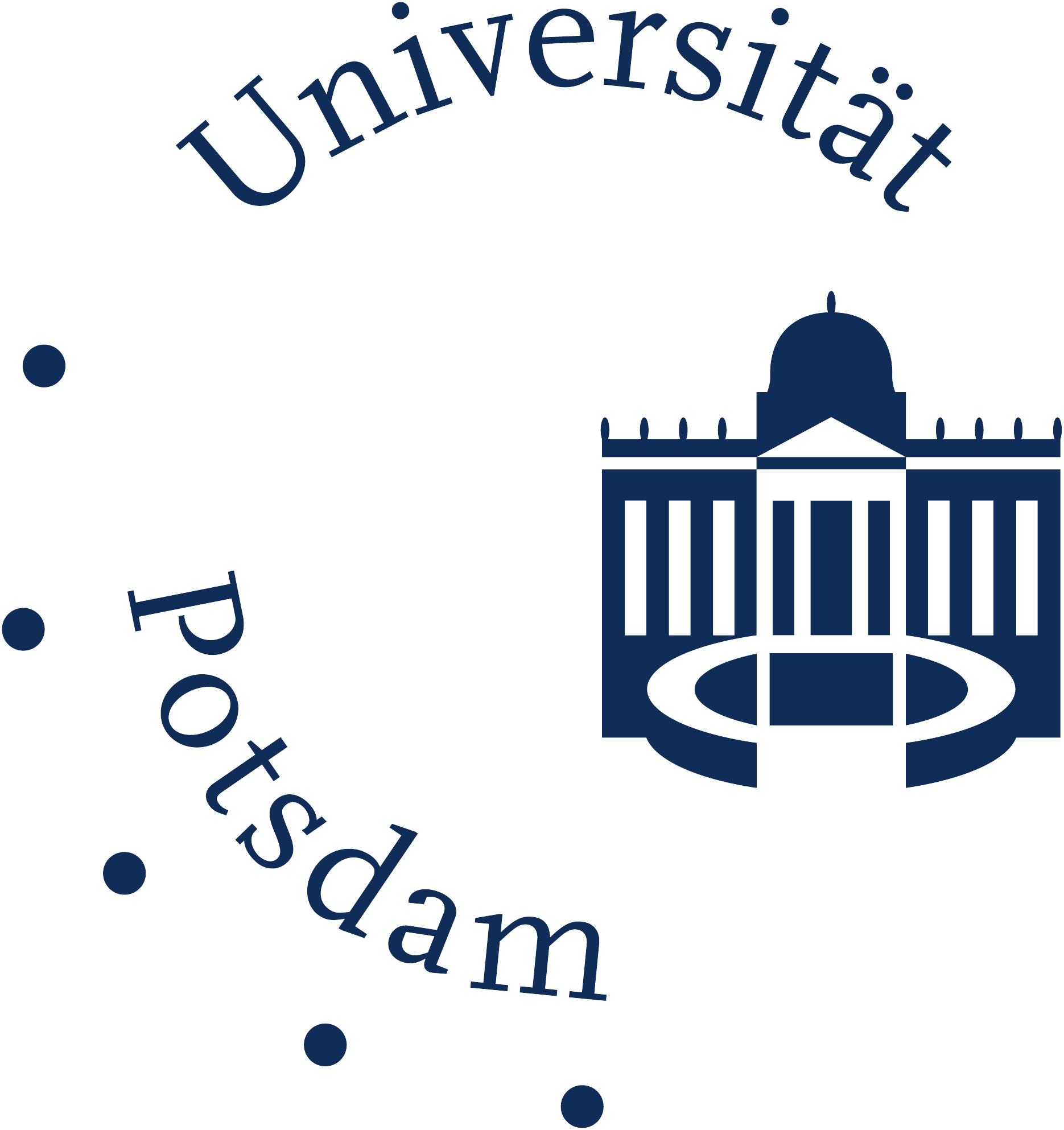}}

	\vspace*{4\baselineskip}
	{\usekomafont{subject}\hpitype}\par
	
	\vfill
	{\usekomafont{title}\hpititle\par}
	\vspace*{\baselineskip}
	{\usekomafont{subtitle}\hpititleother}\par
	
	\vfill
	{\textbf{\iflanguage{ngerman}{von}{by}}\\ 
		\smallskip\usekomafont{author}\hpiauthor}\par
	
	\vfill
	{\textbf{\iflanguage{ngerman}{Betreuung}{Supervisors}}\\ 
		\usekomafont{publishers}\smallskip\hpisupervisor\\ \textit{\hpichair}\\ \smallskip\textbf{\normalfont\hpiexternalsupervisor}\\ \textit{\hpiexternal}}
	
	\vfill
	{\usekomafont{date}\iflanguage{ngerman}{Hasso-Plattner-Institut an der Universität Potsdam}{Hasso Plattner Institute at the University of Potsdam}}\par
	\vspace*{\baselineskip}
	{\usekomafont{date}\hpidate}\par

	\setcounter{page}{1}

\end{titlepage}

	\ifisbook\cleardoubleemptypage\fi

\null\vfil
\begin{otherlanguage}{english}
\begin{center}\textsf{\textbf{\abstractname}}\end{center}

\noindent Physical data layout is an important performance factor for modern databases.
Clustering, i.e., storing similar values in proximity, can lead to performance gains in several ways.
We present an automated model to determine beneficial clustering columns and a clustering algorithm for the column-oriented, memory-resident database Hyrise.
To automatically select clustering columns, the model analyzes the database's workload and provides estimates by how much certain clustering columns would impact the workload's latency.
We evaluate the precision of the model's estimates, as well as the overall quality of its clustering suggestions.
To apply a determined clustering configuration, we developed an online clustering algorithm.
The clustering algorithm supports an arbitrary number of clustering dimensions.

We show that the algorithm is robust against concurrently running data modifying queries.
We obtain a 5\% latency reduction for the TPC-H benchmark when clustering the \texttt{lineitem} table and a 4\% latency reduction for the TPC-DS benchmark when clustering the \texttt{store\_sales} table.

\end{otherlanguage}
\vfil\null


\null\vfil
\begin{otherlanguage}{ngerman}
\begin{center}\textsf{\textbf{\abstractname}}\end{center}

\noindent Die physische Anordnung von Daten spielt eine wichtige Rolle für die Performance moderner Datenbanksysteme.
Ähnliche Werte zu gruppieren und physisch nahe beieinander zu speichern, kann die Performance einer Datenbank auf viele Arten verbessern.
Der Prozess des Gruppierens und Umordnens der Daten wird auch als Clustering bezeichnet.
Wir präsentieren ein Modell, das die momentane Arbeitslast der Datenbank analysiert, um automatisch vielversprechende Clustering-Spalten vorzuschlagen.
Um Clustering-Vorschläge zu bewerten, schätzt das Modell ihren Einfluss auf die Arbeitslast ab.
Wir analysieren die Genauigkeit der Schätzungen des Modells auf verschiedenen Detail-Ebenen.
Um ein vielsprechendes Clustering umsetzen zu können, haben wir einen Clustering-Algorithmus entwickelt.
Der Algorithmus kann im laufenden Datenbankbetrieb nach beliebig vielen Spalten clustern.

Wir zeigen, dass unser Clustering-Algorithmus im laufenden Datenbankbetrieb clustern kann, obwohl parallel datenschreibende Transaktionen ausgeführt werden.
Für die \texttt{lineitem}-Tabelle des TPC-H Benchmark haben wir ein Clustering gefunden, das die Laufzeit des TPC-H Benchmarks gegenüber dem Standard-Clustering um 5\% reduziert.
Analog haben wir für das TPC-DS Benchmark ein Clustering für die Tabelle \texttt{store\_sales} gefunden, das die Laufzeit des TPC-DS Benchmarks um 4\% reduziert.

\end{otherlanguage}
\vfil\null

	\tableofcontents
	\cleardoublepage

	\pagenumbering{arabic}
	\chapter{Introduction}

Modern database systems can store tables with huge amounts of data.
The physical data layout, i.e., the way data is organized on disk, can impact the database's performance in various ways.
A \emph{clustering} is a special form of physical data layout:
If we say a table is clustered by some column $X$, we refer to a physical data layout where rows of the table with a similar value in $X$ are grouped and stored together.
We call $X$ the \emph{clustering column}.
Tables may have an arbitrary number of clustering columns.

Both real-world enterprise systems, as well as analytical systems such as data warehouses, are dominated by read queries~\cite{DBLP:journals/pvldb/KruegerKGSSCPDZ11}.
Typical requests on analytical databases filter multiple attributes to gather information on the data of interest:
For example, 20 of the 22 queries of the analytical TPC-H benchmark~\cite{DBLP:conf/tpctc/BonczNE13, TpchSpec} and 99 of the 99 queries of the analytical TPC-DS benchmark~\cite{DBLP:conf/vldb/OthayothP06, TpcdsSpec} contain at least one filter predicate.
Vogelsgesang et al.~\cite{DBLP:conf/sigmod/VogelsgesangHFK18} have shown that this applies to 97.8\% of real-world queries as well.

\begin{figure}
     \centering
     \includegraphics[width=\textwidth]{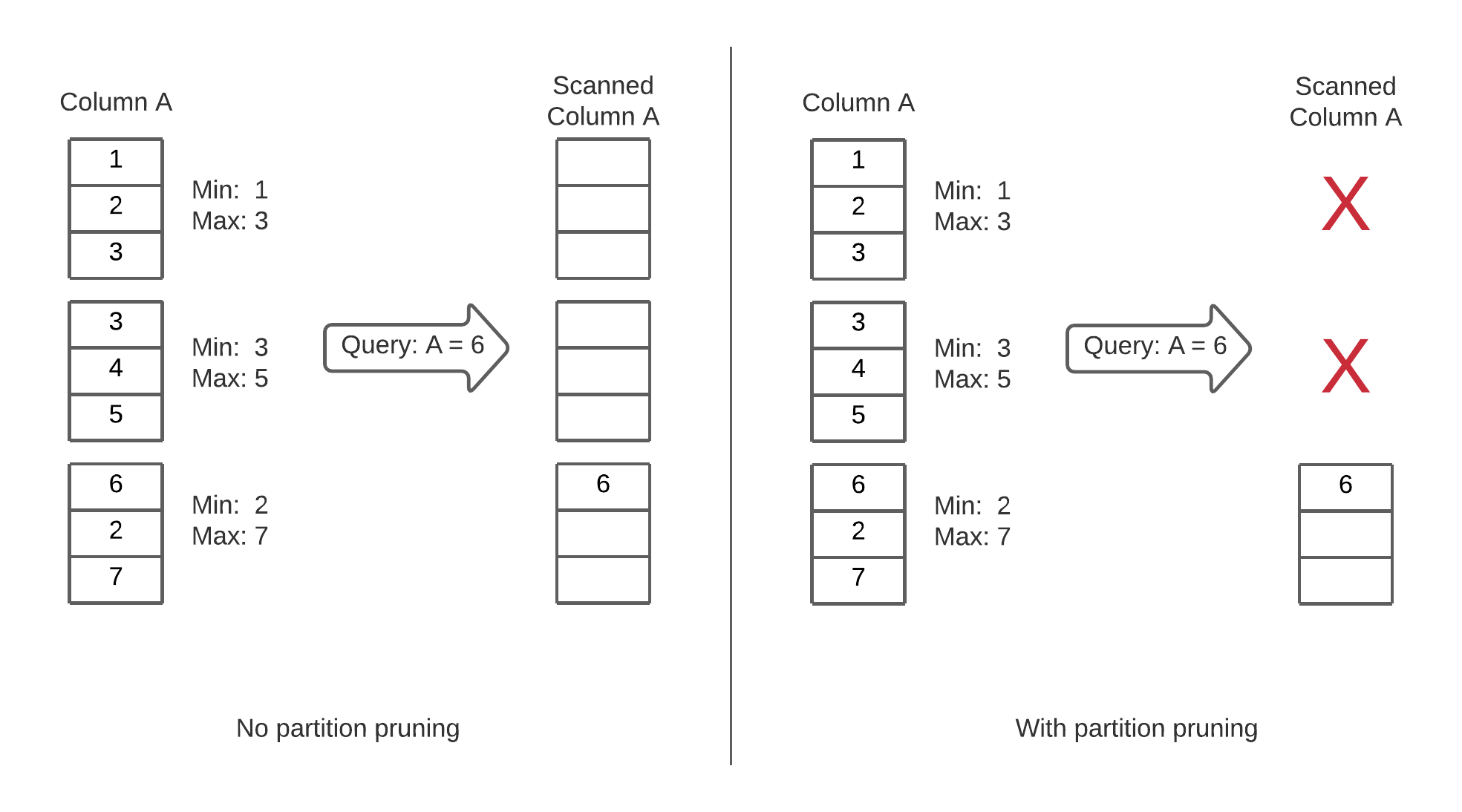}
        \caption{The figure shows a partition pruning example for the predicate $A = 6$. A red crosses indicates a pruned partition. In this example, pruning yields a speed up by factor 3.}
        \label{fig:pruning_example}
\end{figure}

By analyzing the filter predicates and statistics about the stored data, database systems can often tell before executing a query that considerable parts of a table will not qualify, and thus, access to them could be avoided.
To exploit this, many database systems, e.g., Peloton~\cite{DBLP:conf/sigmod/ArulrajPM16}, Hyrise~\cite{DBLP:conf/edbt/DreselerK0KUP19}, HyPer~\cite{DBLP:conf/sigmod/LangMFB0K16}, SAP HANA~\cite{DBLP:journals/pvldb/NicaSACHBG17}, and Quickstep~\cite{DBLP:journals/pvldb/PatelDZPZSMS18}, divide tables into partitions of a fixed number of tuples.
For each of those partitions, database systems can maintain aggregated statistics, such as the minimum and maximum value per column~\cite{DBLP:conf/vldb/Moerkotte98}.
Given a filter predicate, those statistics can be used
to apply \emph{partition pruning}~\cite{DBLP:conf/sigmod/HerodotouBB11}.
Partition pruning is a technique to reduce a table's size early on.
It works by analyzing the partition statistics and  excluding all partitions that cannot contain matching rows from the query execution, without actually accessing the rows.
For example, if a predicate selects only tuples with a value greater than $X$ in a certain column, but the highest value of the respective column in a partition is less than $X$, then no row can qualify for the filter predicate, and the partition can be pruned.
\Cref{fig:pruning_example} visualizes a partition pruning example.

\begin{figure}
     \centering
     \includegraphics[width=\textwidth]{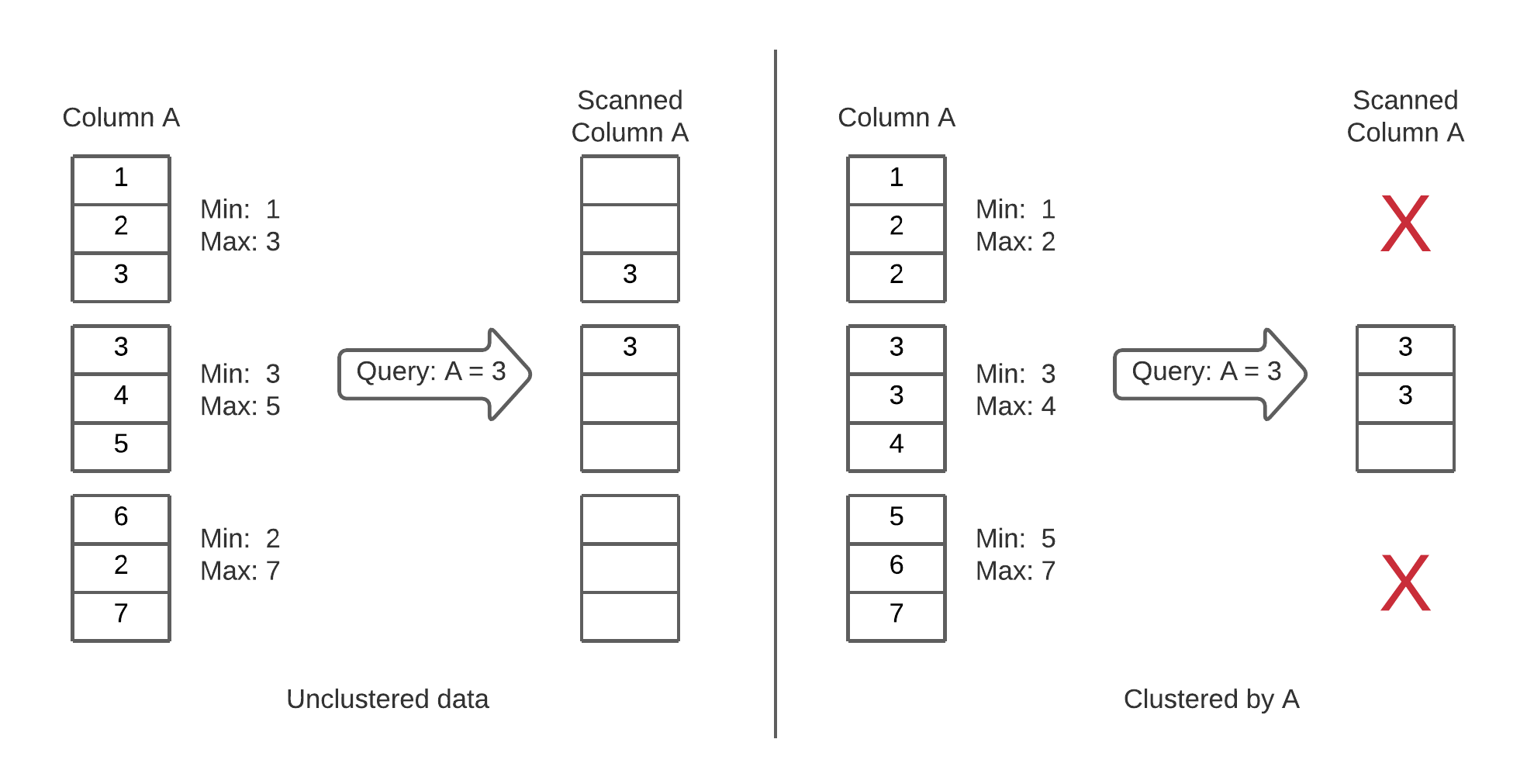}
     
        \caption{The figure shows a how clustering impacts the effectiveness of partition pruning. The red crosses indicate pruned partitions. On clustered data, only a third of the data has to be scanned.}
        \label{fig:pruning_example_clustering}
\end{figure}

Partition pruning is especially effective when the predicate is on a clustering column:
Since the data is clustered along the column, rows that match the filter predicate are stored together, spread over only a few partitions.
The effect of clustering on partition pruning is visualized in \Cref{fig:pruning_example_clustering}.

So far, the examples in our tables had only a small number of columns.
Production databases, however, usually have a large number of columns~\cite{DBLP:conf/sigmod/BianYTCCDM17, DBLP:journals/pvldb/KruegerKGSSCPDZ11}.
Theoretically, it is possible to cluster a table by all of them.
In practice, it is more reasonable to focus on a few
columns that are frequently accessed by expensive operators.
When we restrict ourselves to a few clustering columns, a question arises:
Which clustering columns yield the best clustering, with respect to performance?

In fact, a certain clustering is per se neither good, nor bad.
Whether a clustering is beneficial for the performance always depends on the current workload.
With \emph{workload}, we refer to the set of queries that were recently executed by the database.
In production databases, the workload may change over time~\cite{DBLP:conf/sigmod/DasLNK16, DBLP:conf/sigmod/MaAHMPG18, DBLP:conf/IEEEcloud/RoyDG11}.
Thus, a clustering that is beneficial today may not be suitable anymore in the next week.
Nevertheless, even if the current workload is known, it is still an open question which columns should be used for the clustering.
Traditionally, this question is answered by database administrators, i.e., human experts~\cite{DBLP:conf/sigmod/HilprechtBR20}.
It requires expert knowledge about both the database and the current workload.
This knowledge, however, is expensive and not always available, especially if the workload changes frequently.
As a consequence, we propose an automatic clustering model for the selection of clustering columns.
The model automatically detects which clustering would be beneficial, and thereby reduces the need for human expertise.
Further, it provides an estimate by how much the performance would improve if a certain clustering were implemented.
Database administrators can use those performance change estimates to decide whether the expected performance benefit is worth the time consuming process of implementing a new clustering.

\begin{figure}
    \centering
    \includegraphics[width=\textwidth]{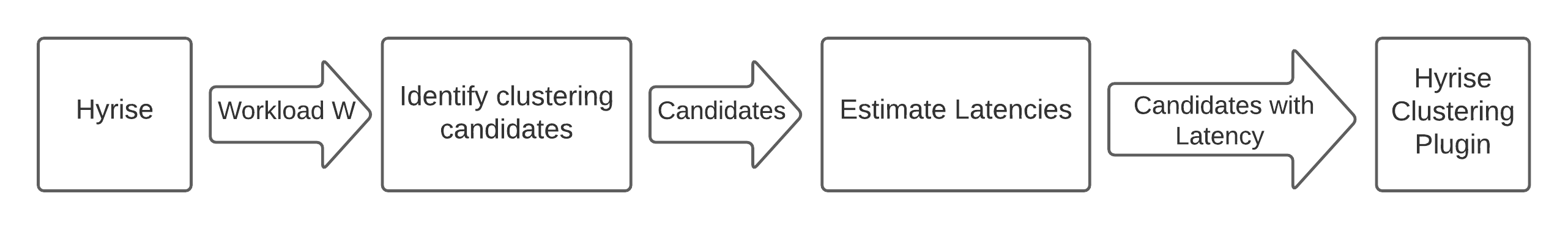}
    \caption{A schematic overview of the model. The model takes a workload as input and returns clustering suggestions, sorted by their expected latency.}
    \label{fig:model_coarse}
\end{figure}

Our model takes a representative workload as input and outputs clustering suggestions.
To obtain the clustering suggestions, we perform the following steps, which are also visualized in \Cref{fig:model_coarse}:
First, we analyze the given workload and extract interesting columns.
Columns are interesting for the clustering if they are used in e.g., filter predicates or joins.
Then, for each clustering candidate, we estimate how such a clustering would affect the total latency of the workload.
The latency estimation process is described in more detail in \Cref{sec:cost_estimation}.
We choose the clustering with the lowest expected latency.

With our model, we can determine a set of clustering columns that yields performance improvements.
However, before we can take advantage of the newfound clustering, we first need a clustering algorithm to implement it, i.e., to change the actual physical data layout accordingly.
Implementing a new clustering is likely to cause massive movements of data, which requires both computation time and memory bandwidth.
If the clustering algorithm is executed during normal database operation, this may cause significant performance drops.
Common approaches to mitigate the impact on the performance include performing the clustering offline, e.g., before bulk loads~\cite{DBLP:journals/pvldb/ZiauddinWKLPK17}; during phases of low load, e.g., at night; or when no data modifying queries are expected.
We consider these approaches valid, but there might also be cases with a steady load:
For example, large online retailers such as Amazon sell all around the world, and orders may be placed at any time.
If there is no phase of low load, the clustering has to be implemented during operation, ideally without affecting the database's performance notably.
Performing the clustering during database operation poses several challenges:
\begin{enumerate}
    \item The clustering algorithm can use only a limited amount of memory and computation power at a time; otherwise, it might cause notable performance drops.
    
    \item The movement of data can be seen as a write-operation. Concurrently running queries might hold locks for some rows, i.e., the rows should not be overwritten nor moved. As a consequence, those rows cannot be clustered at that moment, and have to be clustered later on.
    
    \item The clustering algorithm must obtain locks to indicate to concurrent transactions that it will modify a row.
However, obtaining too many locks, or holding them for a long time, will block other transactions and cause performance drops.

    \item Data modifying statements (e.g., the SQL \texttt{UPDATE} statement) may be executed while the clustering is in progress.
The clustering algorithm must not discard any changes made by such statements.
\end{enumerate}

We present a clustering algorithm that addresses the issues listed above.
The key idea is to perform the clustering in small steps.
This way, we can limit both the memory and the locks necessary at a time.
We implement our algorithm for the research database Hyrise~\cite{DBLP:conf/edbt/DreselerK0KUP19}.

\paragraph{Our contributions}
In this work, we make the following contributions:
\begin{itemize}
    \item We provide a clustering algorithm that can create multi-dimensional clusterings while data modifying queries, e.g., the SQL \texttt{UPDATE} statement, are executed concurrently.
    \item We present a clustering model that analyzes the database's workload to identify suitable clustering columns. Given a hypothetical clustering, the model provides estimations for the workload's run time.
    \item We provide experimental results to assess the precision of the model's estimates. Further, we evaluate our clustering algorithm's robustness against data modifying queries.
\end{itemize}

The rest of this work is structured as follows:
\Cref{sec:background} provides background information about Hyrise.
We describe our online clustering algorithm in \Cref{sec:clustering_algorithm}, and our automated clustering model in \Cref{sec:clustering_model}.
In \Cref{sec:evaluation}, we evaluate both the algorithm and the model.
We discuss related work in \Cref{sec:related_work} and conclude in \Cref{sec:conclusion}.
	\chapter{Background}
\label{sec:background}
All our implementations and experiments are conducted with the research database system Hyrise~\cite{DBLP:conf/edbt/DreselerK0KUP19}.
In this chapter, we provide background knowledge about the Hyrise ecosystem.

\paragraph{Storage layout}
Hyrise is a relational research database system.
It has a non-distributed architecture.
Tables reside in main memory and are stored in a column-oriented fashion.

\begin{figure}
    \centering
    \includegraphics[width=0.4\textwidth]{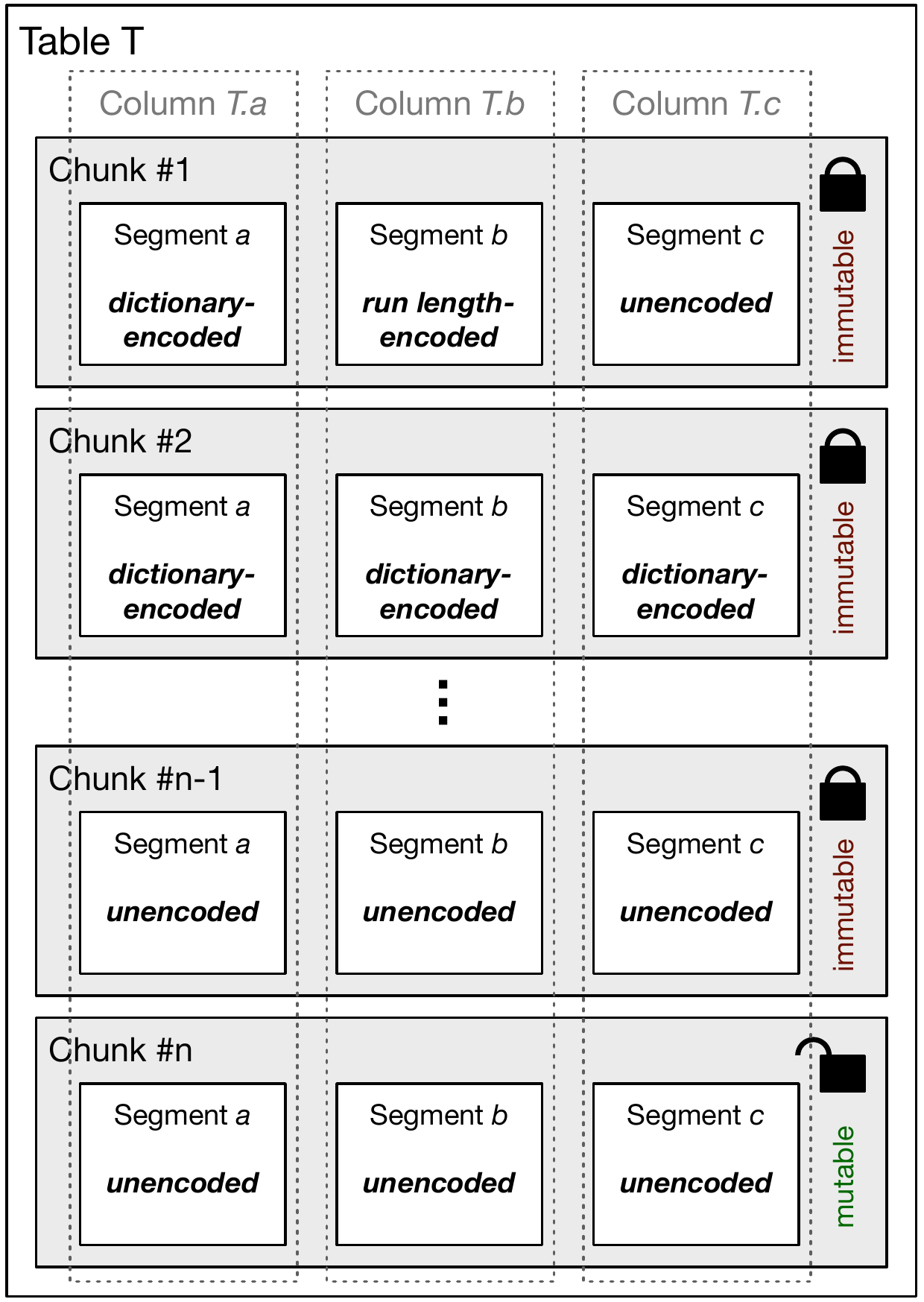}
    \caption{The figure visualizes Hyrise's storage layout. The table has three columns and $n$ chunks. Each chunk contains a segment of each column. Chunks may be immutable, i.e., no more rows can be appended. Segments of immutable chunks can be encoded. This figure is taken from the work of Dreseler et al.~\cite{DBLP:conf/edbt/DreselerK0KUP19}.}
    \label{fig:storage_layout}
\end{figure}

In Hyrise, all tables are implicitly horizontally partitioned.
Those partitions are called \emph{chunks}.
Partitioning the table into chunks yields multiple benefits~\cite{DBLP:conf/edbt/DreselerK0KUP19}.
From a clustering perspective, chunks are particularly interesting because they represent prunable fractions of a table.
Chunks are created with a fixed size.
Due to the horizontal partitioning, a chunk contains a fraction of each column.
These fractions are called \emph{segments}.
\Cref{fig:storage_layout} visualizes Hyrise' storage layout.

Chunks have two states: mutable, and immutable.
After its creation, a chunk is mutable, i.e., rows can be appended (till the pre-allocated capacity is reached).
A chunk becomes immutable when it is finalized, i.e., no new rows may be inserted.
Finalization is usually triggered when the chunk's size reaches its pre-allocated capacity, but may be triggered explicitly, e.g., by our clustering algorithm, at any time.

\paragraph{Encodings}
Hyrise supports different segment encodings.
Data in mutable chunks are never encoded, but stored in \texttt{ValueSegment}s, i.e., the actual value is stored.
For that reason, data in value segments are also considered as \emph{unencoded}.
Segments in immutable chunks can have encodings, such as run-length encoding, or dictionary encoding~\cite{DBLP:conf/sigmod/AbadiMF06}.
Such encoding techniques can reduce a segment's memory consumption, and yield additional optimization potential:
For example, when scanning a dictionary encoded segment, we can perform a dictionary lookup to determine whether the segment contains a certain value, without scanning the actual segment.
This optimization is similar to partition pruning, but allows to check for the actual presence of a given value in the segment, rather than just checking whether the segment could contain the value.

\begin{figure}
     \centering
     \includegraphics[width=\textwidth]{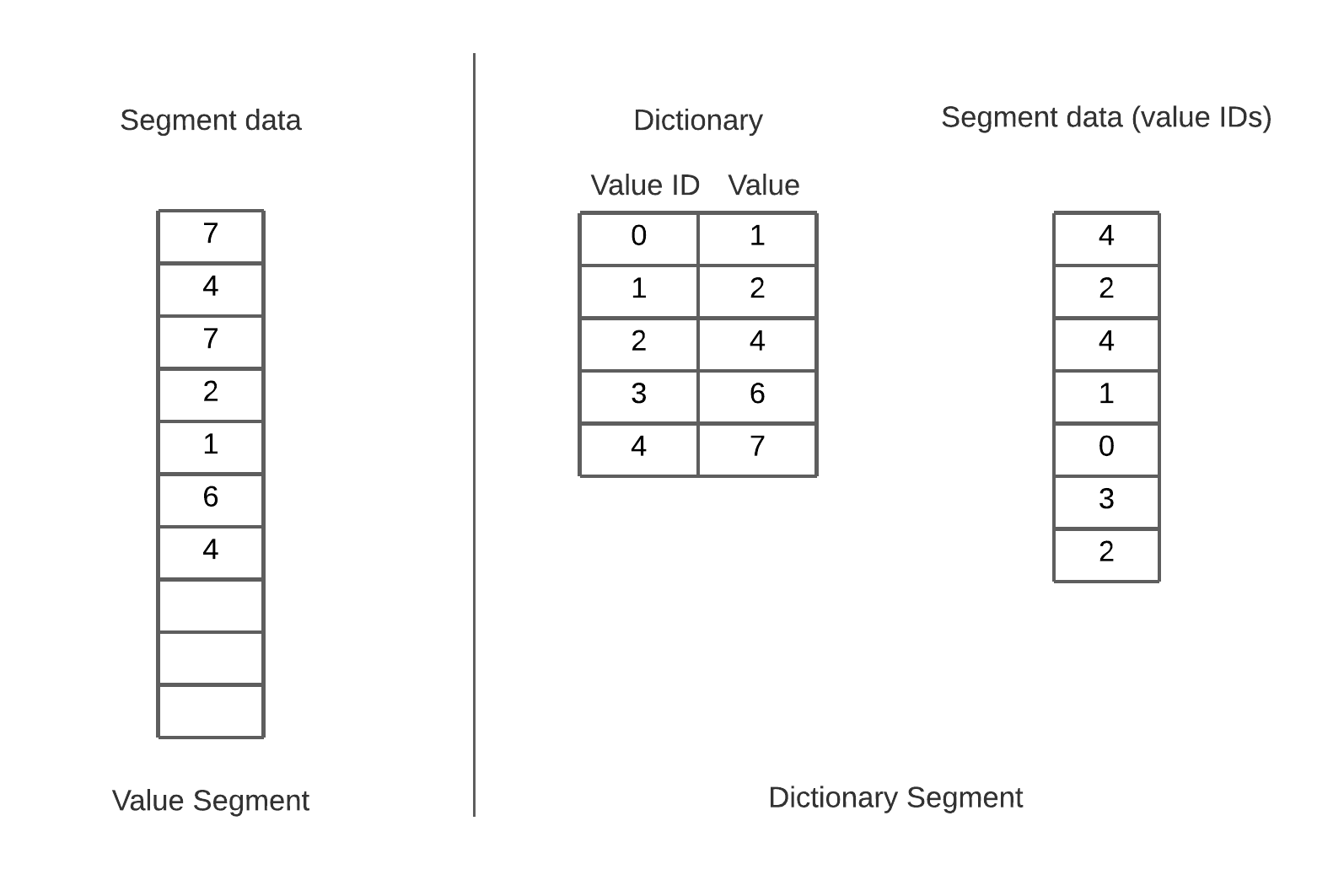}
        \caption{The figure shows a \texttt{ValueSegment} on the left side and a \texttt{DictionarySegment} on the right side. The chunk size is ten.}
        \label{fig:encodings}
\end{figure}

Different segments can have different encodings.
All our experiments were conducted with dictionary segments, which are Hyrise's default encoding for segments in immutable chunks.
A \texttt{DictionarySegment} stores a sorted dictionary that assigns a unique \emph{value id} to each unique value in the segment.
Instead of storing the actual values, a dictionary segment stores value ids.
To access the actual value, a dictionary lookup is performed.
The data type of the value ids depends on the number of unique values in the segment, i.e., 8, 16, or 32-bit integers are used.
The default chunk size in Hyrise is $2^{16}-1 = 65\,535$ rows; this ensures that even if all values in a segment are unique, 16 bits are sufficient to represent all of those values.
Clustering reduces the number of entries in the dictionary (for segments of the clustering columns), which leads to smaller dictionaries, and possibly the choice of a smaller data type for the value ids.
\Cref{fig:encodings} visualizes a dictionary segment.

Unencoded \texttt{ValueSegment}'s for mutable chunks are always created with a fixed size to avoid unexpected re-allocations during some operator's execution.
The other segment encoding types are only used for segments in immutable chunks.
Thus, for encoded segments, we can reserve exactly the required amount of memory.
This exact memory allocation is an important property for us:
Our clustering algorithm may create multiple chunks that end up only partially filled.
Finalizing and encoding those chunks avoids the permanent waste of unused memory.

\paragraph{Data statistics}
Hyrise maintains several statistics on the stored data.
For example, Hyrise maintains statistics on a segment-level.
The segment statistics are generated after the corresponding chunk is finalized: once finalized, the chunk is immutable, and no future inserts can invalidate the statistics.
The statistics include a minimum-maximum-filter, i.e., the minimum and the maximum value present in the segment.
When a predicate filters the respective column, Hyrise uses those statistics to determine whether any value in the segment could qualify, or whether the whole segment (and with that, the whole chunk) can be pruned.

Additionally, Hyrise maintains statistics on table-level.
These table-level statistics include a histogram for each column.
By default, Hyrise uses equal distinct count histograms, i.e., each bin of the histogram covers a similar number of distinct values.
Our clustering algorithm uses these histograms to obtain balanced clusters.
We describe our algorithm and the histograms more detailed in \Cref{sec:choose_cluster_boundaries}.

\paragraph{SQL Pipeline}
Hyrise offers an SQL interface.
SQL queries issued against that interface are processed by Hyrise' SQL pipeline:
First, the query is parsed and transformed into a logical query plan (LQP).
Nodes in the LQP represent logical database operations, such as filter predicates or joins.
In the next step, the optimizer analyzes and optimizes the LQP.
Finally, the optimized LQP is translated into a physical query plan (PQP).
Nodes in the PQP represent concrete operator implementations for the logical database operations, such as, e.g., a hash join, an index join, or a nested loop join.
A physical query plan is a directed acyclic graph (DAG), whose nodes are \emph{operators}.
An operator may have zero, one (e.g., a table scan), or two (e.g., a hash join) inputs, and may be input to an arbitrary number of other operators.

To execute the query, the operators are executed in a bottom-to-top order, i.e., all inputs of an operator are executed before the operator itself is executed.
Query execution usually begins with a \texttt{GetTable} operator, which returns a table and takes no inputs except for the table name.

\paragraph{Reference Segments}
An operator's output is stored in the form of a temporary table.
To keep the memory consumption of those intermediate result tables low, Hyrise often uses \emph{reference segments}.
Reference segments do not store actual data.
Instead, they store a reference to where in the original table the actual data is stored.
The reference consists out of a chunk id, and an offset within the chunk.

\begin{figure}
    \centering
    \includegraphics[width=\textwidth]{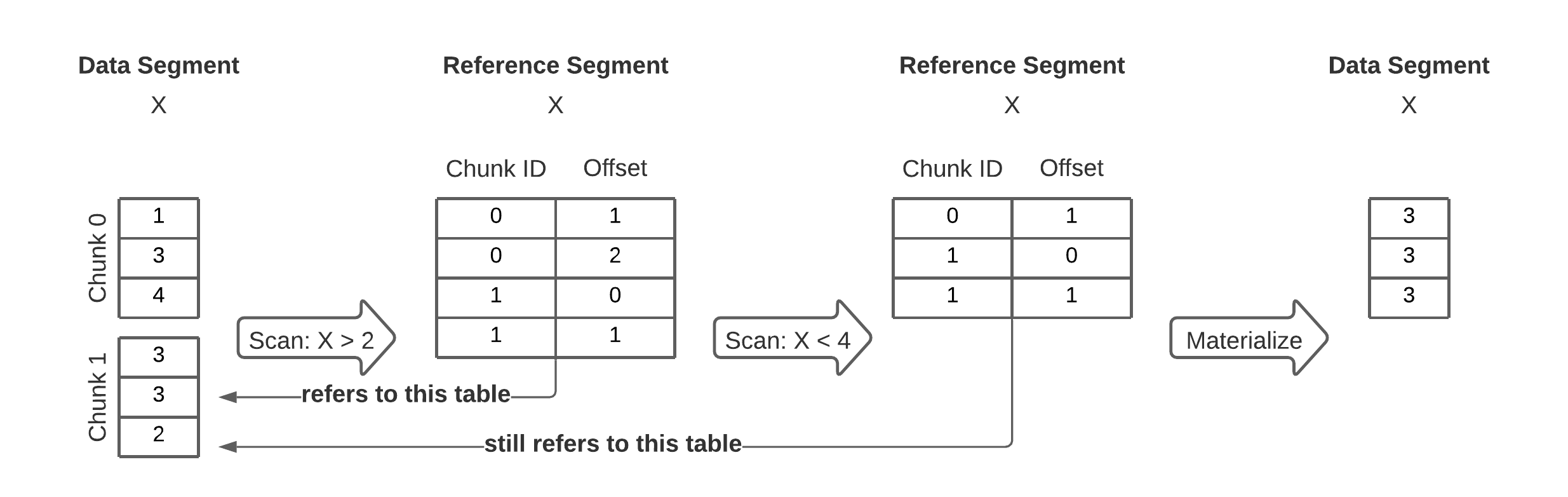}
    \caption{The figure visualizes a column \texttt{X}, on which two scans are executed before the query result is materialized.}
    \label{fig:reference_segment}
\end{figure}

Hyrise allows only one level of indirection, i.e., a reference segment points always to a table that contains actual data, but never to another table containing reference segments.
The process of accessing the actual values behind a reference segment is called \emph{materializing}.
\Cref{fig:reference_segment} visualizes a reference segment.

\paragraph{Operator performance statistics}
Operators have a built-in mechanism to monitor their performance.
For example, each operator stores its execution duration.
The execution of some operators, e.g., the \texttt{JoinHash} operator, involves multiple steps.
Accordingly, those operators provide additional performance statistics, e.g., a step-wise run time breakdown.
In our clustering model, we use the monitored operator performance data to model the workload.
We describe the concrete performance statistics used in \Cref{sec:workload}.

\paragraph{Transaction management}
Hyrise supports both analytical and transactional workloads.
As a means of concurrency control, Hyrise implements Multi-Version Concurrency Control (\emph{MVCC})~\cite{DBLP:journals/tods/BernsteinG83} to obtain snapshot isolation~\cite{DBLP:conf/vldb/SchwalbFWGP14}.
As a consequence, Hyrise follows an insert-only policy:
In MVCC, rows are marked invalid rather than physically deleted.
Updates are never performed in place.
Instead, they are treated as a combination of deletion and insertion, i.e., a new, modified version of the row is inserted, and the old row is marked as invalid.
Older, running transactions will still see the old version of the row, while newer transactions will only see the modified version.
We describe Hyrise's MVCC-implementation in more detail in \Cref{sec:mvcc_consistency}.
	\chapter{MVCC-Aware Clustering Algorithm}
\label{sec:clustering_algorithm}
In this chapter, we present an MVCC-aware clustering algorithm that creates multi-dimensional clusterings and is robust against modifications made by concurrent transactions.

We start by describing Hyrise's status quo regarding clustering, and show the shortcomings of the current approach.
We give an high level overview of our clustering algorithm, before we provide details on its choice of clustering criteria and its interactions with MVCC.
At last, we describe our clustering algorithm's integration into Hyrise.

\paragraph{Status Quo in Hyrise}
At the time of writing, Hyrise has little support for clustering.
Before executing benchmarks, e.g., the analytical TPC-H or TPC-DS benchmark, it is only possible to specify a sorting column for each table.
This yields a one-dimensional clustering, but the rudimentary approach has several shortcomings:
First, the clustering columns are hard-coded.
Hard-coded clustering columns may be sufficient for benchmarks, which have a fixed workload.
In production databases, however, workloads may change over time~\cite{DBLP:conf/sigmod/DasLNK16, DBLP:conf/sigmod/MaAHMPG18, DBLP:conf/IEEEcloud/RoyDG11}, making a different clustering more suitable.
Thus, we consider being able to change the clustering during operation an important property.
Second, the approach supports only one-dimensional clusterings.
We expect a clustering algorithm to support an arbitrary number of clustering dimensions.

\begin{figure}
    \centering
    \includegraphics[width=\textwidth]{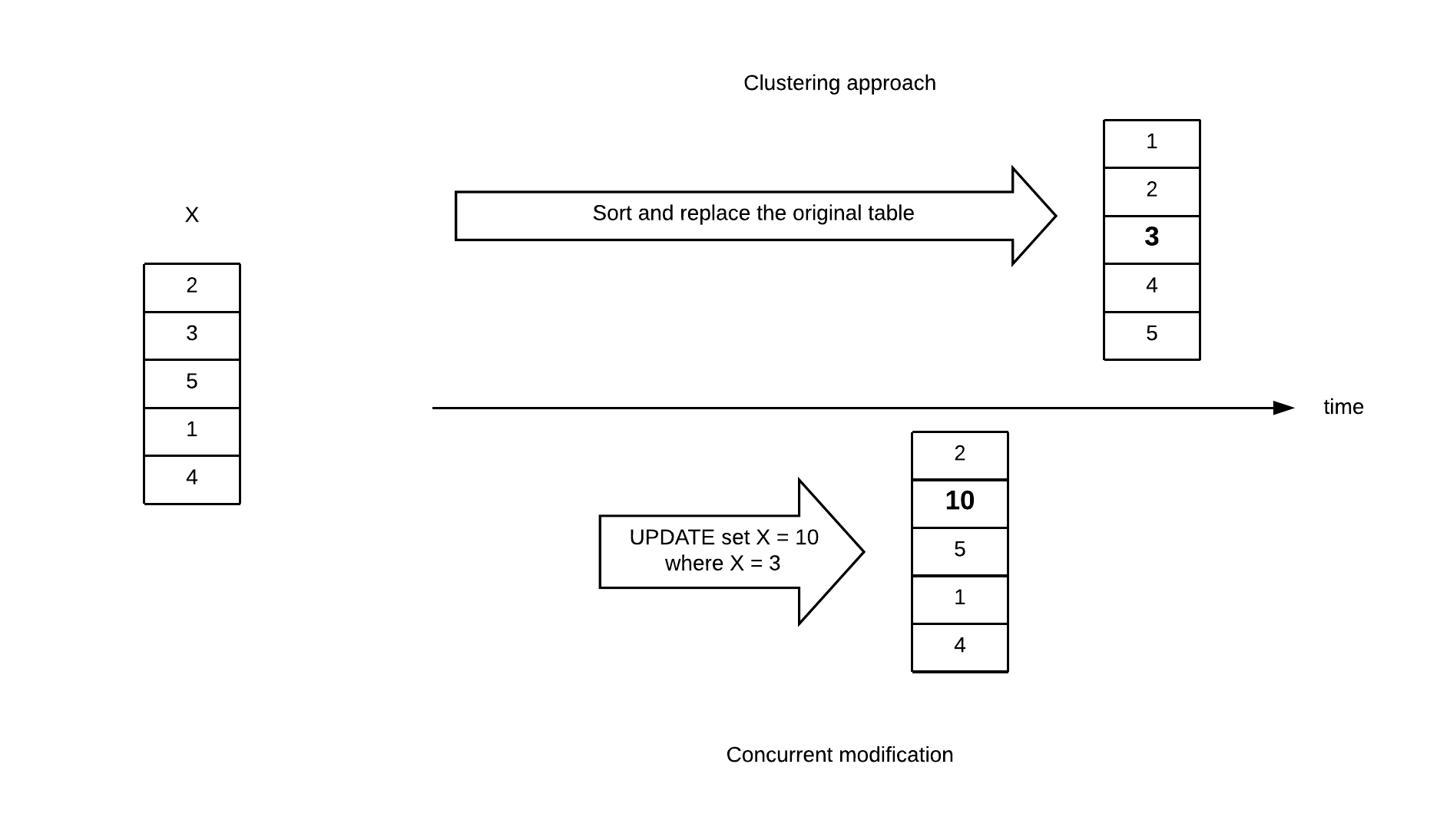}
    \caption{The figure visualizes the impact of a single concurrent modification. The conflicting rows use a bold font.}
    \label{fig:interfering_dml}
\end{figure}

\paragraph{Robustness against concurrent modifications}
Third, and most importantly, Hyrise is designed to perform well for both analytical and transactional workloads.
The rudimentary approach might be sufficient if only analytical (i.e., read-only) queries are executed.
In transactional workloads, however, concurrent modifications produced by, e.g., \texttt{INSERT}, \texttt{UPDATE}, or \texttt{DELETE} statements, are more common than in analytical workloads~\cite{DBLP:conf/cikm/0001MDLMRSZU16, DBLP:journals/pvldb/KruegerKGSSCPDZ11}.
Those concurrent modifications pose a problem:
The rudimentary approach performs an expensive sort operation, during which many concurrent modifications may  occur.
The result of the sort operation, however, reflects the state of the database when the sorting began, i.e., any modifications that occurred during the sort operation are not contained in the sort result.
Discarding the modifications is not an option; and we are not aware of a trivial way to reproduce them on the sort result either.
Thus, we conclude that the current approach lacks robustness against concurrent modifications.
\Cref{fig:interfering_dml} visualizes an example of an interfering concurrent modification.

\section{High-level Overview}
In this section, we describe how our algorithm obtains robustness against concurrent modifications, its intended storage layout, and the phases it undergoes to create a multi-dimensional clustering.

\begin{figure}
    \centering
    \includegraphics[width=\textwidth]{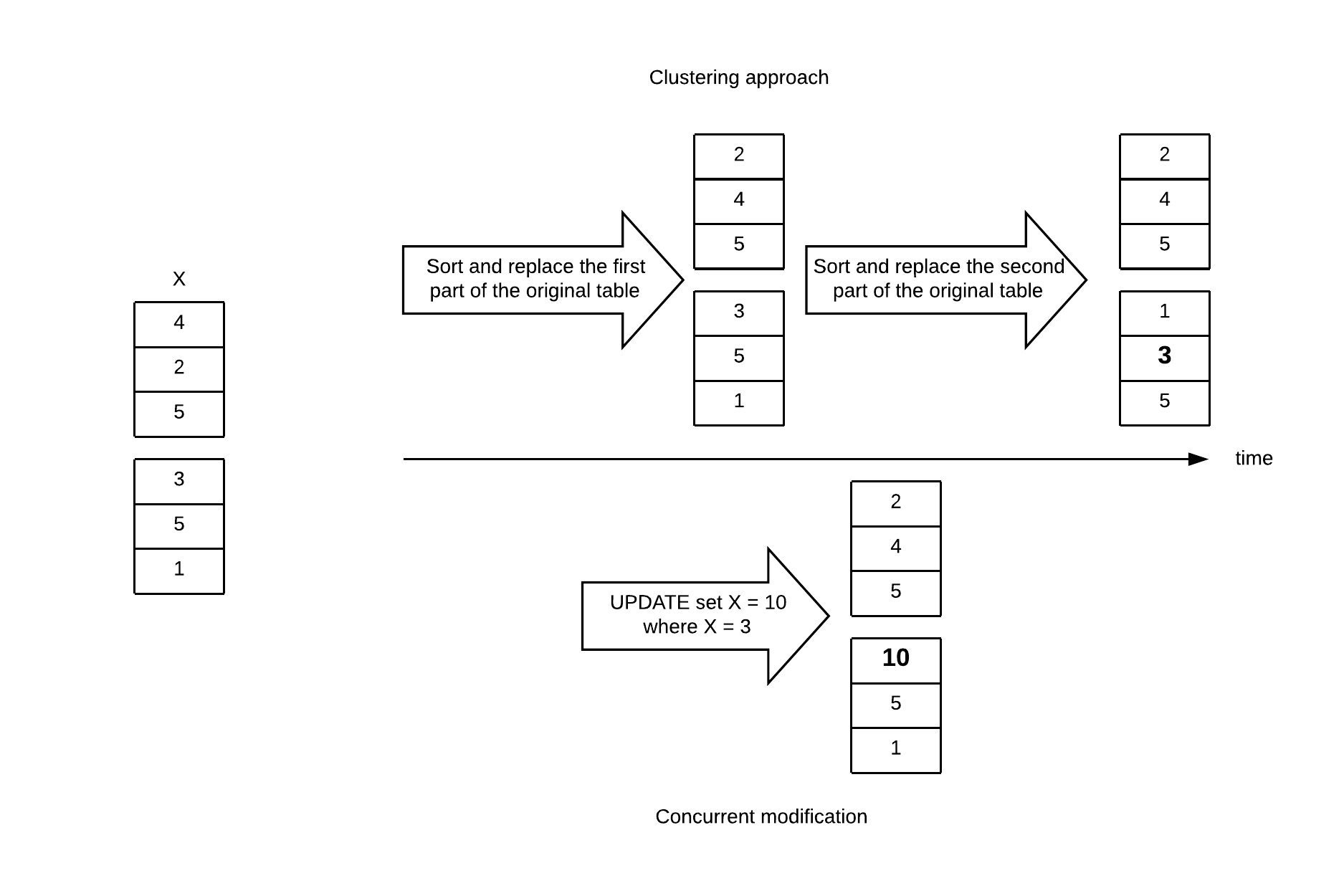}
    \caption{The figure visualizes the impact of a single interfering concurrent modification when performing the clustering in small steps. This time, only the second sort step conflicts. }
    \label{fig:interfering_dml_small}
\end{figure}

\paragraph{Robustness}
Hyrise's current approach for clustering has no robustness against concurrent modifications because it performs an expensive sort operation, during which many modifications may occur.
Consequently, our key idea to obtain robustness is to perform the clustering process in small steps (e.g., per chunk), rather than in a single large step.
Performing clustering in small steps has multiple advantages:
First, the number of interfering DML statements is reduced, as only small parts of the data are being clustered at a time.
DML statements against data that is currently not being clustered cannot cause (clustering-based) lost update problems, and thus do not interfere with the clustering process.
Second, it is sufficient to compute the clustering for a small part rather than the entire table.
This reduces the time required to cluster a certain small part of the data, thereby further reducing the number of interfering concurrent modifications.
Nevertheless, concurrent modifications may still occur, as visualized in \Cref{fig:interfering_dml_small}.
In that case, the computed clustering can be discarded and re-computed.
Since only small parts of the data are being clustered at a time, computing the clustering again is a cheap operation.

\begin{figure}
    \centering
    \includegraphics[width=0.75\textwidth]{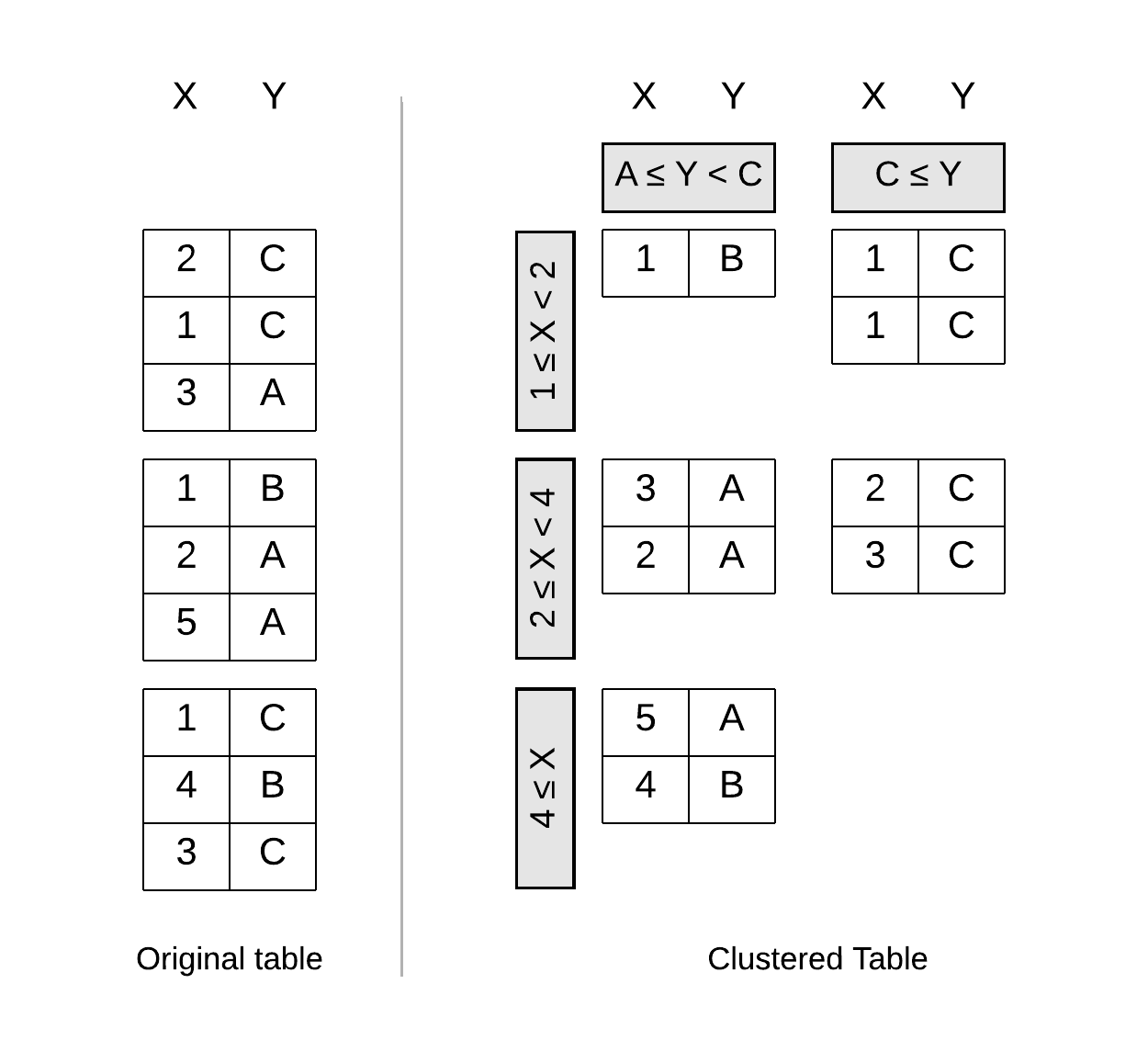}
    \caption{A two-dimensional clustering. The gray boxes show value range constraints the chunks are subject to. Each cluster is individually sorted by column $Y$. In this example, each cluster has exactly one chunk.}
    \label{fig:clustering_result}
\end{figure}

\paragraph{Storage layout}
We propose an algorithm resulting in disjoint clusters, similar to the storage layout described by Lightstone et al.~\cite{DBLP:conf/vldb/LightstoneB04, DBLP:books/mk/Lightstone2007}.
A cluster is a set of one or more chunks with a constrained value range, i.e., all tuples stored in the chunk must satisfy a certain range predicate on the clustered columns.
Clusters can have an arbitrary number of chunks; however, all chunks in a cluster share the same value range constraints.
\Cref{fig:clustering_result} visualizes a two-dimensional clustering example.

For each clustering column, we impose a value range constraint upon all chunks.
The individual value ranges are mutually disjoint.
Combined, they cover the entire value range of the respective clustering column.
As a consequence, each row in the table can be assigned to exactly one cluster.

\subsection{Clustering Phases}
\begin{figure}
    \centering
    \includegraphics[width=.75\textwidth]{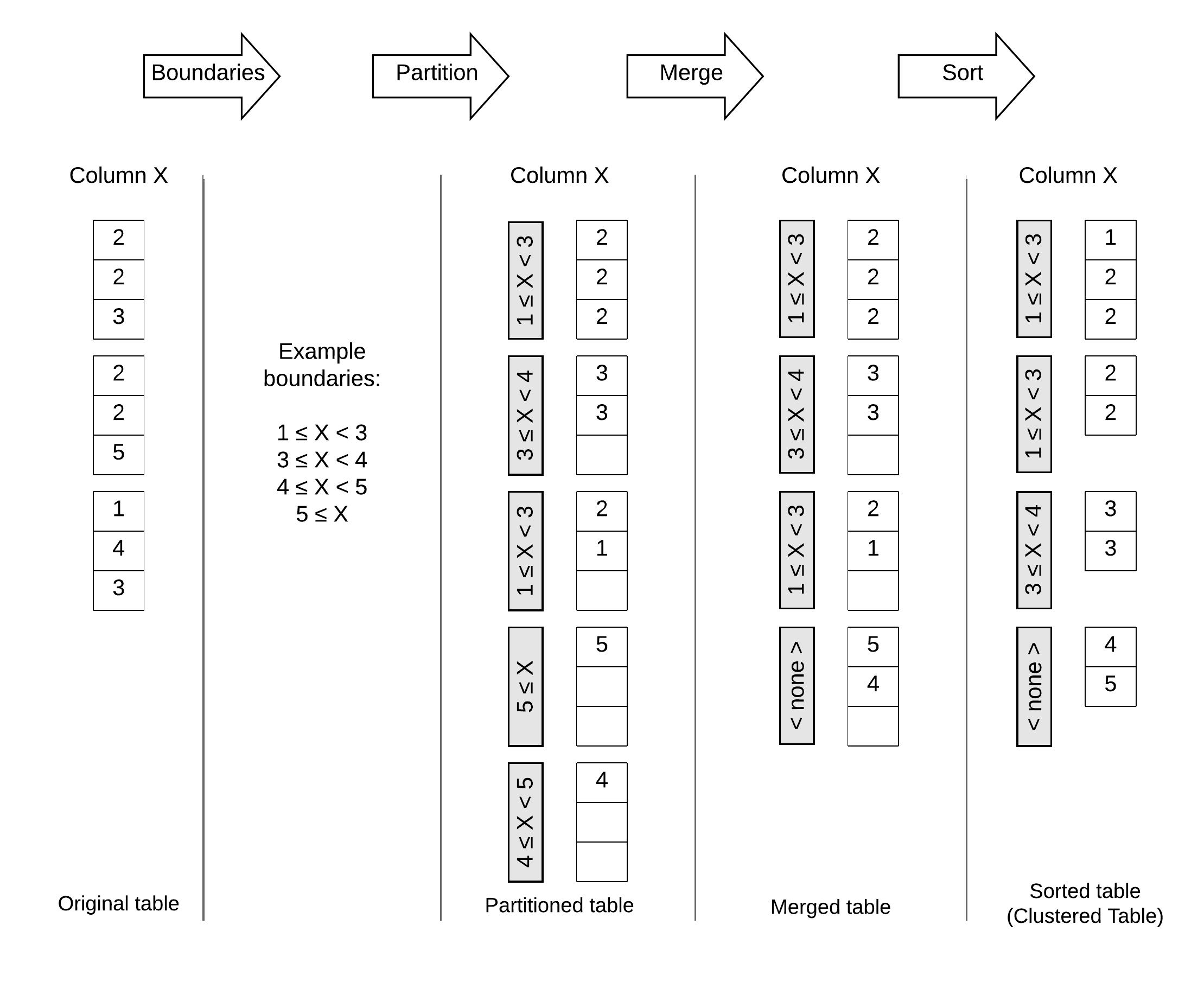}
    \caption{The figure shows the steps performed when clustering the original table (left) by column $X$. The chunk size is three. Chunks with at most one row are considered small and merged. The gray boxes show value range constraints the chunks are subject to. Chunks with the same value range constraints belong to the same cluster.}
    \label{fig:clustering_steps}
\end{figure}

In this section, we describe the phases our clustering algorithm undergoes to create a multi-dimensional clustering.
The phases are described in pseudo-code in \Cref{algo:disjoint_clusters} and visualized in \Cref{fig:clustering_steps}.

\begin{algorithm}[htbp]
\caption{Disjoint Clusters Algorithm}
\label{algo:disjoint_clusters}
    \KwIn{A table, clustering columns and their cluster counts}
    \KwOut{The table, clustered}
    \DontPrintSemicolon
    $B =$ \texttt{ChooseBoundaries}(clustering\_columns, cluster\_counts)\;
    \ForEach{chunk $c$ in the table}{
        \ForEach{row $r$ in $c$}{
            use $B$ to determine the cluster $r$ belongs to\;
            move $r$ to that cluster\;
        }
    }
    
    Optionally, merge all chunks that contain only a small, fixed number of rows\;
    
    \ForEach{cluster in the table}{
        sort the cluster\;
    }
\end{algorithm}

\paragraph{Calculating cluster boundaries}
In the first phase, we calculate the \emph{cluster boundaries}, i.e., the value range constraints the chunks should be subject to.
We describe the calculation process in more detail in \Cref{sec:choose_cluster_boundaries}.
For now, it is sufficient to know that we aim to create balanced clusters, i.e., all clusters have roughly the same size.

\paragraph{Partition phase}
The next phase is called \emph{partition phase}.
The purpose of the partition phase is to group rows with similar values in the clustering columns into clusters.
For that purpose, we perform a \emph{partition step} for each chunk.
In a partition step, we first determine for all rows of the chunk to which clusters they belong.
Since the cluster boundaries are mutually disjoint, there is always exactly one cluster for each row.
Thus, to determine the cluster a row belongs to, it is sufficient to iterate over the cluster boundaries linearly and choose the first (and only) cluster where the row satisfies all constraints.
In the second part of a partition step, we move all rows to their respective clusters.
New chunks are created whenever necessary, i.e., when a cluster does not have a chunk yet, or all chunks of a cluster are full.

\paragraph{Sort phase}
The last phase is called \emph{sort phase}.
Its purpose is to sort the rows within the clusters, so that Hyrise may apply optimizations for sorted data, such as, e.g., binary search.
For that purpose, we perform a \emph{sort step} for each cluster, in which
we sort the whole cluster.
If a cluster contains multiple chunks, they are sorted as a whole rather than each chunk individually. 

Sorting is an expensive operation, during which many concurrent modifications may occur.
However, it is important to recall the fact that we choose the cluster boundaries in a way that yields small, balanced clusters, i.e., clusters of roughly equal size.
Thus, every cluster represents only a fraction of the original table.
As a consequence, sorting an individual cluster is a lot cheaper than sorting the entire table.
Due to the balanced clusters, this remains true even in the case of skewed data.

\paragraph{(Optional) Merge phase}
There is a tradeoff to consider:
The partition phase may create several chunks that contain only a small number of rows.
For example, when a chunk can store $X$ rows, but $X + 1$ rows belong to a cluster, the partition phase would result in two chunks per cluster: one with $X$ rows, the other one with just a single row.
A large number of such small chunks may cause additional processing overhead~\cite{DBLP:conf/edbt/DreselerK0KUP19} while yielding only a marginal pruning benefit.
We introduce an optional \emph{merge phase} (between the partition and the sort phase) that merges those small chunks, thereby trading pruning effectiveness against reduced processing overhead.

In the merge phase, we perform a \emph{merge step} for each small chunk.
In a merge step, the contents of the small chunk are moved to a dedicated merge cluster.
The dedicated merge cluster's chunks do not have value range constraints, and thus can store rows from any cluster.

\section{Choosing Cluster Boundaries}
\label{sec:choose_cluster_boundaries}
The first step of our clustering algorithm is the choice of disjoint cluster boundaries, i.e., the value range constraints the chunks are subject to.
In this section, we describe our algorithm to calculate the cluster boundaries.

\paragraph{Optimization goal}
The cluster boundaries impact the cluster's sizes, i.e., how many rows a given cluster will contain.
There are two different optimization criteria when calculating the boundaries:
We could try to obtain either clusters that contain a similar number of rows, or clusters that contain a similar number of disjoint values.

If skew is present, the latter option might lead to clusters that significantly differ in size.
However, we sort each cluster in the last phase of the clustering algorithm.
When large parts of a table are condensed in just a few clusters, sorting those clusters will be particularly expensive, thereby increasing the likelihood of interfering concurrent modifications from other transactions.
When clusters are balanced equally, i.e., they contain a similar number of rows, there is no such particular expensive sort operation.
For that reason, we consider the first approach superior.

\begin{figure}
    \centering
    \includegraphics[width=0.5\textwidth]{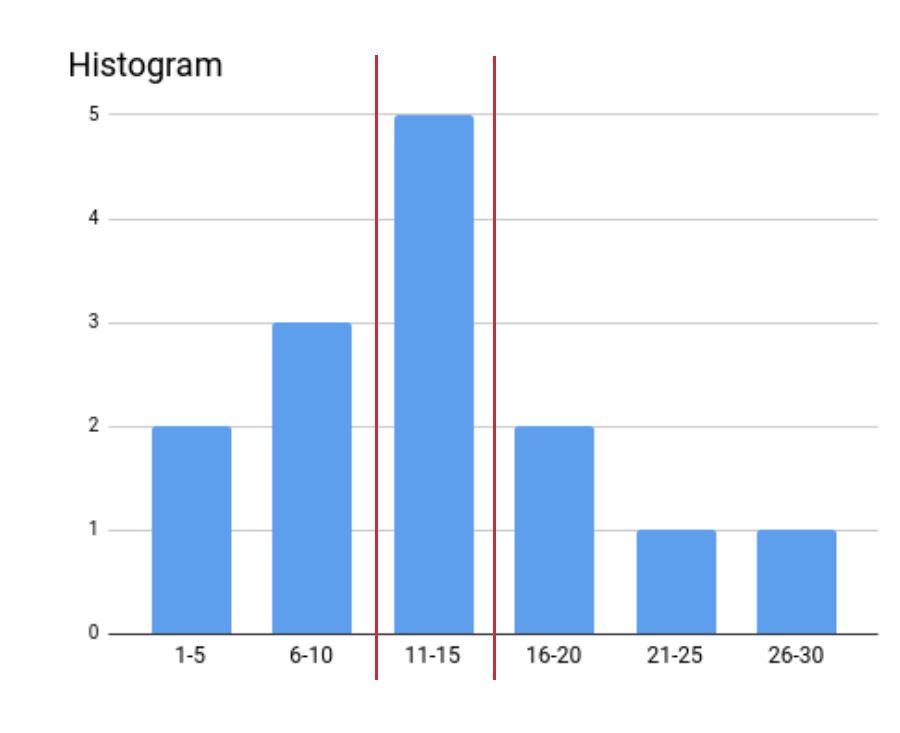}
    \caption{The figure shows an example output when requesting three boundaries. The red lines separate the boundaries.}
    \label{fig:choose_boundaries}
\end{figure}

\paragraph{Algorithm}
Our algorithm takes the clustering columns and their cluster counts as an input to compute boundaries that yield balanced clusters.
A \emph{cluster count} is a measure to quantify the granularity a column should be clustered with.
It describes into how many disjoint sections the clustering column's value range shall be divided.
The higher the cluster count, the more fine-grained a column will be clustered.
\Cref{algo:choose_boundaries} provides a pseudo-code description of the algorithm.
\Cref{fig:choose_boundaries} shows an example of boundary outputs.

\begin{algorithm}[htbp]
\caption{ChooseBoundaries Algorithm}
\label{algo:choose_boundaries}
    \KwIn{Clustering columns and their cluster counts}
    \KwOut{Cluster boundaries}
    \DontPrintSemicolon
    \ForEach{clustering column $X$, its desired cluster count $c$, and its histogram $H$}{
        \If{$X$ is nullable} {
            add null-Cluster for $X$\;
        }
        null\_count = table.size - $H$.size\;
        ideal\_cluster\_size = (table size - null\_count) / $c$\;
        current\_cluster = Cluster(size=$0$, min=null, max=null)\;
        \ForEach{bin $B$ in $H$}{
            \If{current\_cluster.min == null}{
                current\_cluster.min = $B$.min\;
            }
        
            \uIf{current\_cluster.size + $B$.size < ideal\_cluster\_size}{
                current\_cluster.size += $B$.size\;
                current\_cluster.max = next\_bin($B$).min or null\;
            }
            \uElseIf{current\_cluster.size + $B$.size - ideal\_cluster\_size < ideal\_cluster\_size - current\_cluster.size}{
                current\_cluster.size += $B$.size\;
                current\_cluster.max = next\_bin($B$).min or null\;
                cluster\_full = true\;
            }
            \Else{
                cluster\_full = true\;
                In the next iteration, process bin $B$ again\;
            }
            
            \If{cluster\_full}{
                current\_cluster.save\_boundaries\_for($X$)\;
                current\_cluster = Cluster(size=$0$, min=null, max=null)\;
            }
        }
    }
    \Return{all saved cluster boundaries}\;
\end{algorithm}

By default, Hyrise maintains a table-level histogram for each column.
The histogram contains several bins, which cover disjoint value ranges and are sorted in ascending order.
We use the table-level histograms to determine boundaries that yield balanced clusters:
Basically, we follow a greedy approach.
For each clustering column, its histogram, and its desired cluster count, we perform the following steps:
First, we calculate the ideal cluster size, i.e., the size all clusters had if the rows were distributed uniformly.
Then, we iterate over the bins of the histogram and assign all bins we iterate over to an hypothetical cluster.
The iteration stops when including the next bin into the cluster would increase the absolute difference between the estimated current cluster size and the ideal cluster size.
At this point, the cluster's boundaries are known:
The lowest possible value in the cluster matches the lower boundary of the first bin contained in the cluster, i.e., the lower cluster boundary is inclusive.
Analogously, the highest possible value in the cluster is strictly less than the minimum of the first bin that is not included in the cluster, i.e., the upper cluster boundary is exclusive.
An (exclusive) upper boundary that equals the (inclusive) lower boundary of the next cluster ensures that there is no interval between the clusters for which rows cannot be associated to a cluster.

We repeat this process until all bins of the histogram have been processed.
If a column is nullable, we also add a cluster boundary that only allows null values.

\begin{figure}
    \centering
    \includegraphics[width=0.6\textwidth]{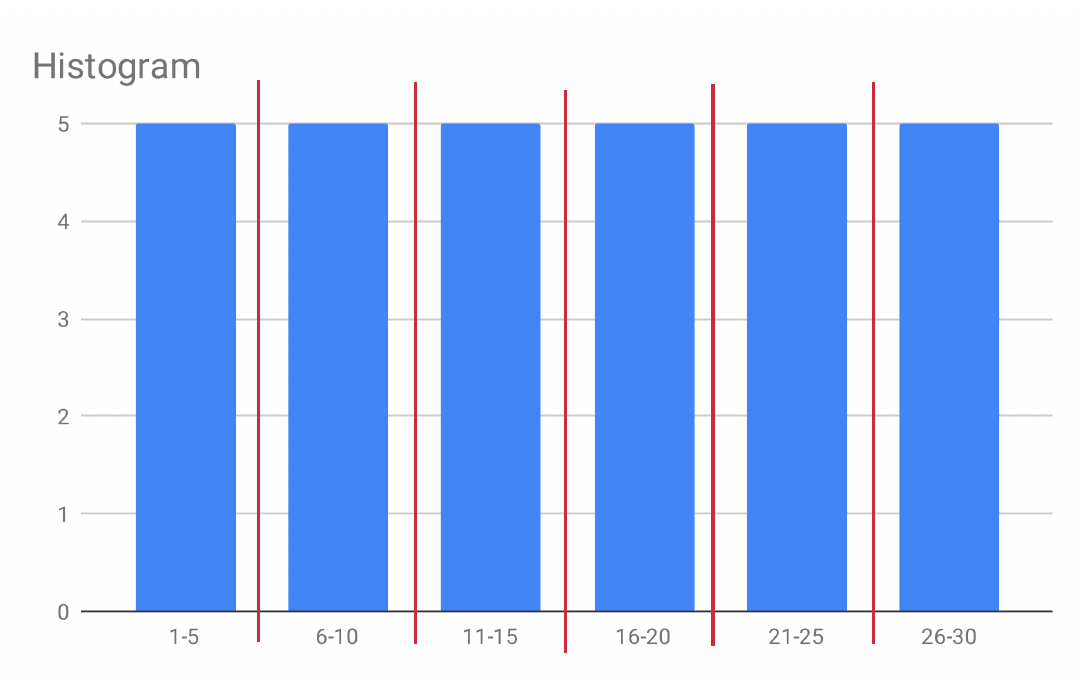}
    \caption{The figure shows an example where \Cref{algo:choose_boundaries} yields more than the requested number of 5 cluster boundaries. The red lines separate boundaries.}
    \label{fig:cluster_count_guess}
\end{figure}

\paragraph{Inaccuracy regarding the obtained number of boundaries}
It is worth noting that our algorithm does not always return exactly the requested number of cluster boundaries, but may return more or fewer boundaries than requested.
Consider the following scenario, which is visualized in \Cref{fig:cluster_count_guess}:
All histogram bins contain the same number of rows (not necessarily the same number of distinct values, though).
The target cluster count $c$ is set in a way such that the ideal cluster size is slightly above the bins' sizes.
In that case, two consecutive histogram bins contain combined almost twice as many rows as an ideal cluster.
Thus, for any reasonable database size, is the number of rows in a single bin closer to the ideal cluster size than the number of rows in two bins.
Consequently, our algorithm creates one boundary for each bin, rather than the requested $c$ boundaries.
However, we consider this inaccuracy as acceptable, as we value balanced clusters more than an exact number of cluster boundaries.

\begin{figure}
    \centering
    \includegraphics[width=0.8\textwidth]{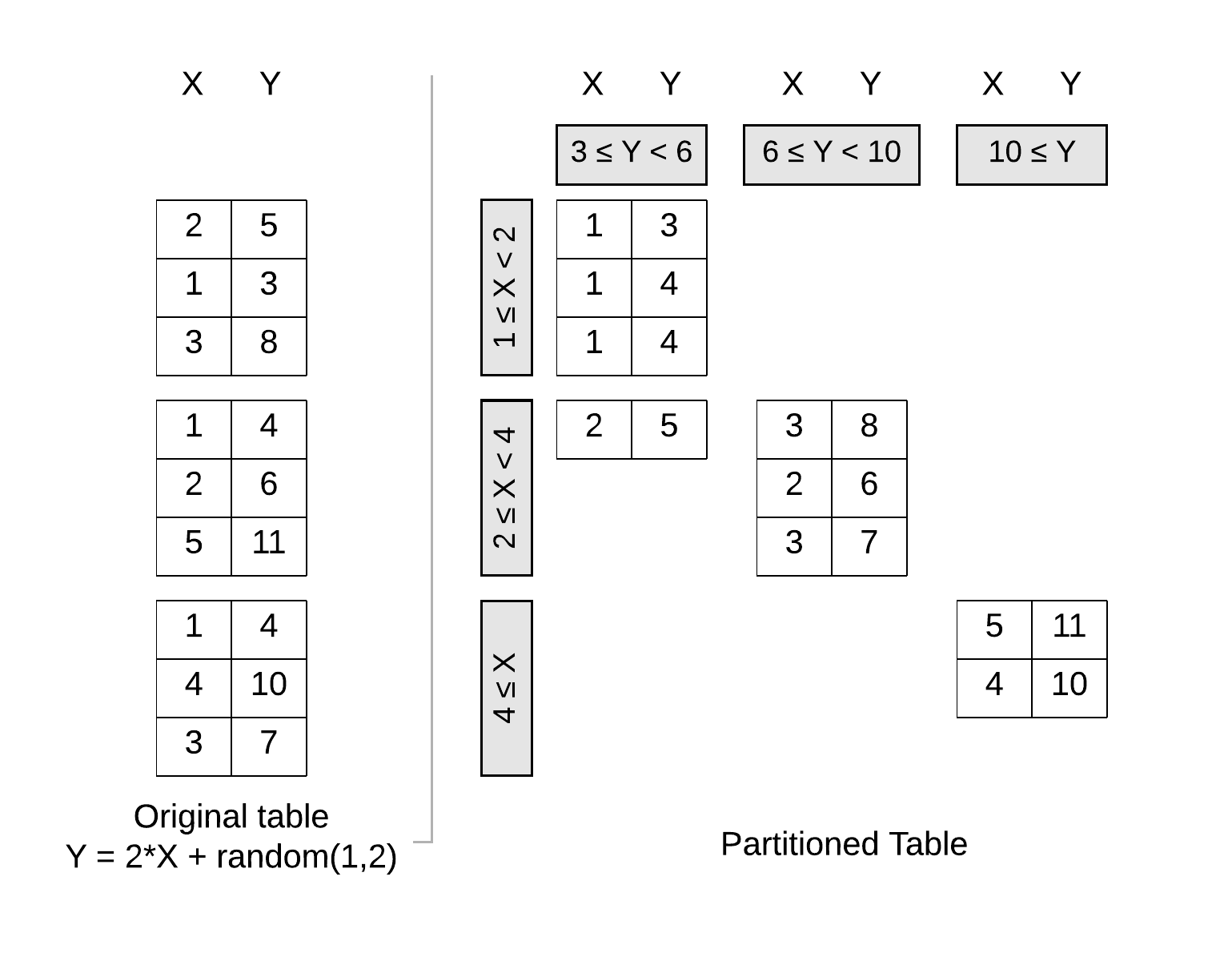}
    \caption{The figure shows an two-dimensional clustering example where the clustering columns are correlated. Five out of nine clusters contain no rows at all, a sixth contains just one row.}
    \label{fig:mdc_correlation}
\end{figure}

\paragraph{Independence assumption}
There is one more issue to consider:
The algorithm aims to create balanced clusters but calculates the boundaries for each clustering column independently.
With that, it indirectly assumes that each combination of cluster boundaries (across all clustering columns) will contain a similar number of tuples.
However, in the presence of correlations or functional dependencies between the clustering columns, this assumption may break.
As a consequence, some clusters may contain no tuples at all, and others might be significantly larger than the average cluster, as visualized in \Cref{fig:mdc_correlation}.
Detecting such situations is difficult, because Hyrise offers only histograms for a single column, not for combinations of columns.
Future work might investigate advanced cluster size estimation techniques, e.g., sampling, which is not in the scope of this thesis.
Instead, for our work, we assume that the clustering model is aware of correlations and does not select two or more correlated columns as clustering columns.

\section{Maintaining Consistency}
\label{sec:mvcc_consistency}
Hyrise follows an append-only policy, i.e., rows are marked as invalid rather than physically deleted; update statements are handled by invalidating all rows affected by the statement, and inserting the modified rows into the table, rather than by updating the values in place.
Analogously, our clustering algorithm cannot perform the clustering in place, but rather has to invalidate the old, unclustered versions, and insert the new, clustered versions of the rows.

During the clustering, it is mandatory to ensure the database maintains a consistent state, i.e., that there is always at most one version of a row visibile to a transaction.
For that purpose, we treat the movement of a row as a special case of an update: an update without modification, but with a dedicated chunk to insert the new version of the row into.
When performing the movement of rows as an update, we can rely on Hyrise's MVCC implentation to ensure that only one version of the row is visible to a each transaction.

In this section, we describe how our algorithm interacts with Hyrise's MVCC implementation to maintain a consistent database state for all transactions.
First, we provide background information about Hyrise's MVCC implementation, and Hyrise's interface for operators that modify rows, the \texttt{AbstractReadWriteOperator}.
Afterward, we describe for each phase of the algorithm its interaction with MVCC and the tradeoffs that should be considered.

\paragraph{Hyrise's MVCC implementation}
Hyrise uses Multi-Version Concurrency Control (MVCC) to handle concurrent transactions~\cite{DBLP:conf/vldb/SchwalbFWGP14}.
In Hyrise's MVCC implementation, every row has a begin commit id (begin cid) and an end commit id (end cid).
The begin commit id represents the id of the commit that inserted the row, the end commit id the id of the commit that invalidated the row.
Invalidating and re-inserting a row within the same transaction thus ensures that exactly one version of the row is visible for any other transaction.
Further, Hyrise maintains a row-level transaction id (tid).
Before modifying a row, transactions have to set the row's transaction id to their transaction id, thereby indicating to other transactions that the row is being modified.
Setting the transaction id of a row that is under modification by another transaction is not allowed.
Given the append-only policy of Hyrise, there are only two reasons for setting the transaction id:
Insertion of a new row, and invalidation of an existing row.
Finally, Hyrise stores for each chunk the number of invalidated rows.

\paragraph{AbstractReadWriteOperator interface}
In our implementation, we introduce two new operators: \texttt{ClusteringPartitioner} and \texttt{ClusteringSorter}.
Both operators are subclasses of \texttt{AbstractReadWriteOperator}, which tells Hyrise that the operators might insert or invalidate rows.
The \texttt{AbstractReadWriteOperator} provides the following interface:
\begin{itemize}
    \item \texttt{\_on\_execute()} - This function is called during the execution of the operator. At this point, it is unknown whether the transaction will commit or abort, so no irreversible actions must be made. We use this function mostly to set the row-level transaction ids, indicating to other transactions that our transaction will modify the row.
    \item \texttt{\_on\_rollback\_records()} This function is called when the transaction is aborted. Any changes performed by the transaction so far must be reverted. We use this function mostly to unset the row-level transaction ids.
    \item \texttt{\_on\_commit\_records(commit\_id)} This function is called when the transaction commits. At this point, the transaction cannot fail anymore (i.e., an action that could fail must be performed earlier) and irreversible actions can be made. We use this function mostly to set begin- and end commit ids.
\end{itemize}

\subsection{Partitioning}
\label{sec:algo_mvcc_partition}
In a partition step, all rows of a chunk are moved to their respective clusters.
Each chunk is partitioned within its own transaction using the \texttt{ClusteringPartitioner} operator.
There are two approaches for the partitioning that differ regarding the point of time the row-level transaction ids are set, which results in different priorities for the partitioning against other transactions:

The first approach sets the transaction ids in \texttt{\_on\_execute()} for all non-invalidated rows.
This prevents operators like \texttt{Delete} or \texttt{Update} to invalidate a row while the partitioning is in progress and causes them to conflict, i.e., it assigns partitioning a high priority.
In case of failure, e.g., because a non-invalidated row is already under modification by another transaction, the whole operator fails.
The actual partitioning work happens in on \texttt{\_on\_commit\_records()}.
Specifically, for each non-invalidated row in the chunk the following actions are performed:
\begin{enumerate}
    \item insert the row, i.e.,
      \begin{enumerate}
        \item look up the target cluster for the row, and get its insertion chunk
        \begin{itemize}
            \item if the cluster does not exist yet, or its insertion chunk is full, acquire the table's append mutex and append a new mutable chunk
        \end{itemize}
        \item set the begin commit id in the insertion chunk
        \item copy the values of the row to the insertion chunk
    \end{enumerate}
    \item invalidate the row by setting the end commit id, and increase the invalid row counter
\end{enumerate}

The second approach performs no action during \texttt{\_on\_execute()}.
In particular, it does not set any transaction ids.
During \texttt{\_on\_commit\_records()}, we first insert all rows analogous to the first approach.
Then, for every row in the chunk, we try to set the transaction id and invalidate the row.
This operation may well fail, e.g., because a row is already under modification by another transaction, or invalidated (Hyrise considers invalidated rows as under modification).
In the case of failure, we do not invalidate the row in the chunk, but rather the corresponding row in the cluster we inserted in the first step.
Setting the transaction ids after all insertions are done reduces the amount of time the rows are under modification, which benefits other transactions.
Overall, the approach causes the partitioning rather than other transactions to cause conflicts, thus assigning it a lower priority than the first approach.

There is a potential downside to the second approach:
Assume that a few rows could not be partitioned because another transaction was already modifying them.
Since we partition only immutable chunks (i.e., no new rows can be inserted), a modification is equivalent to invalidation.
However, if the other transaction aborts, the few rows will remain valid in a chunk that mostly contains invalidated rows.
As a consequence, Hyrise has to execute additional \texttt{Validate} operators, which may reduce the overall performance.
For that reason, we prefer the first approach for setting the transaction ids.



\subsection{Merging}
The partition phase may produce some small chunks.
In the merge phase, those chunks are merged into a single cluster.
Technically, this is implemented as a second partition phase.
For that purpose, we introduce a new, dedicated merge cluster and assign all rows of small chunks to it.
Since the merge step is implemented using the \texttt{ClusteringPartitioner}, all consistency issues have already been addressed.

\subsection{Sorting}
Each cluster is sorted within its own transaction using the \texttt{ClusteringSorter} operator.
Analogous to the \texttt{ClusteringPartitioner}, the actual physical reorganization of the data happens in \texttt{\_on\_commit\_records()}, while \texttt{\_on\_execute()} performs all necessary preparations, i.e., computing the sorted cluster, and setting the transaction id for all non-invalidated rows.

Sorting is an expensive, time consuming operation.
This leads to the question whether the transaction ids should be set in \texttt{\_on\_execute()} before or after the sort operation takes place, i.e., there is a tradeoff between potentially blocking other transactions for a long time, and potentially wasted computation time.

The straightforward way would be to set the transaction ids as the first step.
Setting the transaction ids before the sorting begins ensures that no rows can be invalidated while the \texttt{Sort} operator is executed.
However, keeping the rows under modification for a long time may block other transactions, decreasing overall performance.

The second approach uses the invalid row counters to defer setting the transaction ids:
Every time a row is invalidated, the counter is increased, but it is never decreased.
Thus, it is possible to detect new invalidations by comparing the old and new invalid row counters.
However, there is a race condition to consider:
Invalidating a row and increasing a chunk's invalid row counter are two separate steps.
Consequently, a row could already be invalidated, although the invalid row counter did not increase yet. 
To detect this race condition, we additionally check whether we could set the expected amount of transaction ids.
If we could set fewer transaction ids than expected, we know that some row was invalidated during the sort operation.

Storing and comparing invalid row counters (integers) is a cheap way to detect new invalidations; however, it does not prevent them.
As a consequence, invalidations can occur during the execution of the \texttt{Sort} operator, rendering its result useless.

We consider blocking other transactions for a longer time as the bigger issue and choose the second approach.

Thus, the \texttt{\_on\_execute()} function performs the following steps:
\begin{enumerate}
    \item For each chunk of the cluster, store the invalid row counters
    \item Create a temporary table $T$ that contains only the chunks of the cluster
    \item Execute a \texttt{Validate} operator on $T$
    \item Sort the validated table using the \texttt{Sort} operator. This creates a read-only copy of the cluster chunks but does not change the original table
    \item Set the transaction id for all rows that belong to the cluster (invalidated rows excluded). On any failure, the operator fails
    \item If the invalid row counter of any chunk of the cluster increased since the first step, or we could set fewer transactions ids than expected, the operator fails
\end{enumerate}

The \texttt{Validate} operator takes an input table and excludes all rows that are not visible to the current transaction.
This step is required, because the \texttt{Sort} operator is not MVCC-aware, i.e., rows do not maintain their MVCC state.
In other words, skipping this step would cause the sort operator to - unintentionally - restore invalidated rows.

The \texttt{\_on\_commit\_records()} function replaces the unsorted cluster by its sorted counterpart.
All chunks of the unsorted cluster are completely invalidated (i.e., the end commit id of all non-invalidated rows is set to the commit id of the current transaction).
Next, the original table's append mutex is acquired and all chunks from the sorted table are copied over to the original table, with the current transaction's commit id as their begin commit id.
Performing those two actions in the same transaction ensures that there is exactly one version of the cluster (i.e., sorted or unsorted) visible for a transaction at all times.

\section{Cleanup and Encoding}
In this section, we discuss two aspects of the clustering algorithm that are not directly related to clustering, but are nevertheless important: cleanup and encoding.

\subsection{Cleanup}
During the clustering process, each row is moved and invalidated two or three times (always in the partition and sort phase, and in the merge phase where required).
As a consequence, the clustering algorithm creates a large number of fully invalidated chunks.
Given the append-only policy of Hyrise, it is impossible to perform the clustering in place, i.e., without creating additional chunks.
However, a large number of invalidated chunks poses a performance problem: memory consumption.
Regardless of their rows' invalidity, the chunks consume the same amount of memory as completely valid chunks.
We address this issue by physically deleting invalidated chunks, where possible.
Hyrise offers the \texttt{MvccDeletePlugin}, which periodically consolidates chunks with a high invalidation rate and physically deletes fully invalidated chunks.
The plugin also offers functionality to determine if a chunk can be safely deleted, or whether it might still be referenced by some transaction.
For evaluation purposes, invalidated chunks are explicitly deleted by our \texttt{ClusteringPlugin}.
In the future, the utilization of the \texttt{MvccDeletePlugin} could easily be integrated.

\subsection{Encoding}
By default, Hyrise uses dictionary encoding for segments of immutable chunks.
The partition, merge and sort phase of the clustering algorithm, however, produce unencoded value segments.

Encoding segments benefits Hyrise in multiple ways, e.g., it can reduce the memory consumption, and improve the performance of certain operators, e.g., table scans.
However, calculating a segment encoding is computationally expensive; and we are aware that the clustering algorithm fully invalidates and then removes the outputs of the partition and merge phases as soon as the sort phase is completed, i.e., the effort of encoding the partition and merge phases' outputs only yields a temporary benefit.

As such, there is a tradeoff between memory consumption and temporary query performance on the one side, and clustering run time on the other side:
We could encode the outputs of all clustering phases for a minimum memory consumption and better temporary query performance;
or we could encode only the results of the sort phase for a lower clustering run time.
We evaluate the memory consumption of both options in \Cref{sec:eval_memory_consumption}.
In both cases, we encode the output of the sort phase, i.e., the sorted clusters.

\section{Integration into Hyrise}
Large parts of the algorithm are implemented as a plugin and do not interfere with existing Hyrise code.
This section describes necessary changes to existing Hyrise code and possible interactions.

\paragraph{Insert}
Hyrise provides an \texttt{Insert} operator, which inserts rows into tables, e.g., to process SQL \texttt{INSERT INTO} statements.
Currently, the \texttt{Insert} operator always inserts into the last chunk.
If the last chunk is not mutable or already full, it appends an empty, mutable chunk.

However, this approach does not work well with our clustering algorithm:
While the partition phase is executed, our algorithm appends an empty, mutable chunk whenever a row belongs to a cluster that is either full, or does not have a chunk yet.
These chunks are filled during the partition phase, and only finalized if they reach their maximum capacity, or after all chunks have been partitioned.
In other words: There are multiple mutable chunks at a time, and if left unchanged, the \texttt{Insert} operator is likely to insert into chunks that are used for clustering.
The \texttt{Insert} operator, however, does not consider cluster boundaries, i.e., it might insert an arbitrary value that does not match the cluster's boundaries.
Even a small number of arbitrary values might cause significant harm to the prunability of the chunk:
For example, if a really small and a really large value are inserted, they may prevent the chunk from being pruned, despite 99\% of its values cover only a small, unqueried value range.

To address this issue, we introduce the concept of a dedicated insert chunk.
Instead of always inserting into the last chunk, the \texttt{Insert} operator asks the \texttt{Table} object which chunk should be used for inserts.
To enable this behavior, we add a member field \texttt{ChunkID~\_insert\_chunk\_id} to the \texttt{Table} class.
Further, we added an optional boolean parameter to the member function \texttt{Table::append\_mutable\_chunk()} that determines whether the chunk should be used for inserts.
We use \texttt{true} as the default value, i.e., existing calls to \texttt{append\_mutable\_chunk()} do not have to be modified.
If the parameter is true, the table automatically updates \texttt{\_insert\_chunk\_id}.

In addition to the \texttt{Insert} operator, the \texttt{Update} operator and the \texttt{MvccDeletePlugin} also insert rows.
However, both of them internally use the \texttt{Insert} operator to perform the insertion, and will thus not interfere with the partitioning.

\paragraph{GetTable}
Query execution in Hyrise usually begins with a \texttt{GetTable} operator.
The \texttt{GetTable} operator creates a copy of the original table, but excludes chunks that are, e.g., logically deleted.
Our clustering algorithm produces several fully invalidated chunks, i.e., chunks where all rows are invalidated.
However, we cannot instantly physically delete those chunks, because some running transactions might access them.
Instead, we set the cleanup commit id for these chunks.
Setting the cleanup commit id communicates to the \texttt{GetTable} operator the chunk was logically deleted, i.e., fully invalidated, by the respective commit, and can be safely excluded for transactions that started after the cleanup transaction committed.

\paragraph{Validate}
The \texttt{Validate} operator filters out rows that are invisible to the current transaction.
In \Cref{sec:algo_mvcc_partition}, we described an approach to set transaction ids as late as possible during the partition steps.
If this approach is chosen, the clustering may produce chunks that are mostly, but not fully invalidated.
Fully invalidated chunks can be skipped by the \texttt{GetTable} operator for future queries, but this optimization does not work when chunks still contain valid rows.
As a consequence, Hyrise may have to execute additional \texttt{Validate} operators.

	\chapter{Automated Workload-Based Clustering Model}
\label{sec:clustering_model}
A certain clustering is per se neither beneficial nor harmful for performance:
Whether a clustering is beneficial for performance always depends on the database's workload.
A clustering that is beneficial for a certain workload may harm performance for other workloads.

In this chapter, we present a clustering model that takes a workload as input and outputs multi-dimensional clustering suggestions.
The model is designed to produce clustering suggestions for the algorithm we presented in \Cref{sec:clustering_algorithm}, i.e., it assumes the algorithm will produce disjoint clusters.

\paragraph{Implementation}
Hyrise is implemented in C++.
We believe this is a good incentive to implement the clustering model in C++, too, as it could grant direct access to Hyrise and its statistics.

Nevertheless, for this work, we decided to implement our model in Python.
Python offers a number of advantages that - in our opinion - outweigh the reasons to choose C++:
First of all, Python has a rich set of libraries for data manipulation, such as  \texttt{numpy} and \texttt{pandas}.
Furthermore, it provides plotting libraries such as \texttt{matplotlib} and \texttt{seaborn}, which facilitate the model's evaluation.
Finally, developing the model is an iterative, explorative process.
With interactive tools like Jupyter, we consider Python better suited for such a development process.

If our clustering model is used outside of this work, it might be a wise idea to port it to C++ for a better integration into Hyrise.
The source code is available at Github\footnote{\url{https://github.com/aloeser/clustering_model}}.

\paragraph{Performance metric}
The task of the model is to find a clustering that is, given the recent workload, optimal for the database's performance.
Databases have different performance metrics, such as throughput and latency.
With \emph{throughput}, we describe the number of queries that are executed per second.
Alternatively, throughput can refer to the number of completed transactions per second rather than queries per second.
With \emph{latency}, we refer to the run time of queries.
Latency and throughput are related:
When we improve (i.e., reduce) the latency of a certain unique query, we can execute more queries of that type in the same frame of time, i.e., its throughput will improve, too.
Vice versa, assuming resources such as the number of threads are held constant, an improved throughput implies a reduced latency.

When considering multiple different queries, however, an improved throughput is not equivalent to an improved latency.
For example, when sorting the \texttt{lineitem} table of the analytical TPC-H benchmark by \texttt{l\_shipdate}, throughput improves about 17\% across all queries; at the same time, the total latency (i.e., the time required to execute each query once) increases too, by 11\%.
\Cref{tab:throughput_latency} lists the throughput changes and latencies in the old and the new clustering for each query.
While the clustering also reduces the throughput for some queries (e.g., for TPC-H 9 and 18), the geometric mean of the individual throughput changes increases.
However, the decreased throughput occurs on queries with a big latency.
As a consequence, the total latency increases, too.

In a nutshell: minimum latency and maximum throughput are different optimization goals.
For our model, we decide to optimize the total latency.
With slight modifications, our model could optimize for throughput instead.


\begin{table}[]
    \centering
    \begin{tabular}{ c|c|c|c}
        TPC-H Query & Old latency (ms) & New latency (ms) & Throughput Change (\%)\\
        \hline
        01 & 17\,694 & 16\,433 & +8\\
        03 & 4\,828 & 4\,076 & +18\\
        04 & 3\,722 & 4\,737 & -21\\
        05 & 7\,712 & 8\,652 & -11\\
        06 & 404 & 105 & +285\\
        07 & 2\,762 & 2\,022 & +37\\
        09 & 15\,943 & 19\,069 & -16\\
        12 & 2\,315 & 1\,461 & +58\\
        14 & 1\,055 & 505 & +109\\
        15 & 501 & 231 & +117\\
        18 & 10\,826 & 20\,137 & -46\\
        20 & 1\,120 & 420 & +167\\
        21 & 12\,417 & 15\,576 & -20\\
        \hline
        Total & 113\,322 & 125\,329 & +17\\
    \end{tabular}
    \caption{The table lists the changes in latency and throughput when switching from an \texttt{l\_orderkey} clustering (old) to an \texttt{l\_shipdate} clustering (new). The last row contains the sum for the latency columns and the geometric mean for the throughput column. Queries whose throughput changed by less than $5$\% are not listed. The TPC-H tables were generated with scale factor ten.}
    \label{tab:throughput_latency}
\end{table}

\paragraph{Independence Assumption}
Our model calculates the clustering suggestions for each table separately.
Thereby, it implicitly assumes that the optimal clustering for a table $A$ does not depend on the clustering of or the operators executed on another table $B$.
This assumption is not always correct:
For example, the clustering approach proposed by Ziauddin et al.~\cite{DBLP:journals/pvldb/ZiauddinWKLPK17} allows pruning across different tables.
The same applies to data induced predicates (DiPs)~\cite{DBLP:journals/pvldb/OrrKC19}.
Further, the database might choose a different implementation, e.g., a sort-merge join instead of a hash join, if the join columns from both tables were  sorted.

At the time of writing, Hyrise neither supports DiPs, nor the clustering approach from Ziauddin et al., so we do not need to consider those cases yet.
However, the choice of a different operator implementation might occur.
Thus, by assuming independence, we sacrifice some optimization potential.
On the other hand, the independence assumption leads to a drastic reduction of the search space of potential clustering configurations:
Without the assumption, we had to consider every combination of all clustering configurations of all tables.
In the scope of this thesis, we value the reduction of the search space more than the potentially sacrificed optimization potential.

\section{High-level Overview}
In the following paragraphs, we provide a high-level description of our clustering model.
The model is also described in pseudo-code in \Cref{algo:clustering_model}.

\paragraph{Input}
Our model takes Hyrise's recent workload as input.
The workload contains statistics about recently executed operators, such as their type, run times, and result set sizes.
We describe the input more detailed in \Cref{sec:workload}.

\paragraph{Identifying clustering candidates}
Given the workload, we perform the following steps for each table:
First, we identify the part of the workload that affects the table, i.e., the operators that are executed on that table.
Thereby, we focus on columns that are used in predicates or joins.
We use those columns to construct clustering candidates.
A \emph{clustering candidate} represents a set of columns that shall be clustered.
Further, a clustering candidate also contains a sort column, which determines by which column the individual clusters will be sorted.
The process of identifying clustering candidates is described in more detail in \Cref{sec:clustering_candidates}.

\paragraph{Choosing clustering granularities}
The clustering candidates do not include granularities, i.e., a measure of how fine or coarse the individual columns should be clustered.
Thus, in the next step, we choose one clustering candidate and determine a set of potential clustering granularities.
We use \emph{cluster counts} as a granularity measure.
For each clustering column, the algorithm presented in \Cref{sec:clustering_algorithm} splits the value range of the clustering column into $n$ disjoint subranges, where $n$ is the column's cluster count.
The subranges are then used as cluster boundaries, i.e., a high cluster count leads to small value ranges and thus to a clustering with high granularity.
Each clustering column has its own cluster count.
To determine the cluster counts, we consider a column's usage (e.g., as join column, or in predicates) as well as its number of unique values.
We describe the process of selecting cluster counts in more detail in \Cref{sec:dimension_granularities}.

\paragraph{Latency estimation}
For each clustering candidate and each of its respective cluster count sets, we assume such a clustering were implemented, and estimate its effect on the latency.
To estimate the effect on the latency, we assess the recent workload:
More specifically, the model estimates how the run times of individual scan and join operators are affected by the new clustering.
We use the currently implemented clustering as a baseline.
The latency estimation process is described in more detail, including its limitations, in \Cref{sec:cost_estimation}.

\paragraph{Output}
When all estimations are completed, we sort the clustering candidates ascending by their expected latency.
We return the first $k$ clustering candidates, where $k$ is a parameter that can be set by the user.

\begin{algorithm}[htbp]
\caption{Clustering Model}
\label{algo:clustering_model}
    \KwIn{A workload}
    \KwOut{Clustering suggestions with latency estimations}
    \DontPrintSemicolon
    \ForEach{table in the workload}{
         clustering\_candidates = \texttt{DetermineClusteringCandidates}(workload, table)\;
        \ForEach{clustering\_candidate}{
            cluster\_count\_sets = \texttt{DetermineClusterCounts}(clustering\_candidate)\;
            \ForEach{cluster\_count\_set}{
                \texttt{EstimateLatency}(workload, clustering\_candidate, cluster\_count\_set)\;
            }
        }
        Sort clustering candidates ascending by their estimated latency\;
        
        \Return{first $k$ clustering candidates with estimated latencies}
    }
\end{algorithm}

\section{Capturing the Workload}
\label{sec:workload}
In this section, we describe how our model captures Hyrise's recent workload.
The workload is the set of queries that were recently executed by Hyrise.
In the following paragraphs, we discuss different representations for workloads and describe how our model accesses the workload and its statistics.

\paragraph{Workload representation}
In this paragraph, we discuss different options to represent a workload.
A workload is a set of queries, so we have to find ways to represent a query, ideally with as fine-grained latency information as possible.
We discuss three ways to represent queries: SQL strings, access counters, and query plans.

The first and most straightforward option to represent a query is its SQL string.
SQL strings are human-readable and the best way to express a query's intent.
However, for an SQL string, there may be numerous query execution plans, with significantly varying run times~\cite{DBLP:journals/pvldb/Neumann14, Neumann2009}.
Furthermore, SQL strings provide only a latency for the entire query, i.e., they do not contain detailed information for which operators and columns we spend the most execution time.

The second option are access counters.
Access counters track how often a certain storage structure, e.g., a segment, is accessed.
Segment access counters show which parts of the data are accessed frequently.
Such statistics might be interesting for discriminating clustering algorithms, i.e., algorithms that restrict the clustering to a certain part of the data.
An example for a discriminating clustering algorithm is the database cracking algorithm proposed by Idreos et al.~\cite{DBLP:conf/cidr/IdreosKM07}.
However, our clustering algorithm is neither discriminating, nor do access counters provide detailed latencies.

The third option are physical query plans.
To execute a query, Hyrise's SQL pipeline transforms an SQL string to a physical query plan.
Physical query plans are directed acyclic graphs, which contain operators as nodes, e.g., hash joins, index joins, or table scans.
Once executed, the operators contain fine-grained statistics such as, e.g., their run time, and the size of their input(s) and output.

We conclude that physical query plans are the best-fitting query representation for our model's needs, because they provide latencies on operator level.

%

\paragraph{Accessing the workload}
Hyrise maintains a cache for physical query plans, which maps SQL strings to their respective physical query plans.
We use Hyrise's PQP cache to obtain a sample of the recent workload:
After executing a benchmark to generate the workload, Hyrise's PQP cache is filled with physical query plans of the executed queries.

Our model is implemented in Python, i.e., it does not have direct access to Hyrise's PQP cache.
We use the \texttt{PlanCacheCsvExporter} plugin\footnote{\url{https://github.com/hyrise/example_plugin/tree/martin/benchmark_to_csv_export}} to export the PQPs to CSV files, which can be processed by our model.
Each operator type is exported to its own CSV file.
The \texttt{PlanCacheCsvExporter} plugin is provided by Hyrise.
For our work, we use a slightly modified version of the plugin, which is available on Github\footnote{\url{https://github.com/aloeser/example_plugin/tree/ma}}.

Not all executed operators are relevant for our model.
For example, we are not interested in scans on temporary tables such as aggregation results, because we cannot optimize for those.
For the same reason, we do not export scans comparing two columns.

We export some additional metadata.
These metadata include row counts for all tables and the table's schemata.
For each column, we also export the number of unique values, whether it can contain \texttt{NULL} values, and whether it is sorted.




\section{Identifying Clustering Candidates}
\label{sec:clustering_candidates}
In this section, we describe our approach to identifying clustering candidates.
A \emph{clustering candidate} has a set of one or more clustering columns and a sort column.
The clustering columns determine which columns are used for clustering, the sort column determines by which column the individual clusters will be sorted.

\begin{figure}
    \centering
    \includegraphics[width=\textwidth]{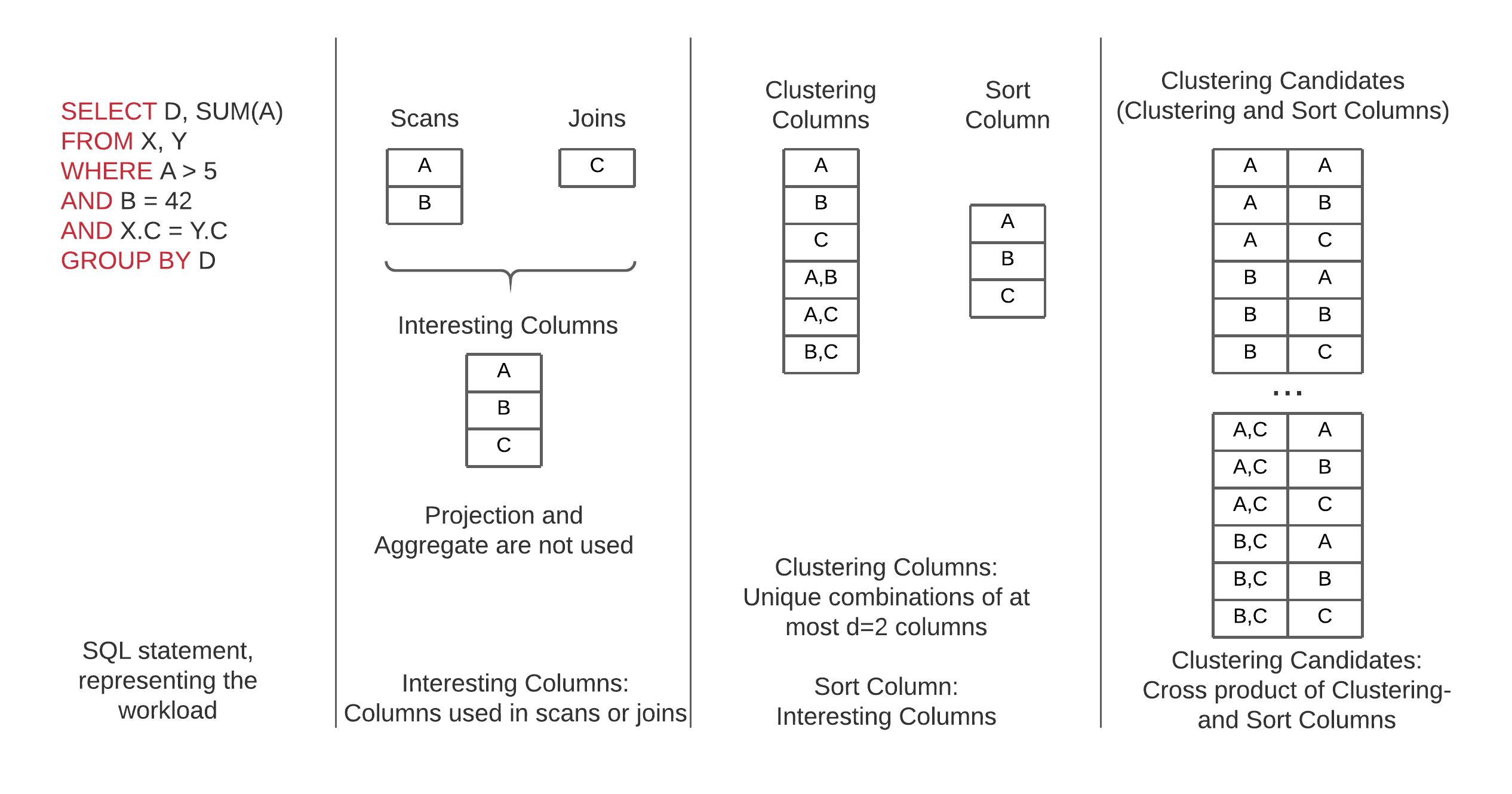}
    \caption{The figure visualizes the steps for identifying clustering candidates for a table $X$. The maximum dimensionality $d$ of the clustering candidates is $2$.}
    \label{fig:clustering_candidates_all}
\end{figure}

In the following paragraphs, we describe how we identify \emph{interesting columns} and how we use them to construct clustering candidates.
\Cref{fig:clustering_candidates_all} visualizes the process of identifying clustering candidates.

\begin{figure}
    \centering
    \includegraphics[width=\textwidth]{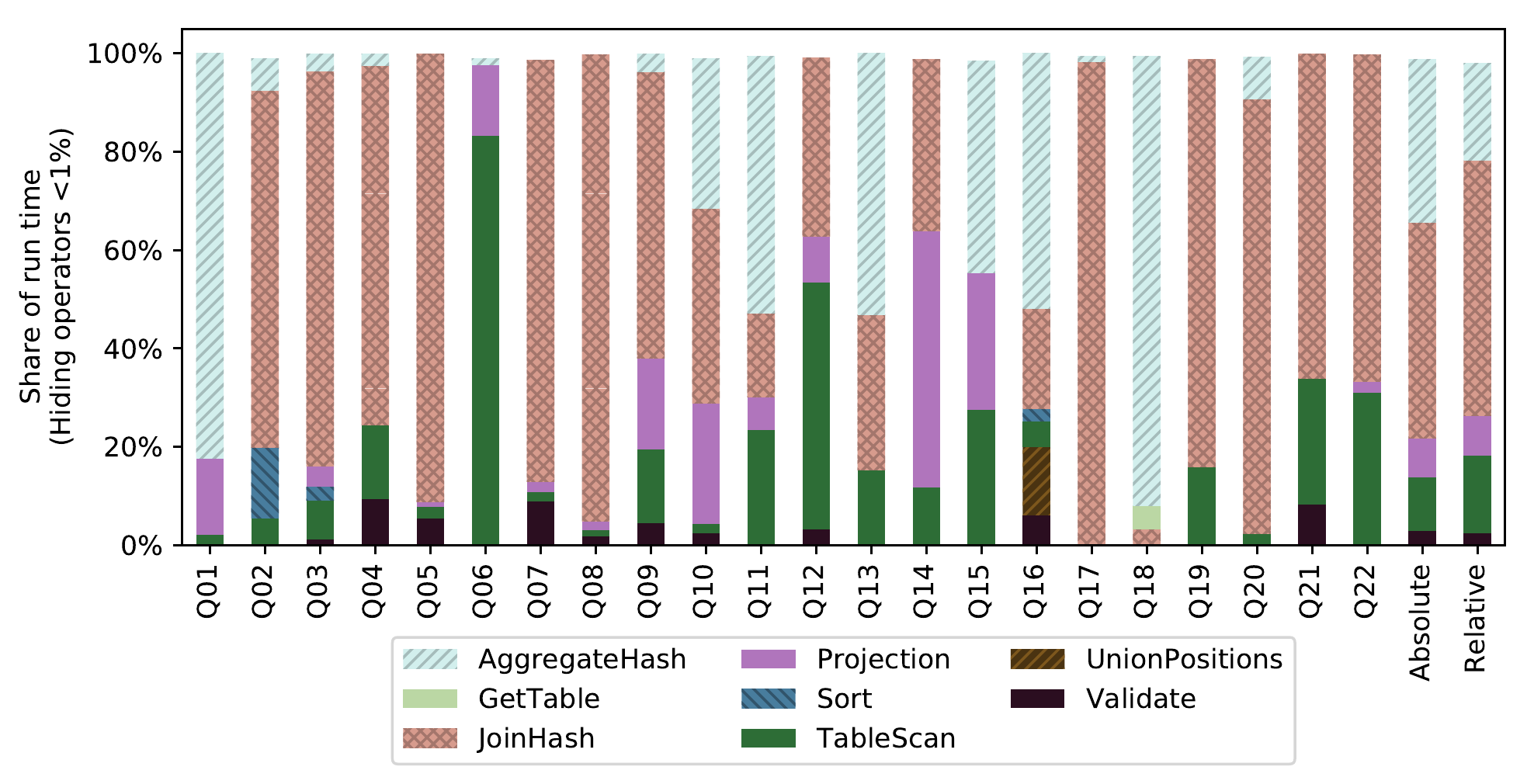}
    \caption{The figure visualizes the share of run time per operator for the queries of the analytical TPC-H benchmark. The TPC-H tables were generated with scale factor ten. All queries were executed once. No clustering was applied, i.e., all tables were left as generated.}
    \label{fig:operator_breakdown_tpch}
\end{figure}

\begin{figure}
    \centering
    \includegraphics[width=\textwidth]{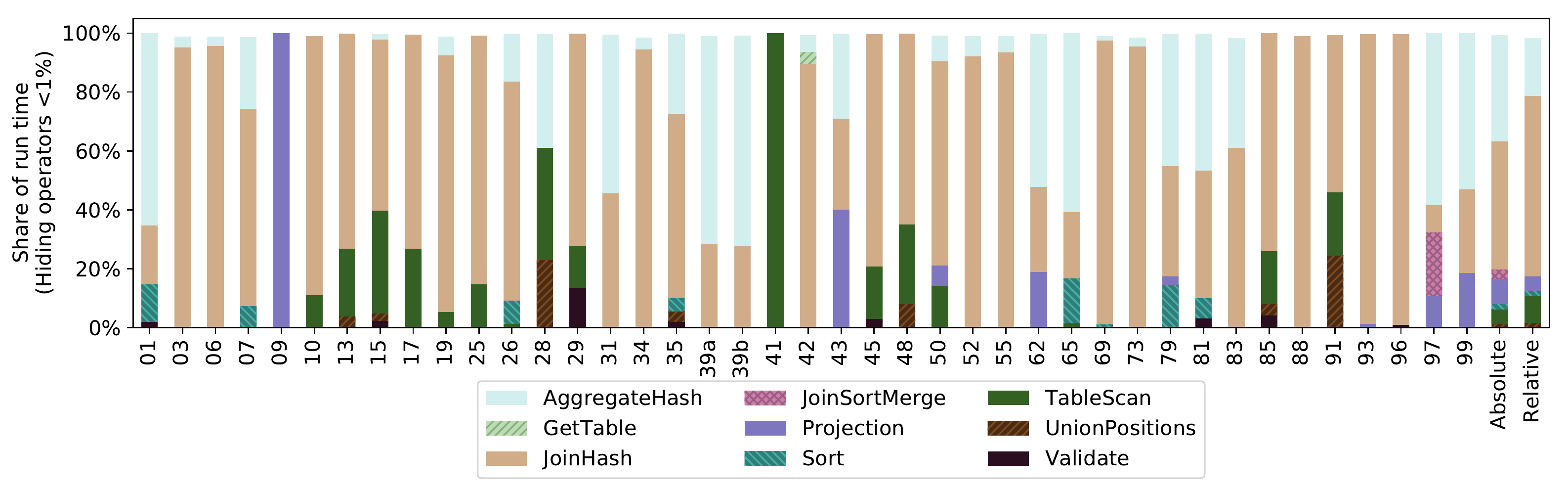}
    \caption{The figure visualizes the share of run time per operator for the queries of the analytical TPC-DS benchmark. The TPC-DS tables were generated with scale factor ten. All queries were executed once. No clustering was applied, i.e., all tables were left as generated.}
    \label{fig:operator_breakdown_tpcds}
\end{figure}

\paragraph{Interesting Columns}
We consider a column \emph{interesting} when it fulfills two criteria:
First, clustering by the column is likely to improve the performance of operators that operate on it.
For example, when a column is used in an, e.g., equality predicate, clustering by that column is likely to enhance pruning abilities, and thereby speed up the respective scan operators on that column.
Second, typical operators executed on that column must consume a notable part of the query execution time to be considered.
We choose this criterion because operators with a small share of the query execution time have a low optimization potential, i.e., even a hypothetical reduction to an execution time of zero nanoseconds for that operator type cannot lead to a notable reduction in total latency.
The operators' shares of query execution time are visualized in \Cref{fig:operator_breakdown_tpch} for the analytical TPC-H benchmark, and in \Cref{fig:operator_breakdown_tpcds} for the analytical TPC-DS benchmark.
For both TPC-H and TPC-DS, the most time-consuming operations are aggregates and joins.
Scans and projections consume less time but are still notable.

Our model focuses on joins and scans:
As mentioned above, clustering can speed up scans mainly by increased pruning opportunities.
Joins, in particular the hash join, can benefit from the clustering because of, e.g., improved cache hit rates.
For that reason, we consider columns that are used in predicates or join conditions as interesting.
Clustering can yield further performance benefits for both operators, which are described in \Cref{sec:cost_estimation_predicates} for scans and \Cref{sec:cost_estimation_joins} for joins respectively.

We do not use aggregates or projections to identify interesting columns.
For most of the time-consuming projections,   (arithmetical) calculations constitute the largest share of their run time.
We believe that it is not obvious how clustering could speed up those calculations, and thus do not use projections as a source of interesting columns.

Aggregates perform the task of aggregating, e.g., calculating a sum or an average, but they also perform grouping, e.g., caused by the SQL \texttt{GROUP BY} clause.
Clustering by the grouped columns ensures that rows that belong to the same group are stored in proximity.
Similar to cache hit rates, an improved "group hit rate" can improve the aggregate's performance.
However, there is a variety of factors to consider when estimating the impact of clustering on aggregate performance.
Due to the limited scope of this thesis, we do not integrate aggregates into the model.
Instead, we only present a list of factors that should be considered when estimating the impact of clustering on aggregates in \Cref{sec:cost_estimation_aggregates}.

\paragraph{Clustering candidates}
We use interesting columns to generate clustering candidates.
Our model has a parameter $d$, which determines the maximum dimensionality of a clustering, i.e., up to $d$ clustering columns are considered per candidate.
In theory, $d$ can be an arbitrary number; in practice, choosing too many dimensions may be counterproductive.
For example, Snowflake~\cite{DBLP:conf/sigmod/DagevilleCZAABC16} recommends choosing at most three to four clustering columns~\cite{SnowflakeMaxDimension}.

We obtain the clustering columns by constructing all combinations of interesting columns.
Each combination has at most \texttt{n} columns, i.e., combinations with fewer than \texttt{n} columns are considered, too.
The algorithm we describe in \Cref{sec:clustering_algorithm} is not sensitive to the order of the clustering columns.
As a consequence, if two combinations only differ in order, but not in columns, we consider them equivalent and emit only one version.

In addition to the clustering columns, a clustering candidate also contains a sort column.
The sort column determines by which column the individual clusters will be sorted.
We use the interesting columns as potential sort columns, i.e., we create clustering candidates for each combination of clustering columns and sort column.

\section{Choosing Cluster Counts}
\label{sec:dimension_granularities}
A clustering candidate  determines which columns should be clustered.
However, it does not specify the desired granularity of the clustering.
For example, a date column could be clustered by day, by month, by year, or even arbitrarily.
In this section, we describe how our model chooses granularities for the clustering candidates.

\paragraph{General approach}
We propose \emph{cluster counts} as a measure for the granularity.
The idea: the value range of a clustering column is divided into $n$ disjoint subranges, where $n$ describes the column's \emph{cluster count}.
The higher the cluster count of a column, the more fine-grained is the resulting clustering; but at the same time, the number of clusters increases.
Each clustering column has its own cluster count, i.e., the granularities of two clustering columns are not directly related.
Without correlations between the clustering columns, the expected number of clusters is the product of all cluster counts.

\begin{figure}
    \centering
    \includegraphics[width=\textwidth]{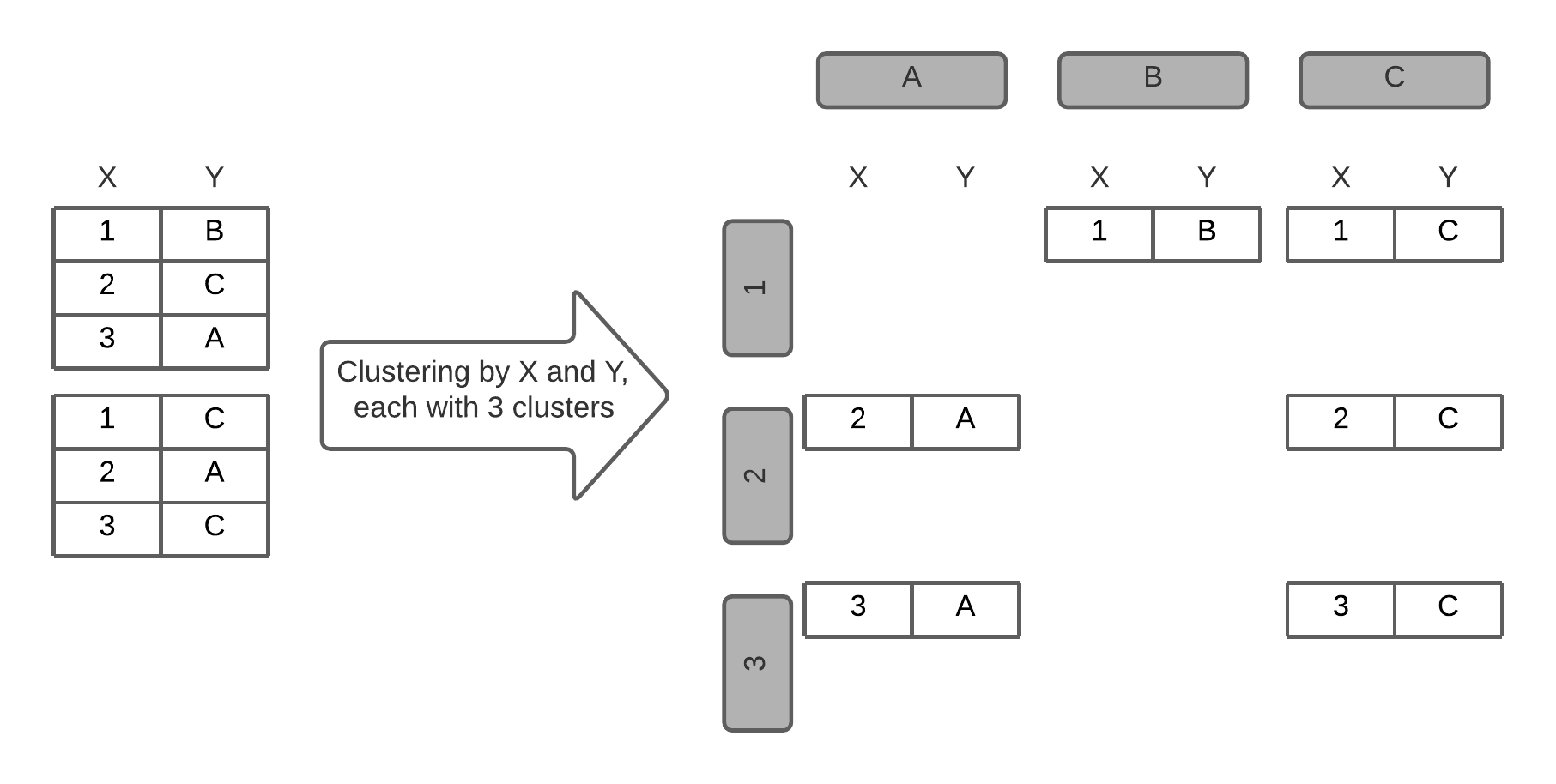}
    \caption{The figure visualizes that too high cluster counts might lead to sparsely populated chunks.}
    \label{fig:cluster_counts_high}
\end{figure}

Theoretically, we can choose for each clustering column a high cluster count, and obtain a fine-grained clustering.
However, choosing too high cluster counts will result in a high number of clusters.
Since each cluster contains at least one chunk, this implies there will also be a high number of chunks, and all only partially filled.
Put differently, we need more chunks to store the same amount of data, which may negatively impact Hyrise's performance ~\cite{DBLP:conf/edbt/DreselerK0KUP19}.
\Cref{fig:cluster_counts_high} visualizes an example of high cluster counts.

In a nutshell: although the cluster counts of two columns are not directly related, choosing a high cluster count for all clustering columns does not necessarily lead to cluster counts that are optimal for performance.
Therefore, we aim for clusters to have at least one (mostly) full chunk.
Due to the limited number of clusters, we must make tradeoffs between the granularities of the clustering columns:
All columns benefit from a fine-grained clustering, but some columns might benefit more from a high cluster count than others.

\begin{figure}
    \centering
    \includegraphics[width=\textwidth]{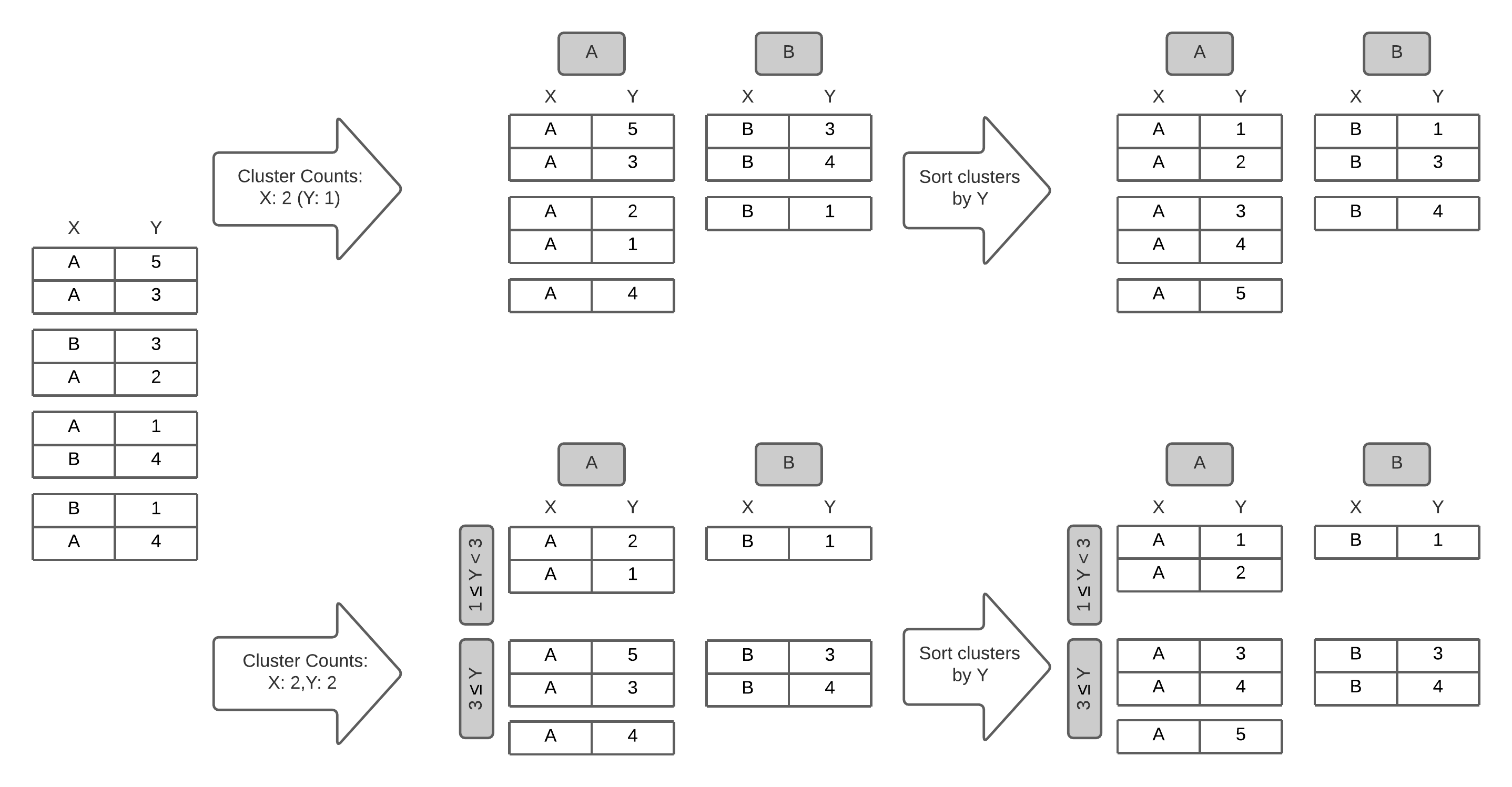}
    \caption{The figure visualizes that the sort column behaves similar an additional clustering column with a cluster count equal to the number of chunks per cluster.
    }
    \label{fig:partition_sort_tradeoff}
\end{figure}

\paragraph{Partition-Sort-Tradeoff}
Clustering candidates also contain a sort column.
Once the partitioning phase of our clustering algorithm is completed, the resulting clusters are sorted by the sort column.
If we expect more than one chunk per cluster, the sort phase can be seen as a special form of partitioning by the sort column.
More specifically, the sort phase behaves similar to an additional clustering column, whose cluster count is equal to the expected number of chunks per cluster.
If the sort column is already clustered, it effectively multiplies the respective cluster count instead.
\Cref{fig:partition_sort_tradeoff} visualizes this behavior.

This enables a tradeoff for our model:
We can obtain a reduced memory consumption during the partition phase of our algorithm, at the cost of an increased duration per cluster during the sort phase, and vice versa.
If the sort column is contained in the clustering columns, we can choose an arbitrary smaller cluster count for the respective clustering column, without impacting the resulting clustering granularities.
During the partition phase, Hyrise maintains a mutable chunk with a fixed size for each cluster.
A reduced cluster count reduces the number of clusters, and thus reduces the temporary memory consumption during the partition phase.
However, a reduced number of clusters increases the average size of the clusters, and with that, the amount of time required to sort a cluster.

\paragraph{Our approach}
Conceptually, our model can consider an arbitrary number of cluster counts sets for each clustering candidate.
However, the model's run time depends linearly on the number of cluster counts sets; for that reason, we generate only one set of cluster counts per clustering candidate.
We describe our algorithm to generate cluster counts in pseudo-code in \Cref{algo:determine_cluster_counts}.

\begin{algorithm}[htbp]
\caption{DetermineClusterCounts}
\label{algo:determine_cluster_counts}
    \KwIn{A clustering candidate}
    \KwOut{Cluster counts}
    \DontPrintSemicolon
    target\_total\_cluster\_count = table.chunk\_count\;
    scan\_ columns = clustering columns that are used in scans\;
    join\_columns = clustering columns that are used in joins\;
    $j = $ len(join\_columns)\;
    
    \uIf{len(scan\_columns) == 0}{
        cluster\_counts $= ceil\left(\sqrt[j]{\text{target\_total\_cluster\_count}}\right)$ for each of the $j$ join\_columns\;
    }
    \Else{
        assign each join column a cluster count of 3\;
        available\_scan\_clusters = target\_total\_cluster\_count / $3^j$\;
        distribute the available scan clusters to the scan columns, proportional to the logarithm of their unique values\;
    }
    
    max\_cluster\_count $= 100$\;
    \ForEach{column in clustering\_candidate.clustering\_columns}{
        column.cluster\_count $=$ min(column.cluster\_count, column.num\_unique\_values, max\_cluster\_count)\;
    }
    
    \Return{cluster\_counts}
\end{algorithm}

Our model tries to generate about as many clusters as the table has chunks, i.e., we expect each cluster to contain one full chunk.
The intent behind that choice is to avoid large clusters because large clusters require more time to be sorted.
Thus, when deciding on cluster counts, our model aims to find cluster counts whose product equals approximately the number of chunks.
The cluster counts shall correlate with the number of unique values in the respective clustering column:
columns with a low number of unique values receive a low cluster count and vice versa.

In the analytical TPC-H and TPC-DS benchmarks, join columns such as \texttt{l\_orderkey} usually have a high number of unique values, compared to their table's row count.
We expect that a high cluster count is required to achieve performance improvements through clustering.
As a consequence, our model handles two cases:
First, it divides the clustering columns into join and scan columns.
For the TPC-H and TPC-DS benchmarks, there are no columns that are used for both scans and joins.
If the clustering columns contain only join columns, we distribute the clusters equally among them.
If the clustering columns contain at least one scan column, we assign a low, constant cluster count to the join columns.
The remaining available clusters are distributed among the scan columns.
The number of unique values of the clustering columns can differ by multiple orders of magnitude, e.g., there are 11 unique values in the \texttt{l\_discount} column, and 2\,526 unique values in the column \texttt{l\_shipdate}.
As a consequence, the cluster counts are distributed proportionally to the logarithm of a column's number of unique values, rather than the plain number of unique values.

\begin{figure}
    \centering
    \includegraphics[width=\textwidth]{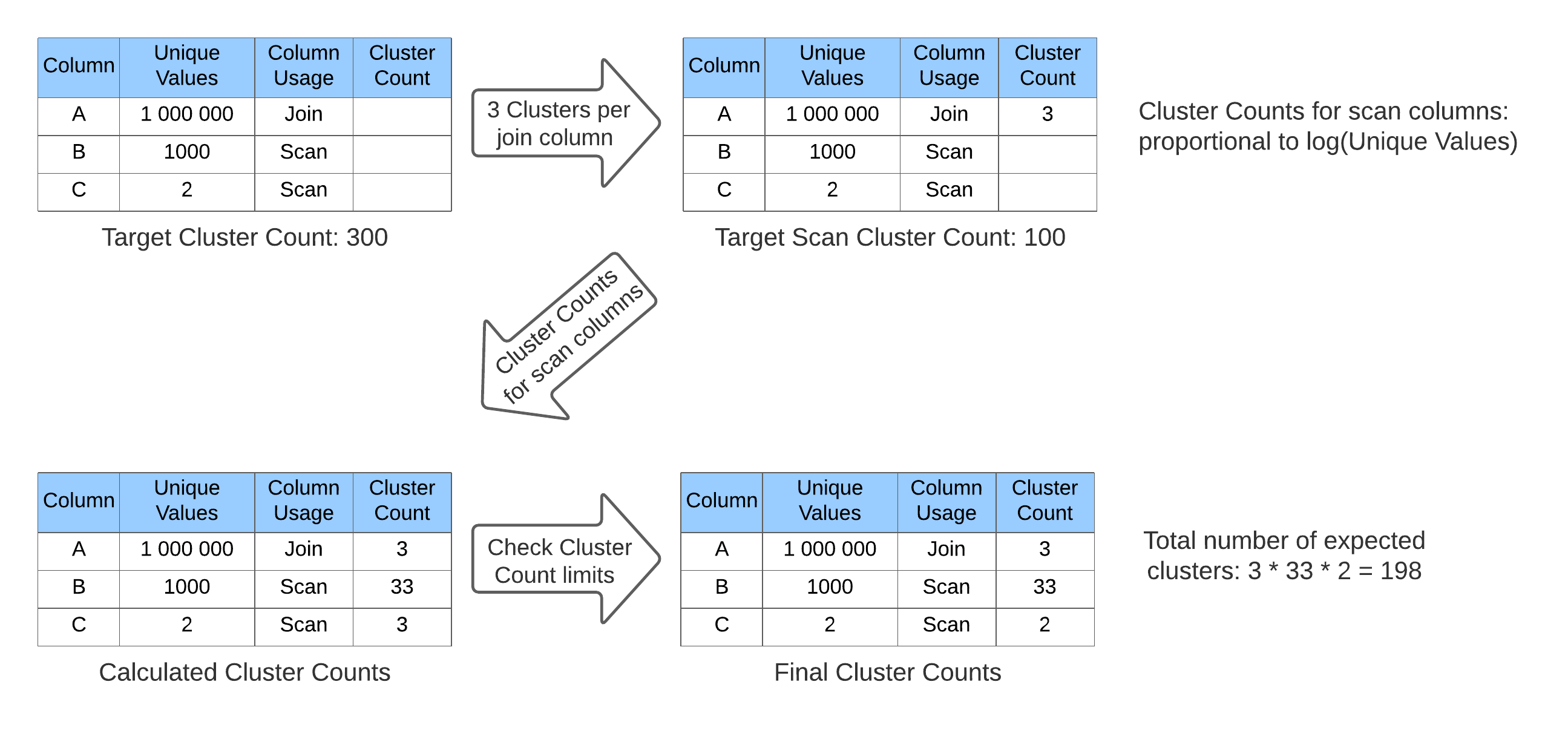}
    \caption{The figure visualizes the process of choosing cluster counts. For the third column, the chosen cluster count is higher than its unique value count. As a consequence, we obtain only two thirds of the requested clusters.
    }
    \label{fig:choose_cluster_counts}
\end{figure}

Finally, we make two restrictions regarding the maximum values of the cluster counts.
First, no cluster count must be higher than the respective column's number of unique values.
Without this restriction, there are at least two clusters that share a section of their value range, i.e., we would lose the property that each row belongs to exactly one cluster.
The second restriction arises from \Cref{algo:choose_boundaries}, i.e., how we choose the cluster's value ranges.
Hyrise maintains histograms for each table.
Each histogram has 5 to 100 bins, depending on the table's size.
At the time of writing, our algorithm assumes that each cluster will cover the entire value range of at least one bin.
Consequently, the global maximum cluster count is 100.
To increase this limit, future work could use bin splitting techniques, which are already implemented in Hyrise.
The upper limits on the cluster counts may lead to a smaller number of clusters than intended.
\Cref{fig:choose_cluster_counts} visualizes an example where the number of clusters is a third lower than intended.

\section{Latency Estimation Process}
\label{sec:cost_estimation}
Estimating the impact of clustered data on the workload's latency is a crucial part of the model.
In this section, we provide details on the latency estimation process.

The base of our estimations is the recent workload, i.e., the executed operators and their run times.
For scans and joins, we identify factors that influence their run times.
We compose these factors in a rule-based system to estimate latencies for scans and joins, given a hypothetical clustering.
Due to the scope of this thesis, we do not provide concrete latency estimates for aggregates, but only point out some factors that should be considered when estimating aggregation latencies.
Thus, for aggregates and all other operator types, our model assumes that they are unaffected by the clustering, i.e., their latency will not change.
As a consequence, the optimization goal of minimizing latency is equivalent to minimizing the combined latency of scans and joins.

%
%

\paragraph{Static physical query plan assumption}
There is a tradeoff to consider:
Even if we assume that the workload's individual SQL queries do not change, the clustering may influence the resulting physical query plan (PQP), e.g., via pruning statistics.
To detect a changed PQP, we could call Hyrise's optimizer with mocked statistics every time we evaluate a clustering candidate.
Alternatively, we can assume that the PQP remains unchanged, i.e., all logical database operations (e.g., a join) are translated into the same physical operators (e.g., a sort-merge join, or a hash join) as before, and that all operators retain their position within the PQP.
The first option is likely to yield more precise latency estimates.
However, a call to the optimizer is not without computational costs~\cite{DBLP:journals/pvldb/DashPA11, DBLP:conf/vldb/PapadomanolakisDA07}.
The latency estimation happens in the model's hot loop, i.e., there would be a large number of calls to the optimizer.
For that reason, we choose the latter option and assume the PQP remains identical.
We discuss the inaccuracies caused by this decision in \Cref{sec:eval_scans_changing_pqp}.

\subsection{Scan Operator}
\label{sec:cost_estimation_predicates}
Hyrise offers two scan implementations: an \texttt{IndexScan}, and a \texttt{TableScan}.
The index scan can only be used when there is an index on the predicate column.
At the time of writing, Hyrise hardly uses indices for analytical workloads because they yield only a marginal benefit~\cite{DBLP:conf/sigmod/KesterAI17}.
The table scan can always be applied and is the dominant scan type.
For these reasons, we restrict our estimations to table scans, i.e., if we encounter any index scans, we assume they are not impacted by the clustering.

The table scan works by scanning the whole table, chunk by chunk.
In the next paragraphs, we describe the two main aspects where the clustering impacts the performance of table scans: sorted columns, and pruning.
\Cref{algo:cost_estimation_scans} visualizes the latency estimation process for scans in pseudo-code.

\begin{algorithm}[htbp]
\caption{EstimateScanLatency}
\label{algo:cost_estimation_scans}
    \KwIn{A scan workload, a clustering candidate, cluster counts}
    \KwOut{A latency estimation for the scan workload}
    \DontPrintSemicolon
    total\_latency $ = 0$\;
    \ForEach{scan\_set in scans.group\_by\_query()}{
        unprunable\_part = \texttt{EstimateUnprunablePart}(scan\_set, clustering candidate, cluster counts)\;
        
        first\_scan = chronologically first scan in scan\_set\;
        
        first\_scan.input\_size = first\_scan.table.size * unprunable\_part\;
        \ForEach{scan in scan\_set}{
            \If{scan.column == clustering\_candidate.sort\_column  and scan.type benefits from sortedness} {
                scan.input\_size $= \log_2{\text{(scan.input\_size)}}$ \;
            }
            latency = time\_per\_row * (scan.input\_size + scan.output\_size)\;
            total\_latency += latency\;
        }
    }
    
    \Return{total\_latency}
\end{algorithm}

\paragraph{Sorted columns}
The clustering algorithm produces clusters that are individually sorted by some column $X$.
To scan segments of $X$, we can use a binary instead of a linear search.
The binary search allows us to skip reading most of the chunk:
Instead of the whole chunk, we only need to read twice (lower and upper bound) $\log_2(65\,535) \approx 16$ rows per chunk.
The output, however, remains unchanged, i.e., we still need to write all matching rows.
Some predicates cannot take advantage of binary searches and always require a linear scan, e.g., SQL \texttt{LIKE} predicates with a wildcard at the beginning: \texttt{X LIKE '\%Customer\%Complaints\%'}.

\paragraph{Pruning}
If a predicate column is clustered, pruning on that column becomes more efficient, i.e., a larger number of chunks can be pruned.
Pruning can speed up table scans because it reduces the scan's input size, i.e., fewer rows have to be scanned.
Vice versa, if a column is currently clustered, choosing a different clustering may prevent pruning on that column, i.e., clustering may also lead to an increased input size (and thus run time) for a scan.

\subsubsection*{Estimating Pruning}
We estimate the effect of the clustering on pruning on a per-query-basis, i.e., all scans that belong to the same query are processed in one batch.
Like Markl et al.~\cite{DBLP:conf/vldb/MarklMKTHS05}, we define a scans's selectivity as the fraction of rows that fulfill the scan's predicate.
For each query, we use the scans' selectivities to estimate the unprunable part of the clustering table.
The \emph{unprunable part} is the fraction of the table that, given a clustering candidate, cannot be pruned, i.e., the part that remains after pruning is completed.

We replace the input size (i.e., row count) of the query's first scan with the size of this unprunable part.
This reduction of the input size impacts the latency estimation.
For all subsequent scans, we assume that their input size does not change significantly.
It is worth noting that the query's first scan in the currently implemented clustering and the currently evaluated clustering candidate are not necessarily the same scans, i.e., the assumption of a static PQP may break here, as can be seen in \Cref{sec:eval_scans_changing_pqp}.

\paragraph{Computing the unprunable part}
We assume that only scans on clustering columns are used for pruning.
To compute the unprunable part, we multiply the selectivities of those scans.
The multiplication result represents a lower bound for the unprunable part of the table.

\begin{figure}
    \centering
    \includegraphics[width=0.8\textwidth]{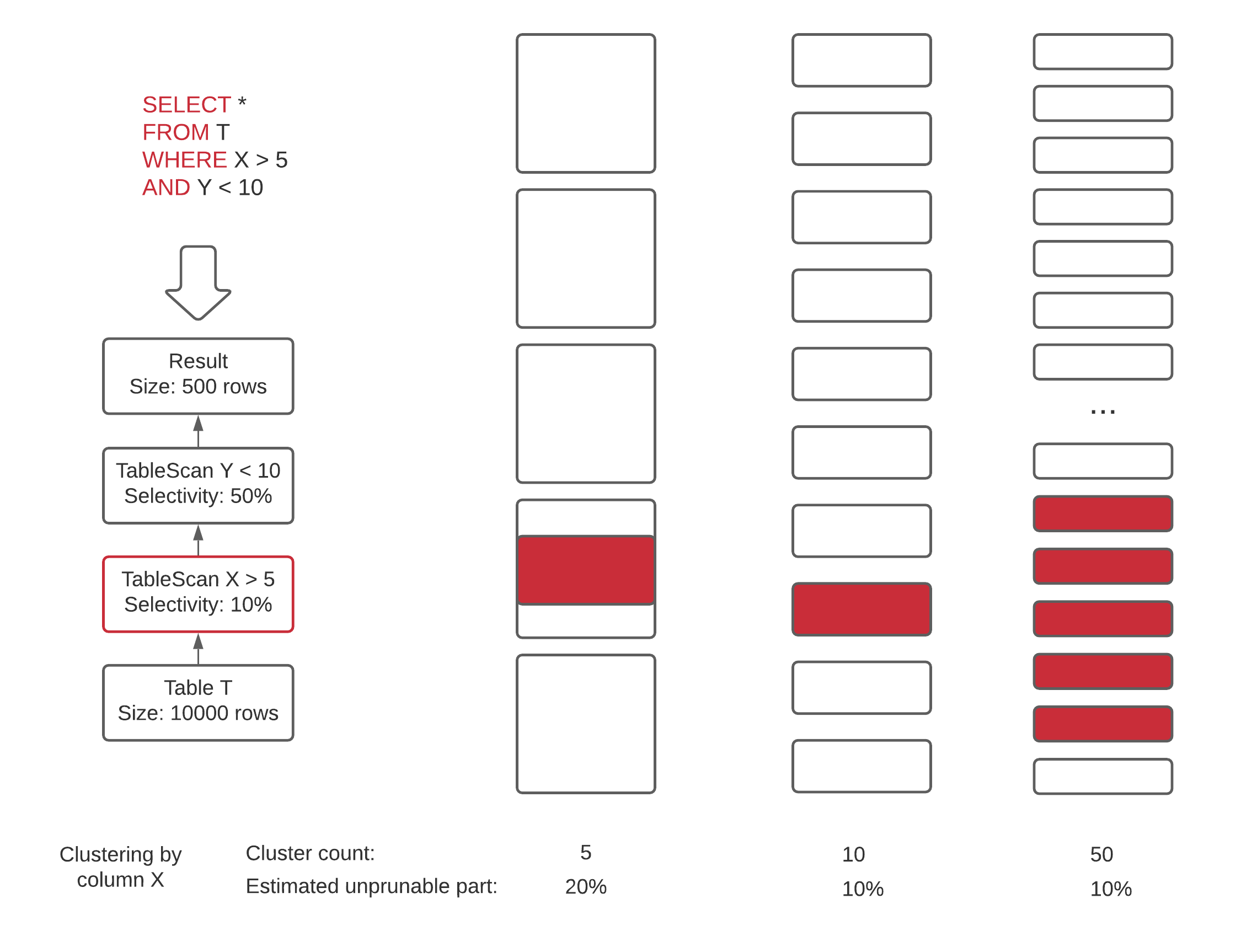}
    \caption{The figure visualizes the estimated unprunable part of the table for different cluster counts. The white boxes represent clusters. The red boxes represent parts of a cluster that qualify.}
    \label{fig:unprunable_parts}
\end{figure}

For a more precise estimate of the unprunable part, we consider that our algorithm splits the value range of a column into $n$ sections of roughly equal size, where $n$ is the cluster count.
If one cluster qualifies for a predicate, we expect all chunks of the cluster to qualify.
Consequently, to obtain a better estimate of the unprunable part, we round up the selectivities to the next multiple of a cluster's size, i.e., $\frac{1}{\text{cluster count}}$, before multiplying them.
\Cref{algo:unprunable_parts} describes the estimation of the unprunable part in pseudo-code.
\Cref{fig:unprunable_parts} visualizes the impact of the cluster count on the estimated unprunable part.

.

\begin{algorithm}[htbp]
\caption{EstimateUnprunablePart}
\label{algo:unprunable_parts}
    \KwIn{A set of scans, a clustering candidate, cluster counts}
    \KwOut{An estimate for the unprunable part}
    \DontPrintSemicolon
    unprunable\_part = $1$\;
    \ForEach{scan in scan\_set}{
        \If{scan.column in clustering\_candidate.clustering\_columns and scan.type is useful for pruning}{
            clusterwise\_selectivity = round up the scan's selectivity to the next multiple of $\frac{1}{\text{cluster\_counts[scan.column]}}$\;
            unprunable\_part *= clusterwise\_selectivity\;
        }
    }
    
    \Return{unprunable\_part}
\end{algorithm}

\paragraph{Unprunable predicates}
Hyrise does not utilize all predicates for pruning.

Some predicates are inherently unsuitable for pruning.
For example, \texttt{LIKE} predicates, e.g. \texttt{X~LIKE~'\%Customer\%Complaints\%'}, cannot be used for pruning:
Due to the wildcard at the beginning, we must check each row individually.
Another example are predicates on temporary tables, such as the \texttt{HAVING} predicate that is performed after an aggregation.

Other predicates could theoretically be used for pruning, but are not considered promising enough to justify the additional implementation complexity.
Examples for such predicates are column vs. column predicates, \texttt{X~IN~(...)} predicates with a list of explicit values, or inequality predicates.

Finally, Hyrise does not use predicates in \texttt{OR}-conditions for pruning, even if the predicate itself could be used for pruning.

In all those cases, the predicates' selectivities are not considered when computing the unprunable part.

\paragraph{Correlations}
Tables may contain correlated columns.
For example, the \texttt{lineitem} table of the analytical TPC-H benchmark contains the columns \texttt{l\_shipdate} and \texttt{l\_receiptdate}.
The value for \texttt{l\_receiptdate} is obtained by adding a random number of days (between 1 and 30) to the value of \texttt{l\_shipdate}~\cite[p.~87]{TpchSpec}.
The \texttt{l\_shipdate} column has a value range of more than 2500 days, i.e., almost two orders of magnitude more than the maximum difference between \texttt{l\_shipdate} and \texttt{l\_receiptdate}.
As a consequence, clustering by \texttt{l\_shipdate} will indirectly also cluster by \texttt{l\_receiptdate}, and thus yield a notable performance improvement for scans on \texttt{l\_receiptdate}.

Our model is not capable of detecting correlations on its own.
However, if the model is made aware of correlations detected by some external tool, it can reflect them in the estimation of the unprunable part (and thus, in the scan's latency estimate).
If a scan occurs on a column that is correlated to a clustered column, our model pretends that the column was clustered, too.
To reflect that the clustering for the correlated column is not as fine-grained as the clustering for the clustering column, we multiply the scan's selectivity with some factor.
Due to the limited scope of this thesis, we use a constant as factor.
More advanced techniques could replace the constant by, e.g.,  a measure for how strong two columns are correlated.

\subsection{Hash Join Operator}
\label{sec:cost_estimation_joins}
In this section, we describe how our model estimates the impact of clustering on joins.
Hyrise has four join implementations: \texttt{JoinHash}, \texttt{JoinSortMerge}, \texttt{JoinNestedLoop}, and \texttt{JoinIndex}.
We describe latency estimations for the hash join, which is used by default.
Some join modes, e.g., full outer joins, are not supported by the hash join.
In that case, Hyrise resorts to other join operators.
We assume that those operators are unaffected by the clustering and do not estimate their latency.

\paragraph{Hash Join}
In this paragraph we first provide a general description of hash joins, before we describe Hyrise's hash join implementation in detail.

\begin{figure}
    \centering
    \includegraphics[width=\textwidth]{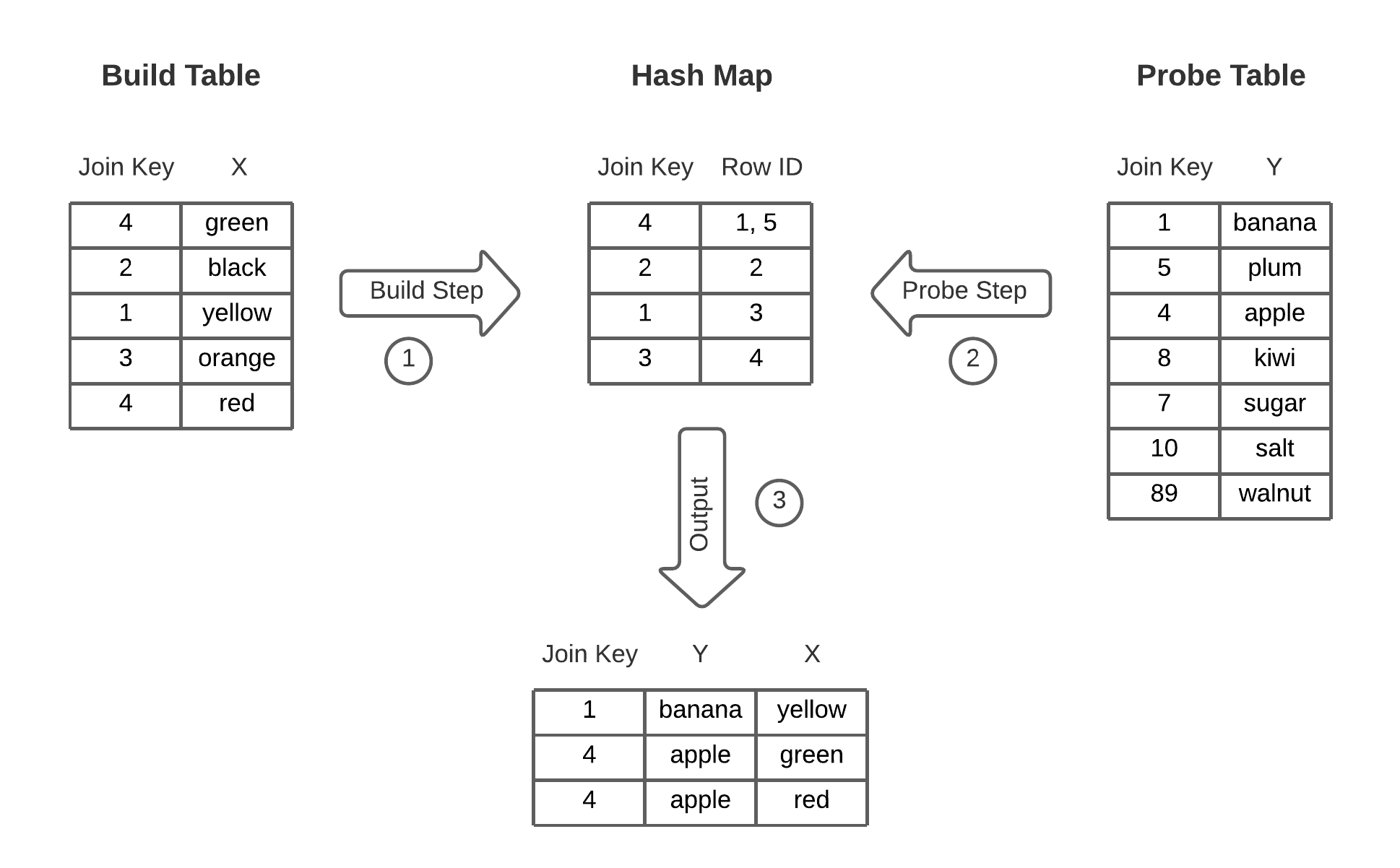}
    \caption{The figure visualizes the hash join's functionality, at the example of an inner join.}
    \label{fig:hash_join_general}
\end{figure}

Hash joins can perform only equi-joins, i.e., joins where equality is the primary predicate.
This is because they use a hash map to find matching tuples:
The key idea of a hash join is to insert the values of all join keys of one input relation, called the \emph{build} relation, into a hash map.
To check for matching tuples, the join keys of the other input relation, called the \emph{probe} relation, are probed against the hash map.
Whether tuples are written depends on the join mode:
For example, a semi-join will only write rows from the probe table, an inner join will combine all matching rows from both tables, and an left (or right) outer join will output tuples even if the probing finds no matches.
\Cref{fig:hash_join_general} visualizes the hash join's functionality.

Hyrise's hash join follows this schema but has additional steps.
In Hyrise, a join's input tables often contain reference segments, and sometimes dictionary encoded segments.
As a preparation for the following steps, the first step of the hash join \emph{materializes} the join columns of the input tables, i.e., we create a list that contains the same values as the join columns, but in an unencoded form.
The hash join maintains separate lists for the build and the probe table.

The \emph{build} and \emph{probe} steps are identical to the general hash join schema.
As the last step, the \emph{output} step, the join output is written.

\begin{figure}
    \centering
    \includegraphics[width=\textwidth]{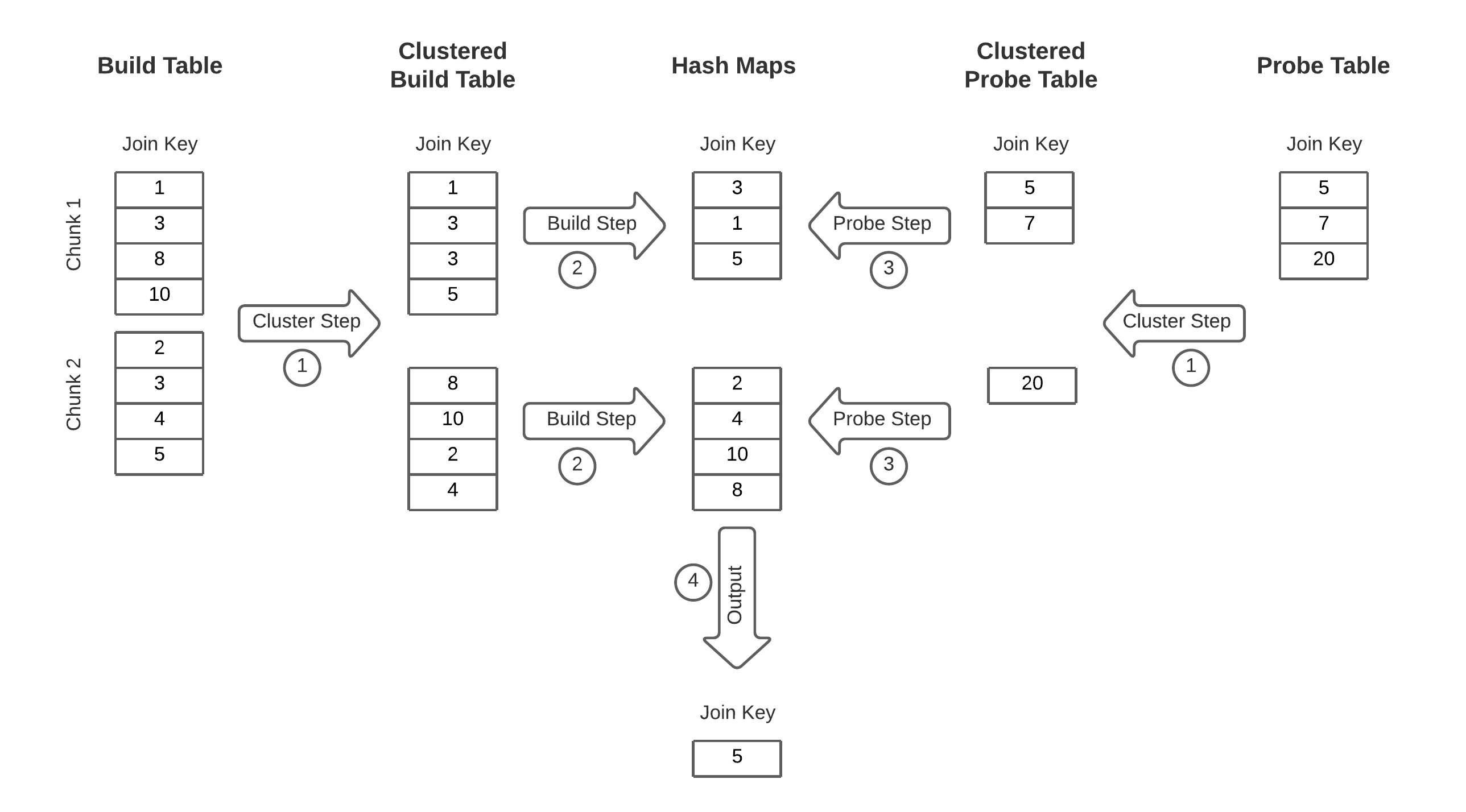}
    \caption{The figure visualizes the execution of a hash join in semi join mode. With this semi join, we want to reduce the right table's size, so the right table must be the probe table, although the left table is larger. We apply a radix clustering before the build step with $r = 1$ radix bit.}
    \label{fig:hash_join_with_clustering}
\end{figure}

Depending on the size of the build table, there may be an additional step between the materialize and the build step, called the \emph{clustering} step.
The purpose of the clustering step is to ensure that the hash map always fits into the L2 cache.
If the expected size of the hash map exceeds the L2 cache size, a radix-clustering as described by Boncz et al.~\cite{DBLP:conf/vldb/BonczMK99} is applied:
We create $2^r$ join clusters and move all join keys to a certain join cluster, which is determined by their last $r$ bits.
The parameter $r$ depends on the build table's size.
This clustering is applied to both the build and the probe table.
If a join key occurs in both the build and the probe table, it will be moved to the same respective cluster, i.e., we can perform the join for each join cluster separately.
As a consequence, the build step is modified to build a hash map for each partition, which reduces the size of an individual hash map.
When probing, we only need to keep the current partition's hash map in the cache, rather than a hash map of all join keys.
\Cref{fig:hash_join_with_clustering} visualizes a semi-join where the clustering step is applied.

Further, it is worth noting that a join cluster will contain rows from different chunks.
As a consequence, choosing a number greater than $0$ for the number of radix bits $r$ will destroy any previously existing sort orders, which has to be considered by our clustering model.

\begin{figure}
     \centering
     \begin{subfigure}[b]{0.48\textwidth}
         \centering
         \includegraphics[width=\textwidth]{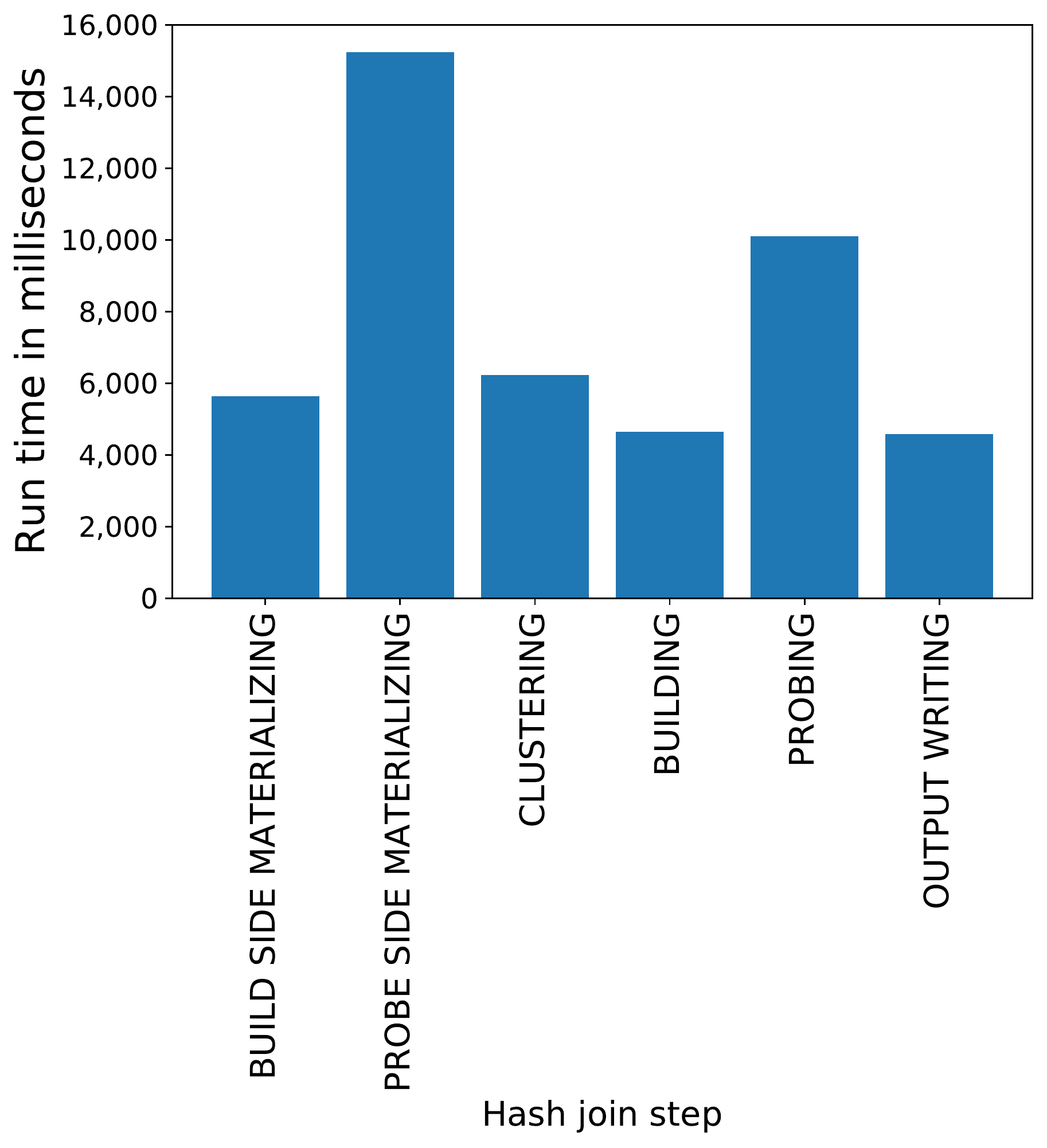}
         \caption{TPC-H}
         \label{fig:hash_join_steps_time_all_tpch}
     \end{subfigure}
     \hfill
     \begin{subfigure}[b]{0.48\textwidth}
         \centering
         \includegraphics[width=\textwidth]{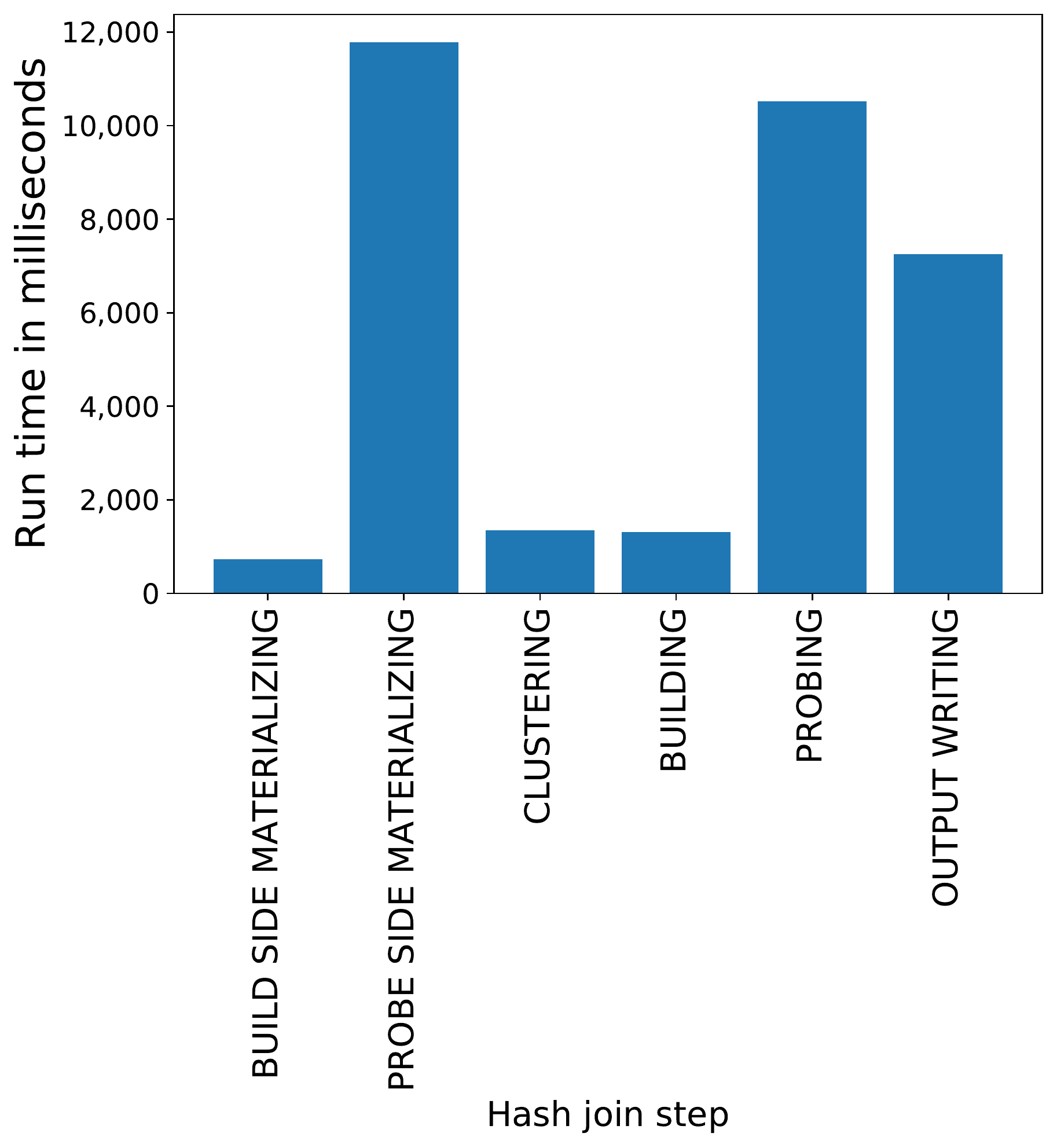}
         \caption{TPC-DS}
         \label{fig:hash_join_steps_time_all_tpcds}
     \end{subfigure}
        \caption{The figure shows the accumulated run times per hash join step across all queries of the analytical benchmarks TPC-H (left) and TPC-DS (right). The benchmark tables were generated with scale factor 10. Each query of the respective benchmarks was executed once.}
        \label{fig:hash_join_steps_time_all}
\end{figure}

\paragraph{Optimization potential}
\Cref{fig:hash_join_steps_time_all} visualizes the run time of the hash join's steps, accumulated across all queries of the analytical TPC-H and TPC-DS benchmarks.
The materialization of the probe table constitutes a large share of the total join run time for both benchmarks, and can benefit from the clustering; thus, we provide latency estimations for the materialize step.
In addition to the materialize step, we also provide a latency estimation for the probing step.
We consider all other join steps as unaffected by the clustering.
The total join latency estimate is thus the sum of estimates for both materialize steps, the probe step, and the current latency of all steps that are unaffected by the clustering.
We describe the estimation process in \Cref{algo:estimate_join_latency}.

\begin{algorithm}[htbp]
\caption{EstimateJoinLatency}
\label{algo:estimate_join_latency}
    \KwIn{A join workload, a clustering candidate, cluster counts}
    \KwOut{An estimated latency for the join workload}
    \DontPrintSemicolon
    total\_latency = $0$\;
    
    \ForEach{join in the workload}{
        total\_latency += \texttt{EstimateMaterializeBuildStep}(join, clustering\_candidate, cluster\_counts)\;
        total\_latency += \texttt{EstimateMaterializeProbeStep}(join, clustering\_candidate, cluster\_counts)\;
        total\_latency += join.clustering\_step\_latency\;
        total\_latency += join.build\_step\_latency\;
        total\_latency += \texttt{EstimateProbeStep}(join, clustering\_candidate)\;
        total\_latency += join.output\_writing\_step\_latency\;
    }
    
    \Return{total\_latency}
\end{algorithm}

\subsubsection*{Estimating Materialization}
Hyrise's hash join has two materialize steps: one for the build table, and one for the probe table.
During the materialization of the build table, a bloom filter~\cite{DBLP:journals/cacm/Bloom70} is built.
However, the construction of the bloom filter constitutes only a small share of the build side materialize step's run time.
As a consequence, we do not take the bloom filter into account and use the estimations described in this section for both materialize steps.

In Hyrise, joins usually operate on reference segments.
In the materialize step, those segments are de-referenced and materialized, i.e., the actual value is accessed and stored.
We have identified three aspects of how this process is affected by the clustering: by the number of unique values per chunk, the sort order, and the average chunk fill grade, also called \emph{chunk density}.

To quantify those aspects, we introduce influence factors, one for each aspect.
The materialize step's run time is multiplied by those influence factors.
An influence factor is always greater than or equal to one.
To take into account that the currently implemented clustering also affects the run time, we calculate the factors for the currently implemented clustering and multiply them with the join's materialize time.
The multiplication result can be seen as the unoptimized run time, i.e., without the benefits granted by the currently implemented clustering.
As the last step, we divide (since influence factors are always $\ge$ 1) the unoptimized run time by the influence factors of the currently evaluated clustering candidate.
The result is the materialize step's estimated run time.
\Cref{algo:estimate_materialize_step} describes our latency estimation process in pseudo-code.

\begin{algorithm}[htbp]
\caption{EstimateMaterializeStep}
\label{algo:estimate_materialize_step}
    \KwIn{A join, a clustering candidate, cluster counts}
    \KwOut{A latency estimate}
    \DontPrintSemicolon
    current\_gains = $1$\;
    \If{join.data\_arrives\_sorted and join.column == current\_clustering.sort\_column}{
        current\_gains *= \texttt{EstimateUniqueValuesFactor}(join, current clustering with cluster counts)\;
        current\_gains *= sort\_factor\;
    }
    current\_gains *= \texttt{EstimateDensityFactor}(join, current clustering with cluster counts)\;
    unoptimized\_run\_time = join.materialize\_step\_latency * current\_gains\;
    
    \;
    
    new\_gains = $1$\;
    \If{join.data\_arrives\_sorted and join.column == clustering\_candidate.sort\_column}{
        new\_gains *= \texttt{EstimateUniqueValuesFactor}(join, clustering\_candidate, cluster\_counts)\;
        new\_gains *= sort\_factor\;
    }
    new\_gains = \texttt{EstimateDensityFactor}(join, clustering\_candidate, cluster\_counts)\;
    \;
    estimated\_run\_time = unoptimized\_run\_time / new\_gains\;
    
    \Return{estimated\_run\_time}
\end{algorithm}

\paragraph{Unique values per chunk}
We have observed that the number of unique values per chunk impacts the run time of the materialize step.
A lower number of unique values per chunk leads to a lower run time and vice versa.
We believe the performance change is due to improved cache hit rates, although we did not investigate whether the cache hit rate actually improved:
By default, Hyrise stores data in segments with dictionary encoding.
Each dictionary segment has an attribute vector that stores value ids, and a dictionary that maps value ids to unique values.
The size of the attribute vector is equivalent to the segment's size, but the size of the dictionary is determined by the number of unique values in the segment, i.e., a lower number of unique values reduces memory consumption.
At the same time, a lower number of unique values increases the likelihood that rows in the reference segment refer to recently accessed values.

We quantify the impact of unique values on the performance of the materializing step by introducing a new influence factor.
To calculate the influence factor, we first estimate the number of unique values per chunk for the given clustering candidate.
When estimating the number of unique values, we assume that each cluster contains the same number of unique values.
This assumption is not correct, especially in the presence of skew.
Our algorithm chooses the cluster boundaries in a way that yields clusters with the same number of rows, not the same number of unique values (see \Cref{sec:choose_cluster_boundaries}).
Nevertheless, this is still our best assumption.
\Cref{algo:estimate_materialize_unique_values_per_chunk} describes the estimation of the unique values per chunk in pseudo-code.

\begin{algorithm}[htbp]
\caption{EstimateUniqueValuesPerChunk}
\label{algo:estimate_materialize_unique_values_per_chunk}
    \KwIn{A column, a clustering candidate, cluster counts}
    \KwOut{The expected number of unique values}
    \DontPrintSemicolon
    
    \uIf{column in clustering\_candidate.clustering\_columns}{
        unique\_values = min(default\_chunk\_size, global\_unique\_values / cluster\_counts[column])\;
    }
    \Else {
        unique\_values = min(default\_chunk\_size, global\_unique\_values)\;
    }
    
    \Return{unique\_values}
\end{algorithm}

We normalize the unique value estimate by dividing by the default chunk size; i.e., a segment with a single unique value would be normalized to 0, a segment that contains only unique values would be normalized to 1.
The factor is then obtained by using the normalized fraction for linear interpolation between a \texttt{LOW} and a \texttt{HIGH} factor.
\texttt{LOW} and \texttt{HIGH} are constants that were experimentally obtained.
\Cref{algo:estimate_materialize_unique_values_factor} describes the estimation process in pseudo-code.

\begin{algorithm}[htbp]
\caption{EstimateUniqueValuesFactor}
\label{algo:estimate_materialize_unique_values_factor}
    \KwIn{A join, a clustering candidate, cluster counts}
    \KwOut{An influence factor}
    \DontPrintSemicolon
    
    unique\_value\_count = \texttt{EstimateUniqueValuesPerChunk}(join.column, clustering\_candidate, cluster\_counts)\;
    
    factor = interpolate(\texttt{LOW}, \texttt{HIGH}, unique\_value\_count / default\_chunk\_size)\;
    
    \Return{factor}
\end{algorithm}

\paragraph{Sort order}
In Hyrise, each chunk can be sorted by a column.
We have observed that the sort order impacts the run time of the materialize step.
More precisely, the materialize step runs faster when the chunk is sorted by the join column.
We believe this run time reduction is due to improved cache hit rates (similar to the effects described by Memarzia et al.~\cite{DBLP:conf/edbt/MemarziaRB19}), although we did not investigate whether the cache hit rates actually improve:
By default, Hyrise stores data in dictionary encoded segments.
The dictionaries of those segments are sorted, i.e., the smallest value id maps to the smallest value, and so on.
If the join column is sorted, the dictionary will be traversed from front to back, rather than randomly accessed.

To quantify the impact of the sort order on the materialize step's performance, we introduce an influence factor.
We choose an experimentally determined constant for this factor, i.e., the number of unique values does not influence the impact of this factor.
Depending on the used hardware, this constant might vary and should be adjusted when the hardware changes.
The influence factor is already integrated in \Cref{algo:estimate_materialize_step}.

\begin{figure}
     \centering
     \begin{subfigure}[b]{0.45\textwidth}
         \centering
         \includegraphics[width=\textwidth]{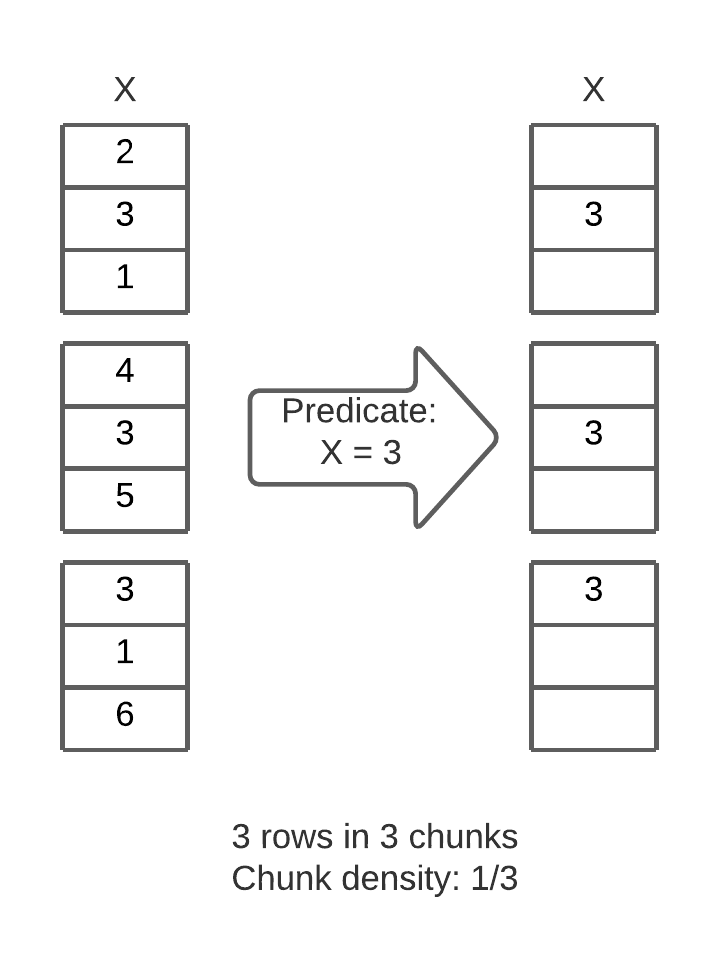}
         \caption{Unclustered table}
         \label{fig:chunk_density_unclustered}
     \end{subfigure}
     \hfill
     \begin{subfigure}[b]{0.45\textwidth}
         \centering
         \includegraphics[width=\textwidth]{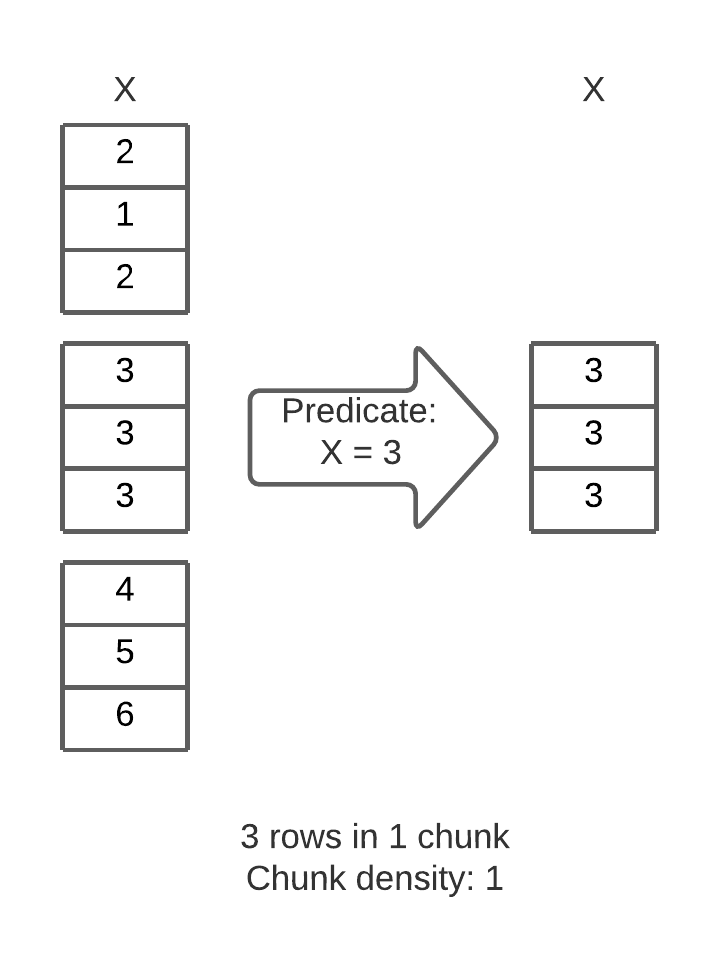}
         \caption{Table clustered by \texttt{X}}
         \label{fig:chunk_density_clustered}
     \end{subfigure}
        \caption{The figure visualizes different chunk densities produced by the scan \texttt{X = 3}.}
        \label{fig:chunk_density}
\end{figure}

\paragraph{Chunk density}
We have observed that the \emph{chunk density} impacts the run time of the materialize step.
With \emph{chunk density}, we refer to the average chunk fill grade:
Some operators, e.g., table scans, scan the table and filter out rows.
The size of a table scan's result does not depend on the clustering; however, the number of chunks of the resulting table does:
For example, a predicate with 10\% selectivity may return all of a table's chunks, which each contain 10\% of their rows, i.e., a chunk density of 10\%.
Alternatively, the same predicate could return only 10\% of the chunks, but all filled to 100\%, i.e., a chunk density of 100\%.
\Cref{fig:chunk_density} visualizes different chunk densities.

A high chunk density is beneficial for the performance of the materialize step because it reduces the number of dictionary segments that have to be accessed (and thus, loaded into memory).
To quantify the impact of the chunk density, we introduce a new influence factor.

As a first step, we estimate the join's expected number of input chunks for a given clustering.
The join's statistics include the input chunk count for the currently implemented clustering.
To estimate the number of chunks for another clustering candidate, we consider the join's corresponding PQP and extract all predicates that operate on the join's table.
We make the following assumption:
All predicates on a clustering column return only chunks with a density of 100\%; and predicates that do not operate on clustering columns filter rows uniformly across all chunks.
In other words: the estimated chunk density is the product of the selectivities of all predicates on unclustered columns.
Based on the estimated chunk density and the input row count of the join, we calculate the expected number of chunks as $\frac{\text{join's input row count}}{\text{default chunk size * chunk density}}$.
The estimation of the number of input chunks is displayed in pseudo-code in \Cref{algo:estimate_chunk_count}.

\begin{algorithm}[htbp]
\caption{EstimateChunkCount}
\label{algo:estimate_chunk_count}
    \KwIn{A join, a clustering candidate, cluster counts}
    \KwOut{The expected number of chunks}
    \DontPrintSemicolon
    
    density $= 1$\;
    \ForEach{scan on scan.table in scan.pqp.scans}{
        \If{scan.column not in clustering\_candidate.clustering\_columns}{
            density *= scan.selectivity\;
        }
    }
    estimated\_chunk\_count $= \min{(\frac{\text{join.input\_size}}{\text{default\_chunk\_size * density}}, \text{scan.table.chunk\_count})}$\;
    
    \Return{estimated\_chunk\_count}
\end{algorithm}

We then normalize the expected number of chunks with the following fraction: $\frac{\text{estimated chunk count - min chunk count}}{\text{max chunk count - min chunk count}}$.
This normalization yields a zero when the minimum number of chunks is expected (i.e., only full chunks), and a one when the maximum number of chunks is expected.
We obtain the influence factor by using the normalized fraction for linear interpolation between a \texttt{LOW} and a \texttt{HIGH} factor.
Analogous to the previous influence factors, the values for \texttt{LOW} and a \texttt{HIGH} are determined experimentally.
\Cref{algo:estimate_materialize_density} displays the factor estimation process in pseudo-code.

\begin{algorithm}[htbp]
\caption{EstimateDensityFactor}
\label{algo:estimate_materialize_density}
    \KwIn{A join, a clustering candidate, cluster counts}
    \KwOut{An influence factor}
    \DontPrintSemicolon
    
    estimated\_chunk\_count = \texttt{EstimateChunkCount}(join, clustering\_candidate, cluster\_counts)\;
    min\_chunk\_count = join.input\_size / default\_chunk\_size\;
    max\_chunk\_count = join.table.size / default\_chunk\_size\;

    fraction\_of\_chunks = $\frac{\text{estimated\_chunk\_count - min\_chunk\_count}}{\text{max\_chunk\_count - min\_chunk\_count}}$\;
    factor = interpolate(\texttt{LOW}, \texttt{HIGH}, fraction\_of\_chunks)\;

    \Return{factor}
\end{algorithm}

\subsubsection*{Estimating Probing}
For the probe step, we use a latency estimation system similar to the materialize step.
However, we include only one influence factor for the probe step:
We have observed that the sort order impacts the performance of the probe step.
If the probe column is sorted, the probe step runs faster.
This seems to apply even if all values in the chunk are unique.
We believe this may be due to an improved cache hit rate for dictionary keys.
As a consequence, we use a constant factor \texttt{SORT FACTOR} that is multiplicative to the probe step's latency.
Analogous to the materialize step, we first calculate the expected unoptimized time, i.e., the expected run time without the benefits granted by the currently implemented clustering.
In the second step, we integrate the benefits granted by the clustering candidate.
The constant \texttt{SORT FACTOR} was determined experimentally.
\Cref{algo:estimate_probe_step} provides a pseudo-code description of the probe step's latency estimation.

\begin{algorithm}[htbp]
\caption{EstimateProbeStep}
\label{algo:estimate_probe_step}
    \KwIn{A join, a clustering candidate}
    \KwOut{A latency estimate}
    \DontPrintSemicolon
    
    factor $= 1$\;
    \If{join.data\_arrives\_sorted and the join.column == current\_clustering.sort\_column}{
        factor *= SORT\_FACTOR\;
    }
    \If{join.data\_arrives\_sorted and join.column == clustering\_candidate.sort\_column}{
        factor /= SORT\_FACTOR\;
    }
    
    \Return{join.probe\_step\_latency * factor}
\end{algorithm}

\subsection{Aggregate Operator}
\label{sec:cost_estimation_aggregates}
In this section, we describe aspects of how the clustering can impact the performance of aggregates.
Due to the limited scope of this thesis, aggregate latency estimation is not integrated into the model.
Hyrise has two aggregate implementations: \texttt{AggregateSort}, and \texttt{AggregateHash}.
We focus on the hash aggregate, which is used by default.

\paragraph{Hash Aggregate}
The hash aggregate operates in multiple steps.
In the first step, the \emph{grouping step}, it assigns each row in the table to a group.
In the second step, the \emph{aggregate step}, the aggregates are calculated.
The aggregates are calculated incrementally, i.e., by iterating over the whole table and updating the respective current aggregate value for each row.
Those two steps are followed by three steps of output writing: for the grouping columns, the aggregates, and the resulting table.
We are only interested in the first two steps for two reasons:
First, those are usually the most expensive steps.
Writing the outputs is comparably cheap, even if the aggregate's output has a high cardinality:
For example, the \texttt{GROUP BY l\_orderkey} aggregate from query 18 of the TPC-H benchmark has an output size of 15\,000\,000 rows, but spends only 10\% of its run time on the output writing.
Second, we do not believe the output steps have a significant optimization potential:
A hash aggregate's cardinality is not impacted by the clustering, i.e., the number of rows we have to write does not depend on the clustering.

\paragraph{Optimization potential}
For the first two steps, we see optimization potential:
For example, clustering the rows in a way that causes rows of a group to be stored consecutively is likely to improve the performance of the first two steps.
Thus, the columns used for grouping might be interesting columns for the clustering.

Interesting estimation metrics could be how far the rows of a group are spread, or the number of different groups every chunk contains.
This estimation is not trivial:
Some aggregates have a low, fixed output cardinality (e.g., the aggregate of TPC-H query 1), others scale linearly with the number of tuples (e.g., the inner aggregate of TPC-H query 18).
Aggregates with a high output cardinality may require a large cluster count, i.e., a more fine-grained clustering, to yield any performance improvement.
Further, multiple columns may be involved in the grouping.
If multiple columns are involved, the number of possible groups, i.e., the number of unique combinations of values of the grouping columns, increases significantly.
At the same time, correlations or functional dependencies between the grouping columns may cause a significantly lower number of groups than expected.


	\chapter{Evaluation}
\label{sec:evaluation}
In this section, we evaluate the above presented model for determining beneficial clustering configurations and the MVCC-aware clustering algorithm.
First, we describe our experimental setup, which is shared between the model's and the clustering algorithm's evaluation.

\paragraph{Experimental setup}
For our evaluation, we perform various experiments with Hyrise.
For all experiments, the last commit from Hyrise's master branch that is included in our work is \texttt{f8e4cd4}\footnote{\url{https://github.com/hyrise/hyrise/tree/f8e4cd4}}.
All experiments were executed on an \texttt{Intel(R) Xeon(R) CPU E7- 8870  @ 2.40GHz} processor with 20 cores, a 32~MB cache per core, and 387~GB of DRAM.
All experiments were performed single-threaded.
Hyrise was compiled with \texttt{gcc 9.2.1} on \texttt{Ubuntu 19.10}.

All experiments were performed with a scale factor of ten, for both TPC-H and TPC-DS.
Unless otherwise specified, all query latencies were obtained by executing each benchmark's queries twenty times and computing the mean latency for each query.
At the time of writing, Hyrise supports only $40$ of the $99$ queries of the TPC-DS benchmark.

Before the clustering algorithm is executed, the clustered table is shuffled to ensure that the latency measurements are not distorted by possible remnants of the clustering produced by the benchmark generator; and dictionary encoding is applied to all segments.

\section{Automated Workload-Based Clustering Model}
In this section, we evaluate the clustering model presented in \Cref{sec:clustering_model}.
First, we evaluate the model's solution quality.
Second, we evaluate the impact of cluster counts.
In the third step, we evaluate the precision of the latency estimations on operator level and analyze common sources of error.
Finally, we evaluate the model's computation time, i.e., the time required to calculate clustering suggestions, for variously sized workloads.

\subsection{Solution Quality}
We consider two workloads to evaluate the quality of our model's clustering suggestions: the analytical benchmarks TPC-H and TPC-DS.
For both benchmarks, we choose individual tables and evaluate the model's top $20$ clustering suggestions for the respective table.
If the model's predictions were perfect, the measured latency would increase from the first to the last suggestion.

\subsubsection*{TPC-H Solution Quality}
\begin{figure}
    \centering
    \includegraphics[width=\textwidth]{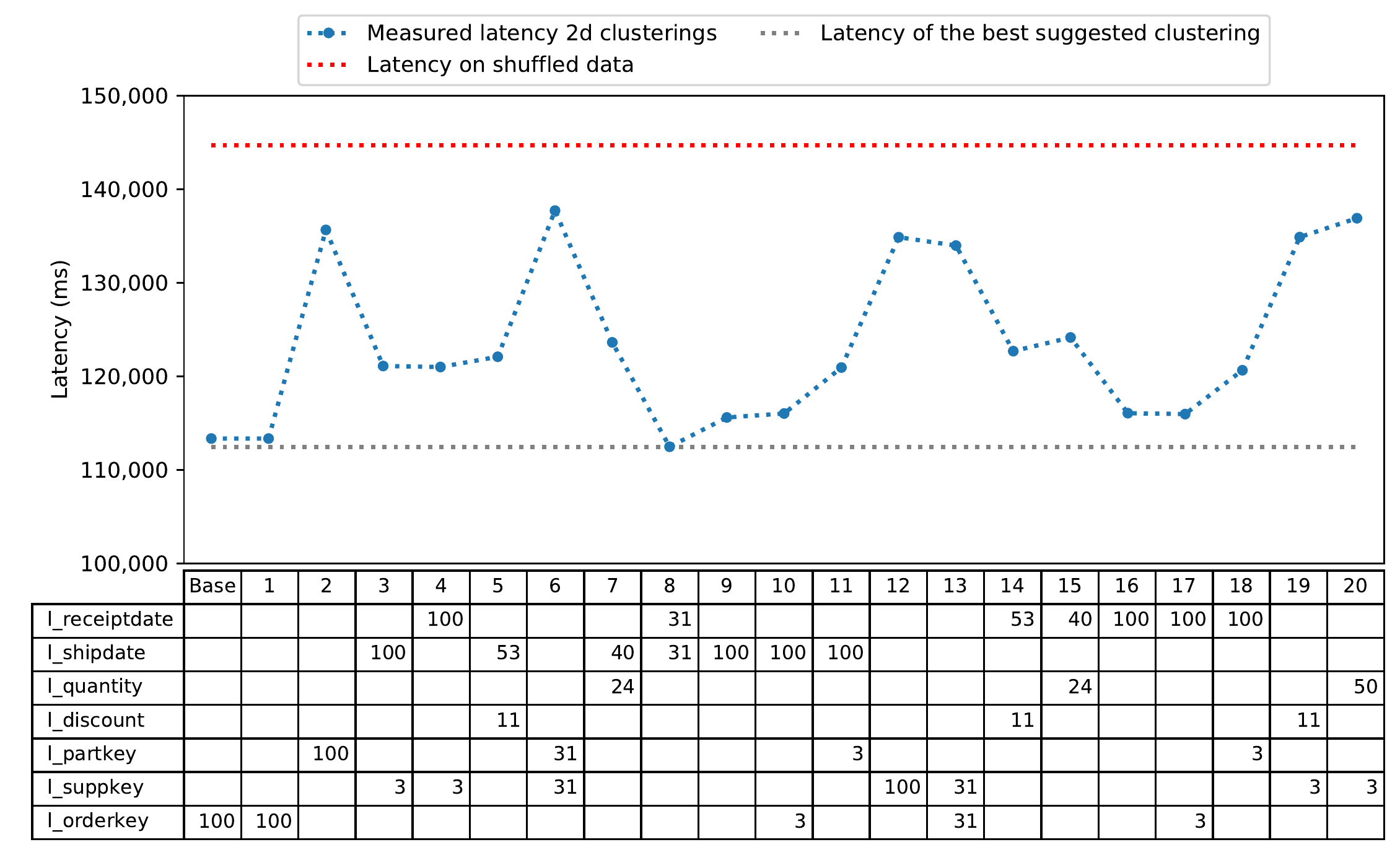}
    \caption{The figure visualizes the measured latency of the model's top $20$ clustering suggestions (blue) for the \texttt{lineitem} table of the TPC-H benchmark. The suggested clusterings have at most two clustering columns. All suggested clusterings use \texttt{l\_orderkey} as sort column. The table displays the clustering columns and cluster counts for each suggestion.}
    \label{fig:top20_tpch_latencies}
\end{figure}

For the TPC-H benchmark, we evaluate only the model's clustering suggestions for the \texttt{lineitem} table.
We choose \texttt{lineitem} for clustering because it is the largest of the TPC-H tables, accounting for $69$\% of all rows.
All other tables were left as generated.
\Cref{fig:top20_tpch_latencies} shows the measured latency of the top $20$ clustering suggestions for \texttt{lineitem} with at most two clustering columns.
All suggested clusterings use \texttt{l\_orderkey} as sort column.

Based on the figure, we make the following observations:
First, the base clustering, i.e., the clustering produced by the benchmark generator (by \texttt{l\_orderkey}), already yields quite good performance.
Our model ranks the base clustering at position 1.
There is only one among the $20$ suggested clusterings that performed slightly better:
A \texttt{l\_shipdate-l\_receiptdate} clustering, which was ranked at position $8$.

Second, there are latency spikes at the ranks $2$, $6$, $12$, and $13$.
Those are exactly the clustering suggestions that focus on the join columns \texttt{l\_partkey} or \texttt{l\_suppkey}.
On the contrary, clusterings on \texttt{l\_shipdate} or \texttt{l\_receiptdate} generally have a comparably low latency.
The majority of the clusterings ($13$ of $20$) cluster by at least one of \texttt{l\_shipdate} and \texttt{l\_receiptdate}.
Those clusterings have a similar latency, from which we conclude that the second clustering column has only a low impact.

Third, most of our model's suggestions have a low latency compared to the workload's latency on shuffled data.
We thus conclude that our model is able to roughly estimate a clustering's impact on the latency, and use those estimates to suggest sensible clusterings.
However, it is still too inaccurate to predict an optimal ranking of the clustering suggestions.

\begin{figure}[H]
    \centering
    \includegraphics[width=\textwidth]{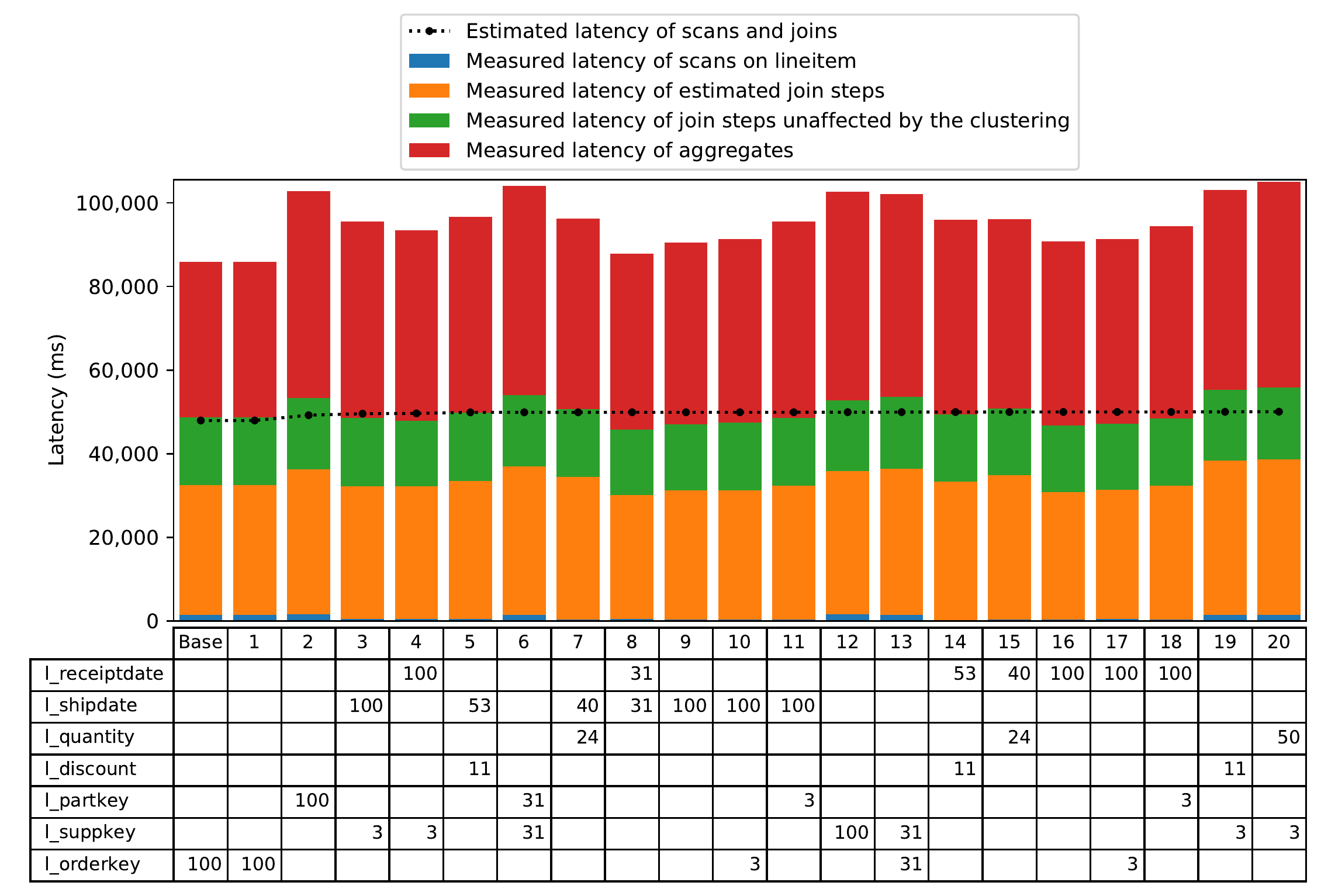}
    \caption{The figure visualizes the measured run times of scans, aggregates and joins for the model's top $20$ clustering suggestions for \texttt{lineitem}. Each TPC-H query was executed once. If the model's estimations were perfect, the black line would be on the same level as the estimated operators, i.e., the lower three bars. The table displays the clustering columns and cluster counts for each suggestion.}
    \label{fig:detailed_estimates_tpch}
\end{figure}

\paragraph{Operator latency breakdown}
The mismatch between estimated and measured workload latencies may be caused by imprecise latency estimations, but also by latency changes of operators that are not included in the model's estimates, such as aggregates.
For a more detailed evaluation, we break down the latencies per operator for each clustering suggestion in \Cref{fig:detailed_estimates_tpch}.

Based on the figure, we make two observations:
First, there is only a difference of $2\,100$ milliseconds between the latency estimate of the first clustering ($47\,947$ milliseconds) and the last clustering ($50\,041$ milliseconds), i.e., the model's estimates vary only marginally.
The measured latencies are spread slightly farther: the minimum combined measured latency for scans and joins is $45\,721$ milliseconds, the maximum $55\,783$ milliseconds.
We observe that the model seems to slightly over-estimate the latency when \texttt{l\_shipdate} or \texttt{l\_receiptdate} is clustered, and to slightly under-estimate the latency otherwise.

Second, the figure shows that the latency spikes can be attributed to aggregates, at least for the largest part.
For example, the aggregate latency of the second clustering suggestion (\texttt{l\_partkey}) is $12\,200$ milliseconds higher than the aggregate of the first suggestion.
Due to the limited scope of this thesis, our model does not estimate the latency of aggregates, but assumes that they are unaffected by the clustering.
As a consequence, the model under-estimates the actual latency of the \texttt{l\_partkey} clustering and ranks it position two.
We conclude that aggregate latency estimations are required to obtain more precise suggestion rankings.

\paragraph{An educated guess regarding aggregate performance}
We observe that clustering by \texttt{l\_partkey} seems to result in a particular high latency for aggregates, compared to, e.g., an \texttt{l\_shipdate} clustering.
While we did not perform experiments on aggregates, we provide an educated guess
why clustering by \texttt{l\_partkey} results in a higher latency for aggregates than, e.g., clustering by \texttt{l\_shipdate}:
TPC-H Query $18$ contains an aggregate operator on \texttt{l\_orderkey} with a result size of fifteen million rows.
This aggregate operator contributes $10$ of the workload's $37$ seconds spent on aggregates when clustering by \texttt{l\_orderkey}, $17$ of $43$ seconds when clustering by \texttt{l\_shipdate}, and $23$ of $49$ seconds when clustering by \texttt{l\_partkey}.
While there is not a direct correlation between \texttt{l\_orderkey} and \texttt{l\_shipdate}, the columns are indirectly related:
According to the TPC-H specification~\cite[p.~87]{TpchSpec}, there may be one to seven rows in \texttt{lineitem} for each order.
Rows with the same \texttt{l\_orderkey} may have different \texttt{l\_shipdate} values; however, \texttt{l\_shipdate} is correlated to \texttt{o\_orderdate}.
This correlation causes the range between the minimum and maximum \texttt{l\_shipdate} per \texttt{l\_orderkey} to be small.
As a consequence, when clustering by \texttt{l\_shipdate}, identical \texttt{l\_orderkey} values are probably not consecutive, but spread over a small number of chunks.
In contrast, there is no such correlation between \texttt{l\_orderkey} and \texttt{l\_partkey}, and identical values of \texttt{l\_orderkey} may be spread over the whole table.
We believe the indirect correlation between \texttt{l\_orderkey} and \texttt{l\_shipdate} may positively impact the aggregate's latency.


\subsubsection*{TPC-DS Solution Quality}
\begin{figure}
    \centering
    \includegraphics[width=0.9\textwidth]{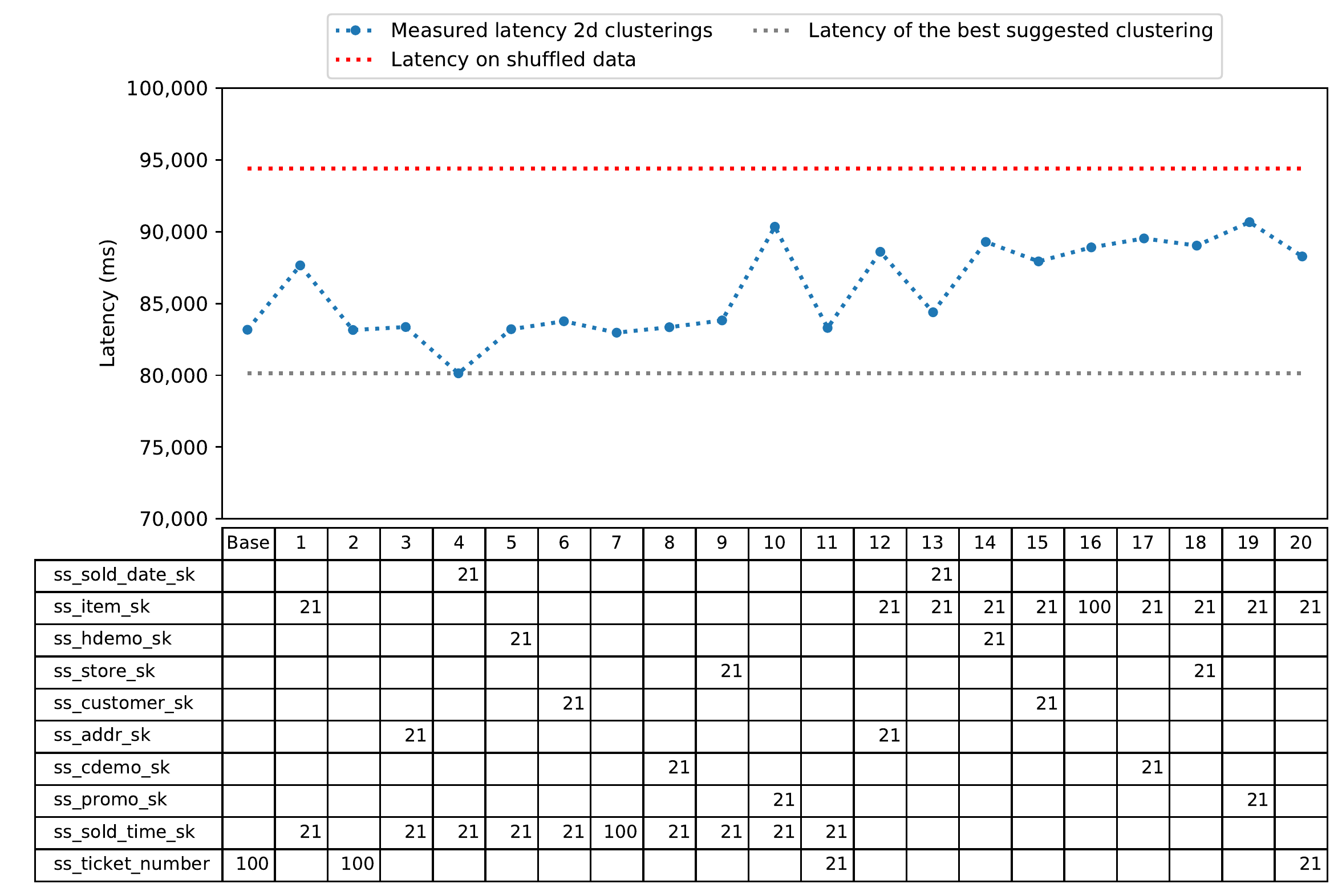}
    \caption{The figure visualizes the measured latency of the model's top $20$ clustering suggestions (blue) for the \texttt{store\_sales} table of the TPC-DS benchmark. The suggested clusterings have at most two clustering columns. All suggested clusterings use \texttt{ss\_ticket\_number} as sort column. The table displays the clustering columns and cluster counts for each suggestion.}
    \label{fig:top20_tpcds_latencies}
\end{figure}

For the TPC-DS benchmark, we evaluate only the model's clustering suggestions for the \texttt{store\_sales} table.
We choose \texttt{store\_sales} for clustering because it is the second largest of the TPC-DS tables.
The \texttt{inventory} table has more than four times as many rows, but is only used in $6$ queries, from which Hyrise can execute only $1$ at the time of writing.
In contrast, \texttt{store\_sales} is used frequently in joins and still contains $28\,800\,991$ rows.
All other tables were left as generated.
\Cref{fig:top20_tpcds_latencies} shows the measured latency of the top $20$ clustering suggestions for \texttt{store\_sales} with at most two clustering columns.
All suggested clusterings use \texttt{ss\_ticket\_number} as sort column.

Based on the figure, we make the following observations:
First, the base clustering, i.e., the clustering produced by the benchmark generator (by \texttt{ss\_ticket\_number}), already yields an acceptable performance.
Our model ranks the base clustering at position 2.
There is only one among the $20$ suggested clusterings that performed significantly better:
A \texttt{ss\_sold\_date\_sk-ss\_sold\_time\_sk} clustering, which was ranked at position 4.

Second, the clustering suggestions are dominated by two columns: \texttt{ss\_sold\_time\_sk}, and \texttt{ss\_item\_sk}.
Ten out of the first eleven suggestions cluster by \texttt{ss\_sold\_time\_sk}, nine out of the last nine suggestions cluster by \texttt{ss\_item\_sk}.
In most cases, clusterings with \texttt{ss\_sold\_time\_sk} achieve a lower latency than clusterings with \texttt{ss\_item\_sk}.
Similar to TPC-H, we conclude that not all columns of a multi-dimensional clustering have the same impact.

Third, the suggestions at ranks $2$ to $9$, $11$ and $13$ have a low latency compared to a clustering on shuffled data.
Thus, analogous to TPC-H, we conclude that our model is able to roughly estimate a clustering's impact on the latency, and use those estimates to suggest sensible clusterings.
However, there is still too much inaccuracy to predict an optimal ranking of the clustering suggestions.

\begin{figure}
    \centering
    \includegraphics[width=0.9\textwidth]{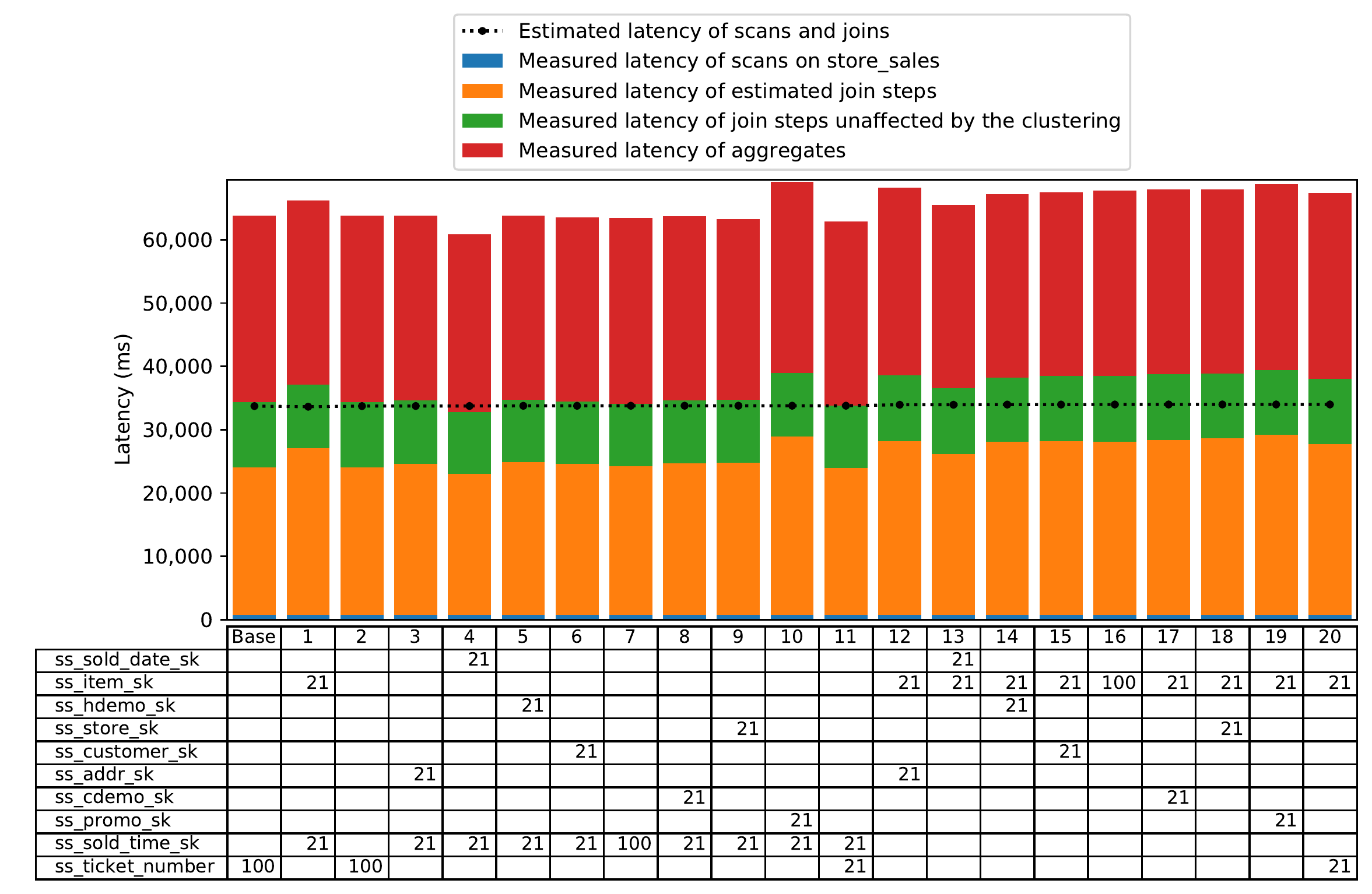}
    \caption{The figure visualizes the measured run times of scans, aggregates and joins for the model's top 20 clustering suggestions for \texttt{store\_sales}. Each TPC-DS query was executed once. If the model's estimations were perfect, the black line would be on the same level as the estimated operators, i.e., the lower three bars. The table displays the clustering columns and cluster counts for each suggestion.}
    \label{fig:detailed_estimates_tpcds}
\end{figure}

\paragraph{Operator latency breakdown}
We break down the latencies per operator for each clustering suggestion in \Cref{fig:detailed_estimates_tpcds}.

We observe that the model's estimates vary only slightly.
The measured latencies of scans and joins can be divided into two groups: clusterings on \texttt{ss\_sold\_time\_sk}, and clusterings on \texttt{ss\_item\_sk}.
Within the groups, the measured latency estimates vary only slightly, too.

For clusterings on \texttt{ss\_sold\_time\_sk}, our model provides precise latency estimates.
However, our model under-estimates the latency when \texttt{ss\_item\_sk} is among the clustering columns (i.e., suggestions at rank 1 and 12-20).

\subsubsection*{Three dimensional clusterings}
So far, we have evaluated only clusterings with at most two clustering columns.
In this section, we compare the measured latencies of the model's top 20 suggestions when using at most two and three dimensions per suggestion, respectively.

\begin{figure}
    \centering
    \includegraphics[width=\textwidth]{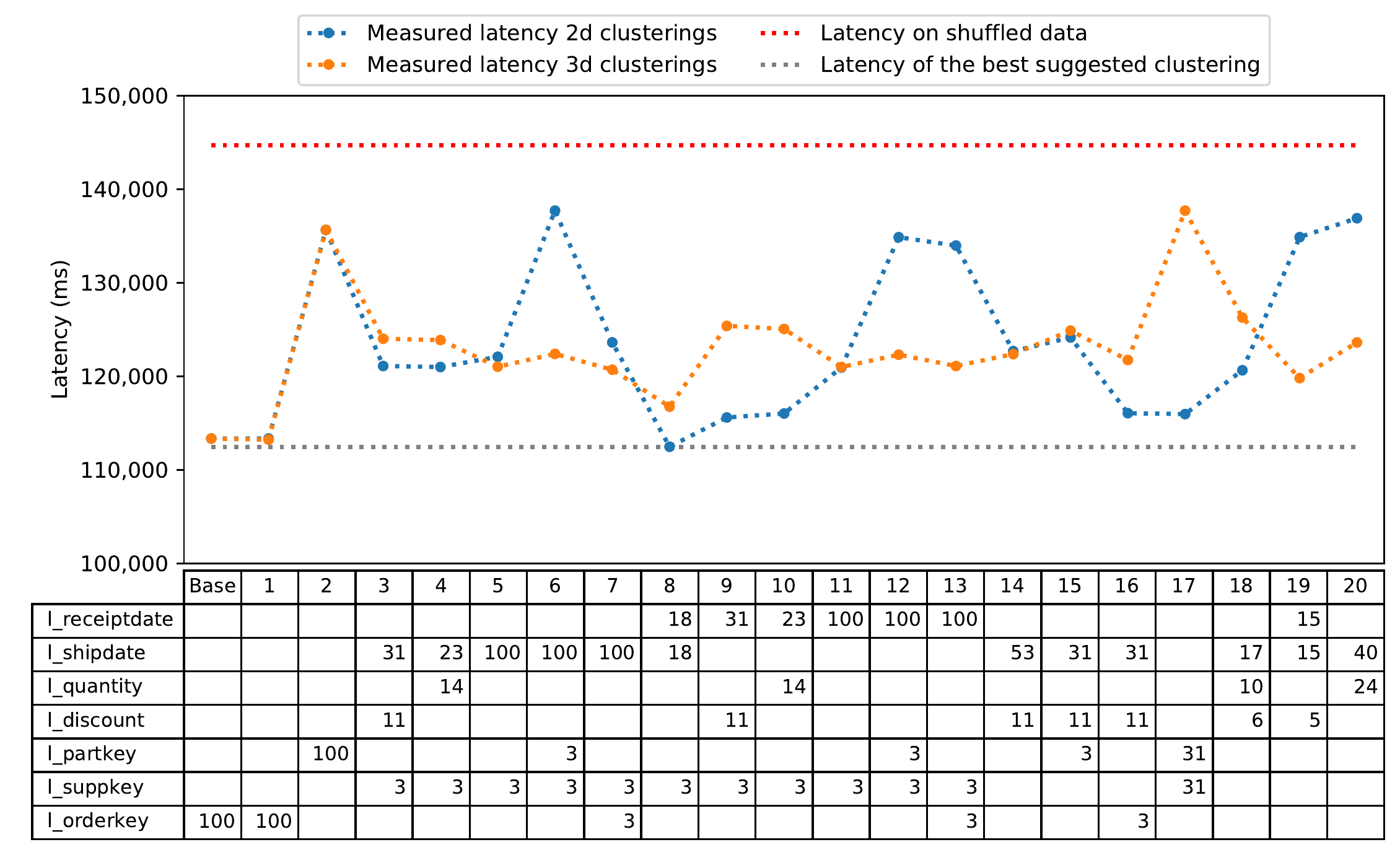}
    \caption{The figure compares the measured latency of the model's top 20 two- and three-dimensional clustering suggestions for the TPC-H table \texttt{lineitem}. All suggested clusterings use \texttt{l\_orderkey} as sort column. The table displays the clustering columns and cluster counts of the three-dimensional clustering suggestions.}
    \label{fig:top20_latency_tpch_3d}
\end{figure}

\paragraph{TPC-H}
\Cref{fig:top20_latency_tpch_3d} compares the measured latencies of the top 20 two- and three-dimensional clustering suggestions for the TPC-H table \texttt{lineitem}.
Based on the figure, we make the following observations:
First, both the two- and three-dimensional suggestions rank the base clustering at position 1, and the \texttt{l\_partkey} clustering at position 2.
From position 3 onwards, the latency of the three-dimensional suggestions is more consistent and has fewer latency spikes than the two-dimensional suggestions.
However, the best three dimensional suggestion has a higher latency than the best two dimensional clustering, i.e., the two-dimensional suggestions find a better clustering.

Second, we observe that \texttt{l\_shipdate} and \texttt{l\_receiptdate} are frequently chosen: 17 out of 20 suggestions cluster by at least one of them.
Further, \texttt{l\_suppkey} is chosen in 12 out of 20 suggestions, while \texttt{l\_orderkey} is present in only 4 out of 20.
We believe this is due to the small table size of the supplier table:
At scale factor 10, it contains 100\,000 rows, i.e., there are only 100\,000 unique values for \texttt{l\_suppkey}.

Our model assumes that number of unique values per chunk impacts the latency of the hash join's materialize step.
When a join column has a low number of unique values, a cluster count of 3 might already yield performance improvements:
For example, with a cluster count of 3 for \texttt{l\_suppkey}, our model expects each chunk to contain $\frac{100\,000}{3}$, i.e, approximately 33\,333, unique values among the 65\,535 rows of a chunk.
On the contrary, \texttt{orders} has fifteen million rows, i.e., there are fifteen million unique values in \texttt{l\_orderkey}.
With a cluster count of 3 for \texttt{l\_orderkey}, our model expects still 65\,535 unique \texttt{l\_suppkey} values per chunk, and concludes that a cluster count of 3 for \texttt{l\_orderkey} yields only little performance improvement.

\begin{figure}
    \centering
    \includegraphics[width=0.9\textwidth]{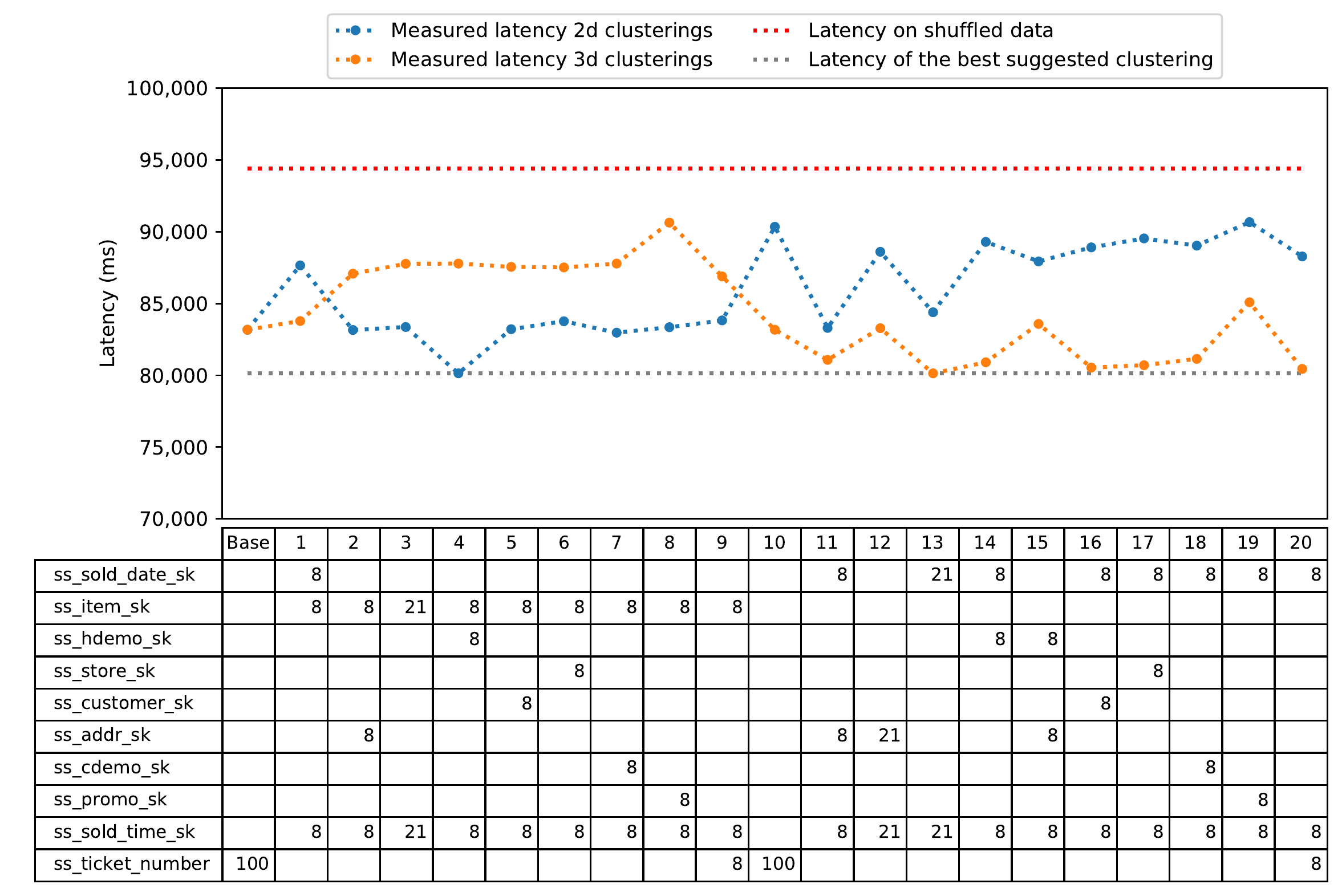}
    \caption{The figure compares the measured latency of the model's top 20 two- and three-dimensional clustering suggestions for the TPC-DS table \texttt{store\_sales}. All suggested clusterings use \texttt{ss\_ticket\_number} as sort column. The table displays the clustering columns and cluster counts of the three-dimensional clustering suggestions.}
    \label{fig:top20_latency_tpcds_3d}
\end{figure}

\paragraph{TPC-DS}
\Cref{fig:top20_latency_tpcds_3d} compares the measured latencies of the top 20 two- and three-dimensional clustering suggestions for the TPC-DS table \texttt{store\_sales}.

Based on the figure, we make the following observations:
First, the three-dimensional version suggests a better clustering at rank 1.
For the ranks two to nine, however, the two-dimensional model yields better results.
Those suggestions all cluster by \texttt{ss\_item\_sk}.
From the tenth rank onwards, the suggestions from the three-dimensional model have consistently lower latencies.
Those suggestions cluster by a combination of \texttt{ss\_sold\_time\_sk} and \texttt{ss\_sold\_date\_sk}.

Second, we observe that three columns dominate the clustering suggestions:
the column \texttt{ss\_sold\_time\_sk} is chosen in 19 out of 20, and  \texttt{ss\_item\_sk} and \texttt{ss\_sold\_date\_sk} are chosen in 9 out of 20 clustering suggestions.
Those columns reference the tables \texttt{time\_dim}, \texttt{item}, and \texttt{date\_dim}, respectively.
Those three tables have a size of 86\,400, 102\'000, and 73\,049 rows, i.e., their corresponding columns in \texttt{store\_sales} all contain a low number of unique values.
Analogous to TPC-H, we conclude that our model prefers join columns with a low number of unique values.

\subsection{Impact of Cluster Counts}
\label{sec:eval_cluster_counts}
\begin{figure}
    \centering
    \includegraphics[width=\textwidth]{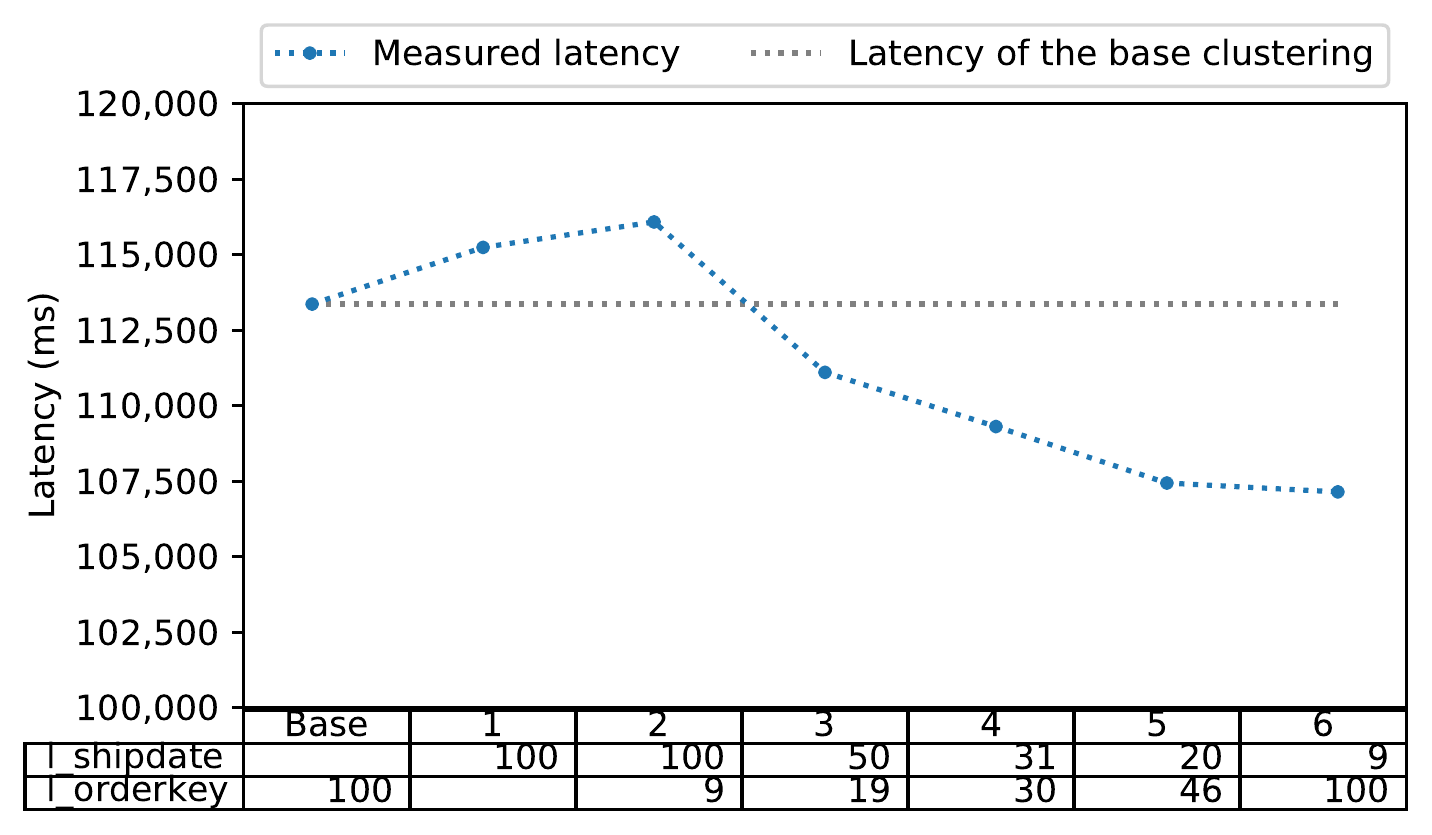}
    \caption{The figure visualizes the measured latencies of \texttt{l\_shipdate-l\_orderkey} clusterings with varying cluster counts. The table describes the clustering columns and their cluster counts. All configurations use \texttt{l\_orderkey} as the sort column.}
    \label{fig:eval_cluster_counts}
\end{figure}

In this section, we evaluate the impact of the cluster counts.
For that purpose, we cluster the \texttt{lineitem} table of the TPC-H benchmark by \texttt{l\_shipdate} and \texttt{l\_orderkey} with varying cluster counts.
We still aim at cluster counts whose product is approximately equal to the number of chunks, i.e., 916 for \texttt{lineitem} on scale factor 10.
The results are displayed in \Cref{fig:eval_cluster_counts}.

So far, our model assumes that join columns such as \texttt{l\_orderkey} require a high cluster count to reduce the number of unique values per chunk, and thus yield latency improvements.
Indeed, the figure shows that the lowest latency was obtained in configuration 6, where the  largest possible cluster count was chosen for \texttt{l\_orderkey}.
Configuration 2 has the lowest cluster count on \texttt{l\_orderkey} among the configurations we evaluated, and yields the highest latency.
However, configurations 3, 4, and 5 show that lower cluster counts on \texttt{l\_orderkey} can still yield latency improvements compared to the base clustering.

Configuration 5 uses a cluster count of 46 for \texttt{l\_orderkey}.
Our model expects $\frac{15\,000\,000}{46} \approx 326\,000$ unique values per cluster, i.e., more unique values than a chunk has rows.
As a consequence, the latency improvement cannot be attributed to a reduced number of unique values per chunk.
We conclude that there might be some additional influence factor beside the number of unique values per chunk, which our model does not yet quantify.

The difference between the configuration with the lowest and highest latency is about 9\,000 milliseconds, i.e., 8\% of the base clustering's latency.
We thus conclude that a choice of useful cluster counts is similarly important as the choice of the clustering columns itself.
Finding the sweet spot of the cluster counts is an interesting opportunity for future work.

\subsection{Precision of Latency Estimates}
In this section, we evaluate the precision of our model's latency estimates for table scans and hash joins.
For that purpose, we generate the TPC-H tables at scale factor ten and execute each TPC-H query ten times.
Based on the resulting workload, we let our model estimate the latencies of the following three \texttt{lineitem} clusterings:

\begin{enumerate}
    \item a clustering by the scan column \texttt{l\_shipdate}. The cluster count of \texttt{l\_shipdate} was set to the maximum, i.e., 100; the clusters were sorted by \texttt{l\_shipdate}. The resulting clustering is equalivalent to sorting the table by \texttt{l\_shipdate}.
    
    \item a clustering by the join column \texttt{l\_partkey}. The cluster count of \texttt{l\_partkey} was set to the maximum, i.e., 100; the clusters were sorted by \texttt{l\_partkey}. The resulting clustering is equalivalent to sorting the table by \texttt{l\_partkey}.
    
    \item a two-dimensional clustering by the scan column \texttt{l\_shipdate} and the join column \texttt{l\_orderkey}. The cluster counts were set to 20 and 50; the clusters were sorted by \texttt{l\_orderkey}.
\end{enumerate}

To evaluate those estimates, we implement the clusterings and compare our estimates against the measured operator latencies.
To quantify the precision of our estimates, we provide a relative error for each operator.
The relative error is calculated by the formula $\frac{\text{measured latency}}{\text{estimated latency}}$, i.e, the relative error is less than one if the estimate was bigger than the actual latency, and vice versa.
The major part of the evaluation analyzes the relative errors and discusses common sources of error in detail.

Additionally, we provide the following aggregated statistics:
the accumulated estimated and measured latencies for scans and join steps, the mean squared error, and the symmetric mean absolute percentage error.
The mean squared error (MSE) is the mean of the squares of the estimation errors: $MSE = \frac{1}{n} \cdot \sum\limits_{i=1}^{n} (y\_true_i - y\_pred_i)^2$, where $y\_true_i$ is the measured latency for the $i$-th operator and $y\_pred_i$ the predicted latency for the $i$-th operator.
Squaring the errors causes large errors to have a bigger impact compared to, e.g., the mean absolute error (MAE).
The symmetric mean absolute percentage error (SMAPE) is a measure based on relative errors.
It is defined as follows: $SMAPE = \frac{100\%}{n} \cdot \sum\limits^{n}_{i=1} \frac{2 \cdot |y\_true_i - y\_pred_i|}{|y\_true_i| + |y\_pred_i|}$, where $y\_true_i$ is the measured latency for the i-th operator, and $y\_pred_i$ is the estimated latency for th $i$-th operator.
By definition, SMAPE yields always a value between 0\% and 200\%.

\subsubsection*{Table scans}
\Cref{tab:aggregated_scan_statistics} lists aggregated statistics for the estimated latencies of scans on \texttt{lineitem}.

\begin{table}[]
    \centering
    \begin{tabular}{ c|c|c|c|c }
        Clustering & \makecell{Sum estimated\\latencies (ms)} & \makecell{Sum measured\\latencies (ms)} &  MSE & SMAPE\\
        \hline
        l\_shipdate & 5\,620 & 7\,580 & 104\,828 & 77.49\% \\
        l\_partkey & 12\,763 & 13\,313 & 243 & 5.21\% \\
        \makecell{l\_shipdate,\\l\_orderkey} & 8\,138 & 4\,121 & 8\'359 & 74.86\% \\
    \end{tabular}
    \caption{The table lists aggregated statistics for the estimated latencies of scans on \texttt{lineitem}. Each TPC-H query was executed 10 times.}
    \label{tab:aggregated_scan_statistics}
\end{table}

We observe that the measured latency of scans on \texttt{lineitem} is almost twice as high when clustering by \texttt{l\_shipdate} compared to the latency when clustering by a combination of \texttt{l\_shipdate} and \texttt{l\_orderkey}.
This may seem confusing, as the generated TPC-H workload contains no scans on \texttt{l\_orderkey}.
As such, it is not intuitive why clustering by \texttt{l\_orderkey} and \texttt{l\_shipdate} yields a significantly lower scan latency than clustering only by \texttt{l\_shipdate}, as the latter clusters by \texttt{l\_shipdate} with a finer granularity.

An in-depth analysis shows that exactly this finer granularity is the reason for the increased latency:
In TPC-H query 1, there is a scan on \texttt{l\_shipdate} that selects almost the entire \texttt{lineitem} table.
With a run time up to 240 milliseconds, this scan is the most expensive scan on \texttt{lineitem} over all TPC-H queries, up to four times slower than the second most expensive scan.
When generating the query execution plans, Hyrise orders predicates ascending by their selectivity, i.e., predicates with a low selectivity are executed first.
Due to the finer granularity of the \texttt{l\_shipdate} clustering, more chunks can be pruned, and the selectivity of the scan increases accordingly.
As a consequence, in two of ten cases, the scan is moved up in the PQP, and thus executed on reference segments rather than dictionary segments.
On reference segments, the scan takes about 2,400 milliseconds, i.e., both scans combined are more than 4\,000 milliseconds slower than before.
In contrast, when using the \texttt{l\_shipdate-l\_orderkey}-clustering, the scan is - due to the lower granularity of \texttt{l\_shipdate} - executed first in ten of ten cases, i.e., on dictionary segments.

\begin{figure}
     \centering
     \begin{subfigure}[b]{0.48\textwidth}
         \centering
         \includegraphics[width=\textwidth]{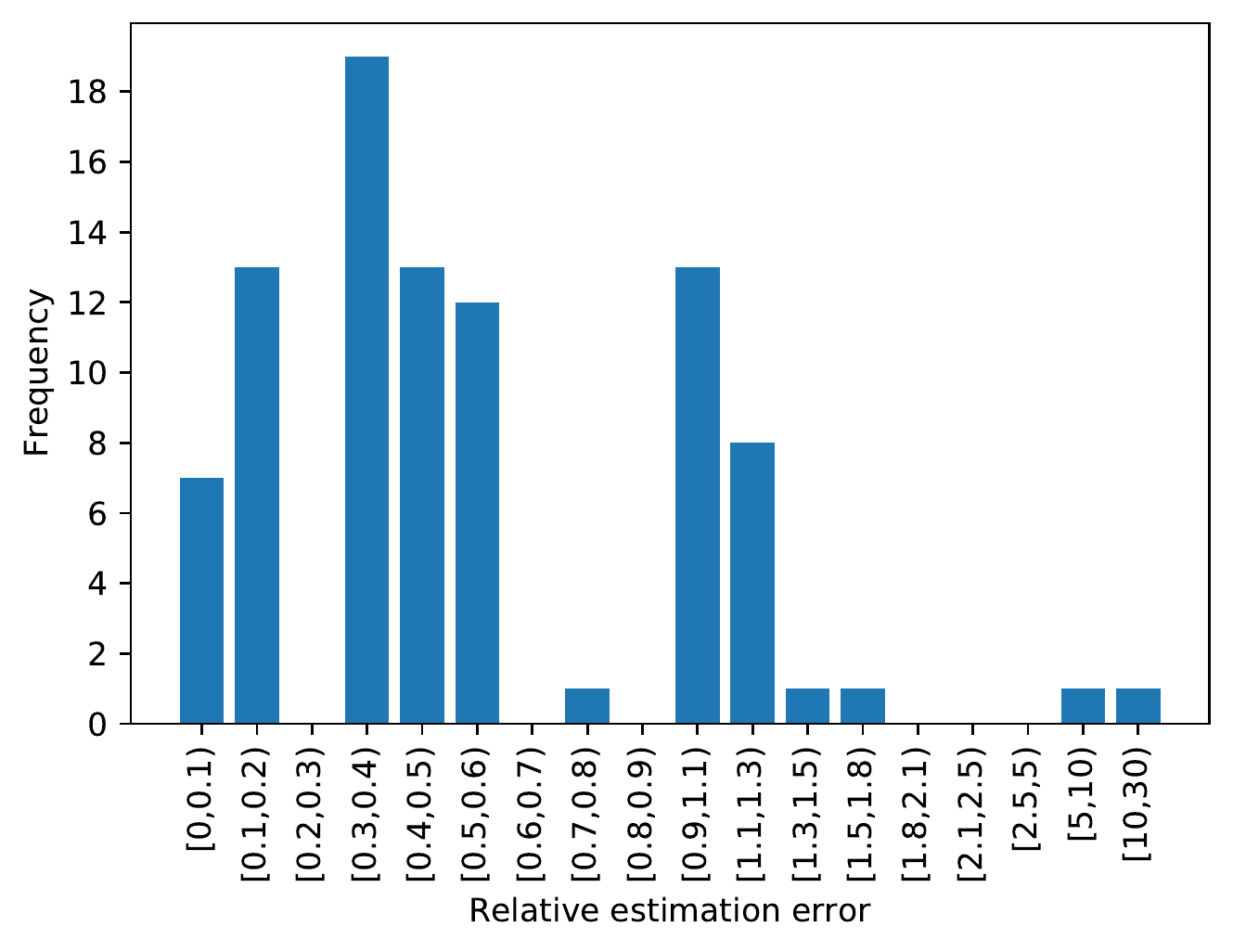}
         \caption{\texttt{l\_shipdate} Clustering}
         \label{fig:relative_errors_histogram_scan_1}
     \end{subfigure}
     \hfill
     \begin{subfigure}[b]{0.48\textwidth}
         \centering
         \includegraphics[width=\textwidth]{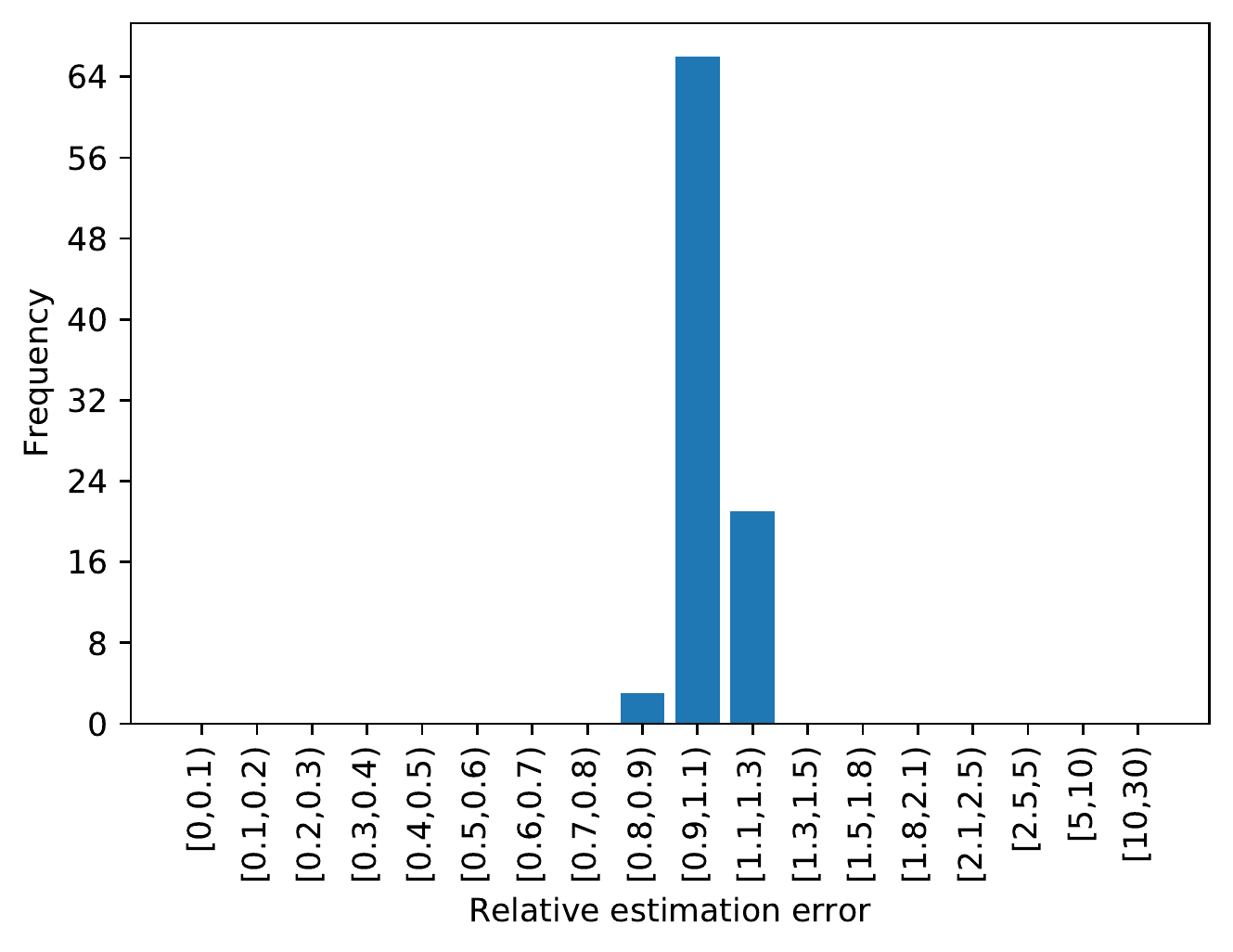}
         \caption{\texttt{l\_partkey} Clustering}
         \label{fig:relative_errors_histogram_scan_2}
     \end{subfigure}
     \hfill
     \begin{subfigure}[b]{0.48\textwidth}
         \centering
         \includegraphics[width=\textwidth]{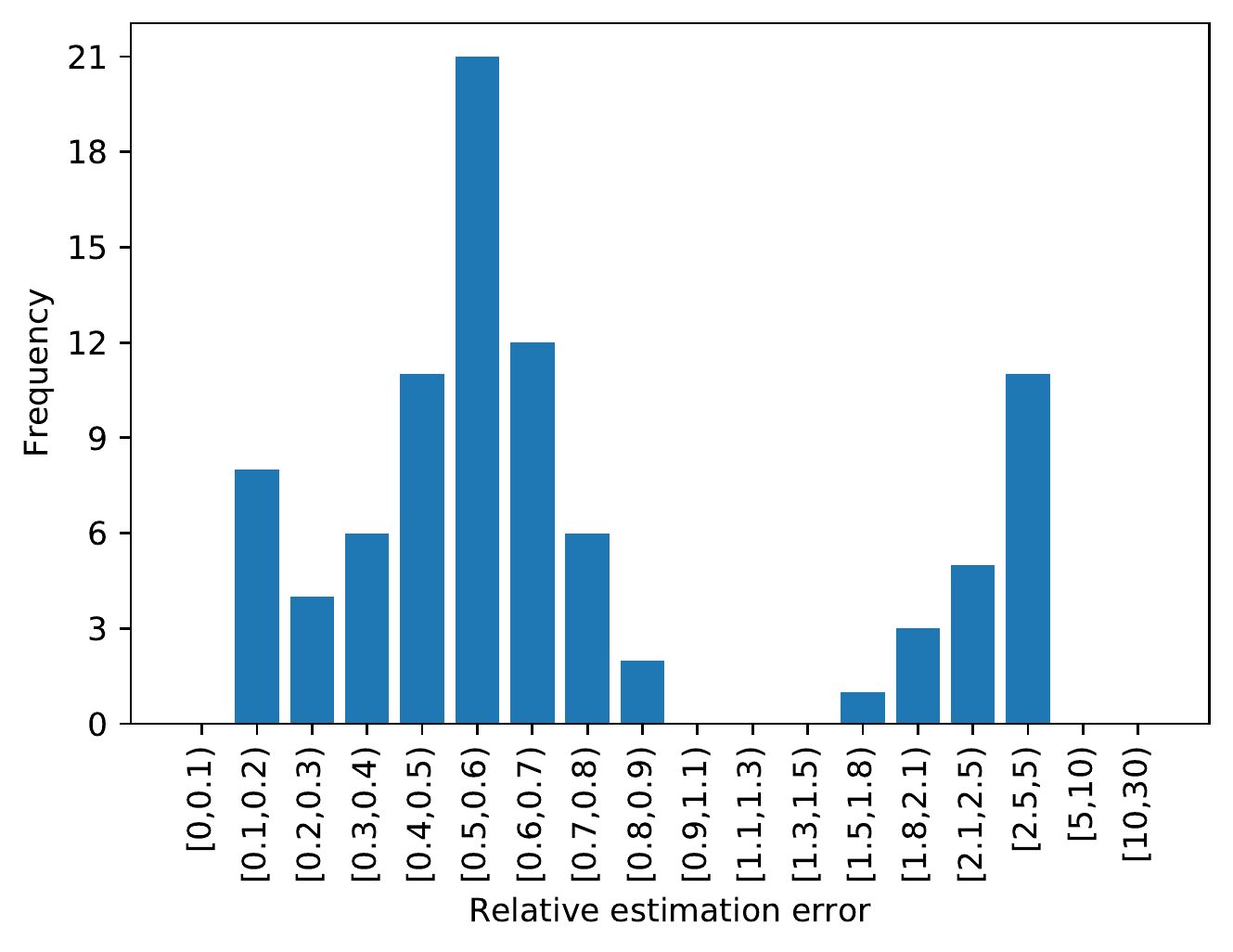}
         \caption{\texttt{l\_shipdate-l\_orderkey} Clustering}
         \label{fig:relative_errors_histogram_scan_3}
     \end{subfigure}
        \caption{The figure visualizes histograms  that describe the distribution of relative errors of our scan latency estimates for different clusterings. The figure contains only relative errors of scans that operate on the \texttt{lineitem} table. Each TPC-H query was executed 10 times.}
        \label{fig:relative_errors_histogram_scan}
\end{figure}

\paragraph{Relative estimation errors}
\Cref{fig:relative_errors_histogram_scan} visualizes the distribution of relative errors of our latency estimates for scans as a histogram, one for each of the three clusterings.
The histograms contain only scans that operate on the \texttt{lineitem} table.

\Cref{fig:relative_errors_histogram_scan_2} indicates that out model produces good scan latency estimates when clustering by \texttt{l\_partkey}:
All measured latencies are at most 30\% lower or higher than their estimate.
We believe that our model performs well in this case because neither \texttt{l\_orderkey} nor \texttt{l\_partkey} are correlated to a scan column.
As a consequence, scans are impacted by the change of clustering, which our model handles correctly.

When a scan column such as \texttt{l\_shipdate} is clustered, our latency estimates are less precise.
This is especially true if the table is clustered by multiple dimensions, as shown in \Cref{fig:relative_errors_histogram_scan_3}.
In some cases, our model over-estimates the scan latency by a factor of ten, in other cases the model under-estimates by a factor of three.
An analysis of those cases yields two sources of error: changes of physical query plans, and correlations.

\paragraph{Changing physical query plans}
\label{sec:eval_scans_changing_pqp}
So far, our model assumes that the physical query plan for a query does not depend on the clustering, i.e., changing the clustering does not change the PQP.
This assumption speeds up the clustering model because we do not have to call Hyrise's optimizer for each clustering candidate we evaluate.

\begin{figure}
    \centering
    \includegraphics[width=0.7\textwidth]{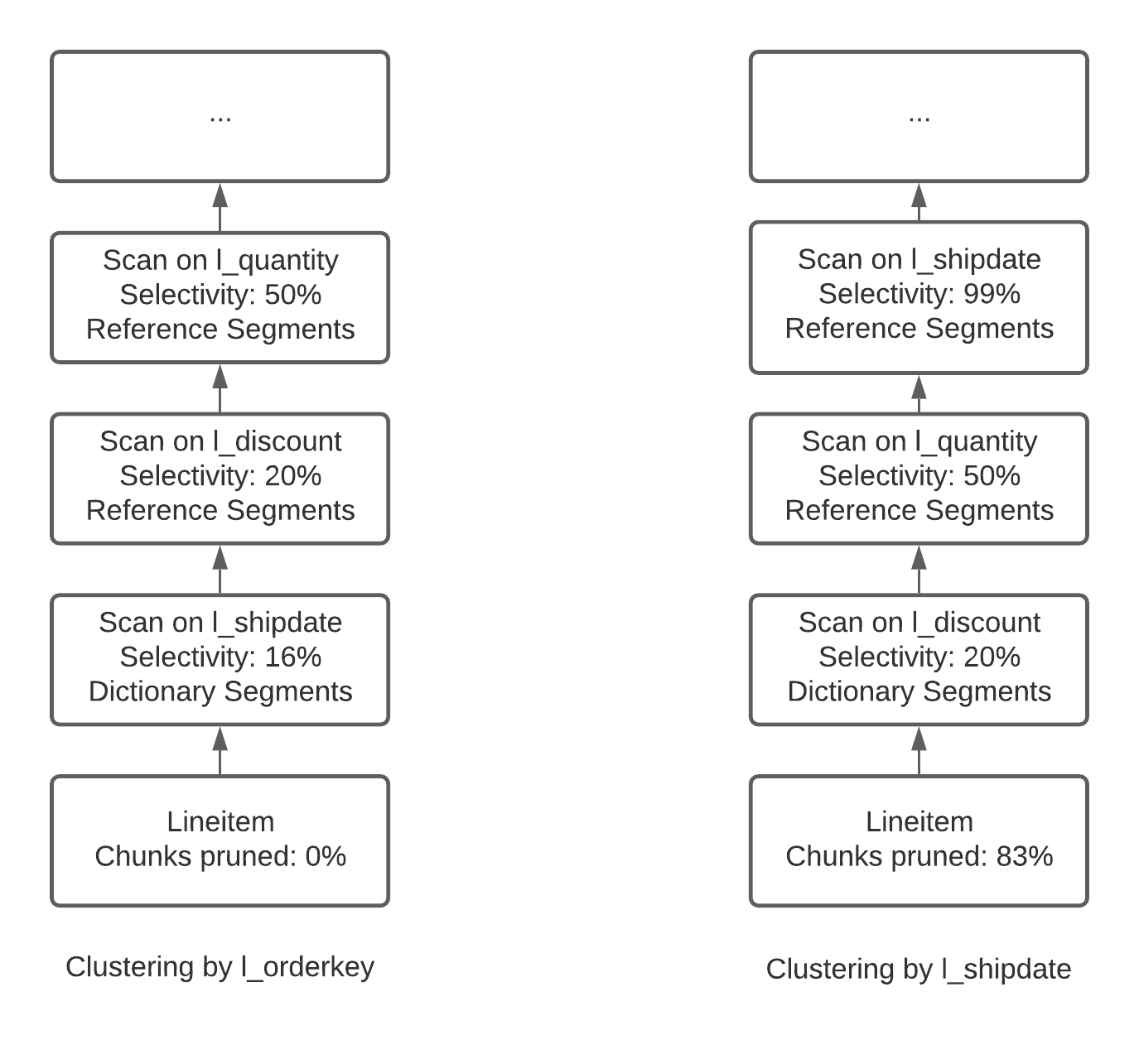}
    \caption{The figure visualizes how a physical query plan may change, depending on the clustering. We use query 6 of the analytical TPC-H benchmark as an example. When clustering by the column \texttt{l\_shipdate} (right), pruning can be applied, which increases the selectivity of the respective scan significantly. As a consequence, the scan is moved from the first to the last position and occurs now on reference segments. The other scans are moved down in the PQP. Since the scan on \texttt{l\_discount} is now the first operator, it occurs on dictionary instead of reference segments.}
    \label{fig:changing_pqp}
\end{figure}

However, the assumption is not always true:
For example, the TPC-H query 6 contains three scans: on \texttt{l\_shipdate}, \texttt{l\_discount}, and \texttt{l\_quantity}.
Hyrise orders these scans ascending by their expected selectivity, i.e., the scan with the lowest selectivity is executed first.
When the \texttt{lineitem} table is clustered by \texttt{l\_orderkey}, no chunks can be pruned, and the scan on \texttt{l\_shipdate} has the lowest selectivity.
If \texttt{lineitem} is clustered by \texttt{l\_shipdate} instead, 83\% of the table can be pruned.
Since most of the mismatching rows were pruned, the selectivity of the \texttt{l\_shipdate} scan is now high, and it is executed last.
This example is visualized in \Cref{fig:changing_pqp}.

This change in the scan order introduces two sources of error:
First, our model assumes that the \texttt{l\_shipdate} scan receives an input size similar to the part of \texttt{lineitem} that could not be pruned.
This were true if the scan was executed first; however, the scan is executed last, i.e., the scan's input size is further reduced by the selectivity of scans that occur beforehand.
In the case of TPC-H 6, this reduces the input by a factor of 10, i.e., exactly the factor by which our model over-estimates.

Second, the (originally) second scan, here \texttt{l\_discount}, is now the first scan.
Its input might increase slightly, but not too much, due to the pruning.
However, since it is now the first scan, it may now operate on dictionary segments rather than reference segments.
As a consequence, the scan may run up to thrice as fast, despite having the same number of input rows.

A similar effect can be observed for, e.g., the TPC-H queries 1 and 15, where a scan on \texttt{l\_shipdate} switches the position with a \texttt{Validate} operator.
\texttt{Validate} operators filter rows that are not visible to the current transaction and are subject to the same selectivity-based reordering as predicates.
The switch of positions causes the scan to operate on reference segments rather than dictionary segments.
Our model does not consider the change of the underlying segment type, which results in an under-estimate by factors of 3 and higher.

To address these issues, future work could assume that scans on a clustered column will always be executed as the last scan.
This assumption might not always be true, but it could help to relax the assumption of a static physical query plan, without the need of calling Hyrise's optimizer.

\paragraph{Correlations}
Another frequent source of errors are correlations.
If a column $X$ is correlated to a clustered column, scans on $X$ may benefit from the clustering as well.
Our model has only rudimentary support for correlations in scans:
If the scan column is correlated to a clustering column, we assume that the scan column is clustered, too, with a slightly worse granularity.
At the time of writing, the granularity is reduced by a constant.
Future work could integrate a measure for how strong two columns are correlated, and adapt the granularity reduction accordingly.

The current approach does not yet yield good estimates, but better estimates than not considering correlations at all.
For example, when clustering by \texttt{l\_shipdate}, the model over-estimates scans on the correlated column \texttt{l\_receiptdate} by a factor of up to six, if it is using our correlation handling approach.
Without our correlation handling approach, the model assumes that the latency of scans on \texttt{l\_receiptdate} does not change, which results in an over-estimation of factor twenty.

\subsubsection*{Hash Join's Probe Side Materialize Step}
\Cref{tab:aggregated_join_materialize_statistics} lists aggregated statistics for the probe side materialize step's latency estimates. 

\begin{table}[]
    \centering
    \begin{tabular}{ c|c|c|c|c }
        Clustering & \makecell{Sum estimated\\latencies (ms)} & \makecell{Sum measured\\latencies (ms)} &  MSE & SMAPE\\
        \hline
        l\_shipdate & 191\,587 & 158\,704 & 125\,600 & 34.22\% \\
        l\_partkey & 170\,005 & 154\,311 & 28\,139 & 30.40\% \\
        \makecell{l\_shipdate,\\l\_orderkey} & 152\,504 & 104\,777 & 118\'656 & 41.29\% \\
    \end{tabular}
    \caption{The table lists aggregated statistics for the probe side materialize step’s latency estimates. Each TPC-H query was executed 10 times.}
    \label{tab:aggregated_join_materialize_statistics}
\end{table}

\begin{figure}
     \centering
     \begin{subfigure}[b]{0.48\textwidth}
         \centering
         \includegraphics[width=\textwidth]{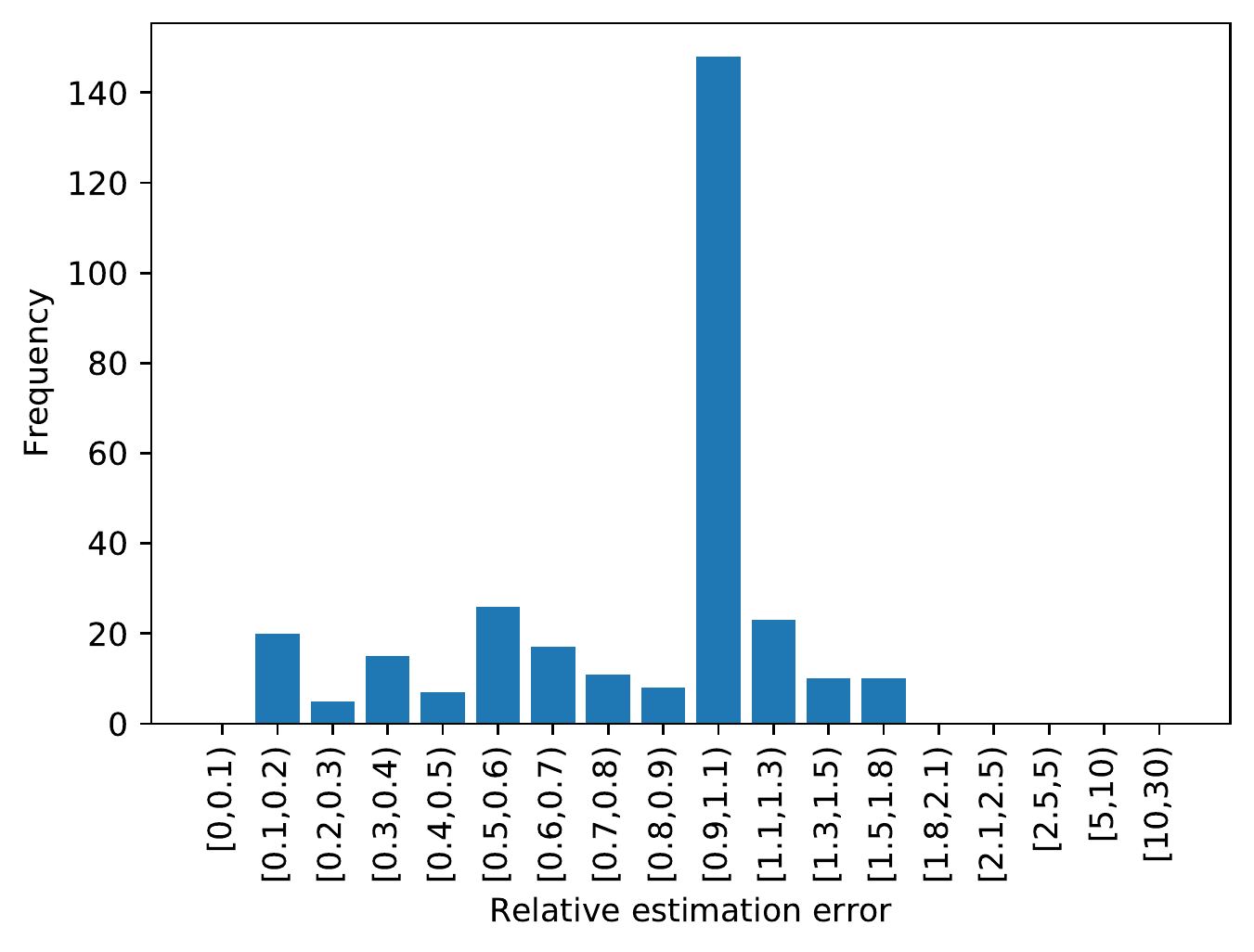}
         \caption{\texttt{l\_shipdate} Clustering - Probe side joins}
         \label{fig:relative_errors_histogram_join_materialize_1}
     \end{subfigure}
     \hfill
     \begin{subfigure}[b]{0.48\textwidth}
         \centering
         \includegraphics[width=\textwidth]{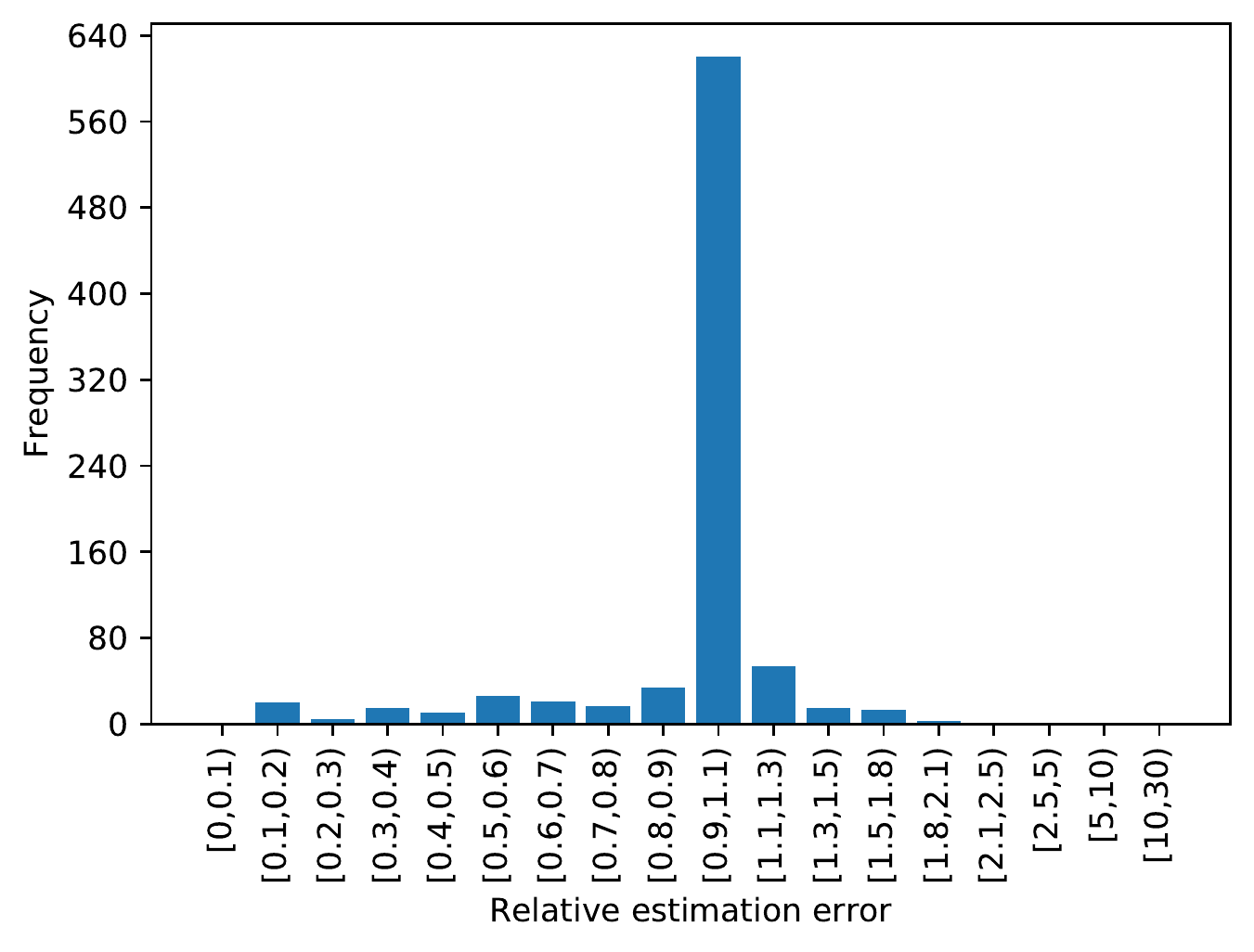}
         \caption{\texttt{l\_shipdate} Clustering - All joins}
         \label{fig:relative_errors_histogram_join_materialize_1_all}
     \end{subfigure}
     \hfill
     \begin{subfigure}[b]{0.48\textwidth}
         \centering
         \includegraphics[width=\textwidth]{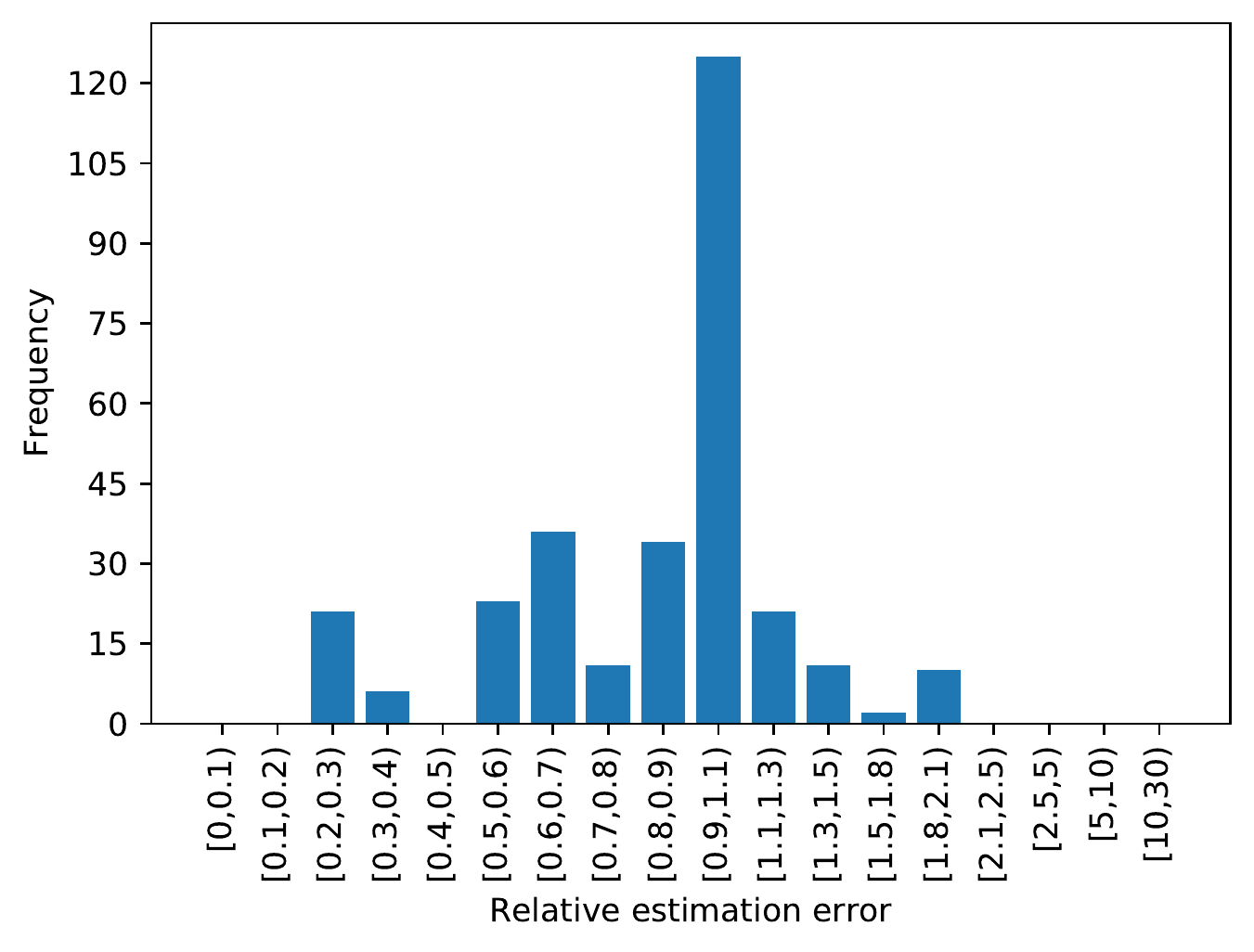}
         \caption{\texttt{l\_partkey} Clustering - Probe side joins}
         \label{fig:relative_errors_histogram_join_materialize_2}
     \end{subfigure}
     \hfill
     \begin{subfigure}[b]{0.48\textwidth}
         \centering
         \includegraphics[width=\textwidth]{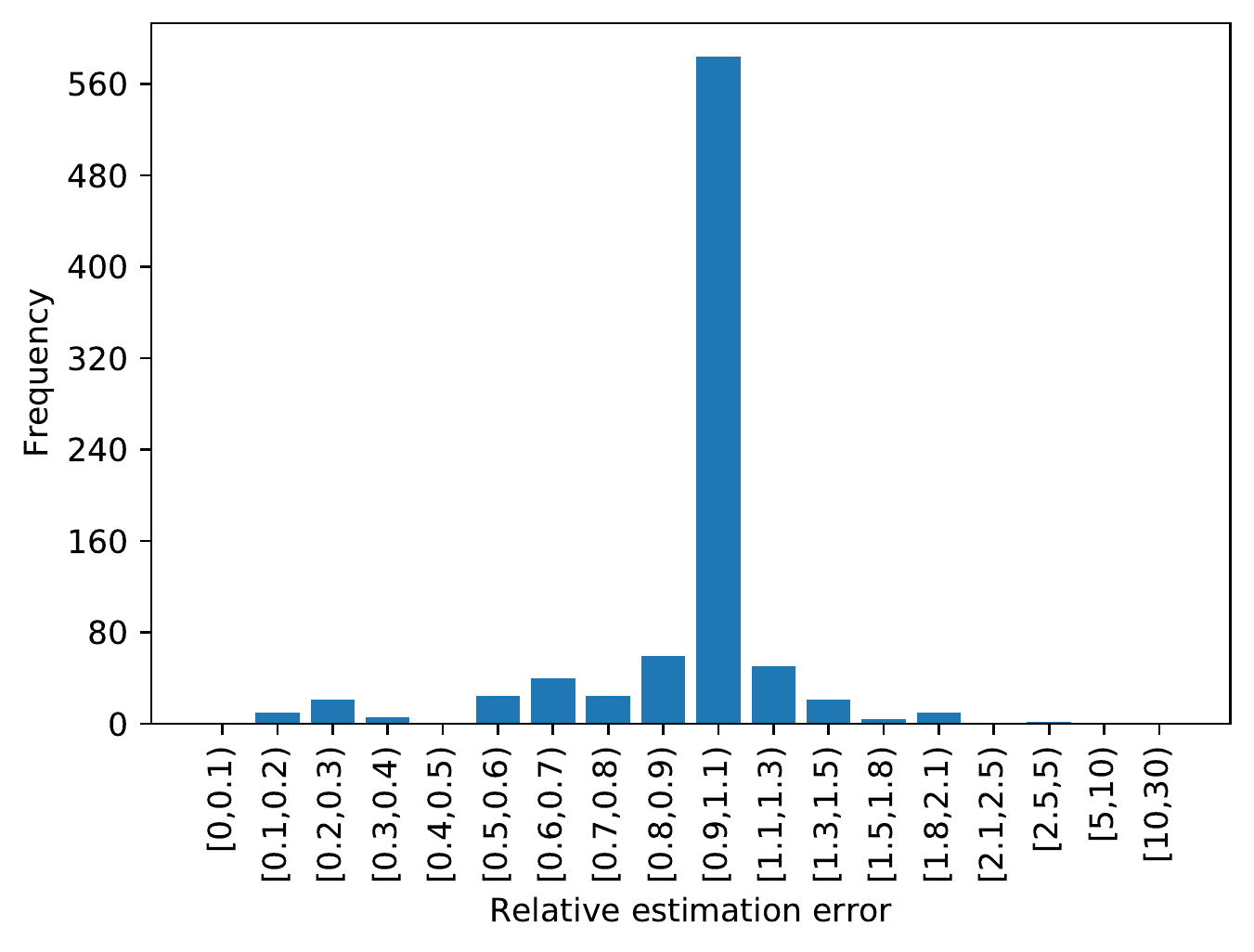}
         \caption{\texttt{l\_partkey} Clustering - All joins}
         \label{fig:relative_errors_histogram_join_materialize_2_all}
     \end{subfigure}
     \hfill
          \begin{subfigure}[b]{0.48\textwidth}
         \centering
         \includegraphics[width=\textwidth]{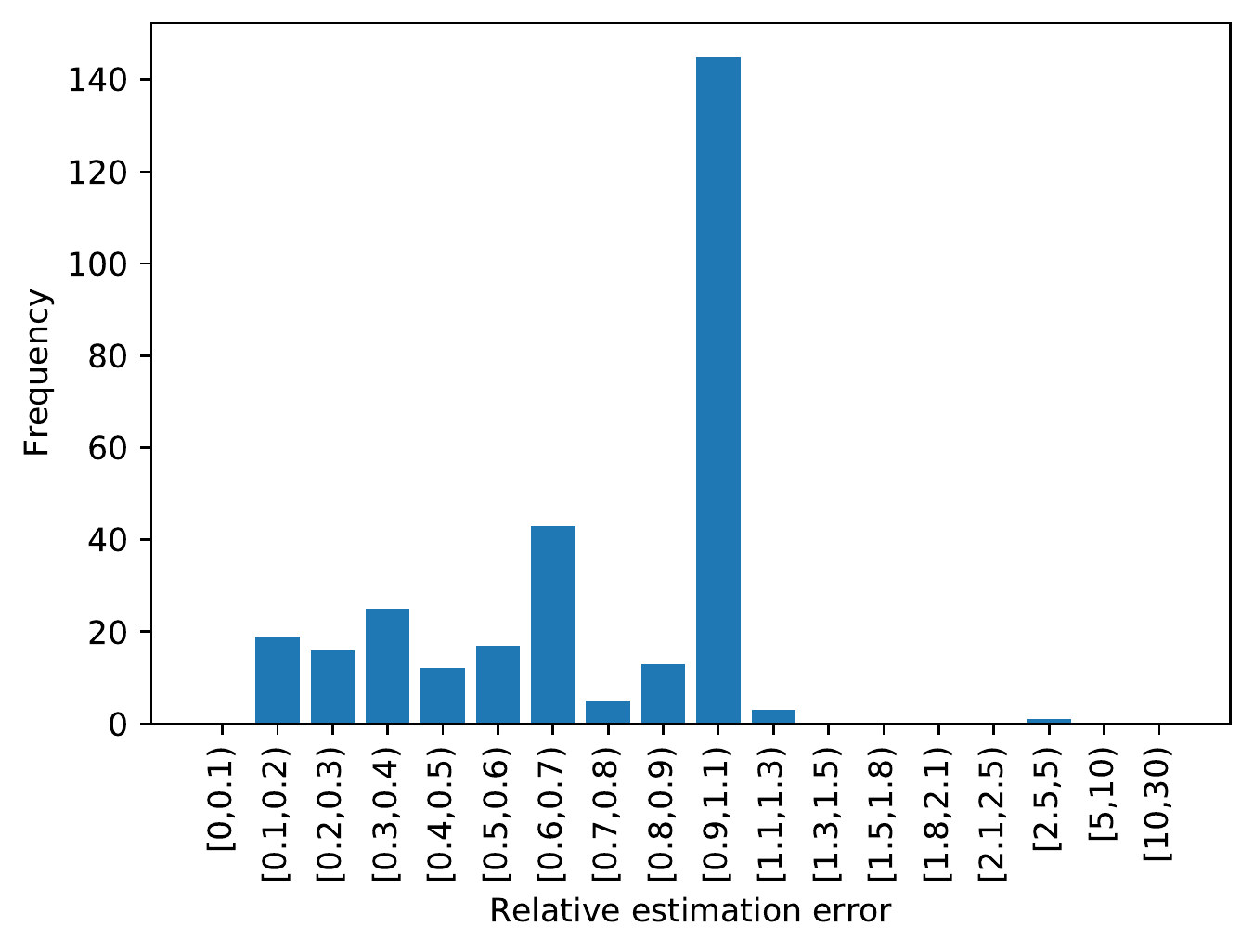}
         \caption{\texttt{l\_shipdate-l\_orderkey} Clustering - Probe side joins}
         \label{fig:relative_errors_histogram_join_materialize_3}
     \end{subfigure}
     \hfill
     \begin{subfigure}[b]{0.48\textwidth}
         \centering
         \includegraphics[width=\textwidth]{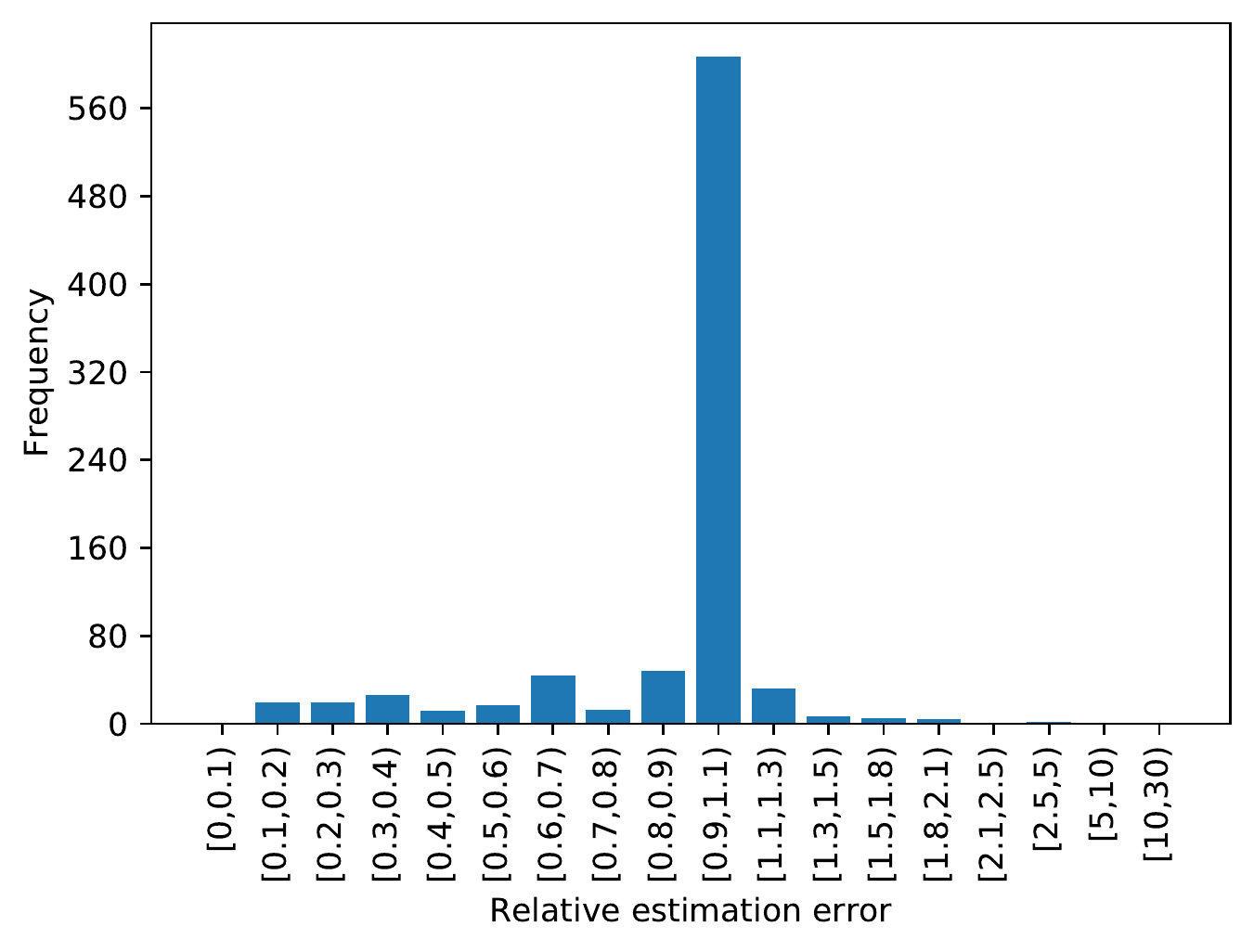}
         \caption{\texttt{l\_shipdate-l\_orderkey} Clustering - All joins}
         \label{fig:relative_errors_histogram_join_materialize_3_all}
     \end{subfigure}
        \caption{The figure visualizes histograms that describe the distribution of relative estimate errors for the hash join's probe side materialize step, for three different clusterings. Each TPC-H query was executed ten times.}
        \label{fig:relative_errors_histogram_join_materialize}
\end{figure}

\Cref{fig:relative_errors_histogram_join_materialize} visualizes the distribution of relative errors of our latency estimates for the hash join's probe side materialize step as a histogram.
The figure shows two histograms per clustering:
The histograms on the left side include only joins where \texttt{lineitem} is used as the probe table.
Histograms on the right side contain relative estimate errors for all joins.

We conclude that our model performs reasonably well at estimating the latency of the probe side materialize step:
The histograms on left side indicate that the measured latency is at most 30\% lower or higher than the estimated latency for the major part of the joins.

The histograms on the right side also include joins where \texttt{lineitem} is not used as the probe table.
Our model assumes that the probe side materialize step of those joins is not affected and predicts the current latency.
Considering the histograms on the right side, we conclude that the assumption is mostly correct:
While there are some joins whose latencies changes significantly, the major part changes by at most 10\%.

Nevertheless, there are cases - especially when \texttt{l\_shipdate} is clustered - where our model over-estimates the latency by factors up to ten.
An analysis of those cases yields three sources of error: predicates above semi joins, bad chunk density estimates caused by correlations, and varying densities of matching rows (i.e., the number of chunks that contain matching rows).

\paragraph{Predicates above semi joins}
When constructing a query execution plan, Hyrise usually pushes scans below joins, i.e., in most cases, scans are executed before joins.
Based on that assumption, we conclude for our model that a join's input size cannot be modified by the clustering:
The clustering may modify the input size of scans (e.g., via pruning); but if all scans are executed first, the join's input size remains constant, as the scans filter a superset of the pruned rows.

\begin{figure}
    \centering
    \includegraphics[width=0.7\textwidth]{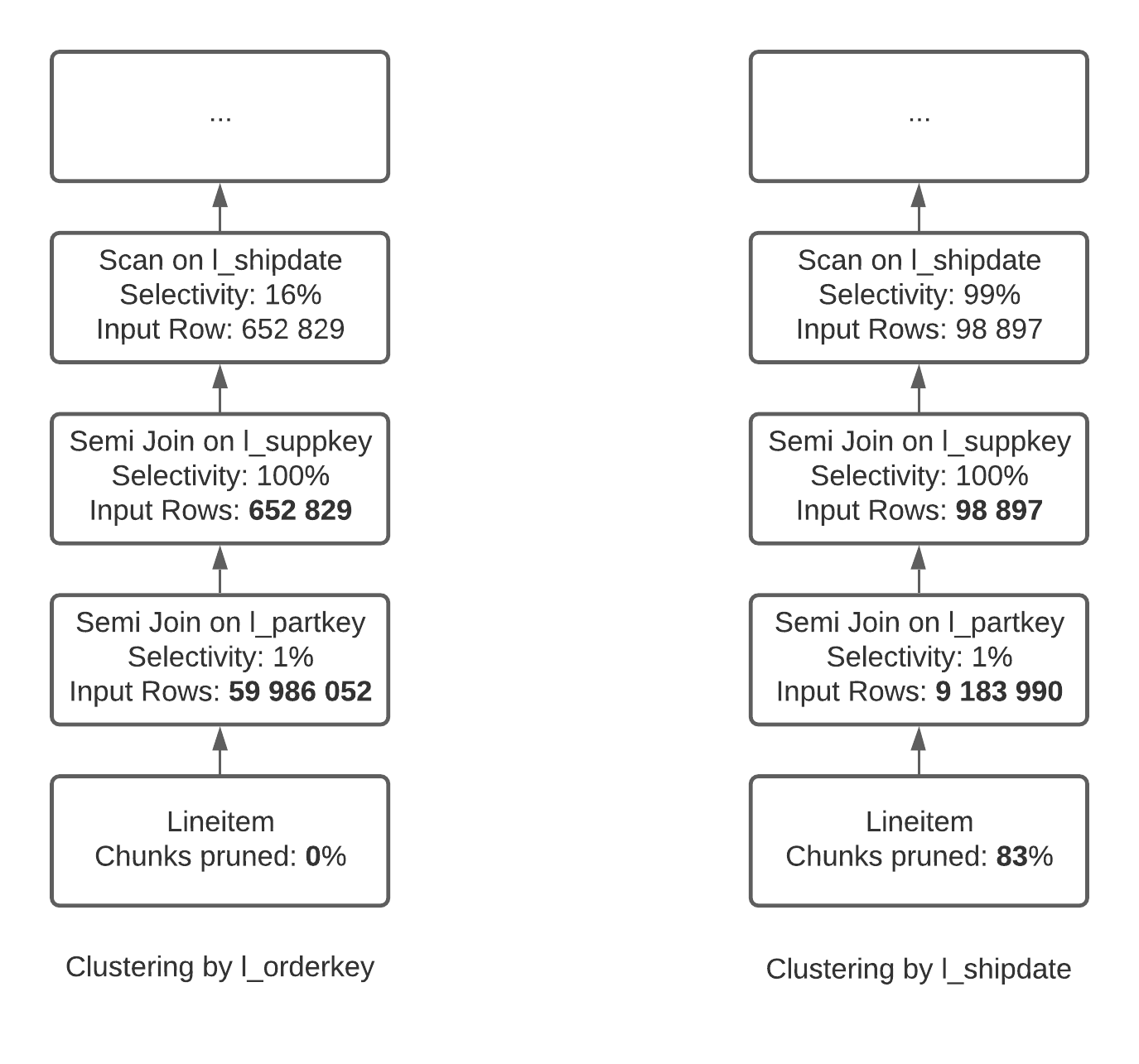}
    \caption{The figure visualizes how predicates above semi joins can impact a semi join's latency. We use query $20$ of the analytical TPC-H benchmark as example.}
    \label{fig:predicate_above_semijoin}
\end{figure}

However, there are cases where semi joins are executed before scans:
Semi joins reduce the number of rows of a single table, and can thus be seen as another predicate with a certain selectivity.
Hyrise orders predicates ascending by the selectivity, i.e., the predicate with the lowest selectivity is executed first.
When reordering the predicates, Hyrise considers semi joins as predicates and may move them below scans.
\Cref{fig:predicate_above_semijoin} visualizes such a case at the example of query 20 of the TPC-H benchmark.

If we cluster by a column that is scanned above semi joins, the input size of semi joins can change:
Even if the PQP remains unchanged, i.e., the scan on the clustering column is still executed after the join, the scan may enable pruning and thereby reduce the join's input size.
In the case of TPC-H 20, the join's input size is reduced by approximately 83\%.
Since there are fewer rows to materialize, the latency of the materialize step decreases accordingly.
Our model does not reflect this reduced number of input rows, which results in an over-estimate by the factor six.
To address such errors, future work could analyze the PQP to identify cases where semi joins are pushed below scans on clustered columns, and apply an input size reduction to the semi joins.

\paragraph{Impact of correlations on chunk density}
In most cases, the clustering will not impact the number of join input rows.
However, the clustering may impact the number of different chunks the input rows belong to.
We have observed that the materialize step runs slower when the same number of rows is distributed about a larger number of chunks.
To describe this behavior, we have introduced the concept of chunk density, which can be interpreted as an average chunk fill ratio.
The chunk density is estimated by multiplying the selectivities of scans on unclustered columns.
Scans on clustered columns are assumed to select whole chunks rather than individual rows, and thus do not affect the chunk density.

This method of estimating the chunk density does not consider correlations.
If two columns are correlated, clustering by one of them also clusters by the other one, with a lower granularity.
Thus, with respect to the chunk density, scans on columns correlated to clustered columns rather behave like scans on clustered columns than on unclustered columns.

For example, consider query $12$ of the TPC-H benchmark:
The query contains a scan on \texttt{l\_receiptdate} that is executed before a join on \texttt{l\_orderkey}.
The join has approximately $300\,000$ input rows.
When using the clustering produced by the benchmark generator (\texttt{l\_orderkey}), those rows are distributed over more than $1900$ chunks; when clustering by \texttt{l\_shipdate}, the same number of rows is distributed over $300$ chunks.

This change of the chunk density is not reflected by our model.
For TPC-H 12, it causes our model to over-estimate the materialize step's latency by factor 2.
To address this source of error, future work could include knowledge about correlations when estimating the chunk density.

\begin{figure}
    \centering
    \includegraphics[width=\textwidth]{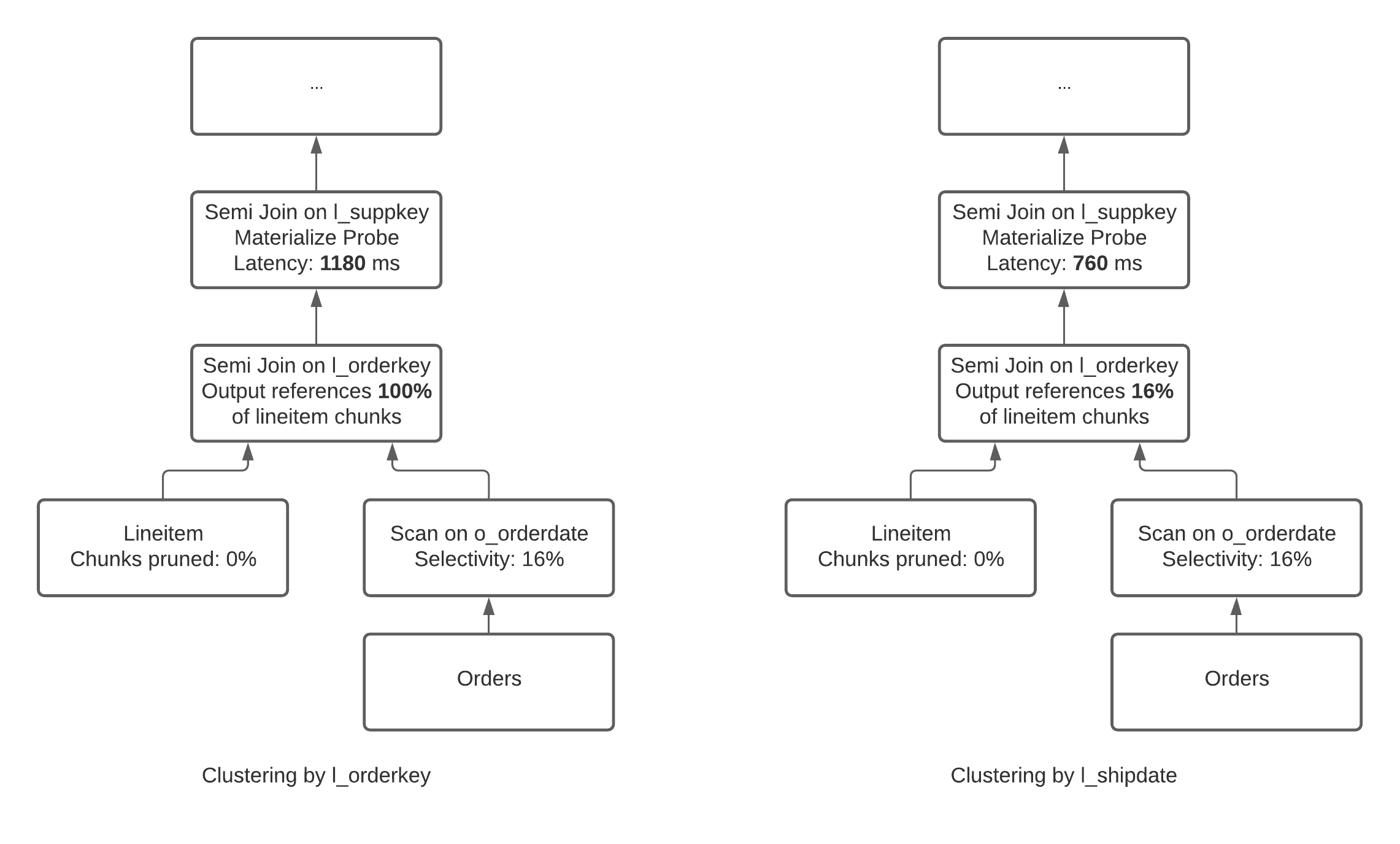}
    \caption{The figure visualizes how clustering can impact the density of matching rows if the other join input is filtered by a correlated column. The figure uses query 5 of the TPC-H benchmark as example.}
    \label{fig:matching_rows_density}
\end{figure}

\paragraph{Varying density of matching rows}
In most cases, the clustering will not impact the number of join output rows.
However, the clustering may impact the number of chunks that contain matching tuples, e.g., when the other join table is filtered on a correlated column before the join.
For example, consider the semi join between \texttt{lineitem} and \texttt{orders} from TPC-H query 5, which is visualized in \Cref{fig:matching_rows_density}.

Before the join is executed, the \texttt{orders} table is scanned on the column \texttt{o\_orderdate}, which is correlated to \texttt{l\_shipdate}.
If we cluster \texttt{lineitem} by \texttt{l\_orderkey}, tuples from all chunks will match in the semi join.
If we cluster by \texttt{l\_shipdate} instead, the matching tuples belong to a smaller number of chunks.
With the \texttt{l\_shipdate} clustering, any join on \texttt{lineitem} that is executed after the join with \texttt{orders} will have to access fewer dictionary segments to materialize its input values.
This holds even if radix partitioning is applied.

Our model does not consider the density of matching rows.
For TPC-H 5, this causes our model to over-estimate the latency by 50\%.
To address this error, future work could introduce a concept of density for rows that match the join condition, and incorporate this density into the latency estimations for joins placed further up in the PQP.

\subsubsection*{Hash Join's Probing Step}
\Cref{tab:aggregated_join_probe_statistics} lists aggregated statistics for the probing step's latency estimates. 

\begin{table}[]
    \centering
    \begin{tabular}{ c|c|c|c|c }
        Clustering & \makecell{Sum estimated\\latencies (ms)} & \makecell{Sum measured\\latencies (ms)} &  MSE & SMAPE\\
        \hline
        l\_shipdate & 87\,130 & 79\,242 & 36\,448 & 28.75\% \\
        l\_partkey & 86\,289 & 94\,339 & 17\,495 & 18.45\% \\
        \makecell{l\_shipdate,\\l\_orderkey} & 73\,163 & 66\,418 & 15\'110 & 24.05\% \\
    \end{tabular}
    \caption{The table lists aggregated statistics for the probing step's latency estimates. Each TPC-H query was executed ten times.}
    \label{tab:aggregated_join_probe_statistics}
\end{table}

\begin{figure}
     \centering
     \begin{subfigure}[b]{0.48\textwidth}
         \centering
         \includegraphics[width=\textwidth]{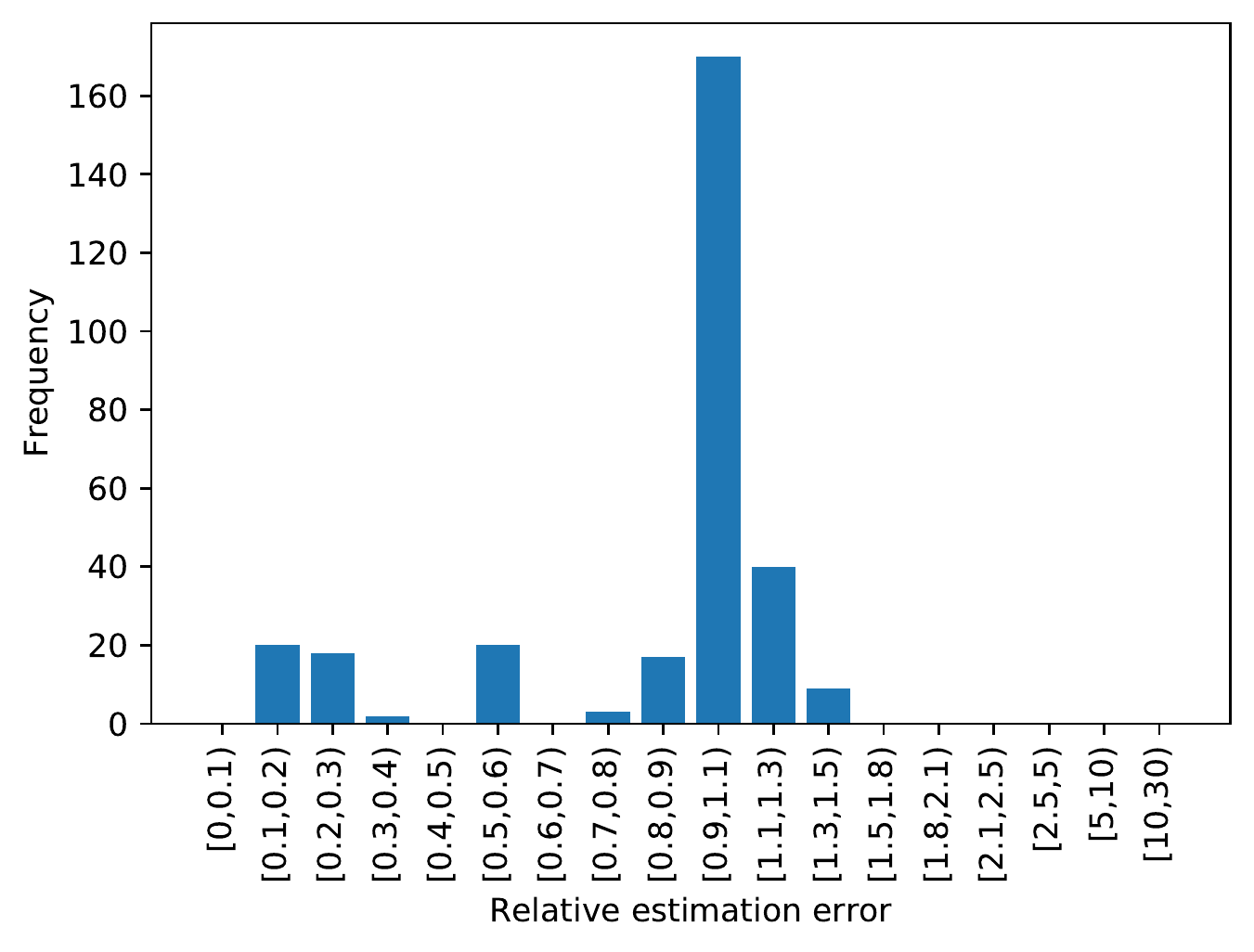}
         \caption{\texttt{l\_shipdate} Clustering - Probe side joins}
         \label{fig:relative_errors_histogram_join_probe_1}
     \end{subfigure}
     \hfill
     \begin{subfigure}[b]{0.48\textwidth}
         \centering
         \includegraphics[width=\textwidth]{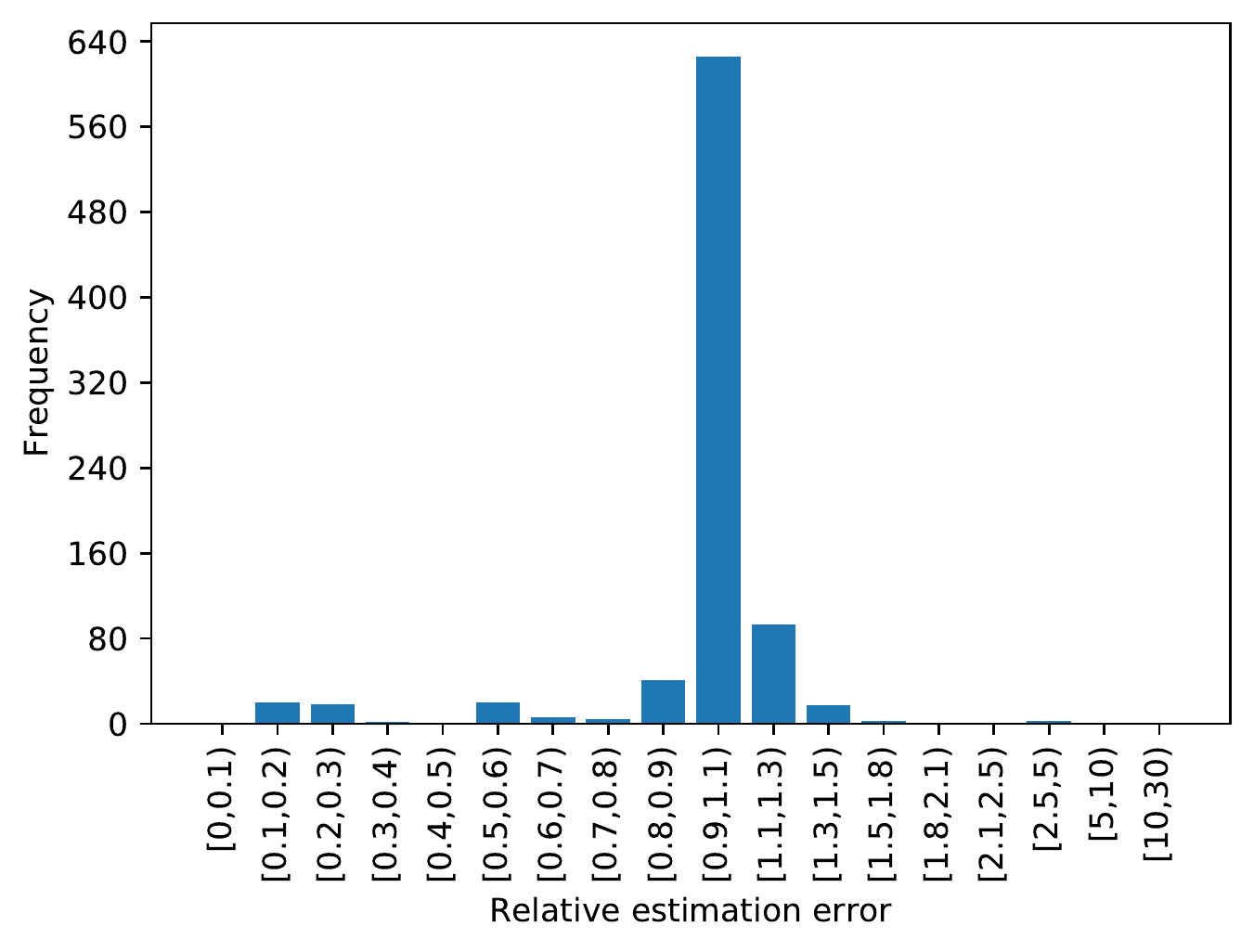}
         \caption{\texttt{l\_shipdate} Clustering - All joins}
         \label{fig:relative_errors_histogram_join_probe_1_all}
     \end{subfigure}
     \hfill
     \begin{subfigure}[b]{0.48\textwidth}
         \centering
         \includegraphics[width=\textwidth]{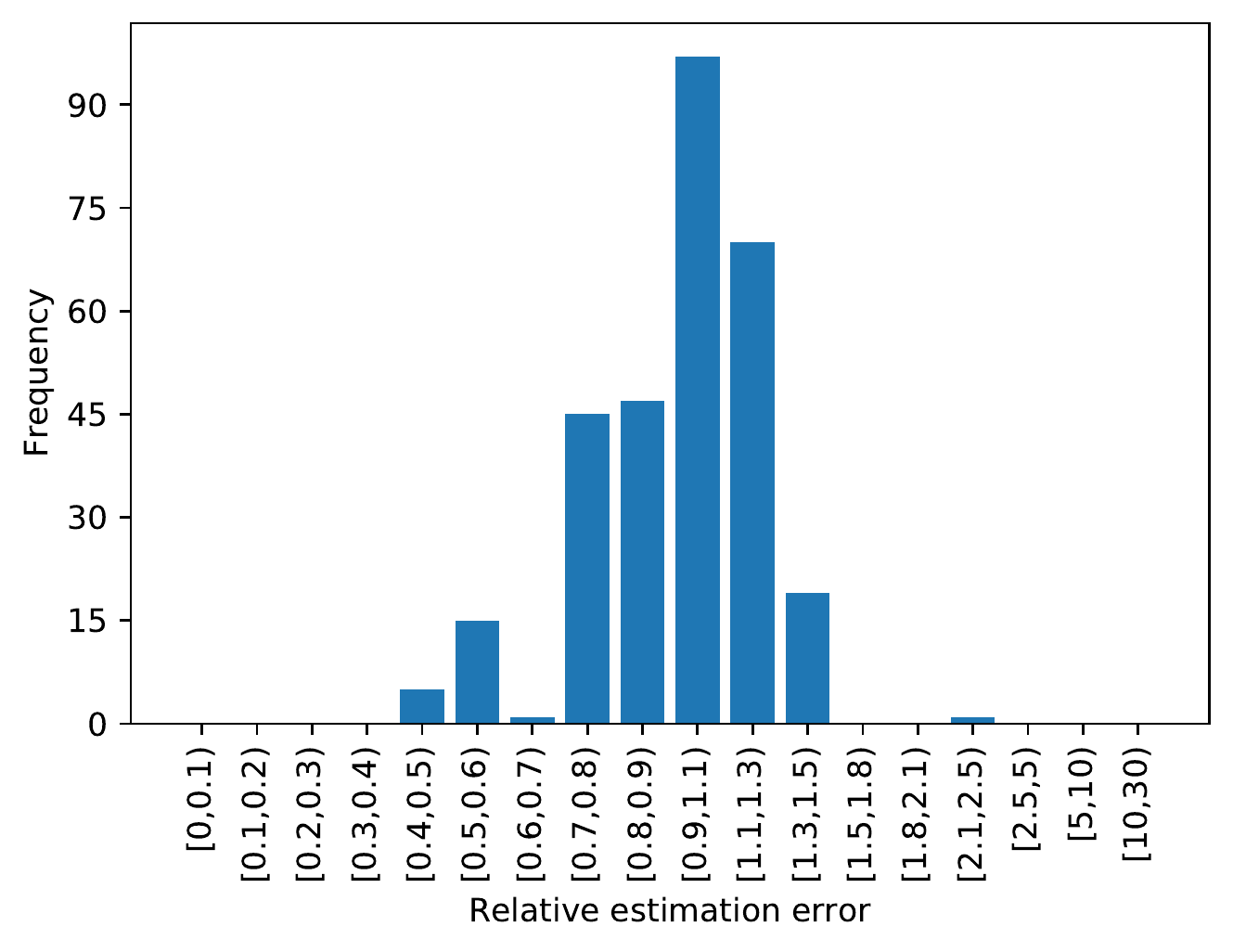}
         \caption{\texttt{l\_partkey} Clustering - Probe side joins}
         \label{fig:relative_errors_histogram_join_probe_2}
     \end{subfigure}
     \hfill
     \begin{subfigure}[b]{0.48\textwidth}
         \centering
         \includegraphics[width=\textwidth]{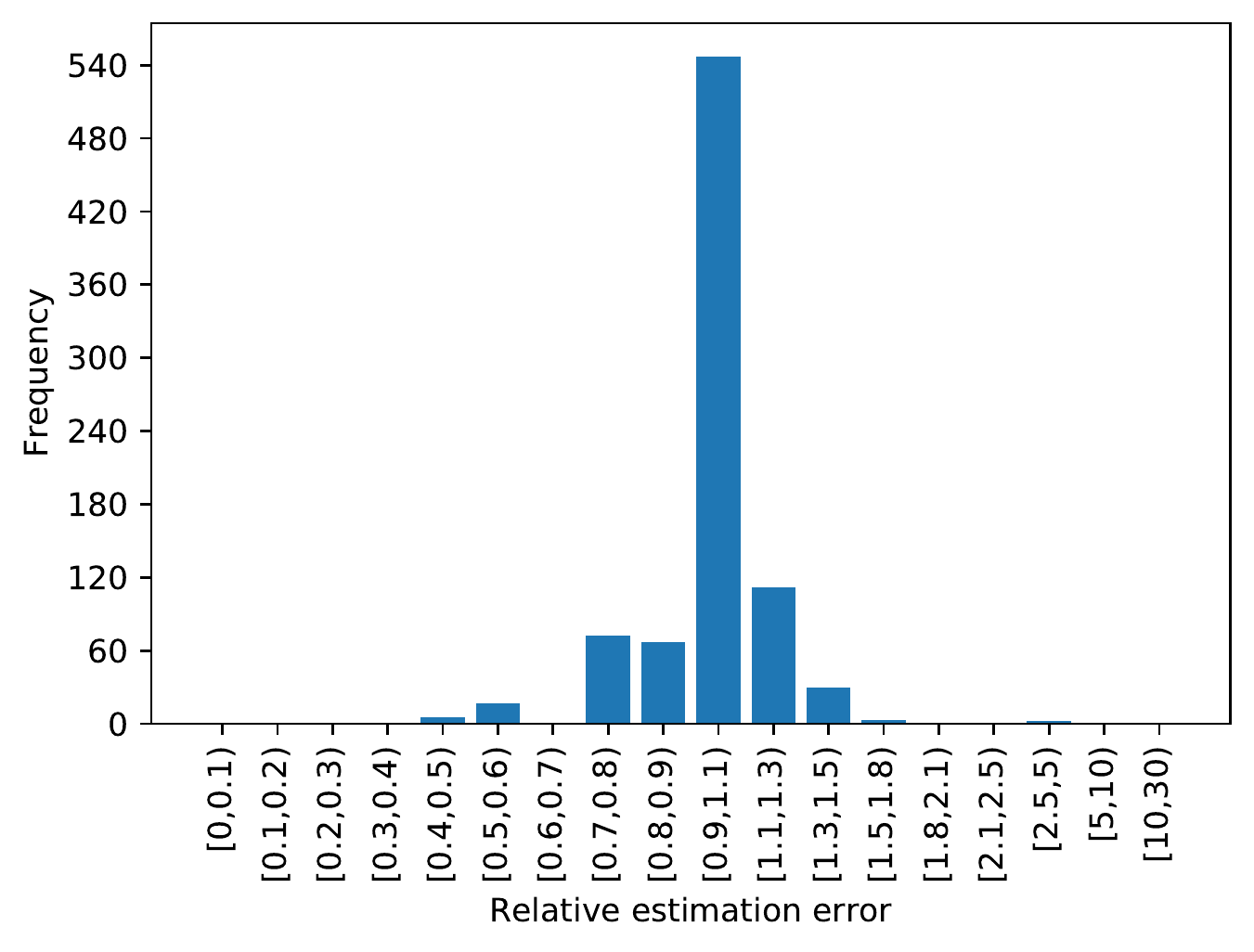}
         \caption{\texttt{l\_partkey} Clustering - All joins}
         \label{fig:relative_errors_histogram_join_probe_2_all}
     \end{subfigure}
     \hfill
          \begin{subfigure}[b]{0.48\textwidth}
         \centering
         \includegraphics[width=\textwidth]{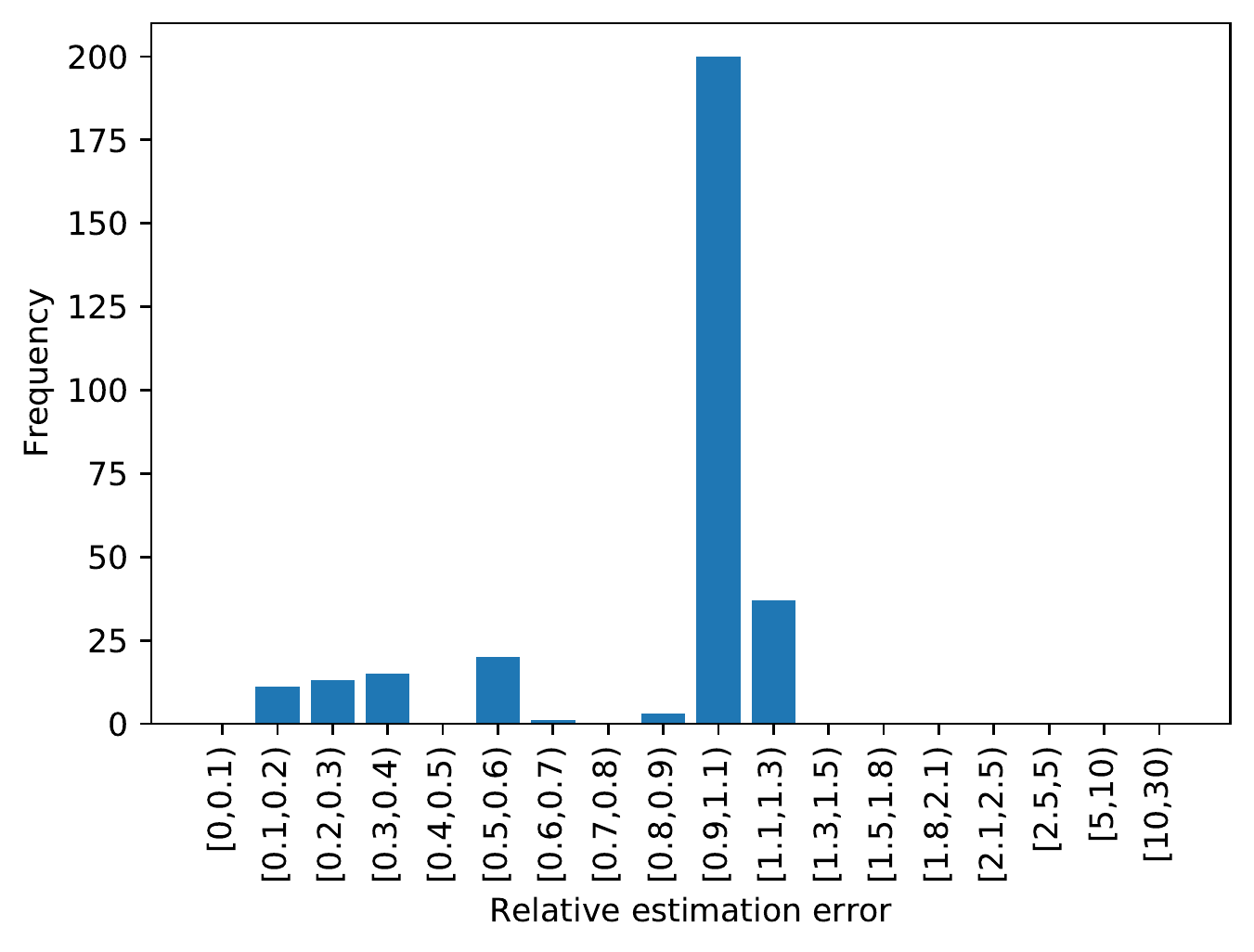}
         \caption{\texttt{l\_shipdate-l\_orderkey} Clustering - Probe side joins}
         \label{fig:relative_errors_histogram_join_probe_3}
     \end{subfigure}
     \hfill
     \begin{subfigure}[b]{0.48\textwidth}
         \centering
         \includegraphics[width=\textwidth]{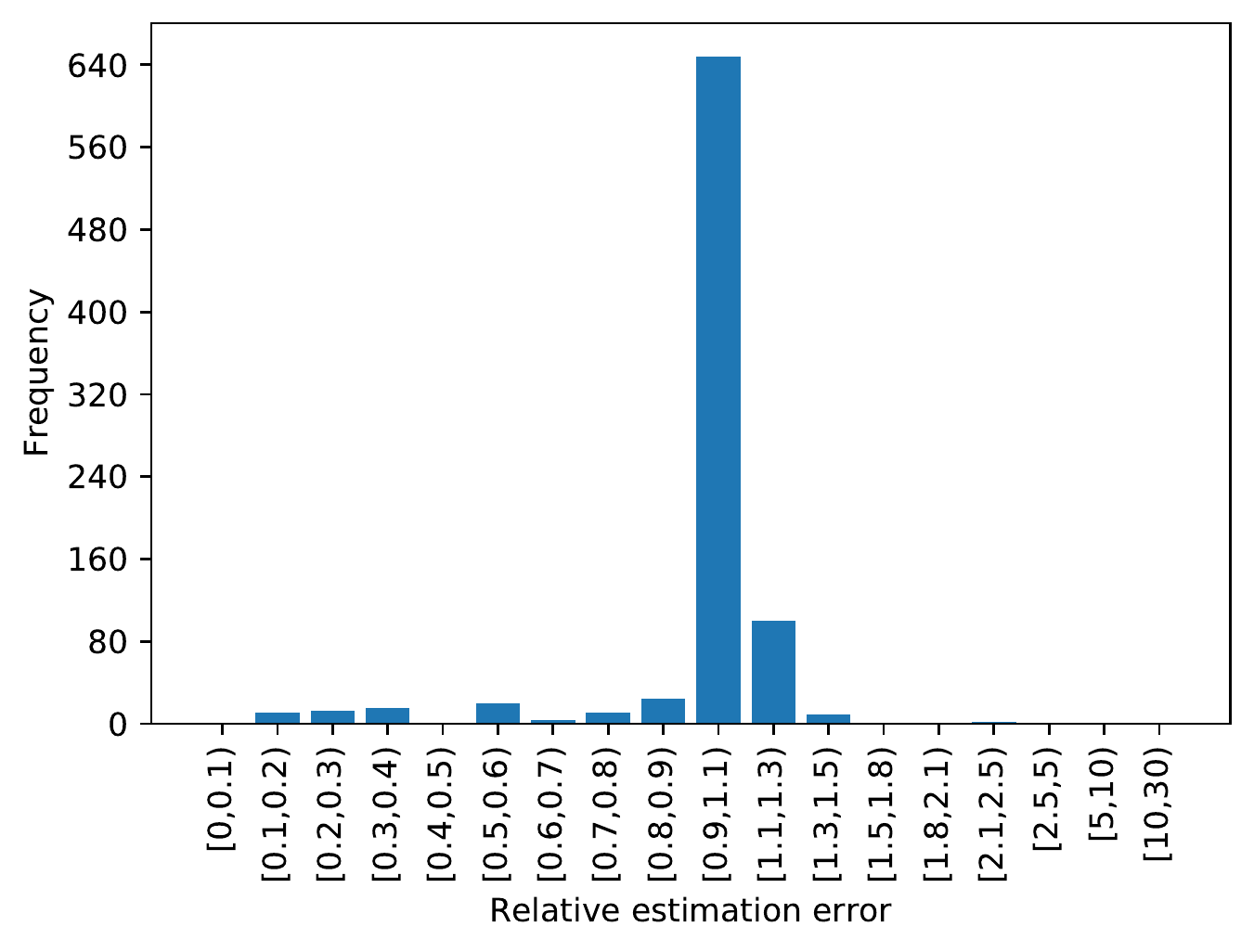}
         \caption{\texttt{l\_shipdate-l\_orderkey} Clustering - All joins}
         \label{fig:relative_errors_histogram_join_probe_3_all}
     \end{subfigure}
        \caption{The figure visualizes histograms that describe the distribution of relative estimate errors for the hash join's probing step, for three different clusterings. Each TPC-H query was executed ten times.}
        \label{fig:relative_errors_histogram_join_probe}
\end{figure}

\Cref{fig:relative_errors_histogram_join_probe} visualizes the distribution of relative errors of our latency estimates for the hash join's probing step as a histogram.
The figure shows two histograms per clustering:
The histograms on the left side include only joins where \texttt{lineitem} is used as the probe table.
Histograms on the right side contain relative estimate errors for all joins.

We observe that for all histograms, the major part of the measured latencies is within $30\%$ of their estimates, which we consider acceptable for the scope of this thesis.
When clustering by \texttt{l\_shipdate}, there are some joins whose run time is over-estimated by factor three to seven.
Those joins belong to the TPC-H queries $7$ and $20$.
An analysis of those joins shows that the over-estimation can be attributed to the predicate-above-semijoin error source, which was already discussed in the evaluation of the materialize step's latency.

\subsubsection*{Unaffected Hash Join Steps}
Our model assumes that the latency of the hash join's clustering, building and output writing steps is not affected by the clustering.
To verify this assumption, we provide histograms that visualize the distribution of the relative estimation errors.
Since our model estimates an unchanged value, the relative error is equivalent to the fraction $\frac{\text{measured latency new}}{\text{measured latency old}}$, i.e., by how much the measured latency changes when the clustering is switched.
The histograms are visualized in \Cref{fig:relative_errors_histogram_join_clustering}, \Cref{fig:relative_errors_histogram_join_build}, and \Cref{fig:relative_errors_histogram_join_output_writing}, respectively.
Although the latency does change for some joins, the major part remains unaffected or changes only slightly.
We thus conclude that it is valid for our model to assume that the clustering, building and output writing step are unaffected by the clustering.

\subsection{Computation Duration}
In this section, we evaluate the model's computation time.
We analyze the model's asymptotic complexity, show measured computation durations for specific instances, and provide suggestions on how to reduce the model's computation duration.

\paragraph{Asymptotic complexity}
Let $c$ be the number of interesting columns (both scan and join columns), and $d$ the maximum dimensionality.
The number of unique combinations of clustering columns of at most size $d$ can be calculated as $\sum\limits_{i=1}^{d} \binom{c}{i}$, which is upper-bounded by the term $c^d$.
For each of those combinations, the model considers $c$ sort orders to obtain clustering candidates.
For each of the clustering candidates, our model considers $s$ cluster count sets.
Overall, our model thus considers $s \cdot c^{d+1}$ clustering configurations.

The most expensive operations used per clustering configuration have a complexity of $\mathcal{O}(n \log(n))$.
Thus, our model has an overall asymptotical complexity of $\mathcal{O}(n \log(n) \cdot s \cdot c^{d+1})$, where $n$ is the number of scans and joins, $d$ the maximum dimensionality, $c$ the number of interesting clustering columns, and $s$ the number of cluster count sets.

\paragraph{Concrete computation durations}
The computation duration depends on various factors, such as the number of interesting columns, the number of unique scans and joins in the workload, and the maximum dimensionality of clusterings to consider.

\Cref{tab:model_computation_time} lists those statistics for the tables \texttt{lineitem} and \texttt{store\_sales}.
The computation duration was measured on an \texttt{Intel(R) Core(TM) i7-4700HQ CPU @ 2.40GHz} processor.
All calculations were performed single-threaded.
The measurements are based on commit \texttt{97fe2f4}\footnote{\url{https://github.com/aloeser/clustering_model/tree/97fe2f4361505d9aac3d0c4a4a1e7b60f167e16a}}.

\begin{table}[]
    \centering
    \begin{tabular}{ c|c|c|c|c|c|c|c  }
        Table & \makecell{Runs per\\query} & \makecell{Interesting\\columns} & \makecell{Unique\\scans} & \makecell{Unique\\joins} & $d$ & \makecell{Confi-\\gurations} & \makecell{Run time\\(min)}\\
        \hline
            lineitem & 10 & 7 & 90 & 785 & 2 & 196 & 12 \\
            lineitem & 10 & 7 & 90 & 785 & 3 & 441 & 29 \\
            lineitem & 100 & 7 & 606 & 5816 &  2 & 196 & 122 \\
            lineitem & 100 & 7 & 606 & 5816 &  2 & 441 & 282 \\
            store\_sales & 1 & 10 & 24 & 347 & 2 & 550 & 14 \\
            store\_sales & 1 & 10 & 24 & 347 & 3 & 1750 & 40 \\
    \end{tabular}
    \caption{The table lists statistics about the model's computation time. In all cases, our model considered only one set of cluster counts per clustering candidate.}
    \label{tab:model_computation_time}
\end{table}

\paragraph{Optimization potential}
During this work, we did not focus on the model's computation time.
Consequently, efficiency of the model was not within the scope of this thesis.
If the model's computation time becomes a bottle neck, its run time could be improved in many ways:
Currently, the model is implemented in Python which is an interpreted scripting language.
Switching to a compiled and natively-executed language, e.g., C++, should yield significant performance improvements.
Further, multi-threading can be used to process multiple clustering configurations in parallel.
Finally, there may be results that can be reused between clustering configurations, without repeated computation.

If the model's computation time is still a bottleneck, search space pruning could be employed, e.g., considering columns only as interesting if operators spend at least a fixed amount of time on them.
Agrawal et al.~\cite{DBLP:conf/sigmod/AgrawalNY04} describe enhanced techniques for search space pruning.

\section{MVCC-Aware Clustering Algorithm}
In this section, we evaluate the clustering algorithm presented in \Cref{sec:clustering_algorithm} regarding the following aspects: its choice of cluster boundaries, its run time, its robustness against concurrent modifications, and its memory consumption.
All benchmarks in this section are based on commit \texttt{11b2d3f}\footnote{\url{https://github.com/aloeser/hyrise/tree/11b2d3f43481f435facaf8a017b9f4b9f56b0471}}.

\subsection{Cluster Boundaries}
In this section, we evaluate our algorithm for determining cluster boundaries, which is described in \Cref{sec:choose_cluster_boundaries}.
The algorithm receives a cluster count as input that determines the requested number of boundaries; however, the algorithm may return more or fewer cluster boundaries than requested.
We evaluate the algorithm's precision regarding the requested and the obtained number of boundaries, as well as its run time.

\begin{figure}
    \centering
    \begin{subfigure}[b]{0.48\textwidth}
         \centering
         \includegraphics[width=\textwidth]{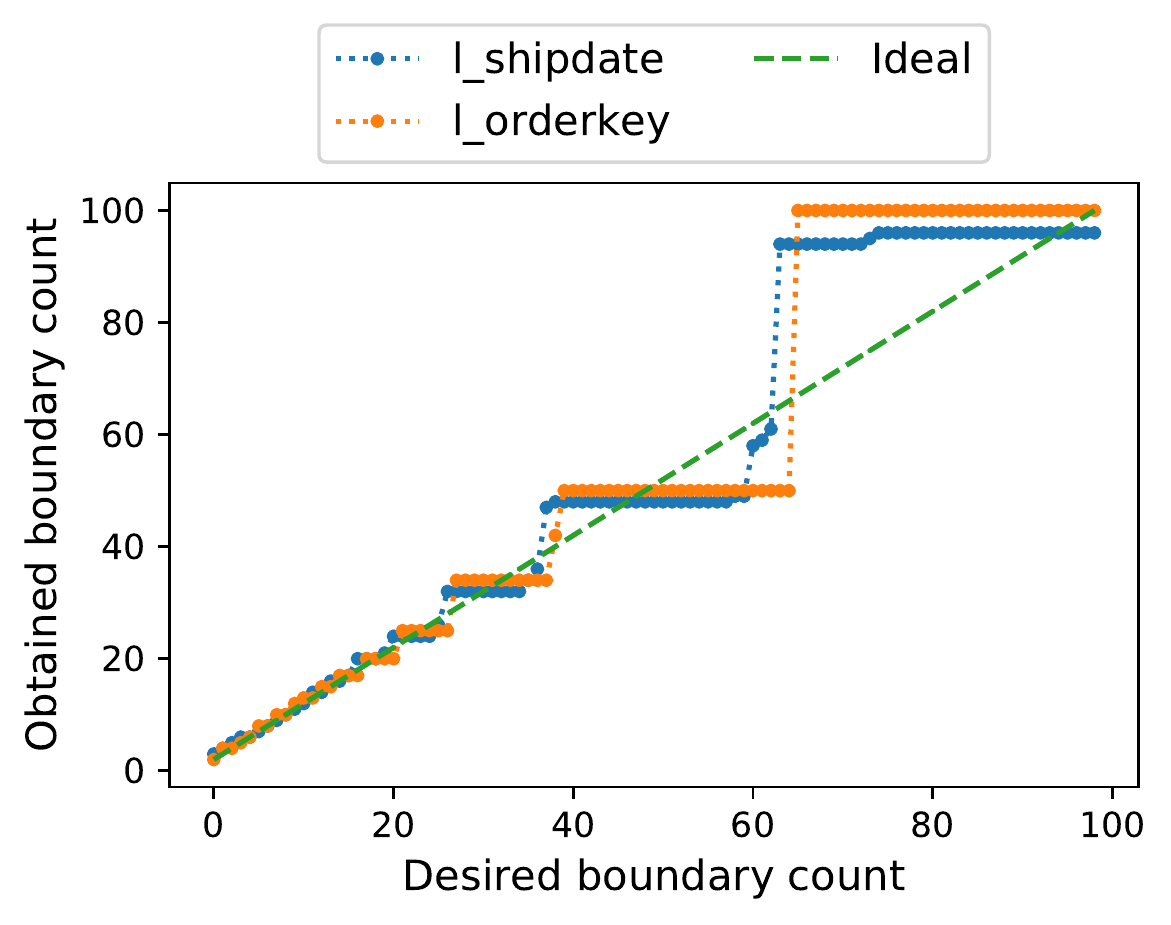}
         \caption{TPC-H \texttt{lineitem}}
         \label{fig:cluster_boundaries_counts_tpch}
     \end{subfigure}
     \hfill
     \begin{subfigure}[b]{0.48\textwidth}
         \centering
         \includegraphics[width=\textwidth]{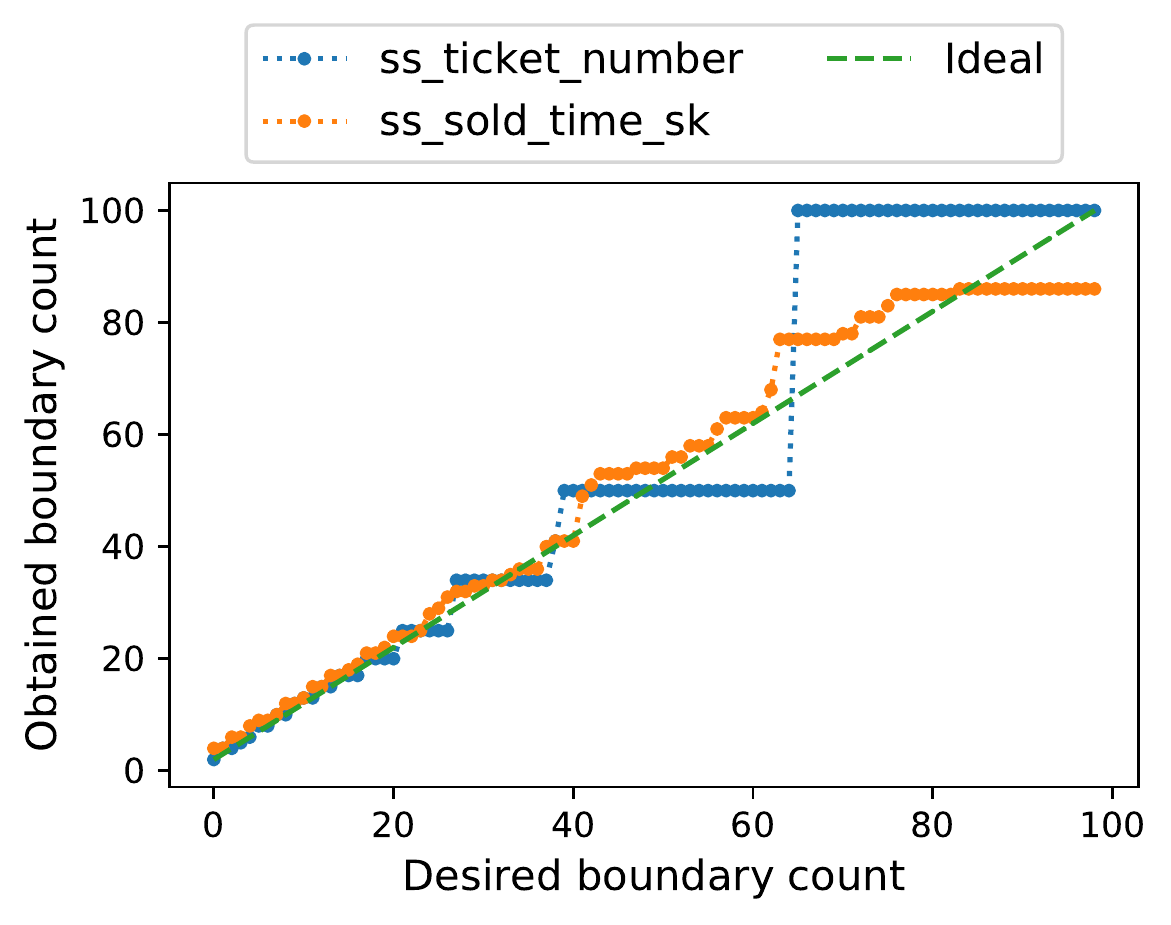}
         \caption{TPC-DS \texttt{store\_sales}}
         \label{fig:cluster_boundaries_counts_tpcds}
     \end{subfigure}

    \caption{Our algorithm for obtaining cluster boundaries does not always yield the requested number of cluster boundaries (which is equivalent to the cluster count of a column). The figure visualizes the number of obtained cluster boundaries when requesting 2 to 100 cluster boundaries for common clusterings columns of the TPC-H table \texttt{lineitem} (left) and the TPC-DS table \texttt{store\_sales} (right).}
    \label{fig:cluster_boundaries_counts}
\end{figure}

\paragraph{Precision regarding the number of boundaries}
\Cref{fig:cluster_boundaries_counts} visualizes the number of requested and obtained cluster boundaries for common clustering columns.
We observe that the figures for TPC-H and TPC-DS look similar:
Up to a cluster count of approximately 40, the number of returned boundaries is close to the requested number of boundaries, and the relative error is close to $0\%$.
For higher cluster counts, the algorithm returns the same number of cluster boundaries for a sequence of cluster counts.
For example, requesting 38 to 66 cluster boundaries on \texttt{l\_orderkey} always yields 50 cluster boundaries, and requesting 67 to 100 boundaries always yields 100 boundaries.
For these sequences, we obtain a relative error of up to 50\%.

We conclude that our algorithm's precision is sufficient for cluster counts up to approximately 40, as well as cluster counts close to $100$.
For cluster counts in between, the algorithm is less precise and might return up to 50\% more cluster boundaries than requested. 
Future work could address this by either using higher resolution histograms (i.e., histograms with more than 100 bins) or by splitting bins of Hyrise's current histograms.

\paragraph{Run time}
In all our experiments, the step of calculating cluster boundaries never exceeded a run time of 1.2 milliseconds.
Thus, we conclude that the algorithm's run time is negligible and does not require further analysis.

\subsection{Clustering Duration}
In this section, we evaluate the run time of the clustering algorithm and its phases.
For that purpose, we measure the run times of the algorithm's phases for the clustering configurations listed in \Cref{tab:algorithm_runtime_clusterings}.

\begin{table}[]
    \centering
    \begin{tabular}{ c|c|c|c }
        Id & Clustering columns & Cluster counts & Sort column \\
        \hline
        1 & l\_orderkey & 2 & l\_orderkey \\
        3 & l\_orderkey & 50 & l\_orderkey \\
        3 & l\_orderkey & 100 & l\_orderkey \\
        4 & l\_shipdate & 2 & l\_shipdate \\
        5 & l\_shipdate & 50 & l\_shipdate \\
        6 & l\_shipdate & 100 & l\_shipdate \\
        7 & l\_shipdate & 100 & l\_orderkey \\
        8 & l\_discount & 11  & l\_discount \\
        9 & l\_shipdate, l\_orderkey & 20, 50 & l\_orderkey \\
        10 & l\_shipdate, l\_orderkey & 50, 50 & l\_orderkey \\
    \end{tabular}
    \caption{The table lists the clustering configurations that are used to evaluate the algorithm's phase run times.}
    \label{tab:algorithm_runtime_clusterings}
\end{table}
 
\paragraph{Partition phase}
The run time of the partition phase is influenced by the number of clustering columns, their data types, and their cluster counts.
\Cref{tab:algorithm_runtime_partition} lists the run times of the partition phase for different clusterings.
 
\begin{table}[]
    \centering
    \begin{tabular}{ c|c|c|c|c|c}
        Id & \makecell{Clustering\\columns} & \makecell{Cluster\\counts} & \makecell{Run time (s)\\(assigning clusters)} & \makecell{Run time (s)\\ (moving data)} & \makecell{Lock time per\\chunk (ms)} \\
        \hline
        1 & l\_orderkey & 2 & 4 & 233 & 254 \\
        2 & l\_orderkey & 50 & 10 & 265 & 289 \\
        3 & l\_orderkey & 100 & 14 & 280 & 306 \\
        4 & l\_shipdate & 2 & 8 & 237 & 259 \\
        5 & l\_shipdate & 50 & 50 & 265 & 289 \\
        6 & l\_shipdate & 100 & 90 & 282 & 308 \\
        8 & l\_discount & 11  & 5 & 240 & 262 \\
        9 & \makecell{l\_shipdate,\\l\_orderkey} & 20, 50 & 35 & 340 & 371 \\
        10 & \makecell{l\_shipdate,\\l\_orderkey} & 50, 50 & 60 & 369 & 403 \\
    \end{tabular}
    \caption{The table lists metrics for the run time of the algorithm's partition phase. Clustering 7 is not contained in this table because it differs from Clustering 6 only in the sort column, which is not considered here.}
    \label{tab:algorithm_runtime_partition}
\end{table}

The partition phase operates in two sub-steps:
First, each row of a chunk is assigned to a cluster, based on its values in the clustering columns.
Second, all rows are moved to their respective clusters.
The table shows that assigning rows to clusters runs faster than the actual movement of data in all cases.

For the first step, we observe that the data type of the clustering columns impacts the run time.
For example, assigning clusters runs 2 to 6 times faster when clustering by \texttt{l\_orderkey}, an integer, instead of clustering by \texttt{l\_shipdate}, a date column, which is internally stored as a 10-character-string (see clusterings 1 to 3 and 4 to 6, respectively).
Furthermore, we observe that the run time is impacted by the cluster counts: a higher cluster count creates more value ranges that must be accessed to find the matching cluster for each row.

For the second step, we observe that the data type of the clustering column does not impact the run time:
The run times of the clusterings 1 to 6 are approximately equal when they use the same cluster count, despite one column stores integers and the other strings.
We believe the run time is not impacted because we need to access all values of a row to move it, not only the clustering columns.
Furthermore, we observe that higher cluster counts lead to higher run time.
This might be because a higher cluster count implies that there are more chunks to insert into, which could result in a worse cache hit rate.

Depending on the clustering, the partition phase has a \emph{lock time} between 254 and 403 milliseconds per chunk.
The \emph{lock time} of a chunk describes the time where no concurrent transactions may modify its rows because they are being moved to their respective clusters, i.e., concurrent transactions need to either wait or abort.
The time required for assigning rows to clusters is not included in the lock time, because there is no need to block other transactions during the first step:
If any rows are invalidated during the first step, they are ignored in the second step, as our algorithm moves only non-invalidated rows.
For the scope of this thesis, we consider the lock times per chunk acceptable.

\paragraph{Merge phase}
\Cref{tab:algorithm_runtime_merge} lists the run times of the merge phase for different clusterings.

\begin{table}[]
    \centering
    \begin{tabular}{ c|c|c|c|c|c}
        Id & \makecell{Clustering\\columns} & \makecell{Cluster\\counts} & \makecell{Merged\\chunks} & Merged rows & Run time (s)\\
        \hline
        1 & l\_orderkey & 2 & 0 & 0 & 0\\
        2 & l\_orderkey & 50 & 0 & 0 & 0\\
        3 & l\_orderkey & 100 & 49 & 456\,687 & 1\,845 \\
        4 & l\_shipdate & 2 & 2 & 9\,090 & 37\\
        5 & l\_shipdate & 50 & 26 & 51\,777 & 206\\
        6 & l\_shipdate & 100 & 0 & 0 & 0\\
        8 & l\_discount & 11 & 0 & 0 & 0\\
        9 & \makecell{l\_shipdate,\\l\_orderkey} & 20, 50 & 9 & 1\,562 & 9\\
        10 & \makecell{l\_shipdate,\\ l\_orderkey} & 50, 50 & 0 & 0 & 0\\
    \end{tabular}
    \caption{The table lists metrics for the run time of the algorithm's merge phase. All chunks with at most 10\,000 were merged.  Clustering 7 is not contained in this table because it differs from Clustering 6 only in the sort column, which is not considered here.}
    \label{tab:algorithm_runtime_merge}
\end{table}
 
We observe that only a low number of chunks are merged:
five out of nine clusterings do not merge any chunks at all.
The other four clusterings merge at most 5\% of the chunks and less than 1\% of the rows of \texttt{lineitem} (clustering 3).
The run time seems to be roughly proportional to the number of merged rows.
Due to the low number of merged rows, the merge phase runs significantly faster than the partition phase.
 
\paragraph{Sort phase}
\Cref{tab:algorithm_runtime_sort} lists the run times of the sort phase for different clusterings.

\begin{table}[]
    \centering
    \begin{tabular}{ c|c|c|c|c|c}
        Id & Sort column & \makecell{Obtained clusters\\(requested clusters)} & \makecell{Chunks\\per cluster} &  \makecell{Run time (s)} & \makecell{Run time per\\cluster (s)}\\
        \hline
        1 & l\_orderkey & 2 & 458 & 173 & 86.5\\
        2 & l\_orderkey & 50 & 18 & 153 & 3.06\\
        3 & l\_orderkey & 100 & 9 & 135 & 1.35\\
        4 & l\_shipdate & 3 (2) & 458 & 270 & 135\\
        5 & l\_shipdate & 48 (50) & 19 & 200 & 4.16\\
        6 & l\_shipdate & 96 (100) & 10 & 173 & 1.8\\
        7 & l\_orderkey & 96 (100) & 10 & 138 & 1.44 \\
        8 & l\_discount & 10 (11) & 91.6 & 126 & 12.6 \\
        9 & l\_orderkey &  1000 & 1 & 117 & 0.12\\
        10 & l\_orderkey & 2400 (2500) & 1 & 114 & 0.05\\
    \end{tabular}
    \caption{The table lists metrics for the run time of the algorithm's sort phase. For experiment 4, we use 2 rather than 3 as the obtained number of clusters because the third cluster is much smaller than the first two, and would distort the average run time per cluster.}
    \label{tab:algorithm_runtime_sort}
\end{table}

We observe that the run time of the sort phase is impacted by the data type of the sort column:
Sorting by \texttt{l\_shipdate}, a date (i.e., string) column, is 30\% to 50\% more expensive than sorting by an integer column such as \texttt{l\_orderkey}.
One could conclude that sorting floats (\texttt{l\_discount}, clustering 8) is faster than sorting integers (clusterings one to three).
However, it is important to consider that \texttt{l\_discount} contains only 11 unique values, which are spread over 10 clusters, i.e., 9 out of 10 clusters contain only a single unique value and are thus already sorted.
Hyrise uses an adaptive sort algorithm, i.e., its run time improves if the input is sorted. 

Furthermore, we observe that the average number of chunks per cluster impacts the run time: the fewer chunks per cluster, the lower the run time (see clusterings 1 to 3, 4 to 6).
This behavior is expected: since sorting has an asymptotic run time of $\mathcal{O}(n \log(n))$, sorting many small sequences should be faster than sorting one large sequence.

Depending on the number of clusters, the sort phase runs between 50 milliseconds and 135\,500 milliseconds per cluster.
This metric is important because it represents the time that rows of a cluster cannot be modified by other transactions without causing the cluster's sort step to fail.
There is a huge gap between 50 and 135\,000 milliseconds.
We conclude that the number of clusters has a significant impact on the sort phase's duration per cluster, and should be chosen with respect to the frequency of data modifying queries.

\subsection{Robustness against Concurrent Modifications}
In this section, we evaluate the algorithm's robustness against modifications made by concurrently running transactions.
Concurrently running transactions may perform three types of modifications: insertion of new data, updates of existing data, and deletion of existing data.
In Hyrise, updates are implemented as a combination of deletion and insertion.
As a consequence, it is sufficient to evaluate the robustness against updates to cover all three modification types.

Due to the low run time of the merge phase (c.f. \Cref{tab:algorithm_runtime_merge}), we evaluate only the robustness of the partition and sort phases.


\paragraph{Robustness metric}
Concurrently executed updates may force the algorithm's clustering steps (partition, merge, and sort) to abort and rollback their changes.
When we talk about the robustness of a certain clustering phase, we refer to the ratio of its clustering steps that finished with a commit.
Consequently, robustness can be expressed as a value between 0\% and 100\%.

Our clustering algorithm can be seen as a concurrently running transaction that executes updates.
As such, the clustering steps may block rows that need to be accessed by another concurrent update operator, forcing the update operator to abort.
Similar to the robustness of the clustering steps, we provide the ratio of successful concurrent updates.

\paragraph{Update statement setup}
We use the \texttt{lineitem} table of the TPC-H benchmark to evaluate our algorithm's robustness against updates.
We use ten threads to issue a varying number of update queries (between $100$ and $10\,000$) per second against the \texttt{lineitem} table.
The updates are distributed uniformly over the entire table.
Each \texttt{Update} operator affects exactly one row.

Unless otherwise specified, we perform exactly one attempt for each clustering steps, i.e., failed clustering steps are not repeated.

\paragraph{Results for the partition phase}
\Cref{tab:algorithm_update_tolerance_partition} lists the results for the partition phase.
For the partition phase, we did not artificially reduce the number of updates per second but rather executed as many updates as possible.
We observe that for all evaluated clusterings almost all updates (at least 99.99\%) are executed successfully.

Furthermore, we observe that in all evaluated cases, the vast majority (at least 99.6\%) of chunks were successfully partitioned.
This high success ratio is particularly interesting because we are executing updates with the highest frequency Hyrise is capable of, i.e., Hyrise is on full load, and 100\% of the workload are updates.
Boissier et al.~\cite{DBLP:conf/cikm/0001MDLMRSZU16} and Krüger et al.~\cite{DBLP:journals/pvldb/KruegerKGSSCPDZ11} have shown that in real-world enterprise resource planning systems the share of updates in the workload may be significantly lower, e.g., between 0.25\% and 5\%.

We thus conclude that the partition phase of our algorithm is sufficiently resistant against updates.
\begin{table}[]
    \centering
    \begin{tabular}{ c|c|c|c }
        \makecell{Clustering columns\\and cluster counts} & \makecell{Updates\\(\% successful)} & \makecell{Updates\\per second} & \makecell{Successful partition steps\\(Robustness)} \\
        \hline
        l\_shipdate: 100  & 3\,779\,304 (99.99\%) &  9\,263 & 913 of 916 (99.67\%)\\
        \makecell{l\_shipdate: 10,\\l\_orderkey: 30} & 1\,107\,947 (99.99\%) & 3\,103 & 914 of 916 (99.78\%)\\
        \makecell{l\_shipdate: 15,\\l\_orderkey: 30} & 1\,513\,157 (99.99\%) & 3\,972 & 914 of 916 (99.78\%)\\
        \makecell{l\_shipdate: 20,\\l\_orderkey: 50} & 2\,482\,043 (99.99\%) & 6\,039 & 906 of 916 (98.91\%)\\
    \end{tabular}
    \caption{The table lists the ratios of successful updates and successfully partitioned chunks during the partition phase of the clustering algorithm.}
    \label{tab:algorithm_update_tolerance_partition}
\end{table}

\paragraph{Results for the sort phase}
\Cref{tab:algorithm_update_tolerance_sort} lists the results for the sort phase.
We observe that for all evaluated clusterings, the vast majority (at least (99.98\%)) of updates executed during the sort phase are successful.

For the sort phase, we evaluated two update frequencies: 1000 updates per second, and 100 updates per second.
When executing 1000 updates per second, we observe that the ratio of successful sort steps varies between 0\% and 89.1\%, depending on the clustering.
If we only execute 100 updates per second, the ratio of successful clustering steps is significantly higher, but still spread between 17.71\% and 98.7\%.
We believe that the number of chunks per cluster heavily impacts the success chances of the sort step: the higher the number of chunks per cluster, the lower the chance of a successful sort step.

For example, the \texttt{l\_shipdate}-clustering contains 9 chunks per cluster, and every sort step fails.
We give an intuition why this behavior is expected:
The \texttt{lineitem} table has 916 chunks.
Thus, when we execute 1000 updates uniformly across those chunks, we expect at least one update per chunk per second.
Since the clusters contain 9 chunks, we expect at least 9 updates per cluster per second.
According to line 6 of \Cref{tab:algorithm_runtime_sort}, sorting a cluster requires 1.8 seconds on average.
As a consequence, we would expect $9 \cdot 1.8 \approx 16$ updates within the cluster while it is sorted.
However, even a single update in one of the cluster's chunks causes the sort step to fail.

We conclude that the robustness of our algorithm's sort phase against concurrent updates heavily depends on the cluster size.
If the clusters are large, the time needed for a sort step is large. Hence, the sort steps have little robustness and fail often.
However, if the clusters are small, sort steps run faster and have a certain robustness against concurrent updates.

\begin{table}[]
    \centering
    \begin{tabular}{ c|c|c|c|c }
        \makecell{Sort\\column}  & \makecell{Chunks per\\cluster} & \makecell{Updates\\(\% successful)} & \makecell{Updates\\per second} & \makecell{Successful sort steps\\(Robustness)} \\
        \hline
        l\_shipdate & 9 & 189\,185 (100\%) & 990 & 0 of 96 (0\%)\\
        l\_shipdate & 9 & 19\,410 (99.98\%) & 100 & 17 of 96 (17.71\%)\\
        \hline
        \makecell{(l\_shipdate,)\\l\_orderkey} & 3 & 143\,358 (99.99\%) & 988 & 118 of 341 (34.60\%)\\
        \makecell{(l\_shipdate,)\\l\_orderkey} & 3 & 14\,760 (100\%) & 100 & 303 of 341 (88.86\%)\\
        \hline
        \makecell{(l\_shipdate,)\\l\_orderkey} & 2 & 144\,779 (99.99\%) & 985 & 348 of 545 (63.85\%)\\
        \makecell{(l\_shipdate,)\\l\_orderkey} & 2 & 14\,600 (100\%) & 100 & 524 of 545 (96.15\%)\\
        \hline
        \makecell{(l\_shipdate,)\\l\_orderkey} & 1 & 139\,800 (100\%) & 991 & 891 of 1000 (89.10\%)\\
        \makecell{(l\_shipdate,)\\l\_orderkey} & 1 & 14\,160 (99.99\%) & 104 &  987 of 1000 (98.70\%)\\
    \end{tabular}
    \caption{The table lists the ratios of successful updates and successfully sorted clusters during the sort phase of the clustering algorithm.}
    \label{tab:algorithm_update_tolerance_sort}
\end{table}

\begin{figure}
    \centering
    \includegraphics[width=\textwidth]{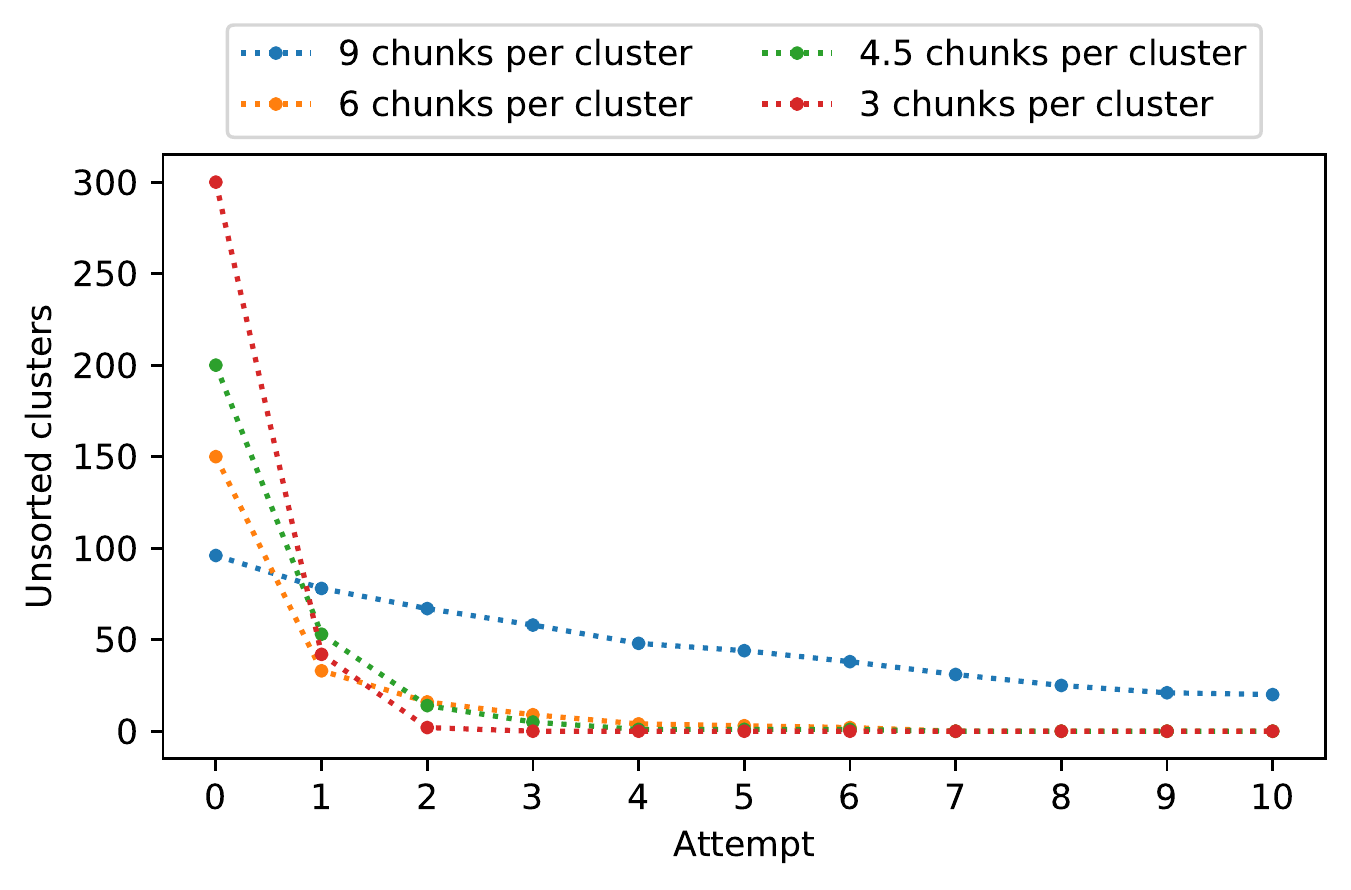}
    \caption{The figure visualizes the number of clusters that could not be sorted successfully after $n$ attempts. During the sort phase, we executed 100 updates per second.}
    \label{fig:failed_steps_after_attempt}
\end{figure}

\paragraph{Effect of multiple attempts}
Some individual steps of our clustering algorithm, e.g., partitioning a chunk, or sorting a cluster, may fail due to concurrent modifications.
To reduce the number of permanently failed steps, our algorithm can perform multiple attempts for each step.
In this paragraph, we evaluate the effect of performing up to ten sort attempts for each cluster.
\Cref{fig:failed_steps_after_attempt} visualizes the number of clusters that could not be sorted successfully after $n$ attempts, for different cluster sizes.
During the sort phase, we executed 100 updates per second.

We observe that for cluster sizes between 3 and 6 chunks the number of unsorted clusters rapidly decreases:
After attempt 4, there are ten or fewer unsorted clusters, after attempt 7 there are none.
With a cluster size of 9 chunks, the number of unsorted clusters decreases too, but slower:
After ten attempts, 20 rather than 79 of 96 clusters are still unsorted.

We conclude that using multiple attempts to sort a cluster can be an efficient way of reducing the number of unsorted clusters if the updates are uniformly distributed.
The number of attempts necessary until a certain share of the clusters is sorted depends on the cluster sizes and the frequency of concurrent modifications (c.f. \Cref{tab:algorithm_update_tolerance_sort}).

In real world databases, updates might not be uniformly distributed, but rather focus on a small part of the table.
For example, there might be legal restrictions that prohibit modifying sales data of previous fiscal years, so updates would focus on the most recent data.
In those cases, most of the sort steps on older data should be successful, while sort steps on recent data are more likely to fail.

\subsection{Memory Consumption during the Clustering Process}
\label{sec:eval_memory_consumption}
In this section, we evaluate how our clustering algorithm affects the memory consumption of a table during the clustering process.
For that purpose, we apply a clustering to the TPC-H table \texttt{lineitem} and estimate its size at regular intervals.

We evaluate three different versions of the algorithm:
\begin{enumerate}
    \item The first version performs all phases sequentially: partitioning, merging, sorting clusters, encoding the sorted clusters, and finally cleaning up (physically deleting chunks that contain only invalidated rows).
    \item During the partition, merge, and sort phases, our algorithm creates a significant number of chunks that contain only invalidated rows, two to three times the original number of chunks. The second version of our algorithm performs the same phases as the first, but executes cleanup steps in the background during the clustering process, i.e., it removes fully invalidated chunks at regular intervals.
    \item The third version is similar to the second version, but also encodes chunks in the background, as soon as they are finalized. In contrast to the first two versions, this version also encodes chunks produced by the partition phase.
\end{enumerate}
 
\paragraph{Experimental Setup}
We evaluate the memory consumption of the TPC-H table \texttt{lineitem}.
As usual, the clustering table was shuffled and dictionary encoding applied to its segments before the measurements begin.
All clustering phases are all performed single-threaded, although the second and third variant use threads to remove or encode chunks in the background.

Hyrise offers an interface to estimate the memory consumption of a table.
We use this interface in regular intervals to track \texttt{lineitem}'s memory consumption.


The second and third version use a thread to remove chunks that are fully invalidated.
The thread runs in the background.
Every second, the thread removes, i.e., physically deletes all fully invalidated chunks that can be safely removed.
The functionality to check whether a chunk can be safely removed, or whether it might still be referenced by some transaction, is taken from Hyrise's \texttt{MvccDeletePlugin}.
Theoretically, we could forgo the self-written thread and simply use the \texttt{MvccDeletePlugin} to delete fully invalidated chunks.
However, for evaluation purposes, we choose the self-written thread, as it grants us more control regarding when and what to delete.

The third version runs an additional thread to encode chunks in the background.
Every second, the thread checks for new immutable chunks created by the partition, merge, or sort phase, and encodes them in the background.
If multiple chunks require encoding, they are processed sequentially, i.e., there is always only a single encoding thread.
The explicit encoding phase of the sorted clusters from the first version is not skipped, but the encoding thread may have encoded some of the sorted clusters by the time the actual encoding phase starts.

\begin{figure}
    \centering
    \includegraphics[width=0.9\textwidth]{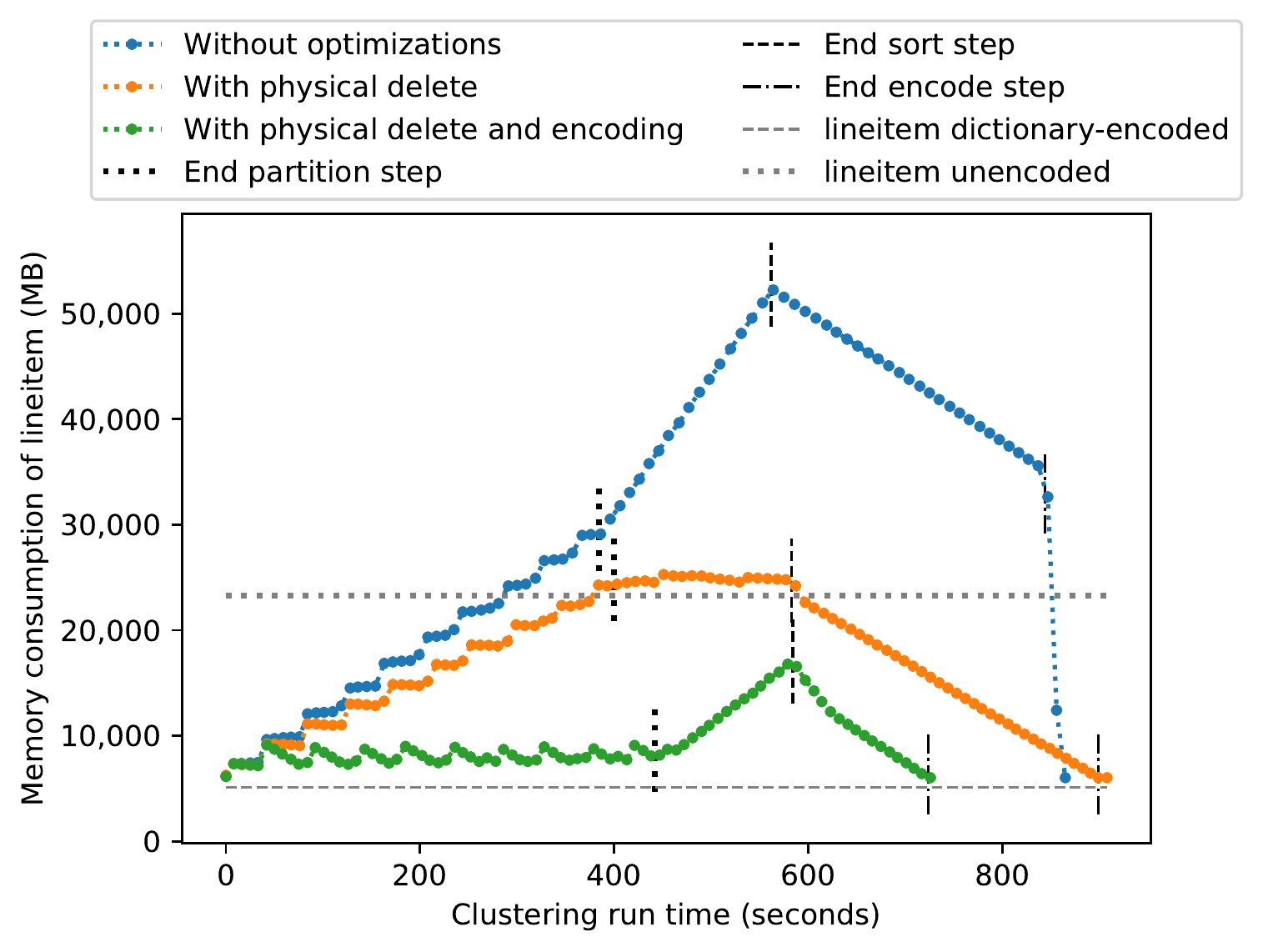}
    \caption{The figure visualizes the memory consumption of the TPC-H table \texttt{lineitem} while it is being clustered by \texttt{l\_shipdate}, with a cluster count of 100. Overall, there are 100 clusters.}
    \label{fig:memory_consumption1}
\end{figure}

\paragraph{Results} 
We evaluate the memory consumption for two clusterings configurations.
The first clustering configuration clusters \texttt{lineitem} by \texttt{l\_shipdate}, with a cluster count of 100.
Once the clustering is complete, there will be 100 clusters, each containing 9 full and one half filled chunk.
\Cref{fig:memory_consumption1} visualizes the memory consumption of \texttt{lineitem} while the clustering is applied.

First of all, we observe that the encoding significantly impacts the memory consumption:
On scale factor 10, \texttt{lineitem} has a size of 5.1~GB if its segments are dictionary encoded, but a size of 23.3~GB if the data is stored unencoded.

\paragraph{First version of the algorithm}
For the first version of our algorithm (blue), we observe that there are 10 plateaus of increasing memory consumption during the partitioning phase.
Both the plateaus themselves and their number are expected: a higher memory consumption in the partition phase occurs when a new chunk for a cluster was added (with a pre-allocated size).
In TPC-H, values of \texttt{l\_shpidate} are distributed uniformly random~\cite[p.~87]{TpchSpec}.
As a consequence, we expect that all clusters are filled at a similar ratio, and thus require a new chunk at approximately the same time, which yields the plateaus.
Since there are 100 clusters, each cluster has 9 full and one a partially filled chunk, which yields 10 chunks per cluster, i.e., 10 plateaus.
During the partition phase, the size of \texttt{lineitem} increases by 23~GB.
This increase is expected, as the entire table was partitioned into chunks with unencoded segments.

The sort phase adds another 23~GB to the table, resulting in a total size of 52~GB.
This increase is expected, too, as the sorted clusters (which contain the entire table) are inserted into the table as unencoded value segments.

During the encoding phase, the sorted clusters are encoded.
The encoding reduces the table size by 18~GB, which is exactly the difference between an encoded and an unencoded version of \texttt{lineitem}.

In the cleanup phase, our algorithm physically deletes chunks that are fully invalidated: the chunks of the original table, and the results of the partition step.
The cleanup phase reduces the table size by 5~GB + 23~GB  = 28~GB, leaving an estimated table size of 6~GB.

\paragraph{Second version with the algorithm}
For the first version of our algorithm, we intentionally perform no cleanup operations until the clustering is complete.
In real-world scenarios, it might be sensible to perform cleanups as soon as they become possible.
The second version of the algorithm (yellow) does just that, and physically deletes fully invalidated chunks as soon as possible.

Both the partition and the sort phase are affected by the intermediate cleanup:
The partition phase partitions chunk by chunk.
Once a chunk is partitioned, it is fully invalidated and can be removed.
As the removed chunk contains dictionary segments, but we add its values to unencoded value segments, we expect that the memory consumption still increases, although not as strong as in the first version.
Indeed, the figure shows that the distance between the plateaus is lower than in the first version of the algorithm.

The sort phase fully invalidates the unsorted chunks of a cluster if the sort operation is successful.
In both the unsorted and the sorted version of the cluster,  data is stored in unencoded value segments.
Thus, when the unsorted chunks of the cluster are replaced by the sorted ones, we do not expect the table's memory consumption to change.
While this is not perfectly true, the figure shows only a small increase in memory consumption during the sort phase.

\paragraph{Third version of the algorithm}
The highest memory consumption during the second version of the algorithm is 25~GB, which is significantly lower than the first version's maximum of 52~GB, but still significantly higher than the initial size of 5~GB.
To further reduce the maximum memory consumption, we use a third version (green) of our clustering algorithm.
Like the second version, it performs cleanup steps as early as possible.
Additionally, it uses a thread to encode the chunks of the partition and sort phases, as soon as they are finalized.

Encoding the chunks significantly reduces the additional memory consumption of the partition phase: instead of 10 increasing plateaus, we now receive 10 spikes, which subside quickly. 
The memory spikes occur in the instant where new chunks are added to the clusters and subside as the old chunks are being encoded.

In the sort phase, the memory consumption does increase again.
The increase occurs because unsorted chunks of the cluster, which are by now encoded, are replaced with the sorted, but unencoded chunks.
The figure also shows that the final encoding phase operates faster than in the first two versions:
This is because the encoding thread already encodes some of the sorted clusters while other clusters are still being sorted.
The maximum memory consumption of our algorithm's third version is 17~GB.

\begin{figure}
    \centering
    \includegraphics[width=0.7\textwidth]{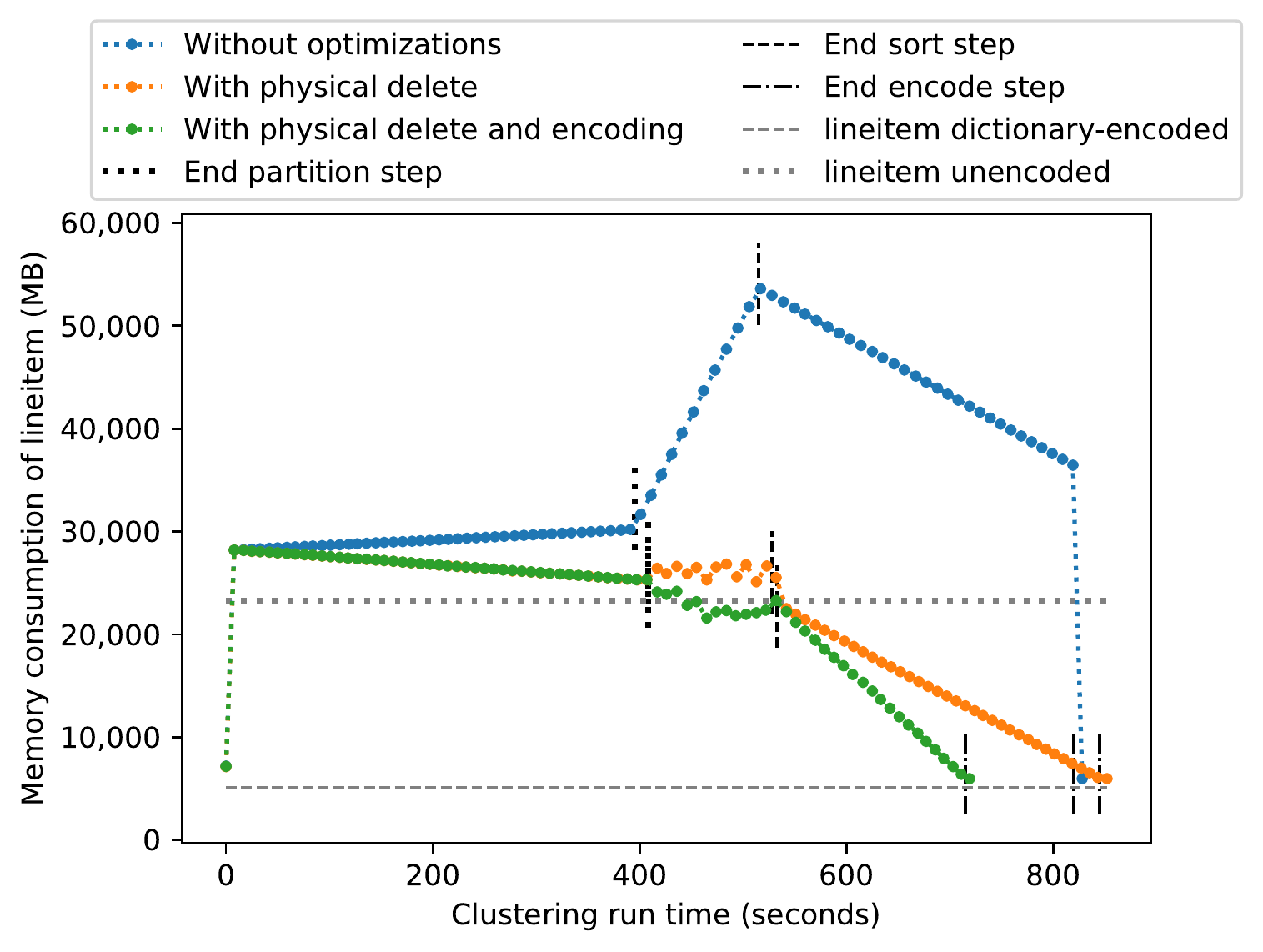}
    \caption{The figure visualizes the memory consumption of the TPC-H table \texttt{lineitem} while it is being clustered by \texttt{l\_shipdate} and \texttt{l\_orderkey}, with cluster counts of 20 and 50. Overall, there are 1000 clusters. The blue and yellow lines begin at the same position as the green line, and are merely overlaid.}
    \label{fig:memory_consumption2}
\end{figure}

\paragraph{Effect of a high cluster count}
The second clustering configuration clusters \texttt{lineitem} by \texttt{l\_shipdate} and \texttt{l\_orderkey}, with cluster counts of 20 and 50, respectively.
Once the clustering is complete, there will 1000 clusters, each containing one mostly full chunk.
\Cref{fig:memory_consumption2} visualizes the memory consumption of \texttt{lineitem} while the clustering is applied.

We observe that the partition phase differs from \Cref{fig:memory_consumption1}:
Instead of several increasing plateaus, we obtain only a single plateau, at the maximum memory consumption reached during the partition phase.
This behavior is expected:
For each of the 1000 expected clusters, the partition phase creates a chunk with a pre-allocated segment size.
Since there are 1000 clusters, we expect only one chunk per cluster (as the original \texttt{lineitem} table has 916).
As the plateaus are caused by the simultaneous adding of new chunks for clusters, we do not expect a second plateau of increased memory consumption.

Furthermore, we observe that performing cleanups (version 2 and 3) reduces the table's sizes, but encoding (version 3, green) does not.
cleanups reduce memory consumption because partitioned chunks can still be removed.
Encoding does not reduce memory consumption because it can be only performed on finalized chunks.
The output chunks of the partition phase are only finalized when they reach their maximum size, or the partition phase is over.
Since we expect at most one chunk per cluster, we do not expect chunks to be finalized before the partition phase ends.
However, the effects of the encoding can be seen in the sort and encode phases.
With all optimizations applied, the maximum memory consumption of the partition phase of \Cref{fig:memory_consumption2} is thrice as large as the memory consumption of \Cref{fig:memory_consumption1}.

\paragraph{Conclusion}
Without any optimizations, the memory consumption of \texttt{lineitem} increases during the clustering by a factor of ten.
The number of clusters determines how fast the memory consumption increases during the partition phase: the more clusters, the faster increases the memory consumption.
The memory consumption of the final clustering result is approximately equal to the memory consumption of the initial table.

We conclude that performing frequent cleanup steps can reduce the maximum memory consumption by a factor of 2.
Frequent cleanups can be implemented efficiently, as identifying chunks that can be deleted is computationally cheap: we can simply track which chunks are fully invalidated by our clustering algorithm.

Applying dictionary encoding to value segments can significantly reduce the segment's memory consumption,  in our case by a factor of 4.5.
However, calculating a dictionary encoding is computationally expensive $(\mathcal{O}(n \log( n)))$, and it has little effect when there is only a low number of chunks per cluster.
We thus conclude that applying dictionary encoding is a useful optimization when the number of chunks per cluster is high, and when memory consumption is more limited than computational power.

	\chapter{Related Work}
\label{sec:related_work}

Clustering algorithms have been researched since the 1980's~\cite{DBLP:conf/sigmod/ChangF80}.
Today, there is a multitude of research on clustering algorithms and the automated selection of clustering columns.
Database vendors such as Oracle~\cite{DBLP:journals/pvldb/ZiauddinWKLPK17}, Microsoft~\cite{DBLP:conf/sigmod/AgrawalNY04} and IBM~\cite{DBLP:conf/vldb/LightstoneB04} have acknowledged the importance of clustering and provide features to optimize partition pruning in their database systems.

We present an overview of clustering algorithms, clustering models and latency estimation models in \Cref{sec:related_work_algos}, and further reading in \Cref{sec:related_work_future}.

\section{Clustering Algorithms, Models, and Latency Estimation}
\label{sec:related_work_algos}
Lightstone et al.~\cite{DBLP:conf/vldb/LightstoneB04} propose a workload-based model for the automated selection of clustering columns.
Further, they propose a storage layout for multi-dimensional clustering.
Both the model and the storage layout were implemented in the row-oriented, disk-based database system DB2 (IBM).
Their storage layout is similar to the disjoint clusters concept:
Each unique combination of values in the clustering dimensions is assigned its own storage block.
Similar to chunks in Hyrise, storage blocks are containers for rows and are allocated with a fixed size.
However, fixed sizes might lead to some very sparsely populated storage blocks, and thus to increased memory consumption.
To counter this effect, the authors use different granularity levels.
For example, date columns could be clustered by week or month, rather than day.
Their model considers different clustering configurations, and at different granularity levels.
For each clustering configuration, sampling is used to estimate the number of tuples for each cluster.
Due to the sampling, their algorithm implicitly considers correlations between the clustering columns, i.e., it is aware that some value combinations of the clustering columns  might occur rarely.
Thus, for two or more dimensions, their approach to estimate the cluster sizes has a significant advantage over our histogram-based approach.
Based on the cardinality estimates obtained by sampling, the model uses optimizer estimates to assess the expected benefit of a given clustering configuration.
Further comparisons of the latency estimation approach are difficult, as the authors do not provide details on the factors included in the optimizer's latency estimation process.
However, the model proposed in this thesis additionally considers different sorting orders for chunks, which can yield a significant speedup.
In a previous paper on IBM's DB2~\cite{DBLP:conf/sigmod/PadmanabhanBMCH03}, the authors describe how to maintain the clustering during bulk loads and inserts.
In contrast to this work, they do not consider how to change the clustering columns while data modifying queries are running.

Ziauddin et al.~\cite{DBLP:journals/pvldb/ZiauddinWKLPK17} present a clustering design specialized for star/snowflake schema databases that enables pruning over multiple tables.
They provide an implementation in Oracle RDBMS.
The authors argue that dimension tables would be filtered more often than fact tables.
As a consequence, they propose to cluster a fact table not (only) by its own columns.
Instead, they suggest to cluster by frequently filtered attributes of dimension tables.
Clustering by dimension table attributes is expensive, because it involves an unfiltered join between the fact and dimension table.
The join result is then clustered by the dimension table's columns.
Once the clustering of the fact table is done, the superfluous dimension table columns are discarded, but a min-max filter of their values is kept.
When a dimension table is both filtered by a fact table's cluster attribute and joined to the fact table, then both dimension and fact table can be pruned.
They provide an SQL-like syntax to define a clustering, i.e., the clustering has to be defined manually.
The actual clustering process can be initiated explicitly, by bulk inserts, or during maintenance.
Ziauddin et al. achieve significant performance gains on star schema databases.
However, they do not provide benchmark results for other database architectures.
Hyrise is designed to work with both analytical and transactional workloads.
Thus, their work is a promising future work approach for star schema architectures, but cannot be applied to all of Hyrise's use cases.

Idreos et al.~\cite{DBLP:conf/cidr/IdreosKM07} present a concept for automated continuous physical reorganization, called database cracking for MonetDB.
The novel idea of their approach is to perform index construction and maintenance as a part of query processing.
Whenever the database receives a range query on a column $X$, it considers the query as an advice to crack $X$ into smaller pieces.
This way, individual columns are continuously physically reorganized.
In contrast to the approach presented in this thesis, their approach is discriminating, i.e., only the requested parts of the column are reorganized, while unqueried parts of the data remain untouched.
Technically, their approach works by creating a copy of the clustering column.
When a range predicate occurs, only the copied column is physically reorganized, leaving the original column unchanged.
As a result, their approach allows to have one clustering per column.
The downside, however, is the additional memory required to store the duplicated columns.
Depending on the number and size of these columns, the additional memory requirements may not be tolerable for memory-resident systems.
Further, their approach relies on range predicates to decide on what to crack.
However, according to Dreseler et al.~\cite{DBLP:journals/pvldb/DreselerBRU20} (Figure 2) the most time consuming operators in Hyrise are joins and aggregates.
We have shown in this thesis that not only predicates, but e.g., joins are affected by clusterings, too.
Join columns, however, are usually joined rather than range-filtered, and might thus be under-represented in the cracking approach.

Hilprecht et al.~\cite{DBLP:conf/sigmod/HilprechtBR20} present a clustering model for cloud databases.
They argue that obtaining precise latency models/optimizer estimates is a difficult task in general; and even harder for cloud databases due to heterogeneous hardware used.
Consequently, they suggest a model based on deep reinforcement learning, that learns the actual performance impact of a clustering.
Before the model can be used, it has to undergo two phases of training.
For every training step, a clustering is chosen, and its influence is measured.
To speed up the first training phase, clusterings are only simulated, not actually implemented; and their impact is estimated rather than measured.
Being in a distributed cloud environment, the authors argue that network transfer latencies are likely to dominate actual computation latencies, and are thus used as primary measure of latency estimation.
For example, assume that two tables are co-partitioned on their common join column, i.e., a single machine stores all rows from both tables where the join column values are within a certain value range.
In that case, the partitions of the two tables can be joined locally, without transmitting them over the network.
Local joins can greatly reduce the amount of transferred data, thus reducing the expected query run time.
The second phase of the training is optional and is used for fine-tuning of the model.
For that purpose, a small sample of the database is generated.
This time, clusterings are implemented, and their actual effect is measured.
Once the model is trained, it suggests a clustering, based on the given workload.
Their approach is not directly comparable to the approach presented in this thesis.
Hyrise is not intended to run as a distributed system.
Thus, there are no distributed joins, and no network transfer latencies.
Consequently, the offline phase cannot be directly applied to Hyrise.

Estimating the latency of operators and queries is a complex task, yet crucial for the approach presented in this thesis.
Usually, latency estimates can be obtained from physical cost models.
At the time of writing, Hyrise' physical cost models are still under development.
Thus, we developed a rule-based estimation approach, which focuses on the impact of the clustering, i.e., the physical data layout.
The rule-based approach covers the most important aspects for predicting an operator's run time.
However, there are still complex interactions that are difficult to express in a rule-based way.
An additional downside of the rule-based approach is the focus on a specific hardware:
The performance benefit of, e.g., partition pruning depends on the specific hardware used.
When some part, e.g., the processor, or main memory, is exchanged, those performance benefits must be re-evaluated to keep the model calibrated.
More sophisticated work tackles this challenge by using deep neural networks.
For example, Marcus et al.~\cite{DBLP:journals/pvldb/MarcusP19} define a neural model for each operator type, and combine them in a query plan-like structure.
However, to the best of our knowledge, previous work on query latency estimation does not focus on the impact of the clustering.

To sum up, in this chapter we have presented previous work on the automated selection of clustering dimensions, as well as previous work on algorithms to implement a clustering during operation.
They all differ from our approach in one way or the other:
Lightstone et al~\cite{DBLP:conf/vldb/LightstoneB04} propose a storage layout for multidimensional clustering.
The layout is similar to the concept of disjoint clusters, and they use sampling for a superior cardinality estimation.
However, they do not describe an approach how to change the clustering while data modifying queries are executed concurrently.
Ziauddin et al.~\cite{DBLP:journals/pvldb/ZiauddinWKLPK17} present an approach to prune across tables that are joined.
Their approach focuses on star schema layouts.
In Hyrise, we aim to support multiple layouts.
The database cracking approach by Idreos et al.~\cite{DBLP:conf/cidr/IdreosKM07} performs discriminating continuous physical reorganization as a part of query processing, i.e., during operation.
In contrast to our approach, database cracking duplicates the clustered columns, thereby increasing memory consumption.
Another difference is the choice of the clustering columns: the cracking approach relies on range predicates, while our model also considers joins.
Hilprecht et al.~\cite{DBLP:conf/sigmod/HilprechtBR20} present a model based on deep reinforcement learning to choose the clustering dimensions.
Their model assumes a cloud environment, i.e., a distributed database system.
The authors argue that network transmission latencies are the dominant cost factor.
Hyrise, however, is not supposed to run in a distributed manner, thus the assumption does not hold.

\section{Further Reading}
\label{sec:related_work_future}
In addition to research on clustering algorithms and cost models, there is also previous work on related topics.

\paragraph{Search space pruning}
Agrawal et al.~\cite{DBLP:conf/vldb/AgrawalCN00} present a model for automated selection of indexes and materialized views for Microsoft SQL Server.
Later, they published an extended version of the model~\cite{DBLP:conf/sigmod/AgrawalNY04} that additionally selects an vertical and/or horizontal partitioning.
The authors argue that the combination of locally optimal decisions (e.g., the best horizontal partitioning and the best index) does not necessarily lead to a globally optimal solution.
Instead, the authors suggest an integrated approach that considers all these aspects of physical database design.
A downside of an integrated approach is the significant increase of potential physical design choices and thereby elevated run time.
To cope with the huge size of the search space, the authors apply a pruning technique:
Clustering column combinations are only considered if their share of execution time exceeds a predefined threshold.
In Hyrise, tables are stored in columnar format, i.e., the entire table is inherently vertically partitioned.
As a consequence, the work in this thesis focuses on horizontal partitioning only.
However, the search space pruning approach is still applicable, and could reduce the run time of the clustering model presented in this thesis.

\paragraph{Data induced predicates}
Orr et al.~\cite{DBLP:journals/pvldb/OrrKC19} describe an approach for pruning across multiple joined tables.
Their approach does not focus on clustering, but could be seen as a generalization of the star-schema approach presented by Ziauddin et al.~\cite{DBLP:journals/pvldb/ZiauddinWKLPK17}.
The core idea: when a scan predicate is used to prune a table $A$, it does not only reduce the value range of the scanned column, but also the value range of $A$'s join column(s).
Using the data statistics on partition level (e.g., min-max-filter), the authors construct a new, so-called data-induced predicate (DiP).
The DiP selects (at least) the same partitions of $A$ as the original predicate, but filters only on join columns.
As such, it can be used to prune the tables joined to $A$ via their join columns.
Depending on the clustering, the DiP might allow to avoid accessing large parts of the data.
It is worth noting that this entire optimization is done before the query is executed.
At the moment of writing, Hyrise does not support DiPs, so they are not considered in the clustering model.
However, this might be promising future work, as Hyrise already has many small partitions (called chunks) which all maintain min-max filters.

\paragraph{Workload prediction}
The clustering model presented in this thesis solely considers the historic workload.
If a DBMS were aware of upcoming workload changes, it could choose an optimized clustering proactively, yielding improved performance.
While our model does not consider this aspect, there is work on the topic of workload forecasting:
Ma et al.~\cite{DBLP:conf/sigmod/MaAHMPG18} propose a model to predict the future workload.
More precisely, their model is based on machine learning and predicts the query composition of the future workload, i.e., how often each query occurs in the workload.
Further work on workload forecasting includes \cite{ DBLP:conf/cnsm/GongGW10, DBLP:conf/adbis/HolzeR08, DBLP:conf/IEEEcloud/RoyDG11}.

	\chapter{Conclusion and Future Work}
\label{sec:conclusion}
In this chapter, we summarize our most important conclusions and provide suggestions for future work.

In this thesis, we have presented a model that determines beneficial clustering configurations and a clustering algorithm for the in-memory research database Hyrise.
The model analyzes the current workload and yields multi-dimensional clustering suggestions.
The clustering algorithm applies multi-dimensional clusterings while other transactions may modify the data concurrently.
As a result of the clustering algorithm, the data is organized in disjoint clusters.

\paragraph{Clustering model}

Our clustering model evaluates clustering candidates by estimating their impact on the latency of scans and joins.
The evaluation shows that our model's join latency estimates are quite precise:
The majority of estimated latencies are within 20\% of the measured latency.
We conclude that our model yields sufficiently precise join latency estimates, although we have identified common sources of error that could be addressed by future work.

The evaluation further shows that our model's scan estimates are spread farther, but still have a  reasonable precision for the majority of operators.
As the main source of error, we have identified the model's assumption of a static physical query plan (i.e., that the physical query plan of an SQL query is chosen independently of the clustering).
This assumption does not hold in all cases.
We have identified pruning as the main reason for the changes in the PQP:
The clustering may enable pruning for some scans, which in turn increases the scan's selectivity, causing the database's optimizer to reorder the scans.
This reordering may impact the input sizes of the scans, which is not considered by our model.
We conclude that the assumption of a completely static PQP is unsuitable for precise scan latency estimates.
Future work could drop the assumption and consider a what-if optimizer~\cite{DBLP:conf/sigmod/ChaudhuriN98} for query plans.
Alternatively, future work could just slightly relax the static PQP assumption to consider the local reordering of consecutive filter predicates.

Our evaluation shows that correlations are not the most common, but still a notable error source for both scan and join latency estimates.
Our model is not capable of detecting correlations on its own, but it offers a rudimentary approach to incorporate user-provided knowledge about correlations in latency estimations for scans.
To obtain more precise latency estimates, future work could use a more sophisticated approach to incorporate knowledge about correlations in latency estimates for scans and joins.

Overall, our model suggests sensible clusterings for the TPC-H and TPC-DS benchmark.
While it is not capable of strictly ordering the clustering suggestions by their measured latency, most suggestions have a similar latency, and a significantly lower latency than, e.g., a clustering where the data is shuffled.
Besides imprecise latency estimates, we have identified aggregates to be a major source of mismatches between estimated and measured workload latency:
Due to the limited scope of this thesis, our model assumes that the latencies of aggregates are not impacted by the clustering.
However, our evaluation has shown that aggregate latencies can be significantly affected by the clustering.
We thus conclude that the clustering model should include latency estimates for aggregates to obtain a better ranking of the suggestions.

For the TPC-H benchmark, the best clustering we found for \texttt{lineitem} yields a 5\% latency reduction.
For the TPC-DS benchmark, the best clustering we found for \texttt{store\_sales} yields a 4\% latency reduction.
In general, we observed that our model's top twenty two-dimensional clustering suggestions often contain a certain column, e.g., \texttt{l\_shipdate} or \texttt{ss\_sold\_time\_sk}, which is combined with several other columns.
We conclude that in a multi-dimensional clustering, a few columns can dominate the clustering's latency impact, while the majority of potential clustering columns have only a marginal impact.

So far, our model considers only one set of cluster counts for each clustering candidate.
Our evaluation has shown that the cluster counts have a significant impact on the clustering's latency:
For example, when clustering by \texttt{l\_shipdate} and \texttt{l\_orderkey}, clustering with the cluster counts suggested by our model yields a 7\% higher latency than a clustering using the best cluster counts we found so far for this clustering candidate.
We conclude that the choice of proper cluster counts is of similar importance as the choice of the clustering columns itself.
Future work could address this issue by considering multiple sets of cluster counts for each clustering candidate.

\paragraph{Clustering algorithm}
We evaluated our algorithm's robustness against modifications made by concurrent transactions.
We found that the partition phase is highly robust against modifications: Even when executing concurrent updates with the maximum frequency Hyrise is capable of, less than 1\% of the partition steps failed.
The robustness of the sort phase depends on the frequency of concurrent modifications and the number of chunks per cluster.
The more chunks a cluster contains, the more likely its sort step is going to fail.
Consequently, to obtain a high ratio of successful sort steps, we conclude that it is sensible to choose many small rather than a few large clusters.

An evaluation of our algorithm's memory consumption shows that the clustering result has a size similar to the original table.
However, during the clustering process, memory consumption increases by up to factor 10.
This temporary memory consumption can be reduced in two ways:
First, our algorithm produces several chunks which contain only invalidated rows, but still consume memory.
Deleting those as early as possible reduces the maximum memory consumption by factor 2.
Second, the intermediate results produced by the clustering algorithm are not dictionary encoded.
We found that applying dictionary encoding reduces their size by a factor of 4.5.
We also found that encoding intermediate results is the most effective when there is only a small number of clusters.

We conclude that the choice of cluster counts does not only impact the workload's latency, but also the memory required to implement it, and the ratio of successful sort steps.
The ideal number of clusters for the memory consumption is diametrically opposite to the ideal number of clusters for a high ratio of successful sort steps.
We thus conclude that the model's choice of cluster counts should consider both the available memory and the frequency of concurrent modifications.

	\appendix\chapter{\appendixname}

\section{Histograms of Relative Estimate Errors for Unaffected Join Steps}
Our model assumes that the clustering, build and output writing steps of the hash join are not affected by the clustering.
In this section, we present histograms of relative errors of the latency estimates of hash join steps our model considers unaffected by the clustering.

\begin{figure}
     \centering
     \begin{subfigure}[b]{0.48\textwidth}
         \centering
         \includegraphics[width=\textwidth]{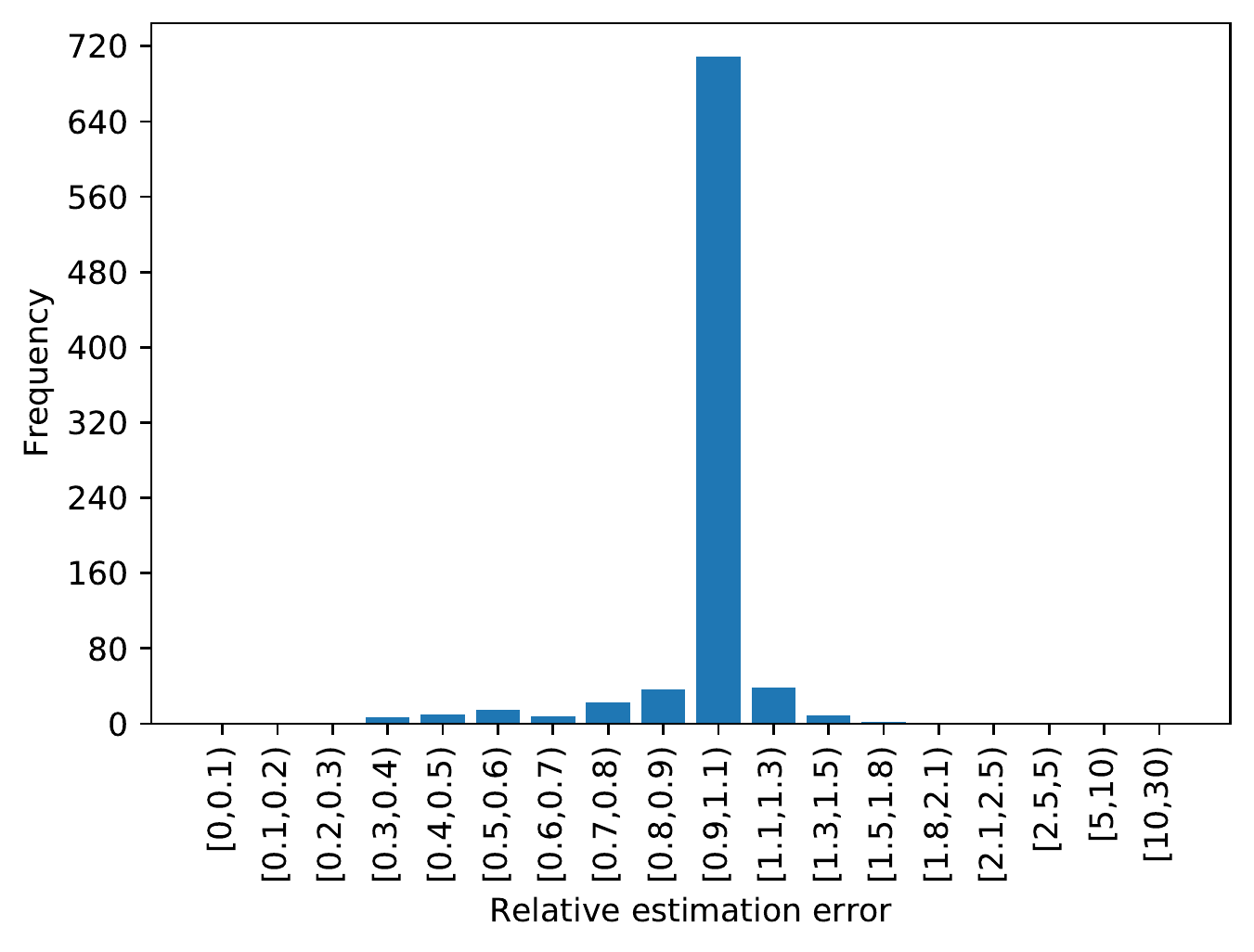}
         \caption{\texttt{l\_shipdate} Clustering - All joins}
         \label{fig:relative_errors_histogram_join_clustering_1}
     \end{subfigure}
     \hfill
     \begin{subfigure}[b]{0.48\textwidth}
         \centering
         \includegraphics[width=\textwidth]{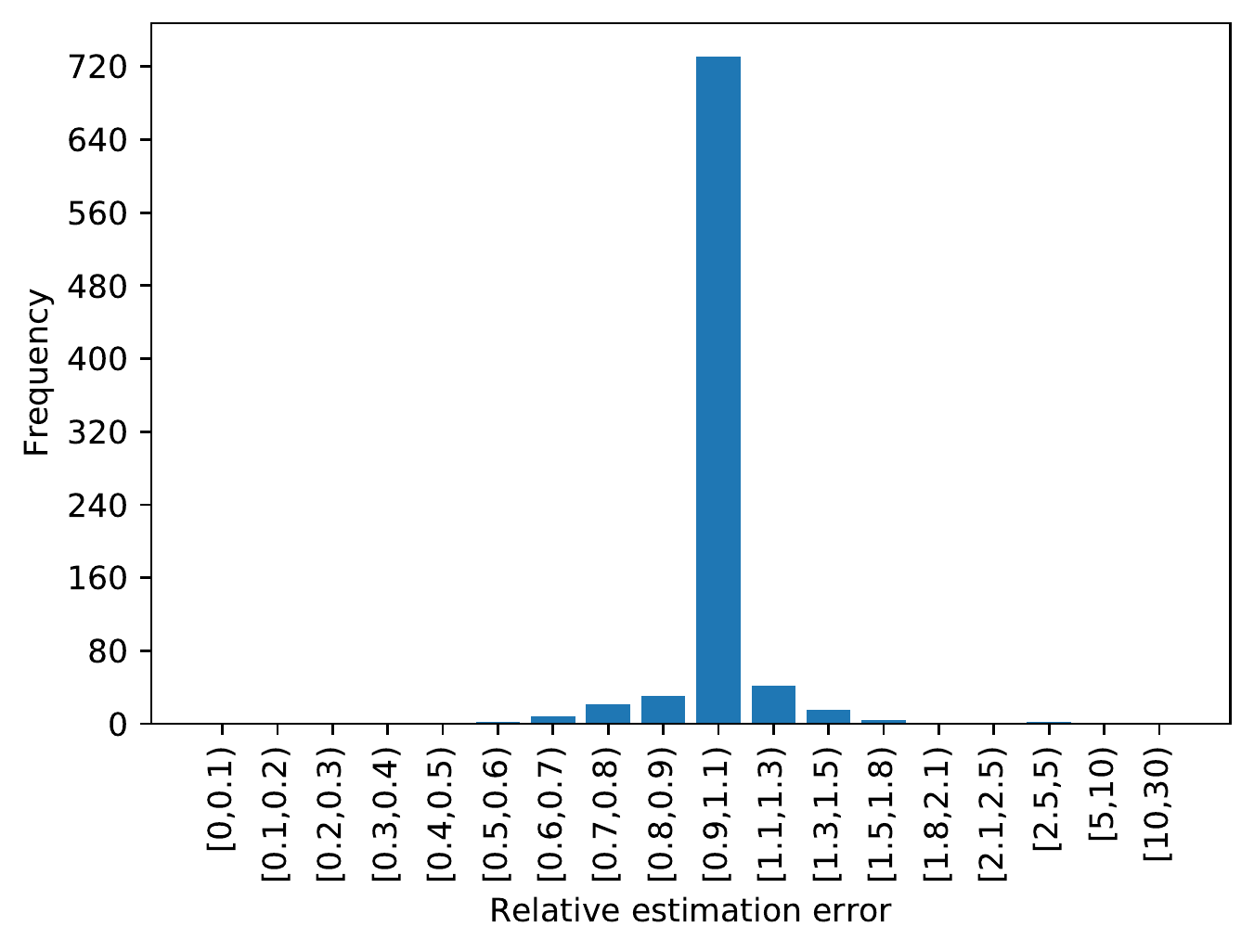}
         \caption{\texttt{l\_shipdate} Clustering - All joins}
         \label{fig:relative_errors_histogram_join_clustering_2}
     \end{subfigure}
     \hfill
     \begin{subfigure}[b]{0.48\textwidth}
         \centering
         \includegraphics[width=\textwidth]{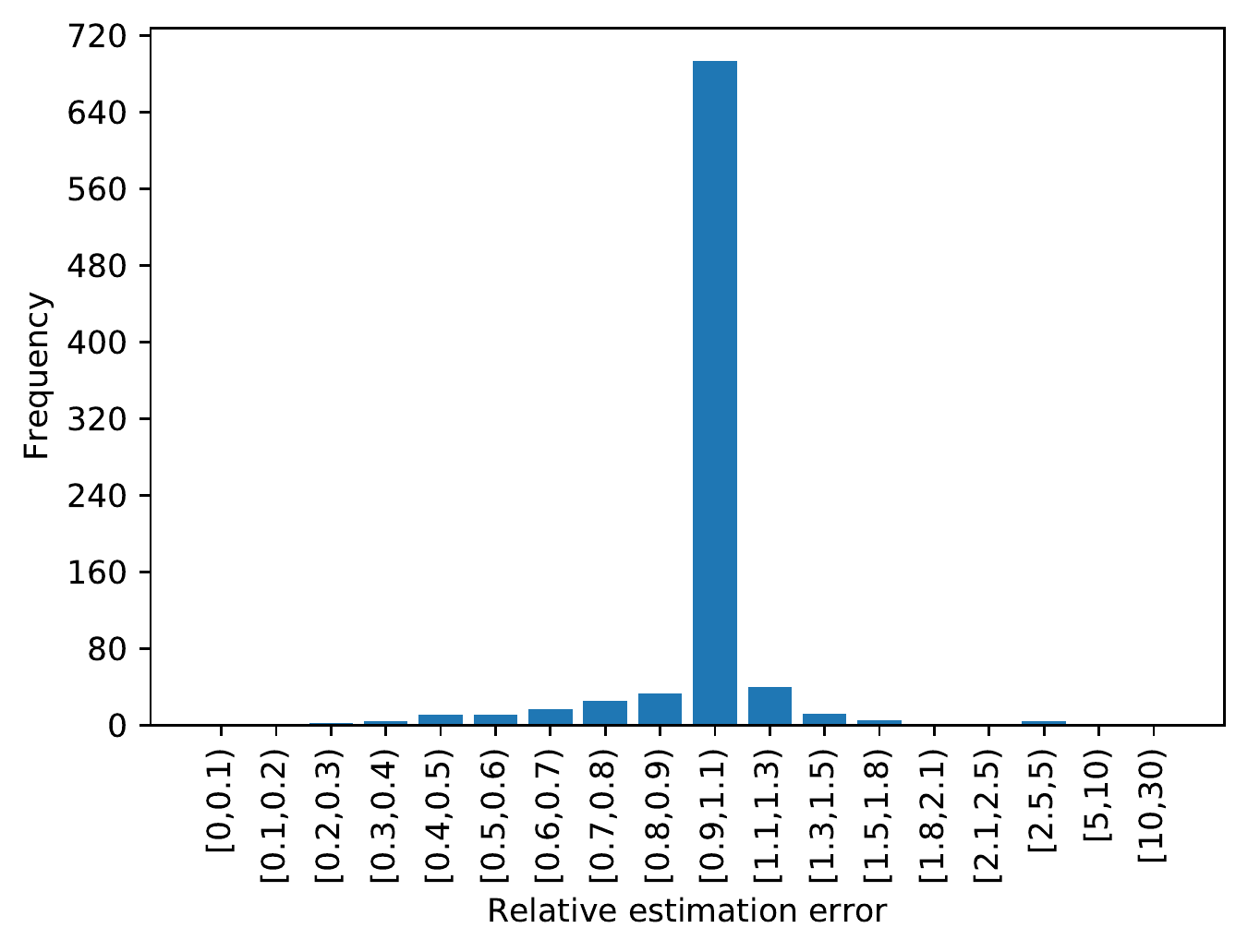}
         \caption{\texttt{l\_partkey} Clustering - All joins}
         \label{fig:relative_errors_histogram_join_clustering_3}
     \end{subfigure}
        \caption{The figure visualizes histograms that describe the distribution of relative estimate errors for the hash join's clustering step, for three different clusterings. The relative error of a latency estimate is the quotient of the measured latency divided by the estimated latency. The relative error is calculated for each join individually and aggregated in the histogram. The histograms on the left contain the relative errors of all joins. Each TPC-H query was executed 10 times.}
        \label{fig:relative_errors_histogram_join_clustering}
\end{figure}

\paragraph{Clustering step}
The distribution of relative errors of the clustering step's latency estimates are displayed in \Cref{fig:relative_errors_histogram_join_clustering}.

\begin{figure}
     \centering
     \begin{subfigure}[b]{0.48\textwidth}
         \centering
         \includegraphics[width=\textwidth]{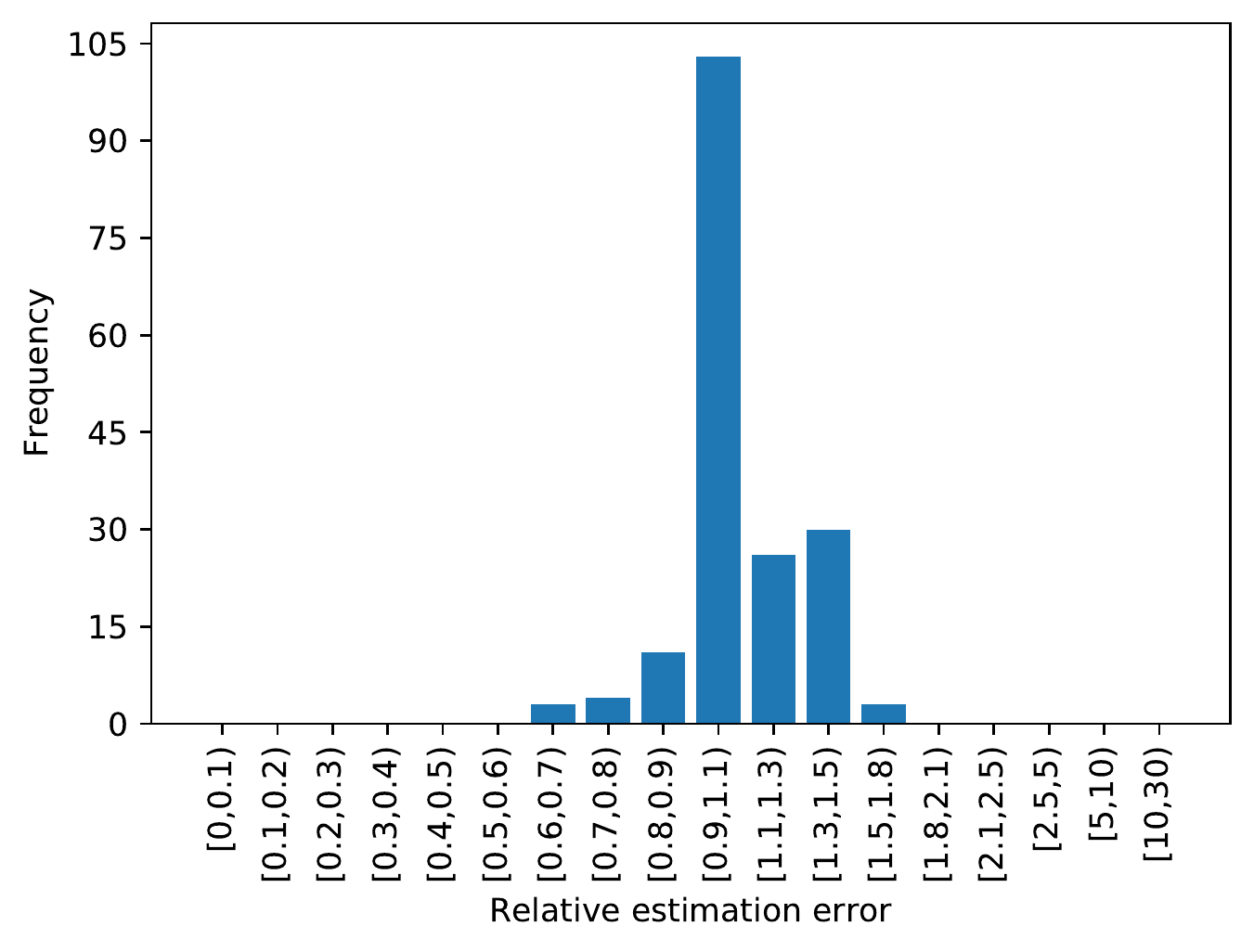}
         \caption{\texttt{l\_shipdate} Clustering - Build side joins}
         \label{fig:relative_errors_histogram_join_build_1}
     \end{subfigure}
     \hfill
     \begin{subfigure}[b]{0.48\textwidth}
         \centering
         \includegraphics[width=\textwidth]{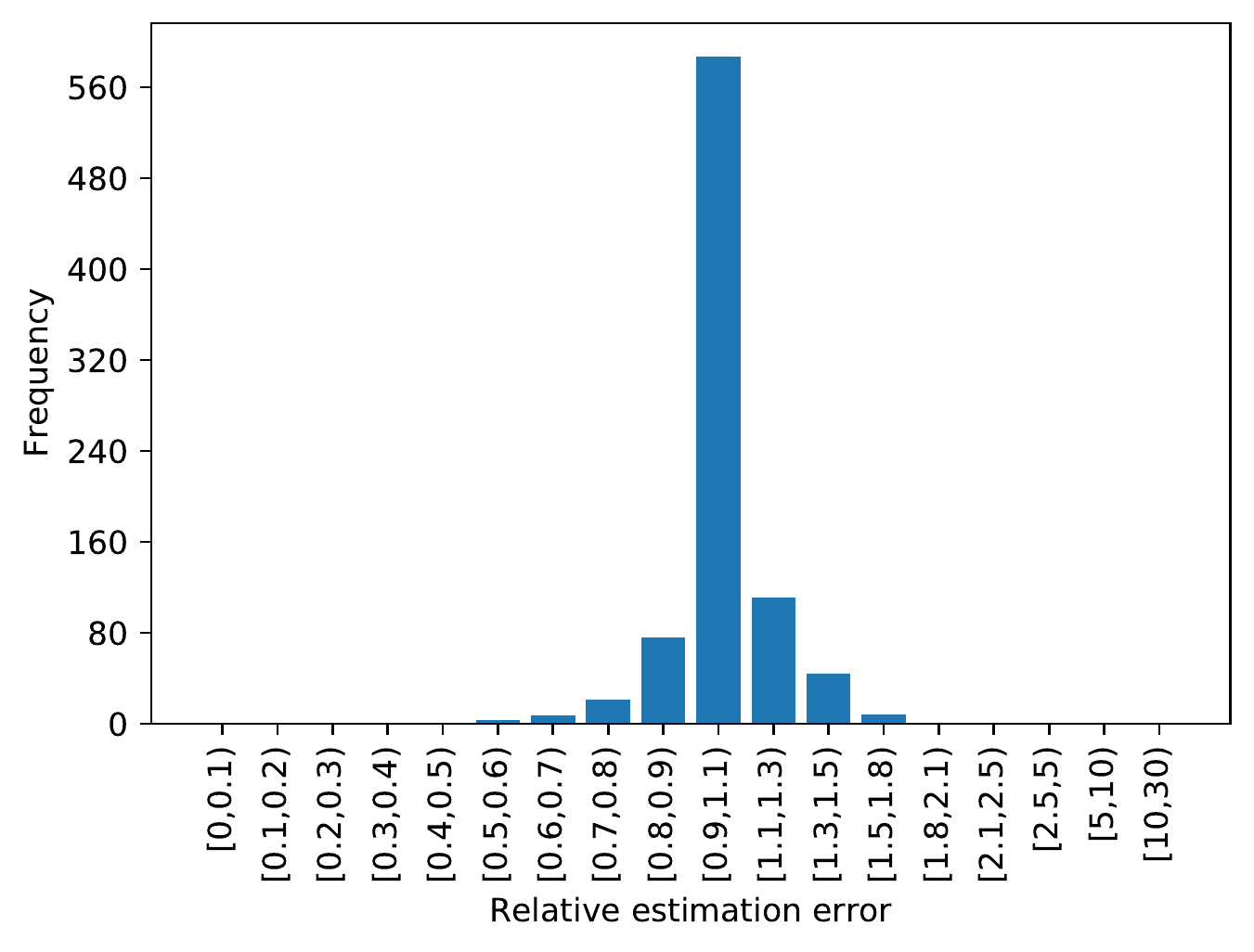}
         \caption{\texttt{l\_shipdate} Clustering - All joins}
         \label{fig:relative_errors_histogram_join_build_1_all}
     \end{subfigure}
     \hfill
     \begin{subfigure}[b]{0.48\textwidth}
         \centering
         \includegraphics[width=\textwidth]{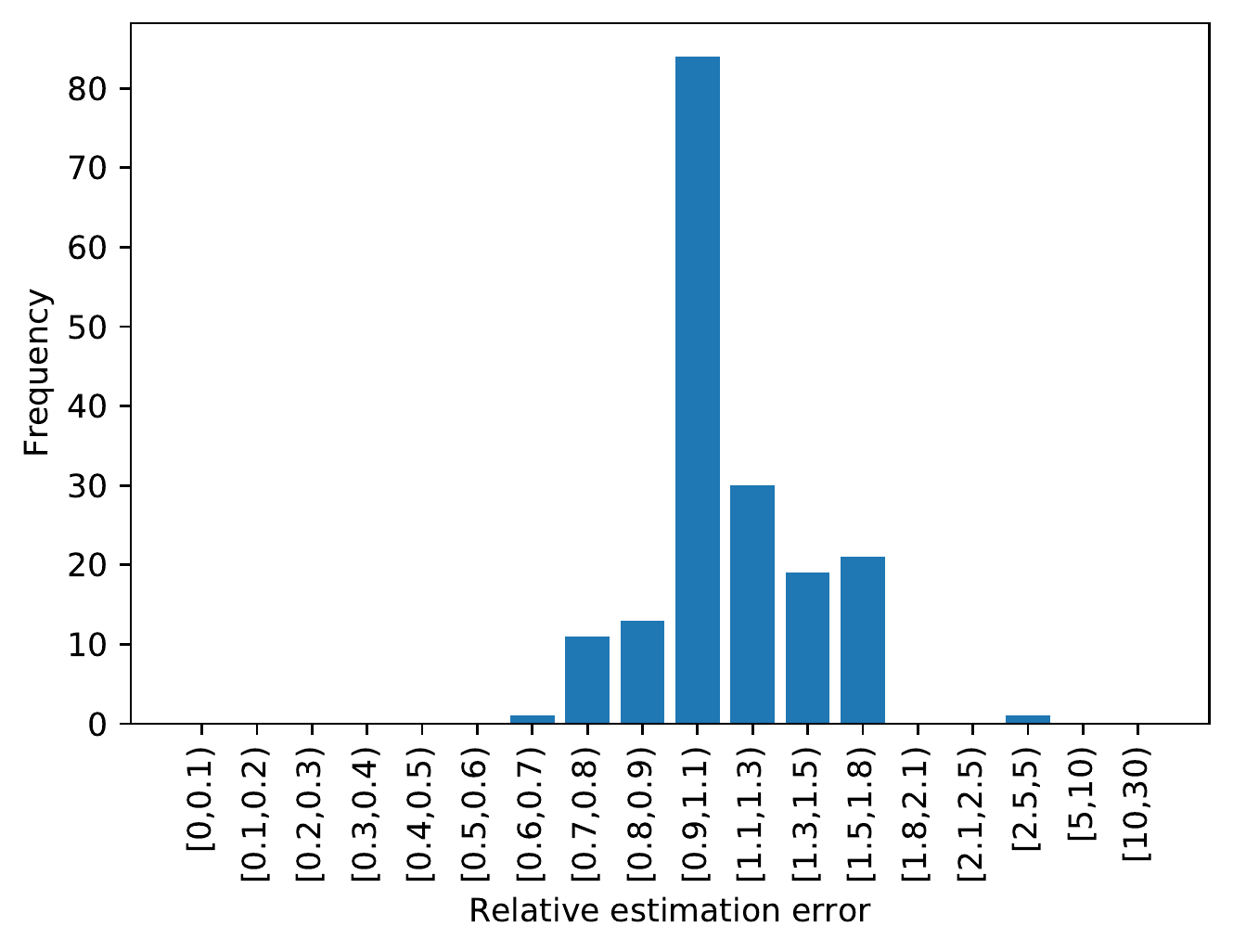}
         \caption{\texttt{l\_partkey} Clustering - Build side joins}
         \label{fig:relative_errors_histogram_join_build_2}
     \end{subfigure}
     \hfill
     \begin{subfigure}[b]{0.48\textwidth}
         \centering
         \includegraphics[width=\textwidth]{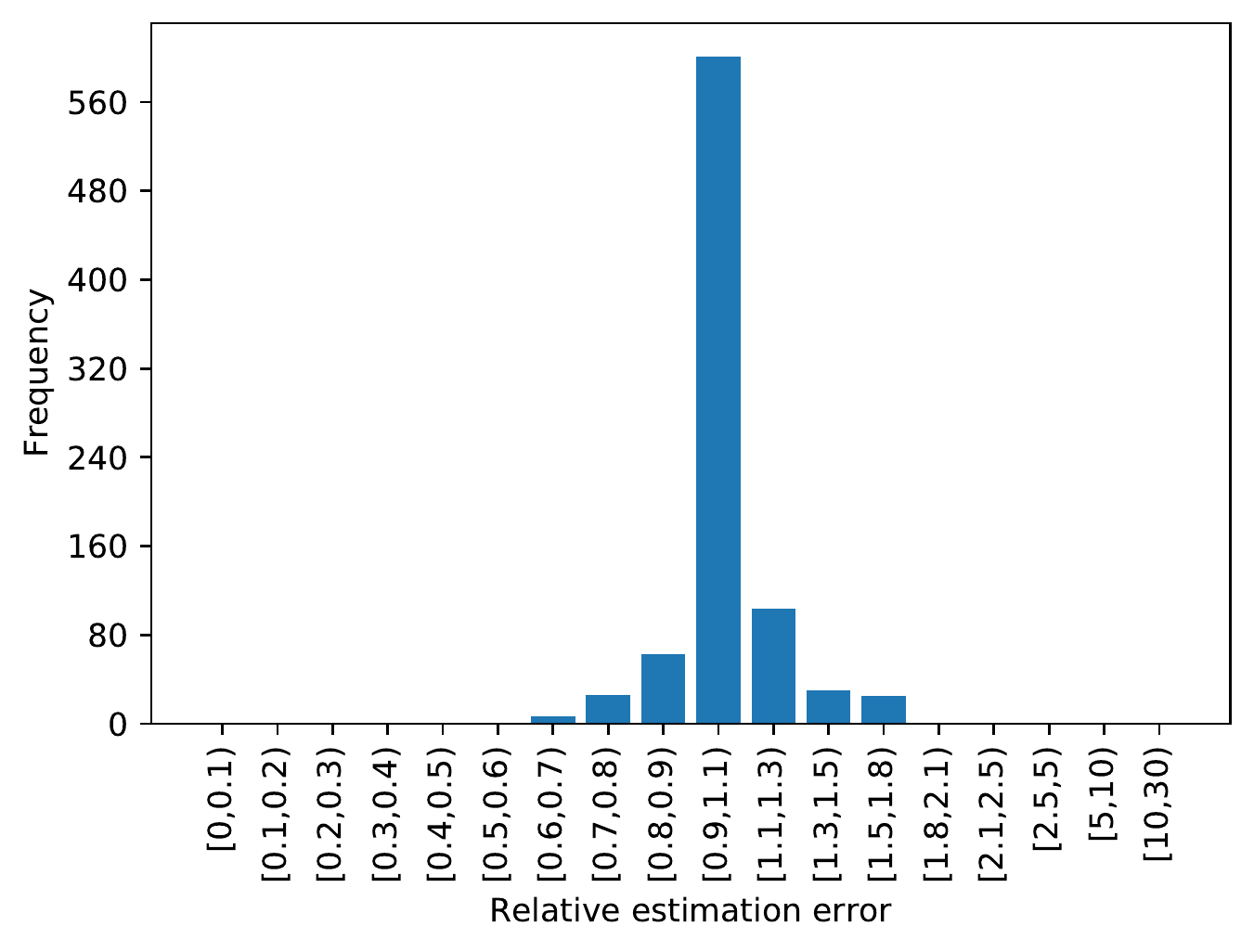}
         \caption{\texttt{l\_partkey} Clustering - All joins}
         \label{fig:relative_errors_histogram_join_build_2_all}
     \end{subfigure}
     \hfill
          \begin{subfigure}[b]{0.48\textwidth}
         \centering
         \includegraphics[width=\textwidth]{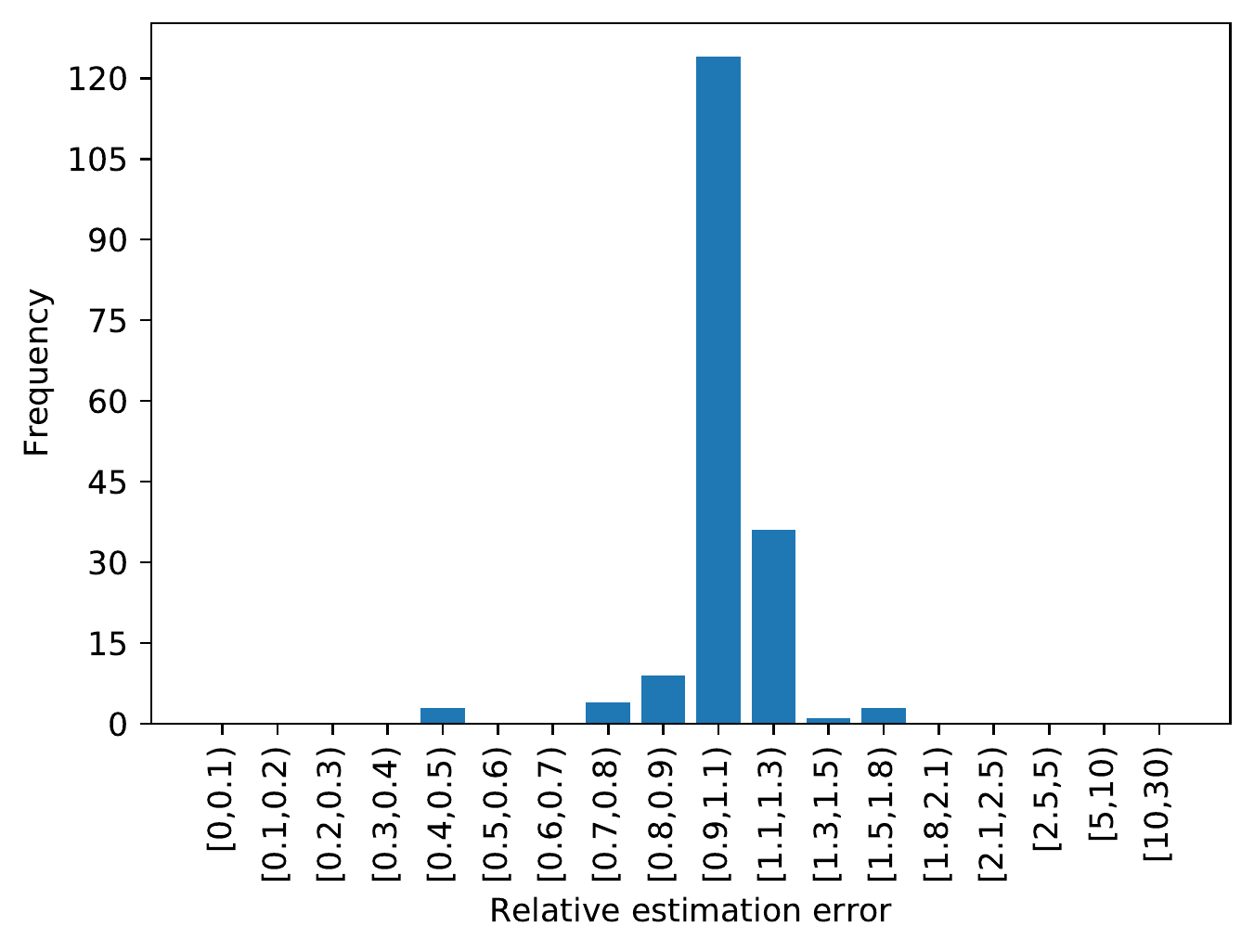}
         \caption{\texttt{l\_shipdate-l\_orderkey} Clustering - Build side joins}
         \label{fig:relative_errors_histogram_join_build_3}
     \end{subfigure}
     \hfill
     \begin{subfigure}[b]{0.48\textwidth}
         \centering
         \includegraphics[width=\textwidth]{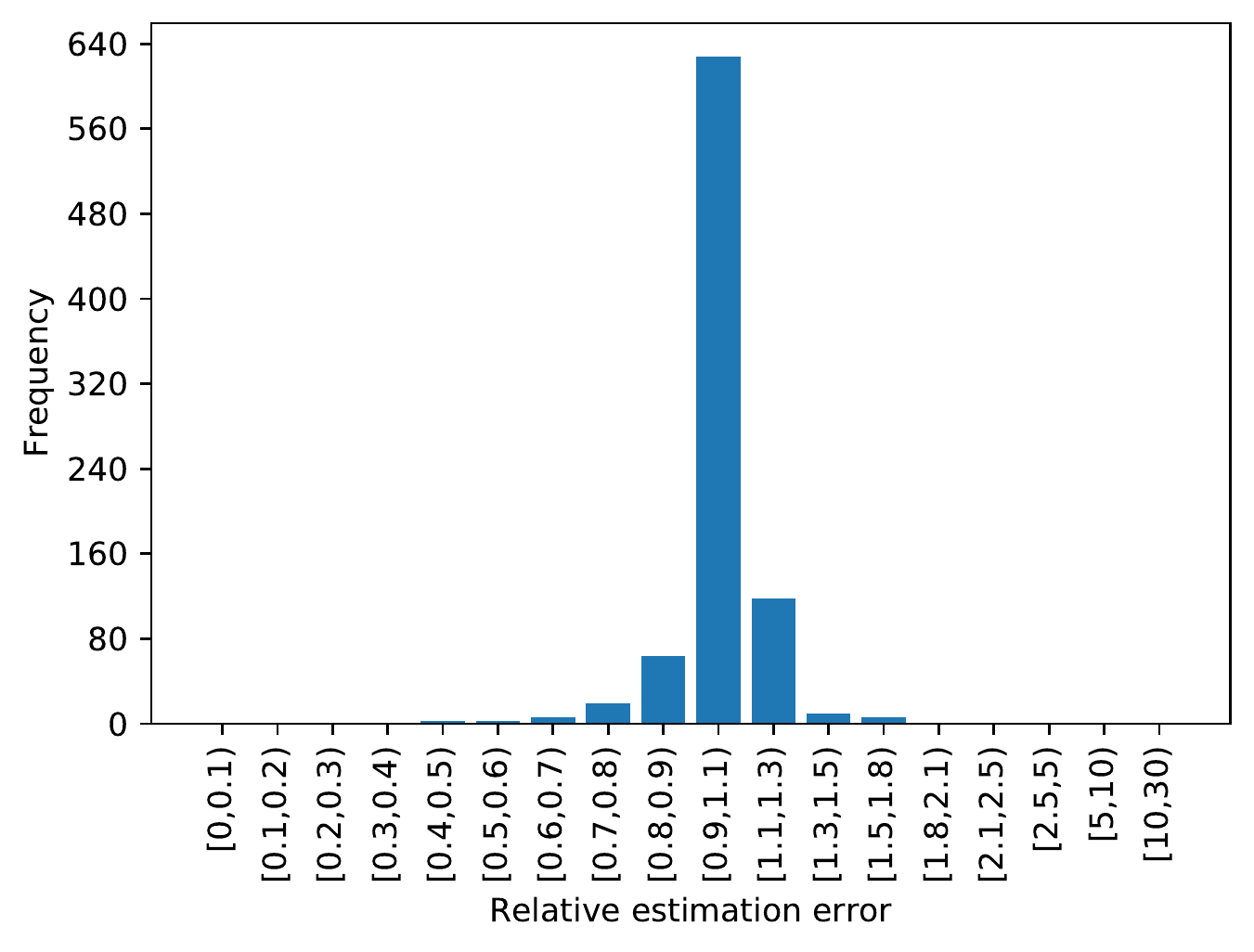}
         \caption{\texttt{l\_shipdate-l\_orderkey} Clustering - All joins}
         \label{fig:relative_errors_histogram_join_build_3_all}
     \end{subfigure}
        \caption{The figure visualizes histograms that describe the distribution of relative estimate errors for the hash join's building step, for three different clusterings. The relative error of a latency estimate is the quotient of the measured latency divided by the estimated latency. The relative error is calculated for each join individually and aggregated in the histogram. The histograms on the left contains only relative errors of joins where \texttt{lineitem} is the probe table, whereas the histograms on the right side contain relative errors of all joins. Each TPC-H query was executed 10 times.}
        \label{fig:relative_errors_histogram_join_build}
\end{figure}

\paragraph{Build step}
The distribution of relative errors of the build step's latency estimates are displayed in \Cref{fig:relative_errors_histogram_join_build}.

\paragraph{Output writing step}

\begin{figure}
     \centering
     \begin{subfigure}[b]{0.48\textwidth}
         \centering
         \includegraphics[width=\textwidth]{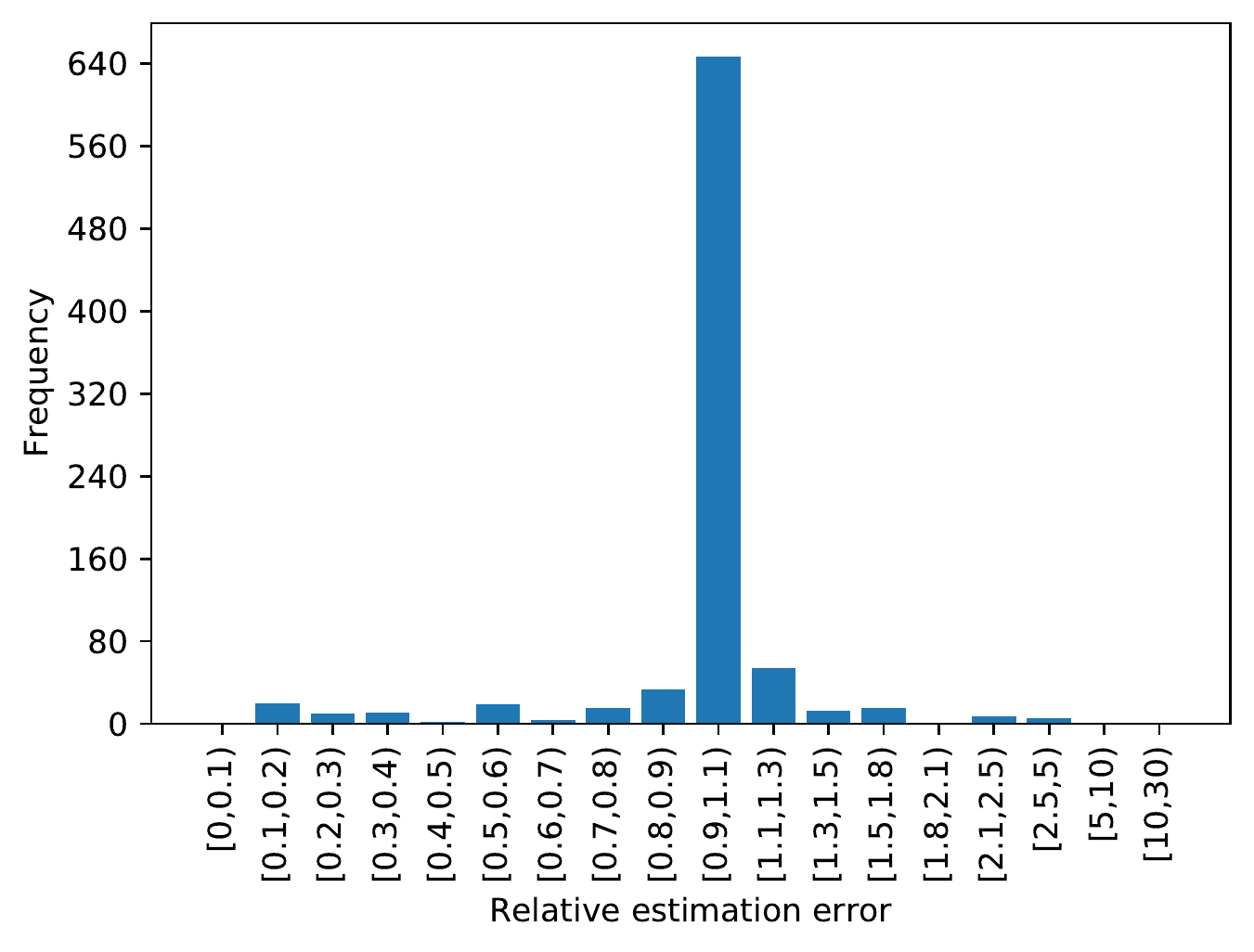}
         \caption{\texttt{l\_shipdate} Clustering - All joins}
         \label{fig:relative_errors_histogram_join_output_writing_1}
     \end{subfigure}
     \hfill
     \begin{subfigure}[b]{0.48\textwidth}
         \centering
         \includegraphics[width=\textwidth]{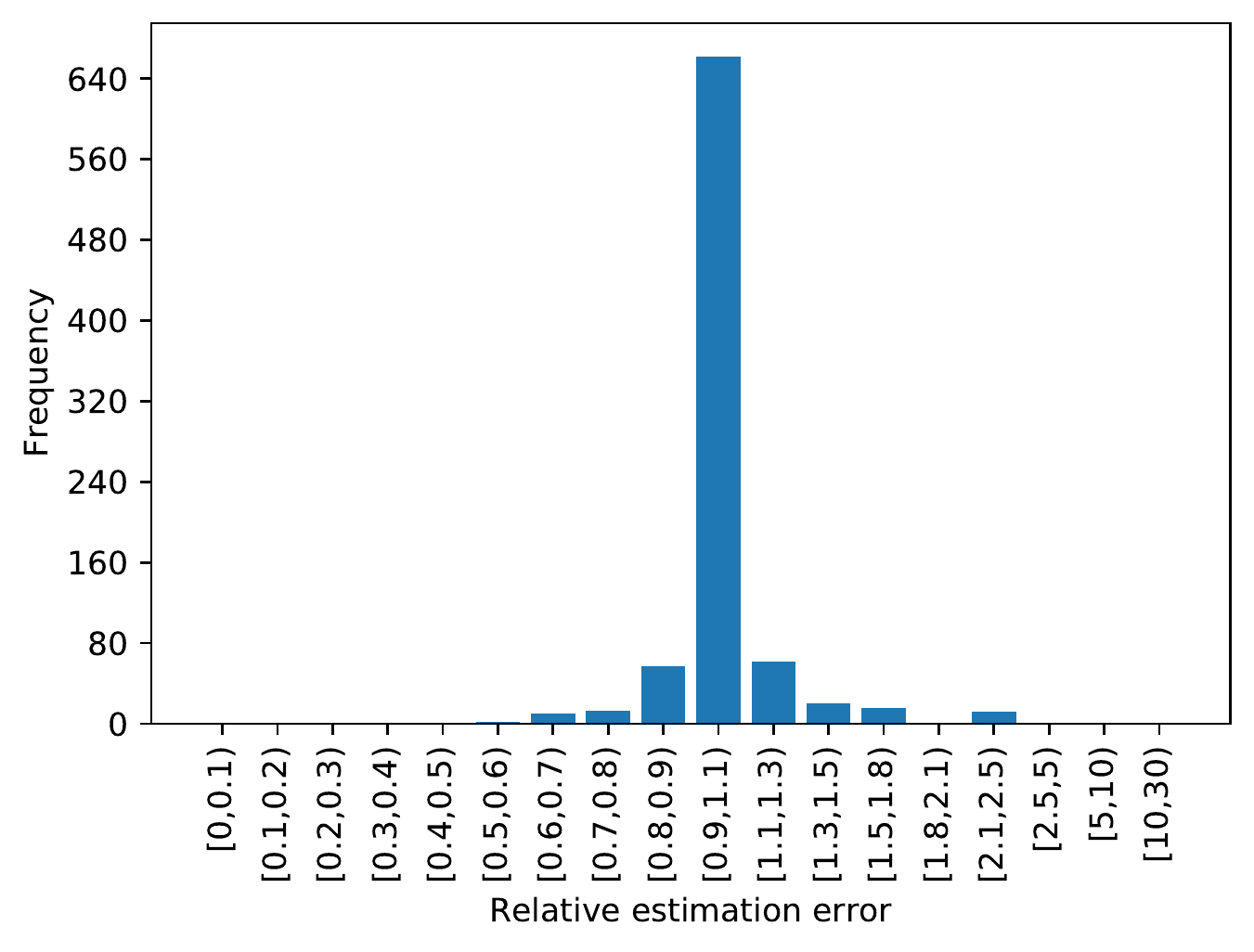}
         \caption{\texttt{l\_shipdate} Clustering - All joins}
         \label{fig:relative_errors_histogram_join_output_writing_2}
     \end{subfigure}
     \hfill
     \begin{subfigure}[b]{0.48\textwidth}
         \centering
         \includegraphics[width=\textwidth]{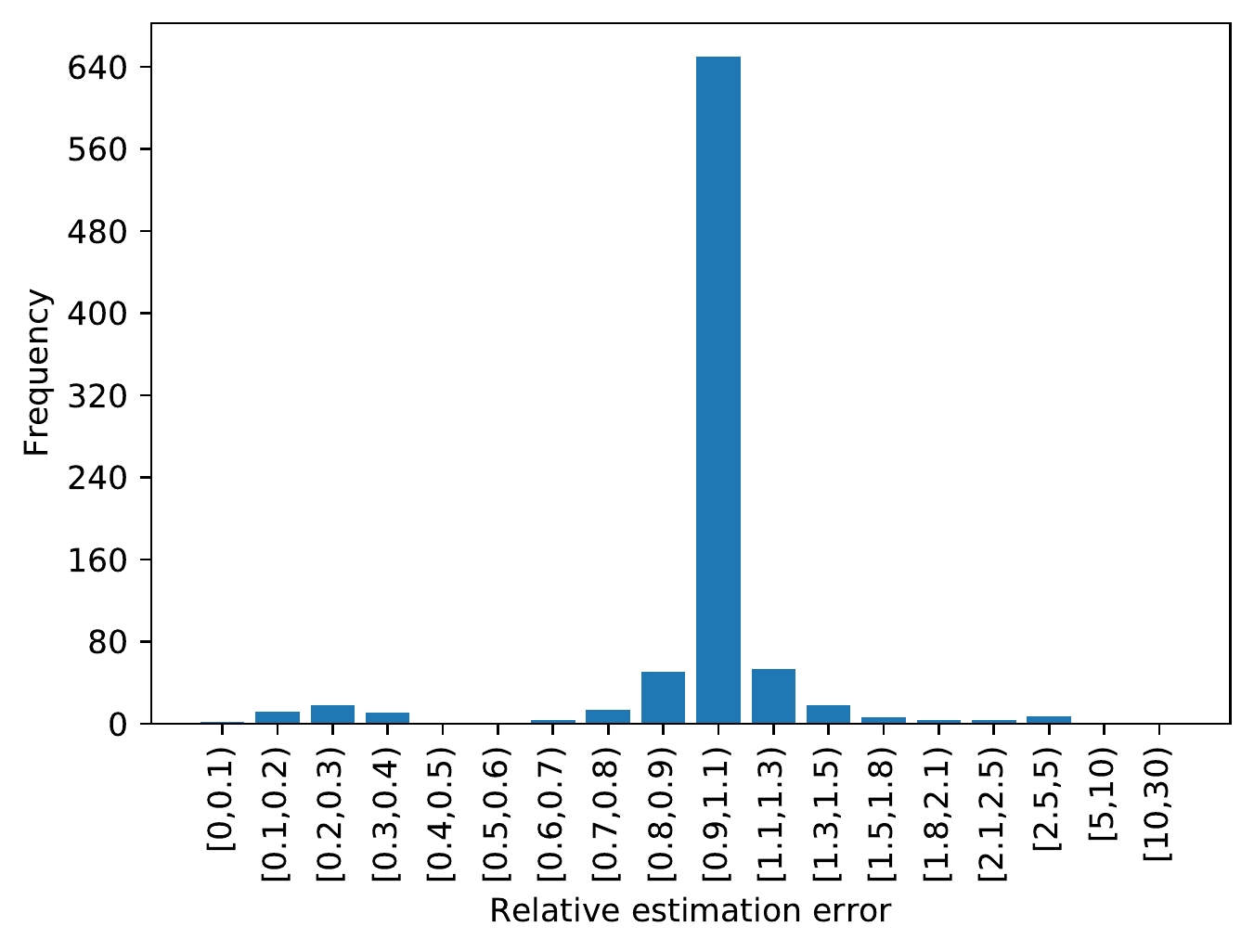}
         \caption{\texttt{l\_partkey} Clustering - All joins}
         \label{fig:relative_errors_histogram_join_output_writing_3}
     \end{subfigure}
        \caption{The figure visualizes histograms that describe the distribution of relative estimate errors for the hash join's clustering step, for three different clusterings. The relative error of a latency estimate is the quotient of the measured latency divided by the estimated latency. The relative error is calculated for each join individually and aggregated in the histogram. The histograms on the left contain the relative errors of all joins. Each TPC-H query was executed 10 times.}
        \label{fig:relative_errors_histogram_join_output_writing}
\end{figure}

The distribution of relative errors of the output writing step's latency estimates are displayed in \Cref{fig:relative_errors_histogram_join_output_writing}.

	\ifisbook\cleardoubleemptypage\fi
	\phantomsection\addcontentsline{toc}{chapter}{\refname}
	\printbibliography[category=cited]

@PREAMBLE{ {\providecommand{\noopsort}[1]{}} }

@STRING{ADBIS = {Advances in Databases and Information Systems (ADBIS)}}

@STRING{CIDR = {Proceedings of the Conference on Innovative Data Systems Research (CIDR)}}

@STRING{CIKM = {Proceedings of the International Conference on Information and Knowledge Management (CIKM)}}

@STRING{EDBT = {Proceedings of the International Conference on Extending Database Technology (EDBT)}}

@STRING{PVLDB = {PVLDB}}

@STRING{SIGMOD = {Proceedings of the International Conference on Management of Data (SIGMOD)}}

@STRING{VLDB = {Proceedings of the International Conference on Very Large Databases (VLDB)}}

@inproceedings{DBLP:conf/vldb/LightstoneB04,
  author    = {Sam Lightstone and
               Bishwaranjan Bhattacharjee},
  title     = {Automated design of multidimensional clustering tables for relational
               databases},
  booktitle = VLDB,
  pages     = {1170--1181},
  year      = {2004},
}

@article{DBLP:journals/pvldb/ZiauddinWKLPK17,
  author    = {Mohamed Ziauddin and
               Andrew Witkowski and
               You Jung Kim and
               Janaki Lahorani and
               Dmitry Potapov and
               Murali Krishna},
  title     = {Dimensions Based Data Clustering and Zone Maps},
  journal   = PVLDB,
  volume    = {10},
  number    = {12},
  pages     = {1622--1633},
  year      = {2017},
}

@inproceedings{DBLP:conf/vldb/AgrawalCN00,
  author    = {Sanjay Agrawal and
               Surajit Chaudhuri and
               Vivek R. Narasayya},
  title     = {Automated Selection of Materialized Views and Indexes in {SQL} Databases},
  booktitle = VLDB,
  pages     = {496--505},
  year      = {2000},
}

@inproceedings{DBLP:conf/sigmod/AgrawalNY04,
  author    = {Sanjay Agrawal and
               Vivek R. Narasayya and
               Beverly Yang},
  title     = {Integrating Vertical and Horizontal Partitioning Into Automated Physical
               Database Design},
  booktitle = SIGMOD,
  pages     = {359--370},
  year      = {2004},
}

@inproceedings{DBLP:conf/cidr/IdreosKM07,
  author    = {Stratos Idreos and
               Martin L. Kersten and
               Stefan Manegold},
  title     = {Database Cracking},
  booktitle = CIDR,
  pages     = {68--78},
  year      = {2007},
}

@article{DBLP:journals/pvldb/OrrKC19,
  author    = {Laurel J. Orr and
               Srikanth Kandula and
               Surajit Chaudhuri},
  title     = {Pushing Data-Induced Predicates Through Joins in Big-Data Clusters},
  journal   = PVLDB,
  volume    = {13},
  number    = {3},
  pages     = {252--265},
  year      = {2019},
}

@inproceedings{DBLP:conf/sigmod/HilprechtBR20,
  author    = {Benjamin Hilprecht and
               Carsten Binnig and
               Uwe R{\"{o}}hm},
  title     = {Learning a Partitioning Advisor for Cloud Databases},
  booktitle = SIGMOD,
  pages     = {143--157},
  year      = {2020},
}

@inproceedings{DBLP:conf/sigmod/MaAHMPG18,
  author    = {Lin Ma and
               Dana Van Aken and
               Ahmed Hefny and
               Gustavo Mezerhane and
               Andrew Pavlo and
               Geoffrey J. Gordon},
  title     = {Query-based Workload Forecasting for Self-Driving Database Management
               Systems},
  booktitle = SIGMOD,
  pages     = {631--645},
  year      = {2018}
}

@article{DBLP:journals/pvldb/MarcusP19,
  author    = {Ryan C. Marcus and
               Olga Papaemmanouil},
  title     = {Plan-Structured Deep Neural Network Models for Query Performance Prediction},
  journal   = PVLDB,
  volume    = {12},
  number    = {11},
  pages     = {1733--1746},
  year      = {2019},
}

@article{DBLP:journals/pvldb/DreselerBRU20,
  author    = {Markus Dreseler and
               Martin Boissier and
               Tilmann Rabl and
               Matthias Uflacker},
  title     = {Quantifying {TPC-H} Choke Points and Their Optimizations},
  journal   = PVLDB,
  volume    = {13},
  number    = {8},
  pages     = {1206--1220},
  year      = {2020}
}

@inproceedings{DBLP:conf/edbt/DreselerK0KUP19,
  author    = {Markus Dreseler and
               Jan Kossmann and
               Martin Boissier and
               Stefan Klauck and
               Matthias Uflacker and
               Hasso Plattner},
  title     = {Hyrise Re-engineered: An Extensible Database System for Research in
               Relational In-Memory Data Management},
  booktitle = EDBT,
  pages     = {313--324},
  year      = {2019},
}

@inproceedings{DBLP:conf/adbis/HolzeR08,
  author    = {Marc Holze and
               Norbert Ritter},
  title     = {Autonomic Databases: Detection of Workload Shifts with n-Gram-Models},
  booktitle = ADBIS,
  volume    = {5207},
  pages     = {127--142},
  year      = {2008},
}

@inproceedings{DBLP:conf/IEEEcloud/RoyDG11,
  author    = {Nilabja Roy and
               Abhishek Dubey and
               Aniruddha S. Gokhale},
  title     = {Efficient Autoscaling in the Cloud Using Predictive Models for Workload
               Forecasting},
  booktitle = {{IEEE} International Conference on Cloud Computing, {CLOUD}},
  pages     = {500--507},
  year      = {2011},
}

@inproceedings{DBLP:conf/cnsm/GongGW10,
  author    = {Zhenhuan Gong and
               Xiaohui Gu and
               John Wilkes},
  title     = {{PRESS:} PRedictive Elastic ReSource Scaling for cloud systems},
  booktitle = {Proceedings of the International Conference on Network and Service
               Management, {CNSM}},
  pages     = {9--16},
  year      = {2010},
}

@inproceedings{DBLP:conf/sigmod/PadmanabhanBMCH03,
  author    = {Sriram Padmanabhan and
               Bishwaranjan Bhattacharjee and
               Timothy Malkemus and
               Leslie Cranston and
               Matthew Huras},
  title     = {Multi-Dimensional Clustering: {A} New Data Layout Scheme in {DB2}},
  booktitle = SIGMOD,
  pages     = {637--641},
  year      = {2003},
}

@inproceedings{DBLP:conf/sigmod/ChangF80,
  author    = {Jo{-}Mei Chang and
               King{-}sun Fu},
  title     = {A Dynamic Clustering Technique for Physical Database Design},
  booktitle = SIGMOD,
  year      = {1980},
}

@inproceedings{DBLP:conf/vldb/SchwalbFWGP14,
  author    = {David Schwalb and
               Martin Faust and
               Johannes Wust and
               Martin Grund and
               Hasso Plattner},
  title     = {Efficient Transaction Processing for Hyrise in Mixed Workload Environments},
  booktitle = {Proceedings of the International Workshop on In Memory Data Management
               and Analytics, {IMDM}},
  pages     = {16--29},
  year      = {2014},
}

@inproceedings{DBLP:conf/sigmod/BianYTCCDM17,
  author    = {Haoqiong Bian and
               Ying Yan and
               Wenbo Tao and
               Liang Jeff Chen and
               Yueguo Chen and
               Xiaoyong Du and
               Thomas Moscibroda},
  title     = {Wide Table Layout Optimization based on Column Ordering and Duplication},
  booktitle = SIGMOD,
  pages     = {299--314},
  year      = {2017},
}

@article{DBLP:journals/pvldb/KruegerKGSSCPDZ11,
  author    = {Jens Kr{\"{u}}ger and
               Changkyu Kim and
               Martin Grund and
               Nadathur Satish and
               David Schwalb and
               Jatin Chhugani and
               Hasso Plattner and
               Pradeep Dubey and
               Alexander Zeier},
  title     = {Fast Updates on Read-Optimized Databases Using Multi-Core CPUs},
  journal   = PVLDB,
  volume    = {5},
  number    = {1},
  pages     = {61--72},
  year      = {2011},
}

@inproceedings{DBLP:conf/sigmod/VogelsgesangHFK18,
  author    = {Adrian Vogelsgesang and
               Michael Haubenschild and
               Jan Finis and
               Alfons Kemper and
               Viktor Leis and
               Tobias M{\"{u}}hlbauer and
               Thomas Neumann and
               Manuel Then},
  title     = {Get Real: How Benchmarks Fail to Represent the Real World},
  booktitle = SIGMOD,
  pages     = {1:1--1:6},
  year      = {2018},
}

@inproceedings{DBLP:conf/cikm/0001MDLMRSZU16,
  author    = {Martin Boissier and
               Carsten Alexander Meyer and
               Timo Dj{\"{u}}rken and
               Jan Lindemann and
               Kathrin Mao and
               Pascal Reinhardt and
               Tim Specht and
               Tim Zimmermann and
               Matthias Uflacker},
  title     = {Analyzing Data Relevance and Access Patterns of Live Production Database
               Systems},
  booktitle = {CIKM},
  pages     = {2473--2475},
  year      = {2016},
}

@inproceedings{DBLP:conf/vldb/Moerkotte98,
  author    = {Guido Moerkotte},
  title     = {Small Materialized Aggregates: {A} Light Weight Index Structure for
               Data Warehousing},
  booktitle = {VLDB},
  pages     = {476--487},
  year      = {1998},
}

@inproceedings{DBLP:conf/sigmod/LangMFB0K16,
  author    = {Harald Lang and
               Tobias M{\"{u}}hlbauer and
               Florian Funke and
               Peter A. Boncz and
               Thomas Neumann and
               Alfons Kemper},
  title     = {Data Blocks: Hybrid {OLTP} and {OLAP} on Compressed Storage using
               both Vectorization and Compilation},
  booktitle = {SIGMOD},
  pages     = {311--326},
  year      = {2016},
}

@article{DBLP:journals/pvldb/NicaSACHBG17,
  author    = {Anisoara Nica and
               Reza Sherkat and
               Mihnea Andrei and
               Xun Chen and
               Martin Heidel and
               Christian Bensberg and
               Heiko Gerwens},
  title     = {Statisticum: Data Statistics Management in {SAP} {HANA}},
  journal   = PVLDB,
  volume    = {10},
  number    = {12},
  pages     = {1658--1669},
  year      = {2017},
}

@article{DBLP:journals/pvldb/PatelDZPZSMS18,
  author    = {Jignesh M. Patel and
               Harshad Deshmukh and
               Jianqiao Zhu and
               Navneet Potti and
               Zuyu Zhang and
               Marc Spehlmann and
               Hakan Memisoglu and
               Saket Saurabh},
  title     = {Quickstep: {A} Data Platform Based on the Scaling-Up Approach},
  journal   = {{PVLDB}},
  volume    = {11},
  number    = {6},
  pages     = {663--676},
  year      = {2018},
}

@inproceedings{DBLP:conf/sigmod/ArulrajPM16,
  author    = {Joy Arulraj and
               Andrew Pavlo and
               Prashanth Menon},
  title     = {Bridging the Archipelago between Row-Stores and Column-Stores for
               Hybrid Workloads},
  booktitle = {SIGMOD},
  pages     = {583--598},
  year      = {2016},
}

@inproceedings{DBLP:conf/sigmod/HerodotouBB11,
  author    = {Herodotos Herodotou and
               Nedyalko Borisov and
               Shivnath Babu},
  title     = {Query optimization techniques for partitioned tables},
  booktitle = {SIGMOD},
  pages     = {49--60},
  year      = {2011},
}

@inproceedings{DBLP:conf/vldb/OthayothP06,
  author    = {Raghunath Othayoth Nambiar and
               Meikel Poess},
  title     = {The Making of {TPC-DS}},
  booktitle = {VLDB},
  pages     = {1049--1058},
  year      = {2006},
}

@online{TpcdsSpec,
  author = {{Transaction Processing Performance Council}},
  year = 2020,
  title = {{TPC-DS Specification}},
  url = {http://www.tpc.org/tpc_documents_current_versions/pdf/tpc-ds_v2.13.0.pdf},
  urldate = {2020-09-28}
}

@inproceedings{DBLP:conf/tpctc/BonczNE13,
  author    = {Peter A. Boncz and
               Thomas Neumann and
               Orri Erling},
  title     = {{TPC-H} Analyzed: Hidden Messages and Lessons Learned from an Influential
               Benchmark},
  booktitle = {Performance Characterization and Benchmarking - {TPC} Technology
               Conference},
  volume    = {8391},
  pages     = {61--76},
  year      = {2013},
}

@online{TpchSpec,
  author = {{Transaction Processing Performance Council}},
  year = 2014,
  title = {{TPC-H Specification}},
  url = {http://www.tpc.org/tpc_documents_current_versions/pdf/tpc-h_v2.17.1.pdf},
  urldate = {2020-09-28}
}

@inproceedings{DBLP:conf/sigmod/DasLNK16,
  author    = {Sudipto Das and
               Feng Li and
               Vivek R. Narasayya and
               Arnd Christian K{\"{o}}nig},
  title     = {Automated Demand-driven Resource Scaling in Relational Database-as-a-Service},
  booktitle = {SIGMOD},
  pages     = {1923--1934},
  year      = {2016},
}

@inproceedings{DBLP:conf/sigmod/AbadiMF06,
  author    = {Daniel J. Abadi and
               Samuel Madden and
               Miguel Ferreira},
  title     = {Integrating compression and execution in column-oriented database
               systems},
  booktitle = {SIGMOD},
  pages     = {671--682},
  year      = {2006},
}

@inproceedings{DBLP:conf/vldb/PapadomanolakisDA07,
  author    = {Stratos Papadomanolakis and
               Debabrata Dash and
               Anastassia Ailamaki},
  title     = {Efficient Use of the Query Optimizer for Automated Database Design},
  booktitle = {VLDB},
  pages     = {1093--1104},
  year      = {2007},
}

@article{DBLP:journals/pvldb/DashPA11,
  author    = {Debabrata Dash and
               Neoklis Polyzotis and
               Anastasia Ailamaki},
  title     = {CoPhy: {A} Scalable, Portable, and Interactive Index Advisor for Large
               Workloads},
  journal   = {PVLDB},
  volume    = {4},
  number    = {6},
  pages     = {362--372},
  year      = {2011},
}

@article{DBLP:journals/tods/BernsteinG83,
  author    = {Philip A. Bernstein and
               Nathan Goodman},
  title     = {Multiversion Concurrency Control - Theory and Algorithms},
  journal   = {{ACM} Trans. Database Syst.},
  volume    = {8},
  number    = {4},
  pages     = {465--483},
  year      = {1983},
}

@book{DBLP:books/mk/Lightstone2007,
  author    = {Sam Lightstone and
               Toby J. Teorey and
               Thomas P. Nadeau},
  title     = {Physical Database Design: the database professional's guide to exploiting
               indexes, views, storage, and more},
  year      = {2007},
}

@inproceedings{DBLP:conf/vldb/MarklMKTHS05,
  author    = {Volker Markl and
               Nimrod Megiddo and
               Marcel Kutsch and
               Tam Minh Tran and
               Peter J. Haas and
               Utkarsh Srivastava},
  title     = {Consistently Estimating the Selectivity of Conjuncts of Predicates},
  booktitle = {VLDB},
  pages     = {373--384},
  year      = {2005},
}

@inproceedings{DBLP:conf/vldb/BonczMK99,
  author    = {Peter A. Boncz and
               Stefan Manegold and
               Martin L. Kersten},
  title     = {Database Architecture Optimized for the New Bottleneck: Memory Access},
  booktitle = {VLDB},
  pages     = {54--65},
  year      = {1999},
}

@Inbook{Neumann2009,
    author="Neumann, Thomas",
    title="Query Optimization (in Relational Databases)",
    bookTitle="Encyclopedia of Database Systems",
    year="2009",
    pages="2273--2278",
}

@online{SnowflakeMaxDimension,
  author = {{Snowflake}},
  year = 2020,
  title = {{Strategies for Selecting Clustering Keys}},
  url = {https://docs.snowflake.com/en/user-guide/tables-clustering-keys.html#strategies-for-selecting-clustering-keys},
  urldate = {2020-09-30}
}

@inproceedings{DBLP:conf/sigmod/DagevilleCZAABC16,
  author    = {Beno{\^{\i}}t Dageville and
               Thierry Cruanes and
               Marcin Zukowski and
               Vadim Antonov and
               Artin Avanes and
               Jon Bock and
               Jonathan Claybaugh and
               Daniel Engovatov and
               Martin Hentschel and
               Jiansheng Huang and
               Allison W. Lee and
               Ashish Motivala and
               Abdul Q. Munir and
               Steven Pelley and
               Peter Povinec and
               Greg Rahn and
               Spyridon Triantafyllis and
               Philipp Unterbrunner},
  title     = {The Snowflake Elastic Data Warehouse},
  booktitle = {SIGMOD},
  pages     = {215--226},
  year      = {2016},
}

@inproceedings{DBLP:conf/sigmod/KesterAI17,
  author    = {Michael S. Kester and
               Manos Athanassoulis and
               Stratos Idreos},
  title     = {Access Path Selection in Main-Memory Optimized Data Systems: Should
               {I} Scan or Should {I} Probe?},
  booktitle = {SIGMOD},
  pages     = {715--730},
  year      = {2017},
}

@article{DBLP:journals/cacm/Bloom70,
  author    = {Burton H. Bloom},
  title     = {Space/Time Trade-offs in Hash Coding with Allowable Errors},
  journal   = {Commun. {ACM}},
  volume    = {13},
  number    = {7},
  pages     = {422--426},
  year      = {1970},
}

@inproceedings{DBLP:conf/edbt/MemarziaRB19,
  author    = {Puya Memarzia and
               Suprio Ray and
               Virendra C. Bhavsar},
  title     = {A Six-dimensional Analysis of In-memory Aggregation},
  booktitle = {EDBT},
  pages     = {289--300},
  year      = {2019},
}

@article{DBLP:journals/pvldb/Neumann14,
  author    = {Thomas Neumann},
  title     = {Engineering High-Performance Database Engines},
  journal   = PVLDB,
  volume    = {7},
  number    = {13},
  pages     = {1734--1741},
  year      = {2014}
}

@inproceedings{DBLP:conf/sigmod/ChaudhuriN98,
  author    = {Surajit Chaudhuri and
               Vivek R. Narasayya},
  title     = {AutoAdmin 'What-if' Index Analysis Utility},
  booktitle = {{SIGMOD}},
  pages     = {367--378},
  year      = {1998},
}

	\ifisbook\pagestyle{plain}\cleardoubleemptypage
\begin{otherlanguage}{ngerman}

\begin{center}\textsf{\textbf{Eidesstattliche Erklärung}}\end{center}
Hiermit versichere ich, dass meine {\hpitype} \enquote{\hpititle} (\enquote{\hpititleother}) selbständig verfasst wurde und dass keine anderen Quellen und Hilfsmittel als die angegebenen benutzt wurden. Diese Aussage trifft auch für alle Implementierungen und Dokumentationen im Rahmen dieses Projektes zu.\\

\noindent
Potsdam, den \hpidate,
\vspace{2cm}

\begin{center}
\begin{tabular}{C{6cm}}
\hline
{\small({\hpiauthor})}
\end{tabular}
\end{center}

\end{otherlanguage}

\fi

\end{document}